\documentclass[useAMS,usenatbib]{mn2e}
\usepackage{graphics}
\usepackage{epsfig}
\usepackage{latexsym}
\usepackage{amsfonts}
\usepackage{amsmath}
\usepackage{amssymb}
\usepackage[english]{babel}
\usepackage{graphicx}
\usepackage{url}
\usepackage{textcomp}
\usepackage{epstopdf}
\usepackage[percent]{overpic}
\usepackage{url}
\usepackage[usenames]{color}
\usepackage[T1]{fontenc}
\usepackage{aecompl}

\def\red#1{\textcolor{red}{ #1}}  
\def\blue#1{\textcolor{blue}{ #1}}  

\begin{document}

\title[Fits of pulsar $\gamma$-ray spectra]{A systematic synchro-curvature modelling 
of pulsar $\gamma$-ray spectra unveils hidden trends}

\author[Vigan\`o, Torres, \& Mart\'in]{Daniele Vigan\`o$^1$, Diego F.~Torres$^{1,2}$ \& Jonatan Mart\'in$^1$\\ 
$^1$Institute of Space Sciences (CSIC--IEEC), Campus UAB, Carrer de Can Magrans s/n, E-08193 Barcelona, Spain\\
$^2$Instituci\'o Catalana de Recerca i Estudis Avan\c{c}ats (ICREA) Barcelona, 08010, Spain}

\date{}
\pagerange{\pageref{firstpage}--\pageref{lastpage}} \pubyear{2015}

\maketitle

\label{firstpage}

\begin{abstract}
$\gamma$-ray radiation from pulsars is usually thought to be mostly produced by the synchro-curvature losses of accelerated particles. Here we present a systematic study of all currently reported, good-quality {\it Fermi}-LAT pulsar spectral data. We do so by applying a model which follows the particle dynamics and consistently computes the emission of synchro-curvature radiation. By fitting observational data on a case by case basis, we are able to obtain constraints about the parallel electric field, the typical lengthscale over which particles emit the bulk of the detected radiation, and the number of involved particles. The model copes well with data of several dozens of millisecond and young pulsars.
By correlating the inferred model parameters with the observed timing properties, some trends are discovered. First, a non-negligible part of the radiation comes from the loss of perpendicular momentum soon after pair creation. Second, the electric field strongly correlates with both the inverse of the emission lengthscale and the magnetic field at light cylinder, thus ruling out models with high-energy photon production close to the surface. These correlations unify young and millisecond pulsars under the same physical scenario, and predict that magnetars are intrinsically $\gamma$-ray quiet via syncrhro-curvature processes, since magnetospheric particles are not accelerated enough to emit a detectable $\gamma$-ray flux.
\end{abstract}

\section{Introduction}

Thanks to the increasing number of detections by {\it Fermi}-LAT, the number of known $\gamma$-ray pulsars is currently in excess of 161.\footnote{This number of pulsars appears already in the list  at {\url {https://confluence.slac.stanford.edu/display/GLAMCOG/Public+List+of+LAT-Detected+Gamma-Ray+Pulsars}}, updated  November 2014.} 71 of them are millisecond pulsars (MSPs), with periods between 1.5  and 22 ms, while the remaining 90 are young pulsars (YPs), with periods up to half a second.
Data for 117 of these pulsars were presented in detail in the second {\em Fermi}-LAT pulsar catalog (2PC hereafter, \citealt{2fpc}). We shall draw the targets of our study from this impressive database. These pulsars are among the most energetic, and can be quiet or loud in radio and/or X-rays. 
Systematic studies of the $\gamma$-ray pulsars usually regard their pulse profile, and the constraints it can impose on the magnetospheric geometry under some particular assumptions of a given model for its location \citep{romani96,romani10,pierbattista15}. Here, we present a systematic study and fit the phase-averaged spectra of the 81 pulsars having good-quality data in the 2PC.

This works builds upon the result of a research program about the high-energy synchro-curvature (SC) radiation coming from pulsar
magnetospheres. Within this program, we have earlier reformulated the SC expressions and computed
particle trajectories under this kind of loss \citep{paper0}. We have also studied the assumptions and caveats in the semi-analytical thin and thick outer gap (OG) models \citep{paper1,paper2}. Finally, we have provided phase-averaged and phase-resolved analysis of the $\gamma$-ray spectra of Crab, Geminga, and Vela \citep{paper3}. The results obtained in the latter work have prompted us to undertake this systematic analysis. Our aim here is to test whether the expected SC radiation is compatible with the $\gamma$-ray spectra of the {\it Fermi}-LAT detected pulsar population, and if so, to obtain general conclusions and explore correlations between model parameters and observational data. We tackle these aims by fitting the phase-averaged spectra of a large population of YPs and MSPs.

\begin{table*}
\begin{center}
\caption{Parameters entering in the SC model, definition, range of observed/expected values, reference to further discussion justifying the adopted range or definition, and treatment in the fitting procedure. The first block of parameters have directly observed values, the second block lists the only three parameters varied in the fitting procedure, and the third block collects the less spectra-affecting, virtually unconstrainable parameters.}
\label{tab:parameters}
\begin{tabular}{l l l l l}
\hline
\hline
{\em Parameter} & {\em Definition} & {\em Range} & {\em Ref.} & {\em Treatment} \\
\hline
$P$ & Spin period & 1--500 ms & [1] & measured \\
$\dot P$ & Spin period derivative & $10^{-21}$--$10^{-12}$ & [1] & measured \\
\hline
$E_\parallel$  & Parallel electric field & $10^6$--$10^{11}$ V/m & [2] & fit \\
$x_0/R_{\rm lc}$ & Lengthscale of the bulk $\gamma$-ray emission, Eq.~(\ref{eq:distribution}) & $10^{-4}$--$10^0$ & [2] & fit \\
$N_0$ & Effective number of particles, Eq.~(\ref{eq:distribution}) & $10^{27}$--$10^{34}$ & [2] & fit  \\
\hline
$\eta$ & Radius of curvature position-dependence, $r_c(x) = R_{\rm lc}\left({x}/{R_{\rm lc}}\right)^\eta$ & 0.2--1.0 & [3] & fixed to 0.5 \\
$b$ & Magnetic field position-dependence, $B(x)=B_s \left({R_\star}/{x}\right)^b$& 2--3 & [3] & fixed to 2.5 \\
$x_{\rm in}/R_{\rm lc}$ & Inner gap location & 0.2--1.0 & [4] & fixed to 0.5 \\
$x_{\rm out}/R_{\rm lc}$ & Outer gap location  & 1.0--2.0 & [4] & fixed to 1.5 \\
$\Gamma_{\rm in}$ & Particle Lorentz factor at birth & $10^3$--$10^4$ & [4] & fixed to $10^3$ \\
$\alpha_{\rm in}$ & Particle pitch angle at birth & $0$--$\pi/2$ & [4] & fixed to $\pi/4$ \\
\hline
\hline
\end{tabular}
\begin{minipage}{\textwidth}
References: [1] 2PC; [2] \S2.3 of \cite{paper3}; [3] Appendix of \cite{paper1}; [4] \cite{paper2}.
\end{minipage}
\end{center}
\end{table*}

In \S\ref{sec:spectrum} we summarize the theoretical model used and how we predict the observed spectra (see also Appendix~\ref{app:formulae} and references therein), focusing on the only three parameters used in the spectral fits. In \S\ref{sec:results} we define the sample and present the results of the spectral fitting for each of the pulsars (see also Appendix~\ref{app:fits}). \S\ref{sec:correlations} is dedicated to the exploration of correlations between the inferred physical parameters of the best-fitting models and the timing and pulse profile properties across our sample. In \S\ref{sec:conclusions} we discuss the implications of these results.

\section{The synchro-curvature model}\label{sec:spectrum}

The model, presented in detail in a series of previous works \citep{paper0,paper1,paper2,paper3}, calculates the SC spectra emitted by a particle accelerated along a gap placed in some region of the pulsar magnetosphere (within or outside the light cylinder).  We refer the reader to the most important formulae, quoted in the Appendix~\ref{app:formulae}, and to \cite{paper0,paper3}, which notation we follow, for further explanations, while here we shall only sketch the main conceptual points.

We assume the existence of a parallel component of the electric field in a gap, $E_\parallel$, which is responsible of accelerating particles up to high Lorentz factors, $\Gamma$. We calculate the evolution of the kinematic properties of said particles while accelerating, $\Gamma$ and the pitch angle $\alpha$, by numerically solving the equations of motion along the particle movement when the full SC losses are taken into account. Together with the local (i.e., varying along the gap extent) values of the magnetic field $B$ and of the radius of curvature $r_c$, $\Gamma$ and $\alpha$ regulate the local emission of SC radiation.
The evolution of these parameters along the particle trajectory is consistently followed when computing the radiation-reaction force (given by the SC power) and the spectral energy distribution of the radiated emission, ${dP_{\rm sc}}/{dE_\gamma}$. The latter is
a cumbersome expression (see Eqs. A2 to A10 of Appendix~\ref{app:formulae}) but can be numerically computed. 

Once we have computed the particles trajectories for each given set of parameters, we assume an effective particle distribution, $dN_e/dx(x;x_0,N_0,x_{in},x_{out})$, along the travelled distance on the magnetic field line, $x$, between the assumed boundaries $x_{\rm in}$ and $x_{\rm out}$ (which are parameters in the model, see below). Thus, in order to obtain the detectable radiation we convolve the single-particle SC spectrum with $dN_e/dx $:
\begin{equation}\label{eq:sed_x}
  \frac{dP_{\rm gap}}{dE_\gamma} =  \int_{x_{\rm in}}^{x_{\rm out}} \frac{dP_{\rm sc}}{dE_\gamma}\frac{dN_e}{d x} {\rm d}x~.
\end{equation}
The effective particle distribution represents only the number of particles, per unit of distance, emitting radiation towards us. 
For representing the particle distribution we shall assume
\begin{equation}
  \frac{dN_e}{dx}=N_0\frac{e^{-(x-x_{\rm in})/x_0}}{x_0(1 - e^{-x_{\rm out}/x_0}) }~.
  \label{eq:distribution}
\end{equation}
This functional form can be used to weight differently (according to the value of $x_0$) the number of particles at different places of the gap extent. Put otherwise, it allows to study whether the synchrotron-like emission (i.e., the loss of the initial component of the particle momentum perpendicular to the magnetic field line, which happens close to the pair creation location), or the curvature-like emission (i.e., the loss of the parallel component of the particle momentum, dominating at larger distances) dominates the emitted spectrum.  

\begin{table*}
\centering
\tiny
\caption{Timing parameters, timing-inferred properties, 0.1-100 GeV luminosity, and best-fitting parameters ($E_\parallel$, $x_0$ and $N_0$) to the phase-average spectra for the 59 YPs of our sample. We indicate lower (upper) limits for $E_\parallel$ ($x_0$ and $x_0/R_{\rm lc}$) in the cases with strong degeneracy (see text and Appendix~\ref{app:fits}). When distance is unknown, we use its upper limit (i.e., the distance to the edge of the galaxy in that particular direction) to have an upper limit on $L$, which is used to constrain $N_0$ (which, therefore, has to be taken as an upper limit). The last columns give the value of $\chi^2$ for the best fitting model and the number of considered data points (the number of degrees of freedom ($dof$) is $N_{\rm bin}-3$).}
\label{tab:psr}
\begin{tabular}{l@{$\quad$}c@{$\quad$}c@{$\quad$}c@{$\quad$}c@{$\quad$}c@{$\quad$}c@{$\quad$}c@{$\quad$}c@{$\quad$}c@{$\quad$}c@{$\quad$}c@{$\quad$}r}
\hline
\hline
Pulsar & $P$ & $\log \dot{P}$ & $\log \dot{E}$ & $\log \tau$  & $\log B_s$ & $\log B_{lc}$  & $\log L$ & $\log E_\parallel$ & $\log x_0$ & $\log (\frac{x_0}{R_{\rm lc}})$ & $\log N_0$ & $\chi^2_{\rm min} (N_{\rm bins})$\\
 & [ms] & & [erg/s] & [yr] & [G] & [G] & [erg/s] & [V/m] & [cm] & & & \\
\hline
J0007+7303 &  315.9 & -12.45 &  35.65 &   4.15 &  13.04 &   3.50 &  34.97 & $> 7.87$ & $< 7.09$ & $<-2.09$ & $31.35^{+0.01}_{-0.03}$ &    41.37 $(11)$ \\
J0106+4855 &   83.2 & -15.37 &  34.46 &   6.49 &  11.28 &   3.49 &  34.32 & $> 8.14$ & $< 7.09$ & $<-2.09$ & $31.04^{+0.19}_{-0.37}$ &     5.43 $( 6)$ \\
J0205+6449 &   65.7 & -12.72 &  37.41 &   3.74 &  12.56 &   5.06 &  34.39 & $> 8.50$ & $< 6.58$ & $<-2.60$ & $29.93^{+0.67}_{-0.33}$ &    11.47 $( 7)$ \\
J0248+6021 &  217.1 & -13.26 &  35.32 &   4.80 &  12.54 &   3.50 &  34.39 & $> 7.69$ & $< 6.99$ & $<-2.19$ & $31.61^{+0.05}_{-0.04}$ &     1.60 $( 6)$ \\
J0357+3205 &  444.1 & -13.88 &  33.77 &   5.73 &  12.38 &   2.41 & <35.71 & $ 6.92^{+0.01}_{-0.14}$ & $ 7.93^{+0.30}_{-0.02}$ & $-1.40^{+0.30}_{-0.02}$ & $33.14^{+0.01}_{-0.19}$ &     3.96 $( 7)$ \\
J0534+2200 &   33.6 & -12.38 &  38.64 &   3.10 &  12.58 &   5.96 &  35.79 & $> 8.71$ & $< 6.69$ & $<-2.49$ & $31.65^{+0.01}_{-0.09}$ &    32.53 $( 9)$ \\
J0631+1036 &  287.8 & -12.98 &  35.23 &   4.64 &  12.74 &   3.33 &  33.75 & $> 8.47$ & $< 6.19$ & $<-2.99$ & $30.74^{+0.06}_{-0.05}$ &     5.35 $( 7)$ \\
J0633+0632 &  297.4 & -13.10 &  35.08 &   4.77 &  12.69 &   3.24 & <35.93 & $ 7.72^{+0.22}_{-0.14}$ & $ 7.05^{+0.22}_{-0.30}$ & $-2.10^{+0.22}_{-0.30}$ & $32.83^{+0.13}_{-0.19}$ &     7.66 $( 7)$ \\
J0633+1746 &  237.1 & -13.96 &  34.20 &   5.53 &  12.20 &   3.05 &  34.50 & $ 7.66^{+0.01}_{-0.07}$ & $ 7.15^{+0.10}_{-0.01}$ & $-1.90^{+0.10}_{-0.01}$ & $31.25^{+0.01}_{-0.04}$ &   194.30 $(11)$ \\
J0659+1414 &  384.9 & -13.26 &  34.58 &   5.04 &  12.67 &   2.88 &  32.37 & $ 6.42^{+0.32}_{-0.42}$ & $ 8.06^{+1.30}_{-0.40}$ & $-1.20^{+1.30}_{-0.40}$ & $30.94^{+0.09}_{-0.16}$ &     3.68 $( 5)$ \\
J0734-1559 &  155.1 & -13.90 &  35.11 &   5.29 &  12.15 &   3.54 & <35.85 & $> 7.60$ & $< 7.18$ & $<-2.00$ & $32.95^{+0.12}_{-0.11}$ &    19.25 $( 8)$ \\
J0835-4510 &   89.4 & -12.90 &  36.84 &   4.05 &  12.53 &   4.64 &  34.95 & $> 8.03$ & $< 7.19$ & $<-1.99$ & $31.19^{+0.01}_{-0.11}$ &  1418.00 $(11)$ \\
J0908-4913 &  106.8 & -13.82 &  35.69 &   5.05 &  12.11 &   3.99 &  34.54 & $ 7.22^{+0.23}_{-0.19}$ & $ 7.31^{+0.33}_{-0.30}$ & $-1.40^{+0.33}_{-0.30}$ & $31.80^{+0.11}_{-0.16}$ &     1.72 $( 6)$ \\
J1016-5857 &  107.4 & -13.09 &  36.41 &   4.32 &  12.48 &   4.34 &  34.74 & $> 7.90$ & $< 7.40$ & $<-1.78$ & $30.86^{+0.19}_{-0.54}$ &     3.68 $( 6)$ \\
J1023-5746 &  111.5 & -12.42 &  37.04 &   3.66 &  12.82 &   4.64 & <36.82 & $> 8.94$ & $< 6.08$ & $<-3.10$ & $32.29^{+0.30}_{-0.23}$ &     6.06 $( 9)$ \\
J1028-5819 &   91.4 & -13.79 &  35.92 &   4.95 &  12.08 &   4.17 &  35.20 & $> 8.45$ & $< 6.69$ & $<-2.49$ & $31.30^{+0.04}_{-0.08}$ &     9.05 $(10)$ \\
J1044-5737 &  139.0 & -13.25 &  35.92 &   4.59 &  12.45 &   3.99 & <36.74 & $> 7.93$ & $< 6.99$ & $<-2.19$ & $32.96^{+0.48}_{-0.03}$ &    18.97 $( 8)$ \\
J1048-5832 &  123.7 & -13.02 &  36.30 &   4.31 &  12.54 &   4.23 &  35.25 & $> 7.91$ & $< 7.19$ & $<-1.99$ & $31.60^{+0.07}_{-0.01}$ &    12.09 $(10)$ \\
J1057-5226 &  197.1 & -14.24 &  34.48 &   5.73 &  12.04 &   3.11 &  33.64 & $ 7.40^{+0.01}_{-0.01}$ & $ 7.47^{+0.01}_{-0.01}$ & $-1.50^{+0.01}_{-0.01}$ & $30.23^{+0.01}_{-0.01}$ &     5.89 $( 8)$ \\
J1105-6107 &   63.2 & -13.80 &  36.40 &   4.80 &  12.00 &   4.57 &  35.16 & $> 8.26$ & $< 6.90$ & $<-2.28$ & $31.28^{+0.23}_{-0.54}$ &     2.08 $( 6)$ \\
J1119-6127 &  408.7 & -11.39 &  36.36 &   3.21 &  13.61 &   3.74 &  35.78 & $> 8.31$ & $< 6.29$ & $<-2.89$ & $32.04^{+0.29}_{-0.06}$ &     5.31 $( 8)$ \\
J1124-5916 &  135.5 & -12.12 &  37.08 &   3.46 &  13.00 &   4.58 &  35.23 & $> 9.32$ & $< 5.58$ & $<-3.60$ & $30.48^{+0.20}_{-0.02}$ &     4.73 $( 8)$ \\
J1135-6055 &  114.5 & -13.11 &  36.32 &   4.36 &  12.48 &   4.27 & <36.29 & $> 8.32$ & $< 6.68$ & $<-2.50$ & $32.38^{+0.27}_{-0.26}$ &     1.98 $( 6)$ \\
J1413-6205 &  109.7 & -13.56 &  35.91 &   4.80 &  12.26 &   4.09 & <36.94 & $> 8.06$ & $< 7.20$ & $<-1.98$ & $33.08^{+0.01}_{-0.05}$ &     5.09 $( 8)$ \\
J1418-6058 &  110.6 & -12.77 &  36.69 &   4.02 &  12.64 &   4.47 &  34.97 & $> 8.12$ & $< 6.99$ & $<-2.19$ & $31.23^{+0.07}_{-0.09}$ &     4.88 $( 8)$ \\
J1420-6048 &   68.2 & -13.08 &  37.00 &   4.11 &  12.38 &   4.84 &  35.81 & $> 8.76$ & $< 6.39$ & $<-2.79$ & $31.28^{+0.41}_{-0.39}$ &     4.04 $( 8)$ \\
J1429-5911 &  115.8 & -13.52 &  35.89 &   4.78 &  12.28 &   4.05 & <36.66 & $> 7.76$ & $< 7.29$ & $<-1.89$ & $33.28^{+0.07}_{-0.08}$ &     6.55 $( 7)$ \\
J1459-6053 &  103.2 & -13.60 &  35.96 &   4.81 &  12.20 &   4.14 & <36.88 & $> 8.76$ & $< 6.18$ & $<-3.00$ & $33.00^{+0.14}_{-0.03}$ &     4.89 $( 8)$ \\
J1509-5850 &   88.9 & -14.04 &  35.72 &   5.18 &  11.96 &   4.08 &  35.02 & $> 8.05$ & $< 7.01$ & $<-2.17$ & $31.36^{+0.27}_{-0.21}$ &    45.43 $( 8)$ \\
J1620-4927 &  171.9 & -13.98 &  34.91 &   5.41 &  12.15 &   3.39 & <37.04 & $> 7.82$ & $< 7.19$ & $<-1.99$ & $33.83^{+0.10}_{-0.18}$ &     9.44 $( 9)$ \\
J1709-4429 &  102.5 & -13.03 &  36.53 &   4.24 &  12.49 &   4.43 &  35.93 & $> 8.10$ & $< 7.09$ & $<-2.09$ & $32.09^{+0.06}_{-0.11}$ &   135.00 $(11)$ \\
J1718-3825 &   74.7 & -13.88 &  36.08 &   4.95 &  12.00 &   4.35 &  35.14 & $> 7.65$ & $< 7.49$ & $<-1.69$ & $31.87^{+0.10}_{-0.70}$ &     4.62 $( 6)$ \\
J1732-3131 &  196.5 & -13.55 &  35.18 &   5.05 &  12.38 &   3.46 &  33.94 & $ 7.56^{+0.01}_{-0.04}$ & $ 7.47^{+0.11}_{-0.01}$ & $-1.50^{+0.11}_{-0.01}$ & $30.28^{+0.01}_{-0.05}$ &     0.80 $( 7)$ \\
J1741-2054 &  413.7 & -13.77 &  33.98 &   5.59 &  12.43 &   2.54 &  33.31 & $ 6.92^{+0.10}_{-0.10}$ & $ 7.99^{+0.23}_{-0.20}$ & $-1.30^{+0.23}_{-0.20}$ & $30.54^{+0.15}_{-0.13}$ &     7.42 $( 7)$ \\
J1746-3239 &  199.5 & -14.18 &  34.52 &   5.68 &  12.08 &   3.13 & <36.74 & $> 7.53$ & $< 7.29$ & $<-1.89$ & $34.08^{+0.23}_{-0.22}$ &     6.38 $( 6)$ \\
J1747-2958 &   98.8 & -13.21 &  36.40 &   4.41 &  12.40 &   4.38 &  35.76 & $> 8.02$ & $< 6.98$ & $<-2.20$ & $31.78^{+0.64}_{-0.23}$ &     4.73 $( 8)$ \\
J1803-2149 &  106.3 & -13.71 &  35.81 &   4.94 &  12.18 &   4.05 & <36.85 & $> 8.27$ & $< 6.79$ & $<-2.39$ & $33.09^{+0.16}_{-0.11}$ &     6.49 $( 8)$ \\
J1809-2332 &  146.8 & -13.46 &  35.63 &   4.83 &  12.36 &   3.82 &  35.21 & $> 7.94$ & $< 7.08$ & $<-2.10$ & $31.74^{+0.01}_{-0.01}$ &    11.58 $( 9)$ \\
J1813-1246 &   48.1 & -13.75 &  36.79 &   4.64 &  11.97 &   4.89 & <37.27 & $> 9.22$ & $< 5.98$ & $<-3.20$ & $32.68^{+0.09}_{-0.10}$ &     8.11 $( 8)$ \\
J1826-1256 &  110.2 & -12.92 &  36.56 &   4.16 &  12.57 &   4.40 & <37.45 & $> 7.73$ & $< 7.39$ & $<-1.79$ & $33.98^{+0.07}_{-0.08}$ &    17.43 $( 8)$ \\
J1833-1034 &   61.9 & -12.69 &  37.53 &   3.69 &  12.56 &   5.14 &  35.19 & $> 7.80$ & $< 7.39$ & $<-1.79$ & $31.92^{+0.06}_{-0.10}$ &     1.57 $( 6)$ \\
J1836+5925 &  173.3 & -14.82 &  34.04 &   6.26 &  11.72 &   2.96 &  34.31 & $> 7.70$ & $< 7.29$ & $<-1.89$ & $31.09^{+0.01}_{-0.01}$ &    22.09 $( 9)$ \\
J1838-0537 &  145.7 & -12.33 &  36.77 &   3.70 &  12.92 &   4.39 & <37.11 & $> 8.77$ & $< 6.28$ & $<-2.90$ & $32.55^{+0.40}_{-0.06}$ &     8.43 $( 8)$ \\
J1907+0602 &  106.6 & -13.06 &  36.45 &   4.29 &  12.49 &   4.37 &  35.50 & $> 8.00$ & $< 7.09$ & $<-2.09$ & $31.18^{+0.75}_{-0.03}$ &    24.13 $( 9)$ \\
J1952+3252 &   39.5 & -14.24 &  36.57 &   5.03 &  11.68 &   4.86 &  34.82 & $> 8.07$ & $< 7.39$ & $<-1.79$ & $30.93^{+0.06}_{-0.02}$ &    12.61 $( 9)$ \\
J1954+2836 &   92.7 & -13.67 &  36.04 &   4.84 &  12.15 &   4.21 & <36.63 & $> 8.10$ & $< 7.08$ & $<-2.10$ & $32.85^{+0.08}_{-0.01}$ &     6.57 $( 9)$ \\
J1957+5033 &  374.8 & -14.17 &  33.71 &   5.94 &  12.20 &   2.45 & <35.82 & $ 7.06^{+0.22}_{-0.18}$ & $ 7.65^{+0.33}_{-0.31}$ & $-1.60^{+0.33}_{-0.31}$ & $33.39^{+0.21}_{-0.28}$ &     6.25 $( 7)$ \\
J1958+2846 &  290.4 & -12.67 &  35.53 &   4.34 &  12.90 &   3.47 & <36.57 & $> 7.29$ & $< 7.70$ & $<-1.48$ & $33.47^{+0.07}_{-0.19}$ &    20.68 $( 8)$ \\
J2021+3651 &  103.7 & -13.02 &  36.53 &   4.23 &  12.51 &   4.42 &  36.77 & $> 7.94$ & $< 7.19$ & $<-1.99$ & $33.15^{+0.08}_{-0.01}$ &    24.60 $(10)$ \\
J2021+4026 &  265.3 & -13.27 &  35.04 &   4.89 &  12.58 &   3.28 &  35.41 & $> 7.65$ & $< 7.09$ & $<-2.09$ & $32.47^{+0.07}_{-0.04}$ &    58.07 $(11)$ \\
J2028+3332 &  176.7 & -14.31 &  34.54 &   5.76 &  11.97 &   3.20 & <36.32 & $ 7.70^{+0.21}_{-0.21}$ & $ 7.03^{+0.43}_{-0.30}$ & $-1.90^{+0.43}_{-0.30}$ & $33.13^{+0.17}_{-0.33}$ &     8.56 $( 8)$ \\
J2030+3641 &  200.1 & -14.19 &  34.51 &   5.69 &  12.08 &   3.12 &  34.53 & $ 7.46^{+0.26}_{-0.27}$ & $ 7.68^{+2.30}_{-0.61}$ & $-1.30^{+2.30}_{-0.61}$ & $30.88^{+0.73}_{-0.31}$ &     5.61 $( 7)$ \\
J2030+4415 &  227.1 & -14.19 &  34.34 &   5.74 &  12.08 &   2.98 & <36.23 & $> 7.73$ & $< 6.88$ & $<-2.30$ & $33.98^{+0.06}_{-0.10}$ &     1.33 $( 6)$ \\
J2032+4127 &  143.2 & -13.69 &  35.43 &   5.05 &  12.23 &   3.73 &  35.23 & $ 8.00^{+0.12}_{-0.18}$ & $ 6.93^{+0.41}_{-0.20}$ & $-1.90^{+0.41}_{-0.20}$ & $31.34^{+0.12}_{-0.22}$ &     6.15 $( 8)$ \\
J2055+2539 &  319.6 & -14.39 &  33.70 &   6.09 &  12.08 &   3.51 & <36.18 & $ 7.10^{+0.10}_{-0.05}$ & $ 7.88^{+0.11}_{-0.21}$ & $-1.30^{+0.11}_{-0.21}$ & $33.11^{+0.17}_{-0.06}$ &     3.43 $( 7)$ \\
J2111+4606 &  157.8 & -12.84 &  36.15 &   4.24 &  12.68 &   4.05 & <36.06 & $> 8.05$ & $< 6.99$ & $<-2.19$ & $32.26^{+0.13}_{-0.10}$ &     9.10 $( 9)$ \\
J2139+4716 &  282.8 & -14.74 &  33.49 &   6.40 &  11.86 &   2.47 & <35.75 & $> 7.15$ & $< 7.71$ & $<-1.47$ & $33.45^{+0.29}_{-0.56}$ &     6.83 $( 6)$ \\
J2229+6114 &   51.6 & -13.11 &  37.34 &   4.02 &  12.30 &   5.13 &  34.29 & $ 9.22^{+0.19}_{-0.60}$ & $ 5.29^{+0.60}_{-0.20}$ & $-3.10^{+0.60}_{-0.20}$ & $29.65^{+0.54}_{-0.16}$ &    13.01 $(10)$ \\
J2238+5903 &  162.7 & -13.01 &  35.95 &   4.42 &  12.60 &   3.93 & <36.07 & $> 7.67$ & $< 7.29$ & $<-1.89$ & $32.79^{+0.07}_{-0.15}$ &     7.75 $( 9)$ \\
\hline
\hline
\end{tabular}

\end{table*}

\begin{table*}
\centering
\tiny
\caption{Same as Table~\ref{tab:psr}, but for the 22 MSPs of our sample.}
\label{tab:msp}
\begin{tabular}{l@{$\quad$}c@{$\quad$}c@{$\quad$}c@{$\quad$}c@{$\quad$}c@{$\quad$}c@{$\quad$}c@{$\quad$}c@{$\quad$}c@{$\quad$}c@{$\quad$}c@{$\quad$}r}
\hline
\hline
Pulsar & $P$ & $\log \dot{P}$ & $\log \dot{E}$ & $\log \tau$  & $\log B_s$ & $\log B_{lc}$  & $\log L$ & $\log E_\parallel$ & $\log x_0$ & $\log (\frac{x_0}{R_{\rm lc}})$ & $\log N_0$ & $\chi^2_{\rm min} (N_{\rm bins})$\\
 & [ms] & & [erg/s] & [yr] & [G] & [G] & [erg/s] & [V/m] & [cm] & & & \\
\hline
J0030+0451 &    4.9 & -19.99 &  33.53 &   9.88 &   8.36 &   4.25 &  32.76 & $ 8.66^{+0.14}_{-0.13}$ & $ 5.57^{+0.22}_{-0.21}$ & $-1.80^{+0.22}_{-0.21}$ & $28.53^{+0.13}_{-0.15}$ &    10.78 $( 8)$ \\
J0034-0534 &    1.9 & -20.70 &  34.46 &  10.18 &   8.00 &   5.12 &  32.75 & $> 9.01$ & $< 5.38$ & $<-1.99$ & $28.38^{+0.04}_{-0.07}$ &     1.45 $( 6)$ \\
J0101-6422 &    2.6 & -20.32 &  34.04 &   9.93 &   8.04 &   4.77 &  32.58 & $ 8.70^{+0.19}_{-0.20}$ & $ 5.99^{+2.10}_{-0.51}$ & $-1.10^{+2.10}_{-0.51}$ & $27.51^{+0.81}_{-0.16}$ &     1.99 $( 6)$ \\
J0218+4232 &    2.3 & -19.11 &  35.40 &   8.67 &   8.63 &   5.51 &  34.58 & $> 9.84$ & $< 4.47$ & $<-2.90$ & $29.51^{+0.10}_{-0.01}$ &     1.37 $( 9)$ \\
J0340+4130 &    3.3 & -20.23 &  33.81 &   9.95 &   8.15 &   4.56 &  33.86 & $ 8.84^{+0.42}_{-0.37}$ & $ 5.60^{+2.60}_{-0.70}$ & $-1.60^{+2.60}_{-0.70}$ & $28.97^{+0.61}_{-0.41}$ &     6.12 $( 6)$ \\
J0437-4715 &    5.8 & -19.27 &  34.04 &   9.23 &   8.75 &   4.42 &  31.69 & $> 8.19$ & $< 5.88$ & $<-1.49$ & $28.06^{+0.16}_{-0.14}$ &     4.04 $( 7)$ \\
J0613-0200 &    3.1 & -20.02 &  34.11 &   9.71 &   8.26 &   4.73 &  33.46 & $ 8.96^{+0.34}_{-0.21}$ & $ 5.27^{+0.44}_{-0.50}$ & $-1.90^{+0.44}_{-0.50}$ & $28.84^{+0.26}_{-0.28}$ &     4.99 $( 6)$ \\
J0614-3329 &    3.2 & -19.75 &  34.32 &   9.45 &   8.38 &   4.83 &  34.67 & $> 9.04$ & $< 5.47$ & $<-1.90$ & $29.86^{+0.06}_{-0.09}$ &    12.62 $(10)$ \\
J0751+1807 &    3.5 & -20.11 &  33.86 &   9.85 &   8.23 &   4.56 &  32.40 & $> 9.00$ & $< 5.39$ & $<-1.98$ & $28.02^{+0.24}_{-0.26}$ &     2.22 $( 6)$ \\
J1124-3653 &    2.4 & -20.24 &  34.23 &   9.82 &   8.08 &   4.90 &  33.63 & $> 9.02$ & $< 5.59$ & $<-1.78$ & $28.95^{+0.24}_{-0.27}$ &     1.72 $( 6)$ \\
J1231-1411 &    3.7 & -19.67 &  34.20 &   9.44 &   8.45 &   4.71 &  33.37 & $ 8.92^{+0.08}_{-0.12}$ & $ 5.35^{+0.20}_{-0.11}$ & $-1.90^{+0.20}_{-0.11}$ & $28.76^{+0.05}_{-0.13}$ &    12.81 $( 9)$ \\
J1514-4946 &    3.6 & -19.73 &  34.20 &   9.48 &   8.41 &   4.71 &  33.68 & $> 9.20$ & $< 5.18$ & $<-2.19$ & $29.05^{+0.08}_{-0.10}$ &     1.85 $( 8)$ \\
J1614-2230 &    3.2 & -20.02 &  34.08 &   9.72 &   8.26 &   4.70 &  33.09 & $ 8.58^{+0.18}_{-0.14}$ & $ 6.38^{+1.80}_{-0.73}$ & $-0.80^{+1.80}_{-0.73}$ & $27.96^{+0.92}_{-0.08}$ &     3.29 $( 6)$ \\
J1744-1134 &    4.1 & -20.05 &  33.71 &   9.86 &   8.28 &   4.41 &  32.83 & $> 8.42$ & $< 5.78$ & $<-1.59$ & $29.00^{+0.14}_{-0.14}$ &     2.06 $( 6)$ \\
J1810+1744 &    1.7 & -20.34 &  34.57 &   9.77 &   7.95 &   5.23 &  34.05 & $> 9.41$ & $< 4.87$ & $<-2.50$ & $29.68^{+0.07}_{-0.04}$ &     2.28 $( 7)$ \\
J1902-5105 &    1.7 & -20.05 &  34.86 &   9.48 &   8.08 &   5.37 &  33.56 & $> 9.52$ & $< 4.87$ & $<-2.50$ & $28.86^{+0.04}_{-0.08}$ &     2.54 $( 6)$ \\
J2017+0603 &    2.9 & -20.08 &  34.11 &   9.74 &   8.20 &   4.77 &  33.99 & $ 9.22^{+0.19}_{-0.15}$ & $ 5.14^{+0.33}_{-0.30}$ & $-2.00^{+0.33}_{-0.30}$ & $28.96^{+0.17}_{-0.20}$ &     4.26 $( 8)$ \\
J2043+1711 &    2.4 & -20.24 &  34.20 &   9.82 &   8.08 &   4.90 &  34.00 & $> 9.13$ & $< 5.28$ & $<-2.09$ & $29.47^{+0.14}_{-0.09}$ &     4.60 $( 8)$ \\
J2124-3358 &    4.9 & -19.69 &  33.84 &   9.58 &   8.51 &   4.40 &  32.60 & $ 8.50^{+0.08}_{-0.10}$ & $ 6.37^{+0.61}_{-0.31}$ & $-1.00^{+0.61}_{-0.31}$ & $27.65^{+0.12}_{-0.10}$ &     2.07 $( 7)$ \\
J2214+3000 &    3.1 & -19.82 &  34.30 &   9.51 &   8.34 &   4.83 &  33.97 & $ 8.88^{+0.09}_{-0.11}$ & $ 5.37^{+0.21}_{-0.11}$ & $-1.80^{+0.21}_{-0.11}$ & $29.33^{+0.05}_{-0.11}$ &     3.72 $( 7)$ \\
J2241-5236 &    2.2 & -20.06 &  34.51 &   9.60 &   8.15 &   5.08 &  33.02 & $> 9.04$ & $< 5.58$ & $<-1.79$ & $28.15^{+0.04}_{-0.07}$ &     2.09 $( 7)$ \\
J2302+4442 &    5.2 & -19.88 &  33.57 &   9.79 &   8.43 &   4.24 &  33.79 & $ 8.70^{+0.11}_{-0.14}$ & $ 5.79^{+0.51}_{-0.21}$ & $-1.60^{+0.51}_{-0.21}$ & $29.08^{+0.10}_{-0.28}$ &     4.48 $( 8)$ \\
\hline
\hline
\end{tabular}

\end{table*}

By expanding Eq.~(\ref{eq:sed_x}) above using the details found in Appendix~\ref{app:formulae}, one can see that the model contains eleven parameters. The model parameters are, however, neither all unknown, nor equally relevant. Two of the parameters are in fact the observable spin period $P$ and its derivative $\dot P$, which define in turn the light cylinder radius, defined as
\begin{eqnarray}
 && R_{\rm lc} = \frac{Pc}{2\pi} = 4.77\times 10^9 P{\rm[s] ~ cm} ~,\label{eq:rlc}
\end{eqnarray}
and the inferred value of the magnetic field at the polar surface, $B_s$ (see \S\ref{sec:correlations_timing_derived}). Then, six other parameters regulate the particle trajectory: $E_\parallel$, $\eta$, $b$, $x_{\rm in}$, $\Gamma_{\rm in}$, and $\alpha_{\rm in}$; and three additional parameters describe the effective particle distribution: $N_0$, $x_0/R_{\rm lc}$, and $x_{\rm out}/R_{\rm lc}$. Table \ref{tab:parameters} gives details on these parameters, their considered range, and provide specific references to further discussion. 

In \cite{paper3} we proved that only three parameters are the most relevant in determining the spectrum: the parallel electric field in the gap, $E_\parallel$, the normalization of the effective particle distribution, $N_0$, and the weighting parameter $x_0$.
The other model parameters barely affect the predicted spectral outcome if they vary within the considered range, and we fix them here to fiducial values for all pulsars, as indicated in the third block of Table~\ref{tab:parameters}. Those six fixed parameters, which will not be discussed further here, are the initial Lorentz factor $\Gamma_{\rm in}$, the initial pitch angle $\alpha_{\rm in}$, the boundaries $x_{\rm in}$ and $x_{\rm out}$, and the parameters describing the decreasing of $B$, and the increasing of $r_c$ along the trajectory of the particles. The relative irrelevance of these parameters is, in itself, an interesting result. On the one hand,  it makes the model more robust. On the other hand, it tells that the spectra are not particularly sensitive to details like the precise form of the magnetic field decay in the magnetosphere, that would seem otherwise key. 

\section{Fitting the pulsar spectra }\label{sec:results}

\subsection{Sample definition and fitting}

We consider the publicly available, {\it Fermi}-LAT processed data for the phase-averaged pulsar spectra contained in the 2PC. 
As a good-quality criterion, and not to consider less constraining data,
we select those sources having measured flux (not upper limits) in at least five consecutive energy bins. 
Such selection criterion somewhat favors the YPs, since they are brighter in $\gamma$-rays. Among the 81 sources of our sample (out of the 117 of the 2PC), 59 are YPs (Table~\ref{tab:psr}) and 22 are MSPs (Table~\ref{tab:msp}). 

We apply our models exactly in the same way to each YP and MSP in the sample. We neither make any distinction between radio-quiet or radio-loud. Thus, we constrain the values of the three relevant parameters, $E_\parallel$, $x_0$, and $N_0$, by comparing the SC radiation expected from our models once they assume a particular value
with the observational data. The fitting procedure is described in detail in Appendix~\ref{app:fits}. We span a range of several orders of magnitude in the values of the three mentioned parameters, thus considering a grid of thousands of models for every pulsar before selecting the best fitting set of parameters. 
For each model tested, we compute the particle trajectory in the gap and obtain the gap-integrated spectra according to Eq.~(\ref{eq:sed_x}). To quantitatively compare model and data we compute $\chi^2$ (see Appendix~\ref{app:fits} for details). 

We model only the particles directed to the observer, in an effective way, as stated in \S 2 with $N_0$ being the normalization of the number of particles that are moving in the direction of the observer. The fits show what collection of the three model parameters can fit the data. The gap width is not a parameter in our model, since we have no geometry included.

\subsection{Fitting results and model assessment}

Tables~\ref{tab:psr} and \ref{tab:msp} show the timing and spectral properties of the {\it Fermi}-LAT pulsars in our sample, together with the best-fitting parameters inferred by applying our model. Appendix~\ref{app:fits} presents the case by case results. 
Figs.~\ref{fig:best_fit1}-\ref{fig:best_fit5} show the observed $E^2dN/dE$  spectra (where $dN/dE$ is the spectral photon distribution radiated by the source) 
for each of the 81 pulsars in the sample
together with the corresponding best-fitting models. The theoretical spectra for YPs (MSPs) are plotted there in red (blue), a color code we maintain in all sections. 

The best-fitting models show quantitative agreement with data. With just three parameters, the model copes relatively well with data of dozens of pulsars, disregarding their type (YP or MSP) and spanning several orders of magnitude in any of their main characteristics. 

\begin{figure*} 
\begin{center}
\includegraphics[width=0.42\textwidth]{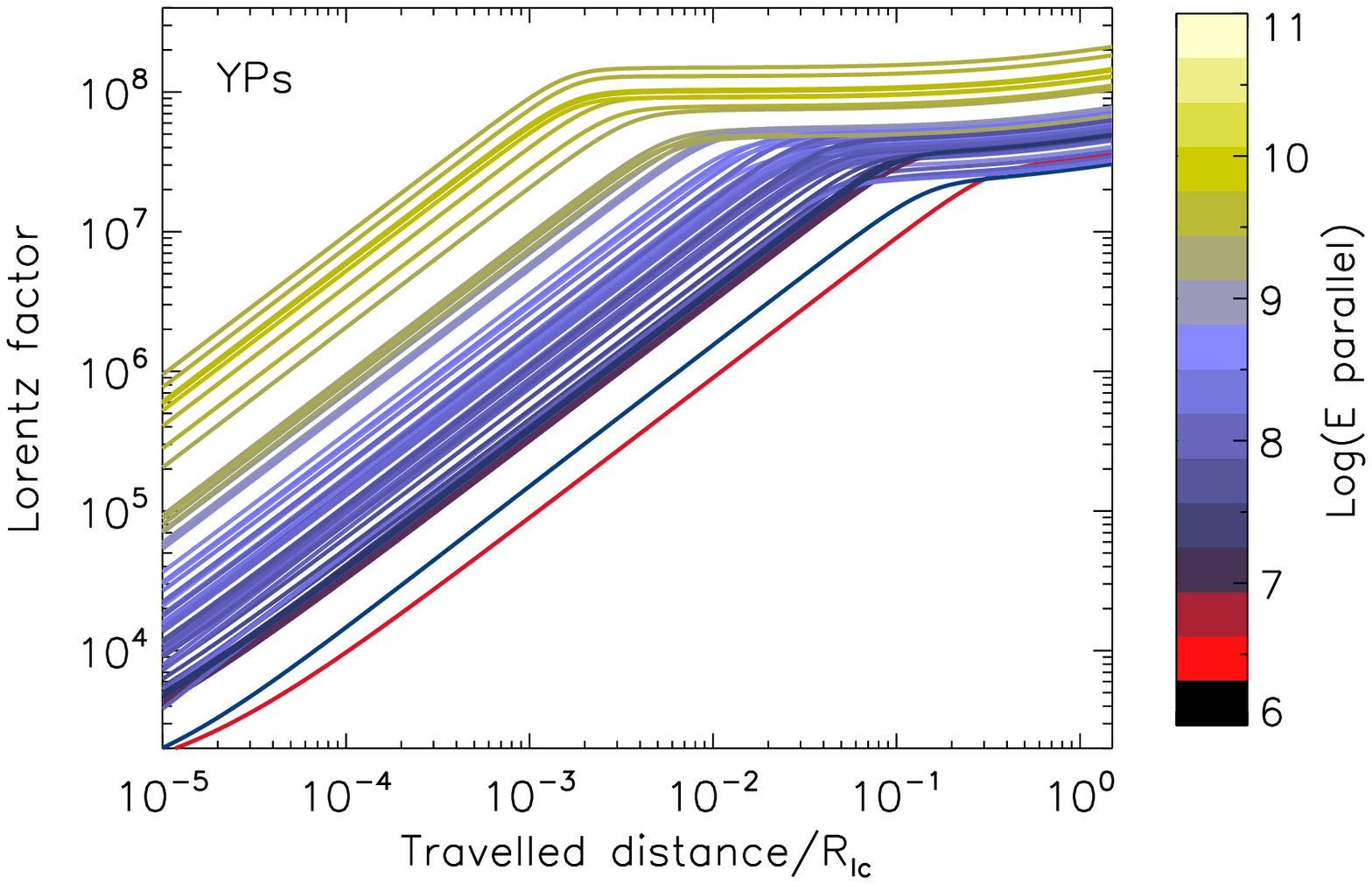}
\includegraphics[width=0.42\textwidth]{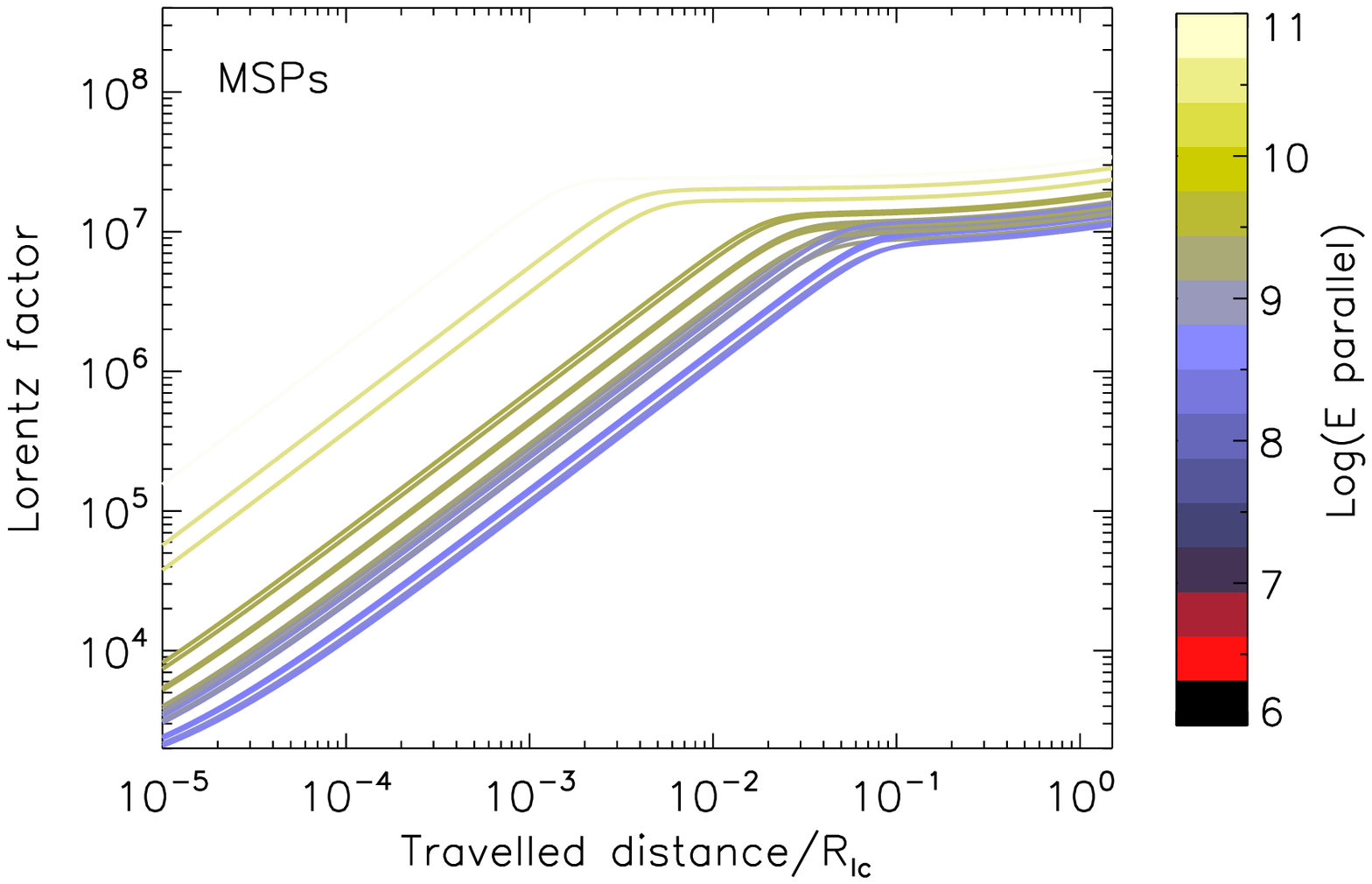}\\
\includegraphics[width=0.42\textwidth]{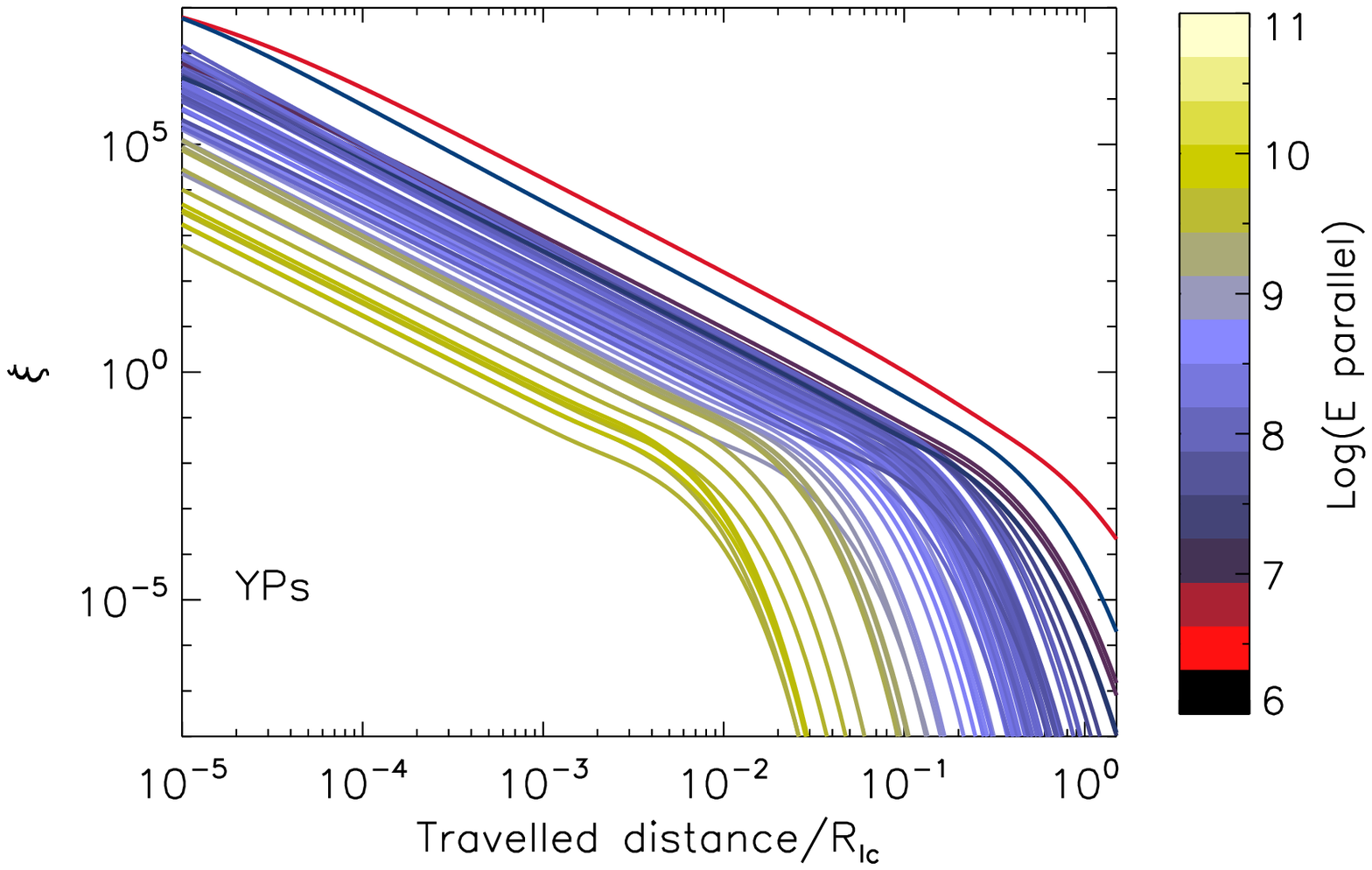}
\includegraphics[width=0.42\textwidth]{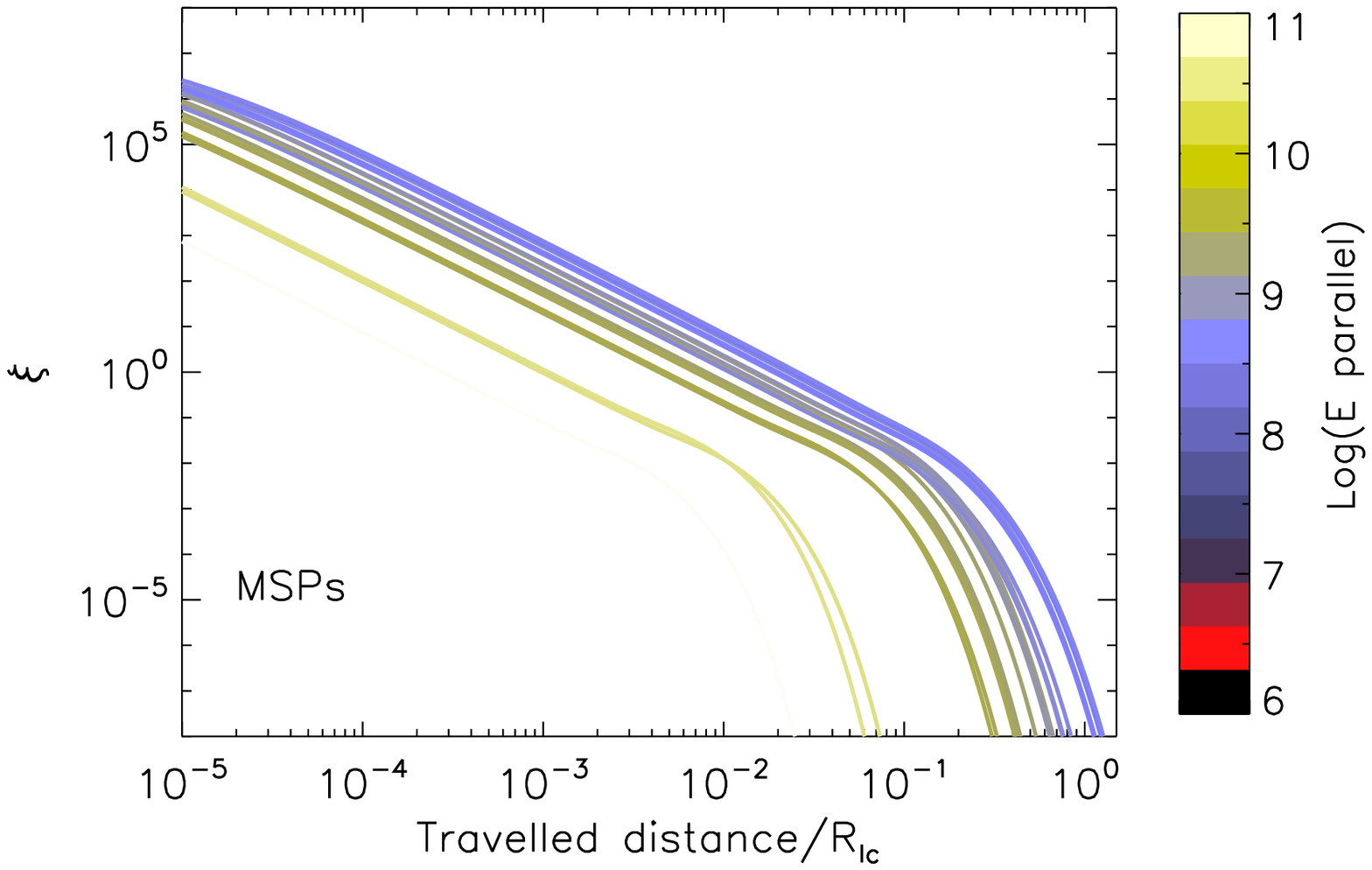}
\end{center}
\caption{Evolution of the trajectory parameters for YPs (left-hand panels) and MSPs (right-hand): Lorentz factor of the particles in the gap, and $\xi$ value, determining the dominance of synchrotron-like ($\xi \gg 1$) or curvature-like ($\xi \ll 1$) radiation along the traveled distance in the gap ($x$), normalized with the corresponding light cylinder radius. The color scheme corresponds to the value of the parallel electric field.}
\label{fig:traj}
\end{figure*}

\subsubsection{Analysis of the residuals}

The highest values of $\chi^2$ occur for a few, very bright YPs: J0835-4510 (Vela pulsar, $\chi^2 > 1000$, $dof=8$), J0633+1746 (Geminga $\chi^2 =194.3$, $dof=8$), and J1709-4429 ($\chi^2 =135$, $dof=8$). They show residuals at high energies, similarly to the Crab pulsar ($\chi^2 = 32.53$, $dof=6$).
There are two main reasons producing the large values of $\chi^2$ in these cases. On the one hand, being these the brightest pulsars, the relative error bars at the low-energy part of the spectra are extremely small (better than $1\%$), and this entails a relatively large deviation from theoretical curves even when the overall description as shown in the corresponding spectral plot of Appendix \ref{app:fits}  is appropriate. 
On the other hand, the high-energy end of the spectra of these pulsars is decaying slower than what the model predicts ({\it Fermi}-LAT data is better described with a power-law with a sub-exponential cutoff, dubbed PLEC in the 2PC). This may, in fact, be a spurious effect of considering phase-averaged spectra. 

The latter is the case for Geminga, where the set of fine-binned phase-resolved spectra has shown the absence of sub-exponential decays at all phases \citep{abdo10a}, i.e., being described by power-law with exponential cut-off models (PLEC1 in the 2PC). The phase-averaged, harder Geminga spectrum arises as a consequence of summing up phase-resolved contributions that can be individually well described by a SC model
\citep{paper3}. Two more cases are similar to Geminga, with which they share similar values of $P$ and $\dot P$: J0007+7303 in CTA1 ($\chi^2 =41.37$, $dof=8$) and J2021+4026  ($\chi^2 =58.07$, $dof=8$). For both, the phase-resolved spectra can always be fitted by PLEC1 models \citep{abdo12,allafort13}. 

Crab and Vela are different, however, since they show pulsed emission above tens of GeV \citep{aliu08,leung14}, and, in at least one phase, their phase-resolved spectra is harder than an exponentially decaying one. Such high-energy tails are very likely a signature of inverse Compton processes contributing to the total yield and cannot be described well by just a SC model, particularly at high energies. We have discussed the cases of Geminga, Crab, and Vela pulsars in more detail using phase-resolved spectra in \cite{paper3}.

In some cases, the first couple of data points show non-negligible residuals within a few $\sigma$, or upper limits apparently incompatible with our model. For instance, in J1907+0602 and J1958+2846, the low-energy slope is apparently harder than what the SC radiation of the best-fitting model produces. If these discontinuous points were real (e.g., the first point in J1509-5850), then the phenomenology of $\gamma$-ray emission is more complicated than what this SC model is able to describe, and perhaps include more than one component. We note, however, that diffuse model uncertainties can especially affect the first couple of low-energy data points. In this regard, we hope that pulsar results using Pass 8 in analyzing {\it Fermi}-LAT data will bring additional insight.

\subsubsection{Trajectories and spectra}

It is interesting to relate the best-fitting spectra with the corresponding particle dynamics along the gap, shown in  Fig.~\ref{fig:traj} for the YPs (left) and the MSPs (right). We show the evolution of their Lorentz factor, $\Gamma$ (top), and their SC parameter, $\xi$ (bottom) --see Eq.~\ref{eq:xi} and Appendix~\ref{app:formulae} for details--, as a function of travelled distance along the field line. Colors indicate the intensity of the accelerating field: the larger $E_\parallel$, the larger the final $\Gamma$.

We see the overall similarities of models, despite the different spectra they produce. Particle trajectories start being dominated by the loss of perpendicular momentum (initial plateaux in $\Gamma$, $\xi \gg 1$), then the electric acceleration boost them up, till they reach a balance between the electric acceleration, and the SC losses, dominated by the loss of longitudinal momentum (curvature regime, $\xi\ll 1$). The transition between the different $\xi$-regimes happens at about $(10^{-3}$--$10^{-2}) R_{\rm lc}$. The larger is the value of the accelerating electric field, the sooner this transition happens. At this same (corresponding) location (i.e., at physically different location for each of the pulsars, since it depends on the corresponding $R_{\rm lc}$) the Lorentz factor saturates. 

In MSPs, despite having a larger best-fitting value of $E_\parallel$, particles reach lower (or similar) values of $\Gamma$, because the radiative losses are stronger, being the light cylinder smaller (and, consequently also their curvature radius, \citealt{paper1}). The different saturated values of the Lorentz factors between MSPs and YPs forms a continuing set similar to $E_\parallel$ values, discussed in the next section.

If the effective particle distribution were uniform, particles with  saturated values of $\Gamma$ would dominate the spectra \citep{paper3}, since their radiation is more intense (higher flux) and harder (larger energy peak). However, this could not explain the variety of spectral slopes. The flatness of spectra require that less energetic particles, having lower $\Gamma$ and larger $\xi$, to significantly contribute as well. In our model, the parameter $x_0$ regulates the relative weight of this contribution.

\subsubsection{Error estimate and degeneracy}\label{sec:errors}

Figs.~\ref{fig:contours1}-\ref{fig:contours5} of Appendix \ref{app:fits} show the contours of $\chi^2/\chi^2_{\rm min}$ in the plane $\log E_\parallel$-$\log (x_0/R_{\rm lc})$. Again, red and blue labels help distinguish the pulsar class. The contour plots  measure the deviation from the minimum attainable value of $\chi^2$ in the model grid, i.e., when the parameters differ from the best-fitting set. In each of the panels, we mark with a white cross the location of the best-fitting model (i.e., where $\chi^2 = \chi^2_{\rm min}$).

In order to estimate the errors on parameters, we consider the region of the contour plots where $\chi^2/\chi^2_{\rm min}\leq 1.3$ (approximatively, at the boundary between black and red regions). Then, we take the minima and maxima of the three parameters within the enclosed region to estimate the errors. For pulsars having a well-defined region, typical relative errors on the inferred best-value parameters of about $\delta log(E_\parallel) / \log(E_\parallel) \sim 5-10\%$, $\delta \log(x_0) / \log(x_0) \sim 30-50\%$, and $\delta \log(N_0)/\log(N_0) \sim 1\%$. These ranges of uncertainties are acceptable since we explore several orders of magnitude in the three parameters for each of the pulsars. 14 YPs and 10 MSPs have well-constrained parameters, and thus it is possible to study trends in the population using just these (filled circles in all the plots). 

Many pulsars show a very elongated (or linear) shape in their corresponding (log--log) contour plot of Figs.~\ref{fig:contours1}-\ref{fig:contours5}, pointing to a clear anti-correlation between $E_\parallel$ and $x_0$. In these cases, there is a systematic degeneracy in choosing the best-fitting model. This can be understood when one considers that at a larger $E_\parallel$, particles are pushed to larger energies; while on the contrary, a smaller $x_0$ tends to neglect the part of the trajectories where particles attain the largest energies weighting more heavily on the synchrotron dominated part and thus compensating for the previous effect. This anti-correlation, $E_\parallel \sim x_0^{-1}$ for large enough $E_\parallel$, can be analytically proven to hold from the equation of motions  themselves
(see Appendix~\ref{app:formulae} for an approximate derivation), when the electric term dominates over the SC loss term.
It does not hold whenever particles with larger $\Gamma$ are dominating the spectrum (larger $x_0$): in these cases, the degeneracy is partly broken and the value of $E_\parallel$ is more constrained (see the contours with more vertically-oriented shapes, e.g. J0101-6422).

From the trajectories shown in Fig.~\ref{fig:traj}, one can understand that a larger $E_\parallel$ displaces the $\Gamma(x)$ curves to the left; thus, to reach the same observed spectrum, one has to compensate by convolving a particle distribution weighted towards lower values of $x$. In such cases, the exact location of the minimum found by our pipeline is slightly dependent on the resolution and range of the explored grid of parameters: all the pairs of parameters inside the black and red valley in the pulsars $\log E_\parallel$--$\log (x_0/R_{\rm lc})$ contour plots of Appendix~\ref{app:fits}
are closely as good as the ones marked by the white cross.

The degeneracy can be partially solved on physical grounds. In particular, we think that within the constrained range, the lowest values of the allowed range of $E_\parallel$ and the corresponding largest values of $x_0$ are more physically meaningful. As a matter of fact, it is difficult to justify large values of $E_\parallel$ exceeding the local value of magnetic field, $B_{\rm lc}$ by orders of magnitude, and, at the same times, values of $x_0$ of the order of meters (with an accordingly very large particle density). 
In Tables~\ref{tab:psr} and \ref{tab:msp}, the pulsars presenting a stronger degree of 
degeneracy in the selection of the best-fitting values (over two thirds of the sample) are represented by lower (upper) limit on $E_\parallel$ ($x_0$). Such lower/upper limits are also employed in the correlation plots hereafter, identified by empty circles.

\begin{figure*}
\begin{center}
\includegraphics[width=0.4\textwidth]{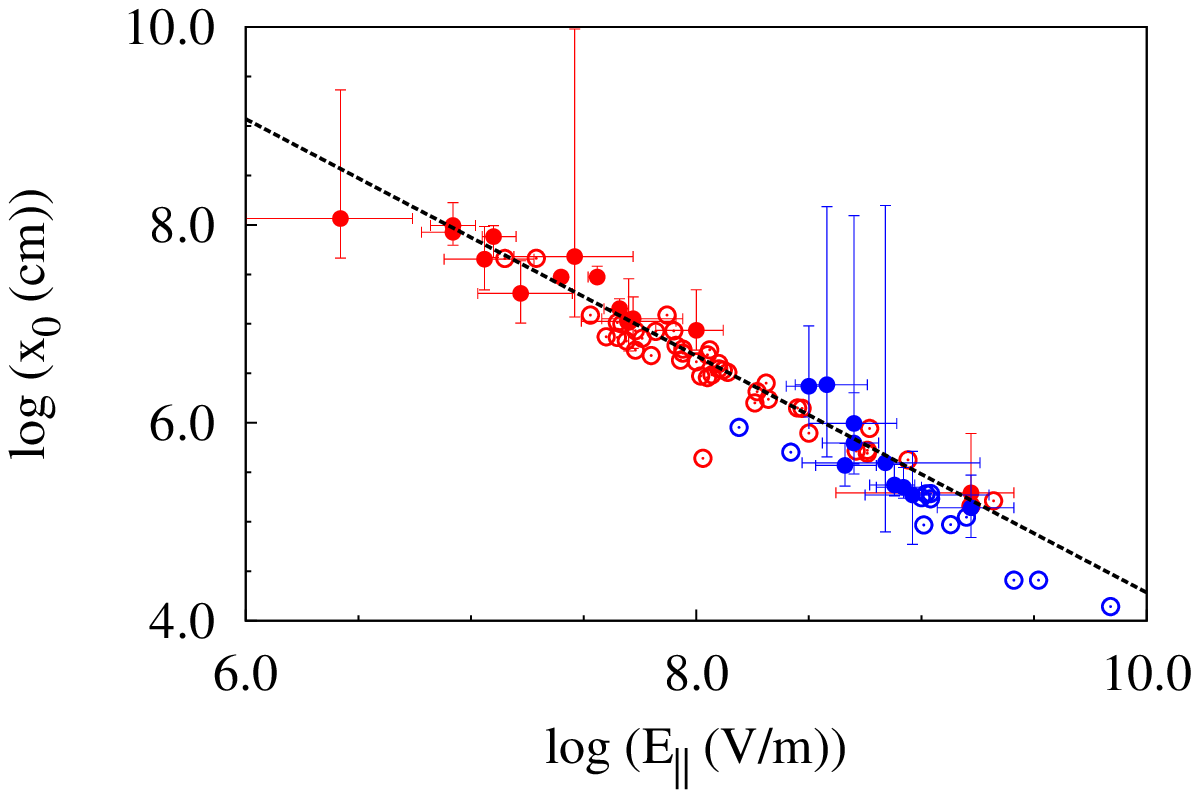} 
\includegraphics[width=0.4\textwidth]{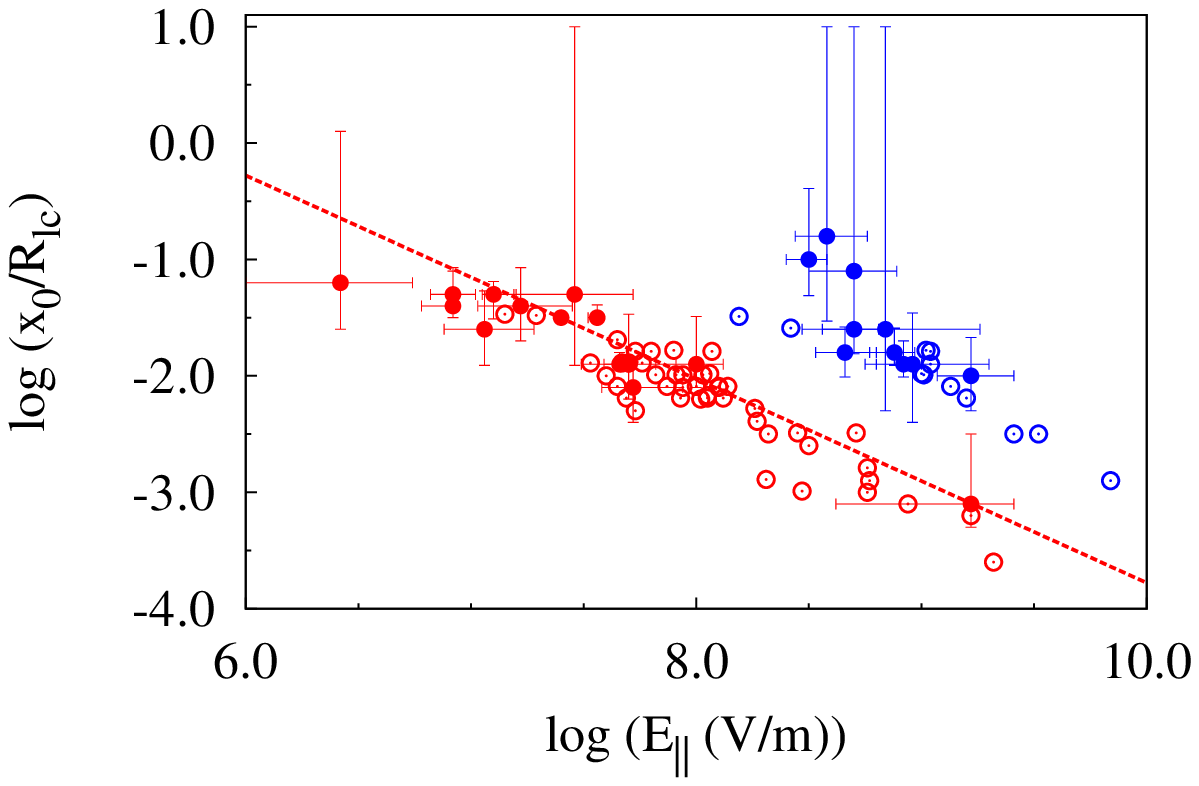}\\
\includegraphics[width=0.4\textwidth]{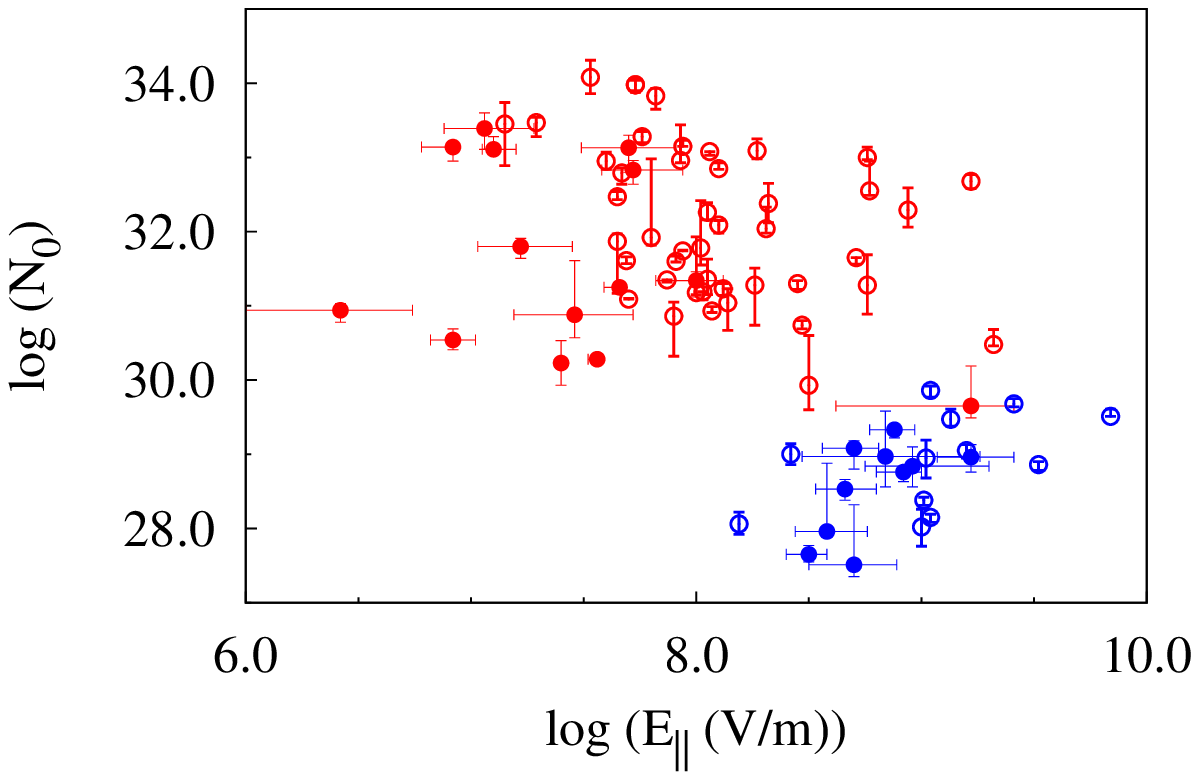}  
\includegraphics[width=0.4\textwidth]{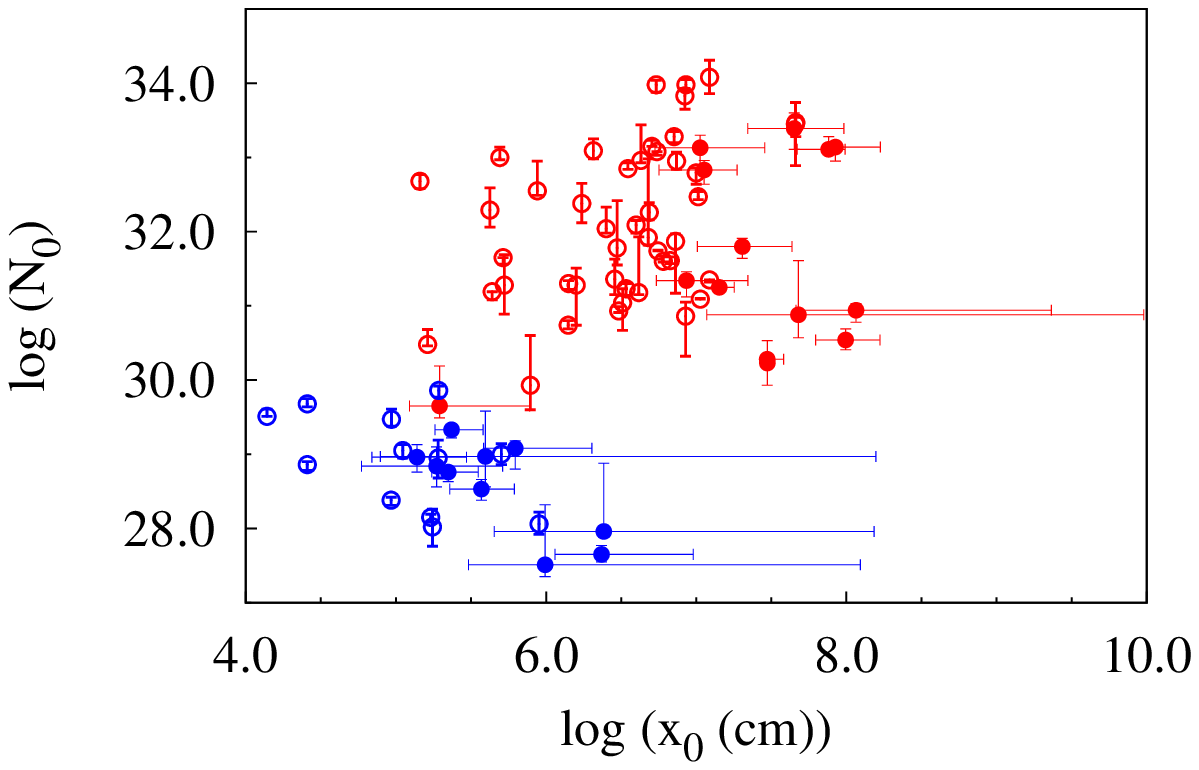}\\
\end{center}
\caption{Distributions and correlations between our best-fitting SC model parameters. Hereafter, YPs and MSPs are marked by red and blue points, respectively, and empty circles represent the cases where $E_\parallel$ and $x_0$ can only be constrained by lower and upper limits, respectively. Therefore, empty points could actually be moved along $E_\parallel \sim (1/x_0)$ towards larger (smaller) values of $E_\parallel$ ($x_0$). See text for a detailed discussion. Lines represent linear fits (shown only when the corresponding Pearson coefficient is $r>0.85$, in black for the whole sample, in red (blue) for YPs (MSPs)) to the non-degenerate pulsar parameters only (filled circles).}
\label{fig:corr_parameters}
\end{figure*}

\section{Correlations and trends}\label{sec:correlations}

In this section we explore correlations between the best-fitting model parameters and the timing and timing-derived properties of pulsars, the best phenomenological fit (PLEC1 model values) as quoted in the 2PC, and (briefly) the pulse profile characteristics. 

\subsection{Values of parameters and self-correlations}\label{sec:self-correlations}

Fig.~\ref{fig:corr_parameters} summarizes the distributions of our best-fitting SC model parameters. 
As Fig.~\ref{fig:corr_parameters} and the Tables~\ref{tab:psr} and \ref{tab:msp} show, MSPs can typically have up to one order of magnitude larger $E_\parallel$ when compared with YPs. Large electric fields, $E_\parallel \sim 10^{8.5}-10^{10}$ V/m (compared with $E_\parallel \sim 10^{6.5}-10^{9}$ V/m for YPs) are required in MSPs in order to compensate for their much smaller radius of curvature (due to the much closer light cylinder) and let particles be energetic enough ($\Gamma \sim 10^7$-$10^8$, see Fig.~\ref{fig:traj}) as to be able to produce $\sim$ GeV photons. 
The lowest (highest) values of $E_\parallel$ are found for YPs (MSPs), respectively.  


Typical values of $x_0/R_{\rm lc}$ are consistently low across both sub-samples of pulsars, $10^{-4}$--$10^{-1}$ (or, equivalently, $x_0\sim 1$--$1000$ km), as was already the case for Crab, Geminga, and Vela pulsars \citep{paper3}. These values are needed to accommodate the many, soft spectral indices of the $\gamma$-ray pulsars below the cutoff energies, as observed by {\it Fermi}-LAT. They indicate a common feature: the $\gamma$-ray radiation is not produced only by particles with saturated values of Lorentz factor. Instead, most photons need to be produced while particles have non-negligible perpendicular momentum, thus radiating mostly synchrotron emission. The lowest (highest) values of $x_0$ are found for MSPs (YPs), respectively. 

The values of $N_0$ clearly separate YPs from MSPs. This is a reflection of the MSPs being less luminous in general, which directly implies that all MSPs have smaller $N_0$ (in the extreme case, for up to six orders of magnitude). Physically, this could be related either to the much reduced magnetospheric volume in MSPs, and/or to the lower rotational energy losses, which represent the ultimate energetic budget partially converted in particle acceleration, and, therefore, in visible radiation.

\begin{figure*} 
\begin{center}
\includegraphics[width=0.32\textwidth]{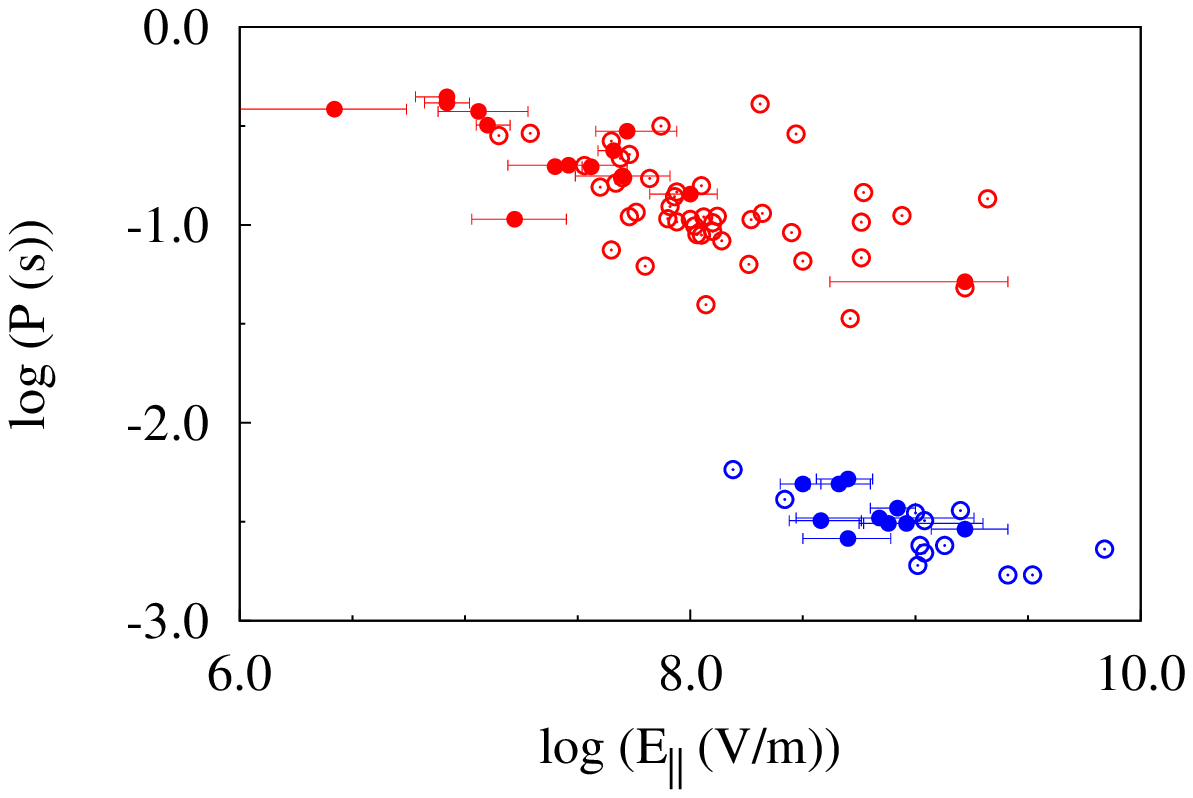}
\includegraphics[width=0.32\textwidth]{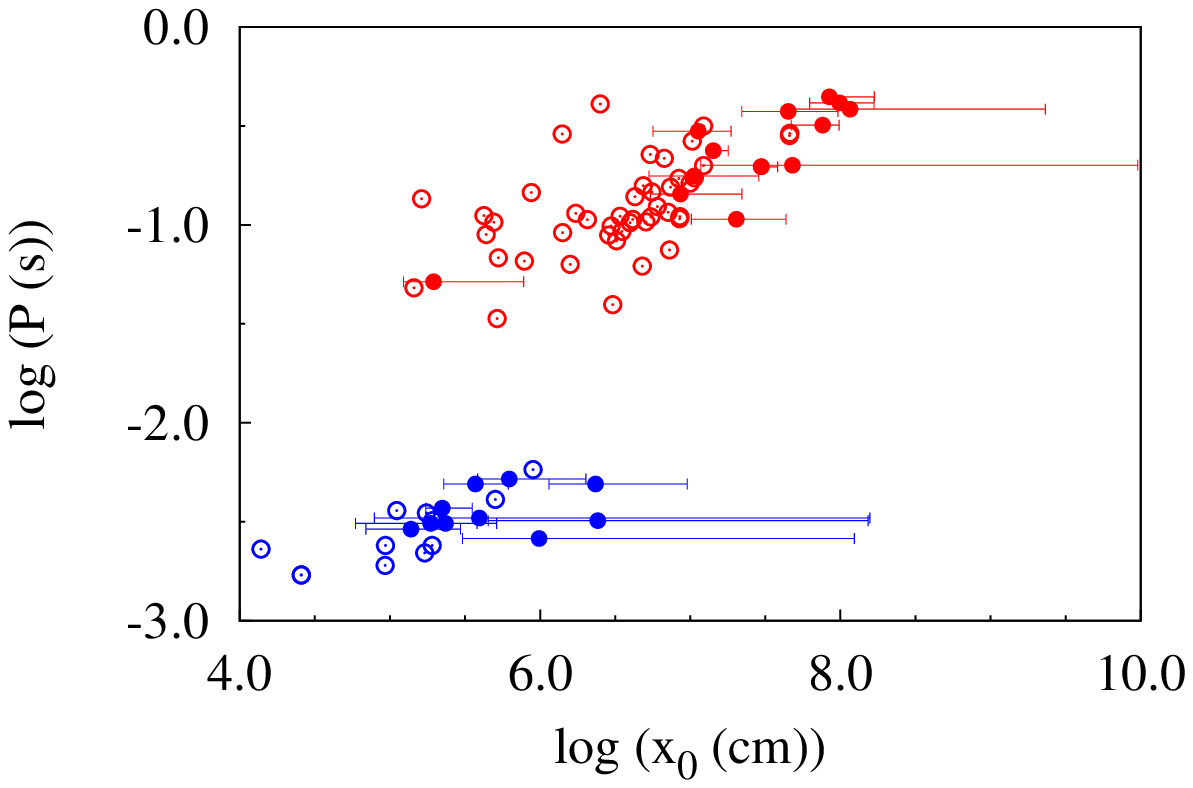}
\includegraphics[width=0.32\textwidth]{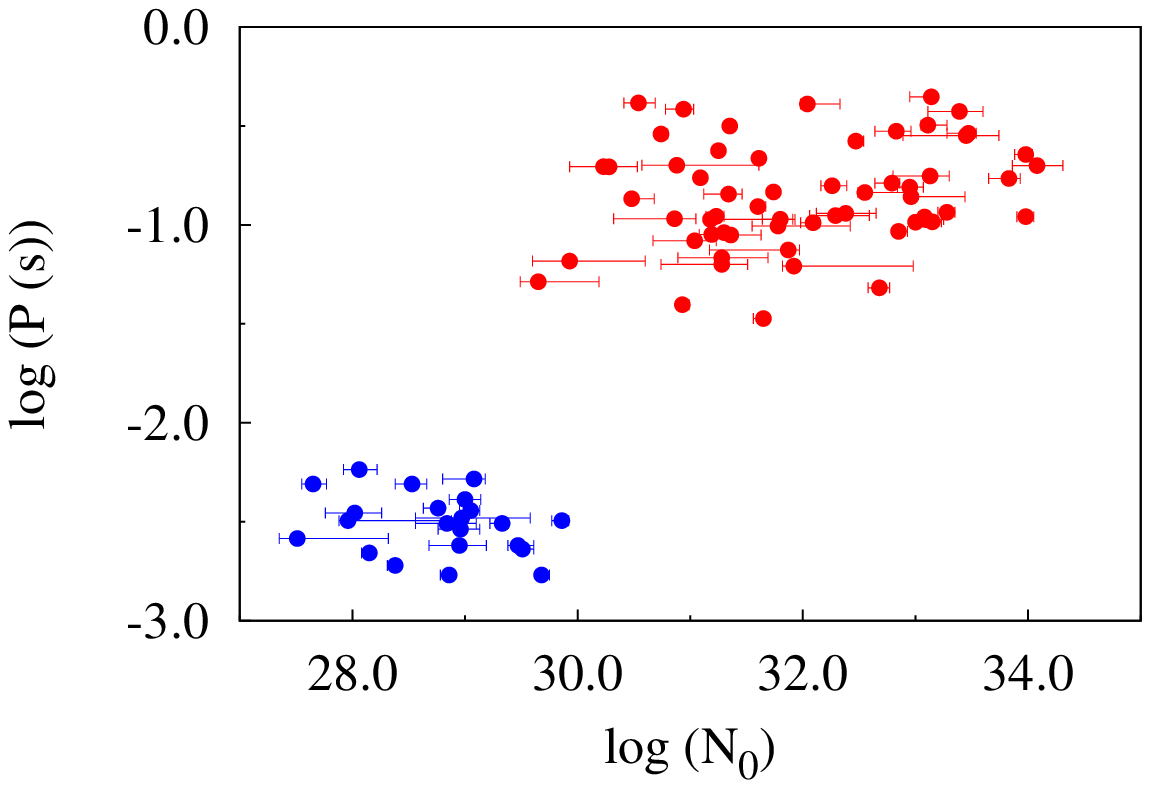}\\
\includegraphics[width=0.32\textwidth]{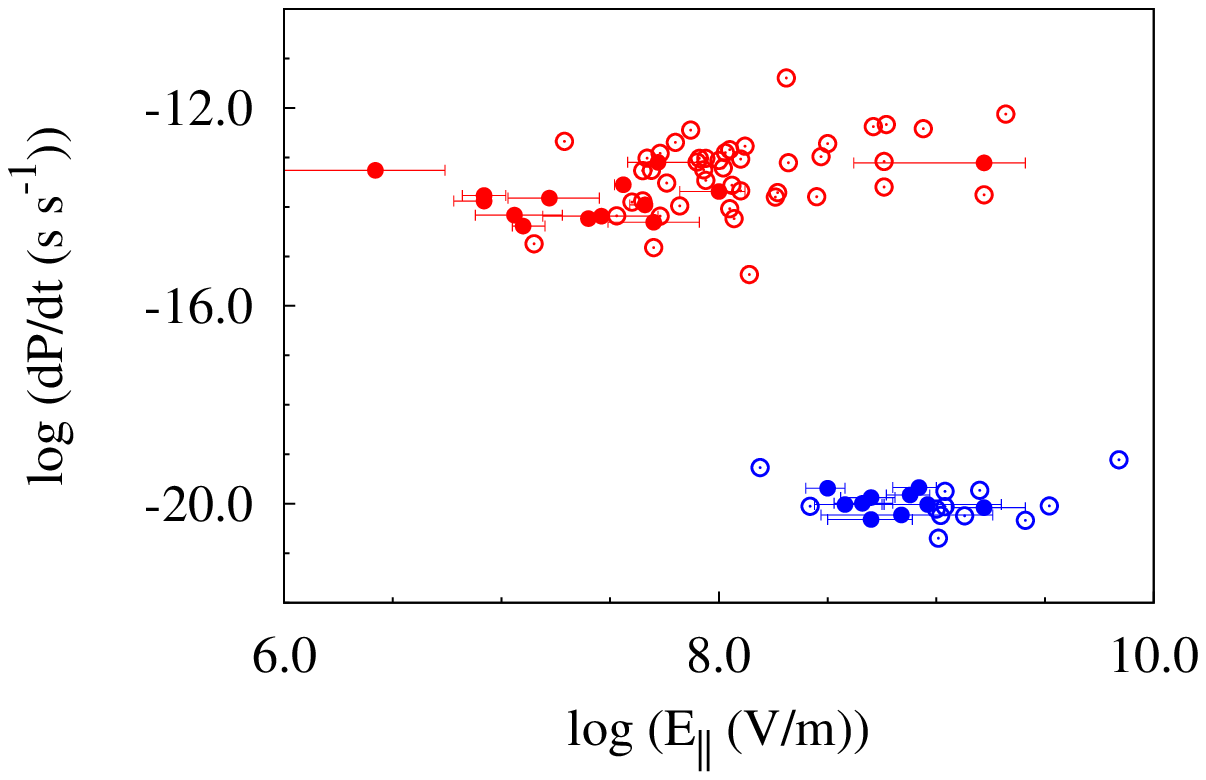}
\includegraphics[width=0.32\textwidth]{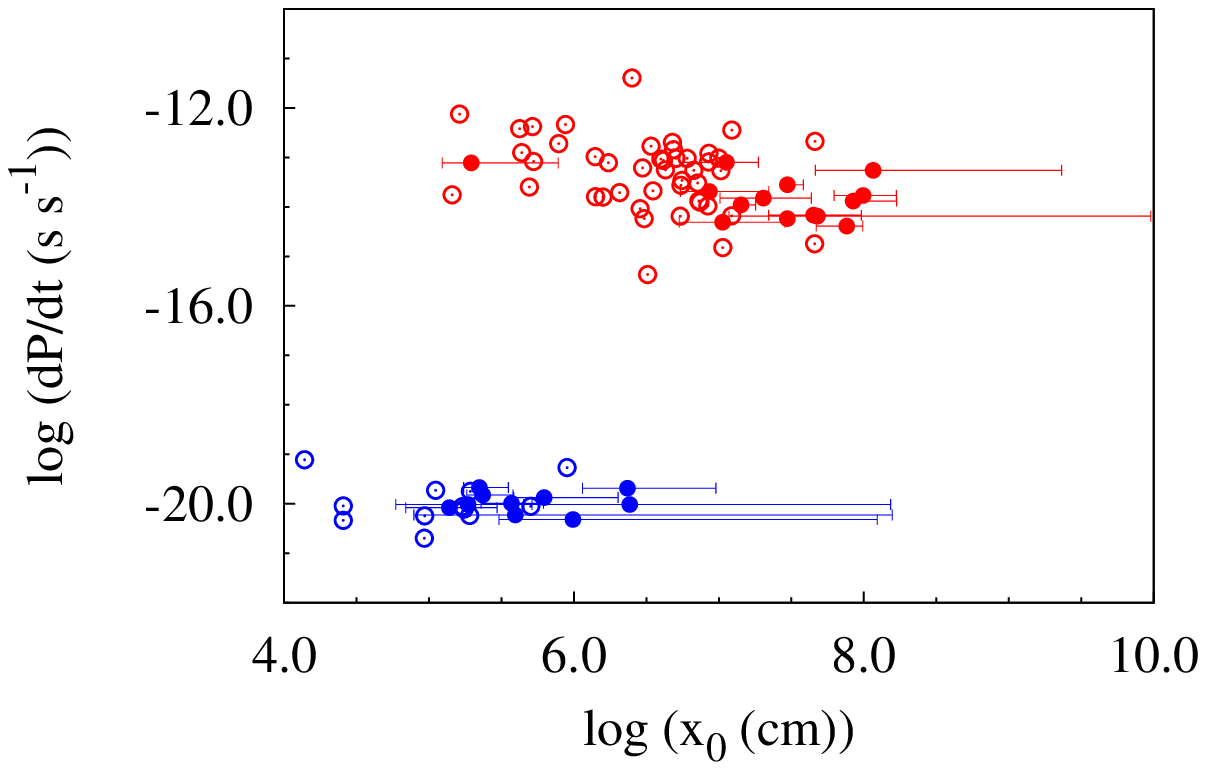}
\includegraphics[width=0.32\textwidth]{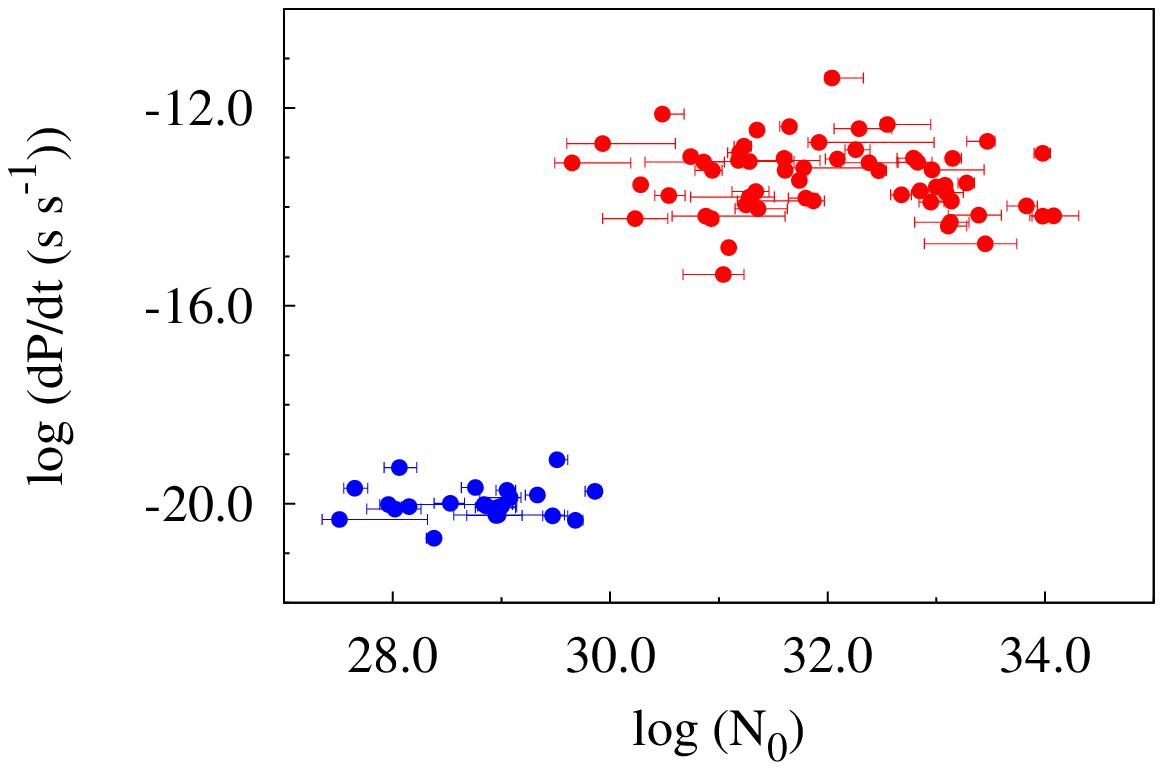}
\end{center}
\caption{Distribution of the three parameters fitted in our models as a function of $P$ (top panels) and $\dot P$ (bottom). Color and empty/filled codings are the same as Fig.~\ref{fig:corr_parameters}.}
\label{fig:sample-dist}
\end{figure*}

We note that, if the number of effective particles we infer were all born at the inner boundary of the gap, then the inferred flux of particle, $(N_0/x_0) c$, is much larger (between a factor of a few  up to 4 orders of magnitude) than the estimated Goldreich-Julian flux coming from the open field line region and hitting the polar cap (see e.g. \citep{zhang97}). Our larger number of particles could be an indication of the oversimplification of our model (e.g., particles are generated in different altitudes), or that the volume occupied by the gap is much larger than thought, or simply that cascading is very active (large multiplicity factor, similarly to pulsar wind nebulae models, e.g., \citealt{bucciantini11,torres14}). Only numerical simulations considering all the magnetosphere could provide detailed answers. How active can the cascading be before starting to significantly screen the accelerating electric field also needs simulations to tell.

Among the three parameters, we find that only $E_\parallel$ and $x_0$ maintain a significant anti-correlation, which is separately visible and continuing across both sub-samples (upper left-hand panel of Fig.~\ref{fig:corr_parameters}). This is confirmed by the Pearson coefficient shown in Table~\ref{tab:correlations}. 
Note that the correlation lines in Fig.~\ref{fig:corr_parameters} and all subsequent ones, as well as the correlation parameters in Table~\ref{tab:correlations}, are obtained considering the filled circles only, i.e., the pulsars for which all $E_\parallel$, $x_0$, and $N_0$ are well-constrained. The degenerate pulsar cases match these correlations very well in all cases. 

The global correlation in the upper left-hand panel of Fig.~\ref{fig:corr_parameters}
shows all the best-fitting solutions together. It should not be confused with the individual correlations found in the same plane for many pulsars in Figs.~\ref{fig:contours1}-\ref{fig:contours5}. Those contours cover different ranges of parameters from pulsar to pulsar (which is natural since each pulsar has different timing properties), but the linear parts of the contours have similar slopes, as already discussed in \S\ref{sec:errors} and in Appendix~\ref{app:fits}. 
Note that the global trend will be maintained despite the degeneracy in the values of parameters in many individual cases, since the latter would move the points along a similar diagonal line in the global plane $\log E_\parallel$--$\log x_0$ of Fig.~\ref{fig:corr_parameters}. In other words, the particular degeneracy in determining the specific pulsar best-fitting model does not affect the overall trend. 
Since we find no reason to expect any a-priori bias, this global trend is a novel result, which can highlight physical implications. 

The correlation also points to the fact that particles are accelerated to similar Lorentz factors, as the rightmost part of Fig. 1 (top-left panel) show, between $\sim 2 \times 10^{7}$ and $\sim 3 \times 10^{8}$. These high values can be interpreted as those needed to start pair cascades, which ultimately support the gap. In fact, the product $x_0 E_\parallel$ stays within 1 order of magnitude around $10^{13}$ V, a value that has been seen as the voltage that is limited by curvature radiation pair cascades, e.g., \citet{harding2002}.



The linear anti-correlation is found also when $x_0$ is rescaled by $R_{\rm lc}$, $E_\parallel \sim (x_0/R_{\rm lc})^{-1}$ (see top right-hand panel), but, in this case, the two sub-samples get separated (due to the very different values of $R_{\rm lc}$). This reinforce the idea that the same underlying microphysical process is at work, being barely dependent on the macrophysical lengthscale of the light cylinder. In other words, the observed spectra sistematically present features which can be explained by the local particle dynamics, more than by the global properties like the magnetospheric size (set by the spin period).

\subsection{Correlations with $P$ and $\dot{P}$}

Fig.~\ref{fig:sample-dist} shows the distribution of the three parameters fitted in our models as a function of $P$ and $\dot P$. The two populations are clearly distinguishable.

The distribution of the accelerating electric field is shown in the left panels. Taking the two sub-samples individually, $E_\parallel$ hints to have increasing values for smaller spin periods.
This is consistently reflected in the middle panel, where the hint is for a larger value of $x_0$ at larger periods.
If true,  such behavior can be understood as follows: the smaller is the period of the pulsar, the closer is the light cylinder to the surface of the neutron star (NS), and the smaller is the range in which the Lorentz factor needs to achieve values high enough as for particles to be able to emit $\gamma$-rays with $E\sim $ GeV.  The two sub-samples are separated, indicating that the period alone is not the main parameter determining the electric field (certain values of $E_\parallel$ can correspond to different values of $P$). No apparent correlation is seen between $E_\parallel$ or $x_0$ and $\dot{P}$. 
and no correlation between $N_0$ and the measured timing properties is seen.
A larger $x_0$ being obtained for a larger $P$ 
probably reflects too the pair cascade physics: a smaller $P$ have larger $E_\parallel$. In our models, the latter requires
a smaller $x_0$ to reach high voltages. Here there is the interesting possibility of an observational bias, since we have only {\it Fermi}-LAT survey data, the peak of the emission of all these pulsars 
are at $\sim$1--few GeV, 
a larger $x_0$ value for a similar $E_\parallel$ would reduce the peak energy to mostly observationally-unexplored regimes in the MeV--few hundreds of keV region.

\begin{figure*}
\begin{center}
\includegraphics[width=0.33\textwidth]{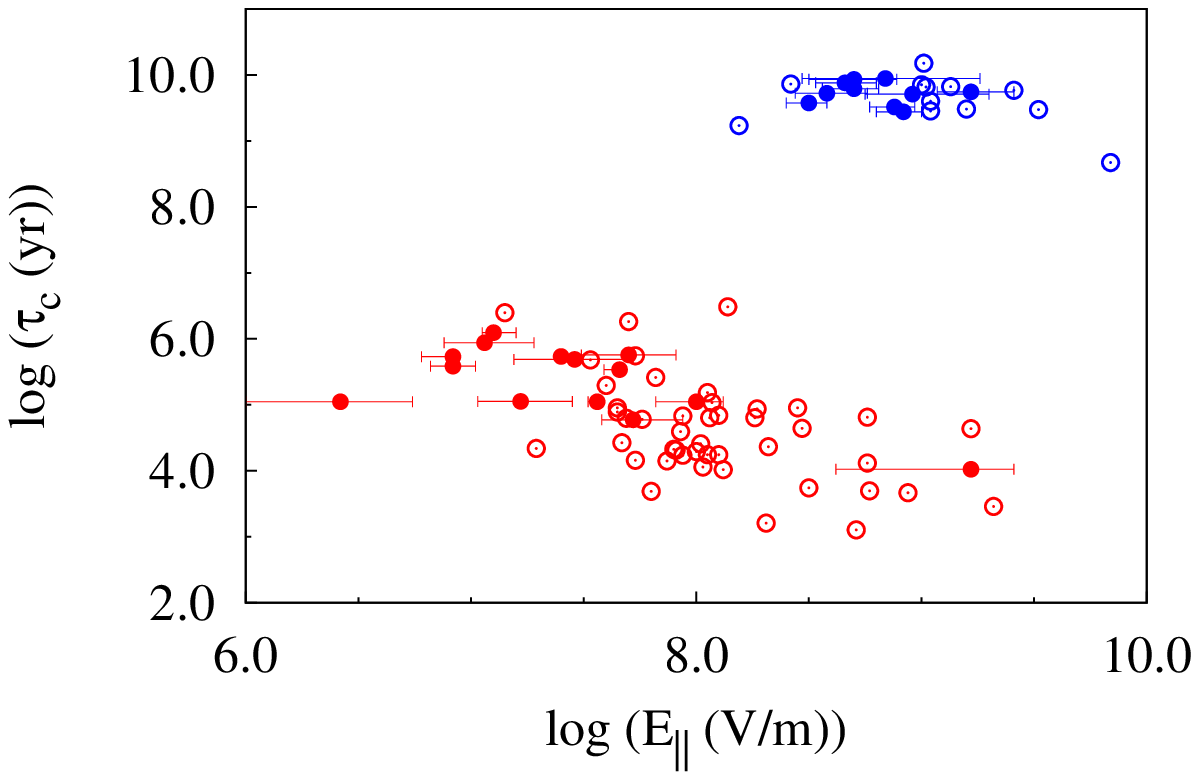}
\includegraphics[width=0.33\textwidth]{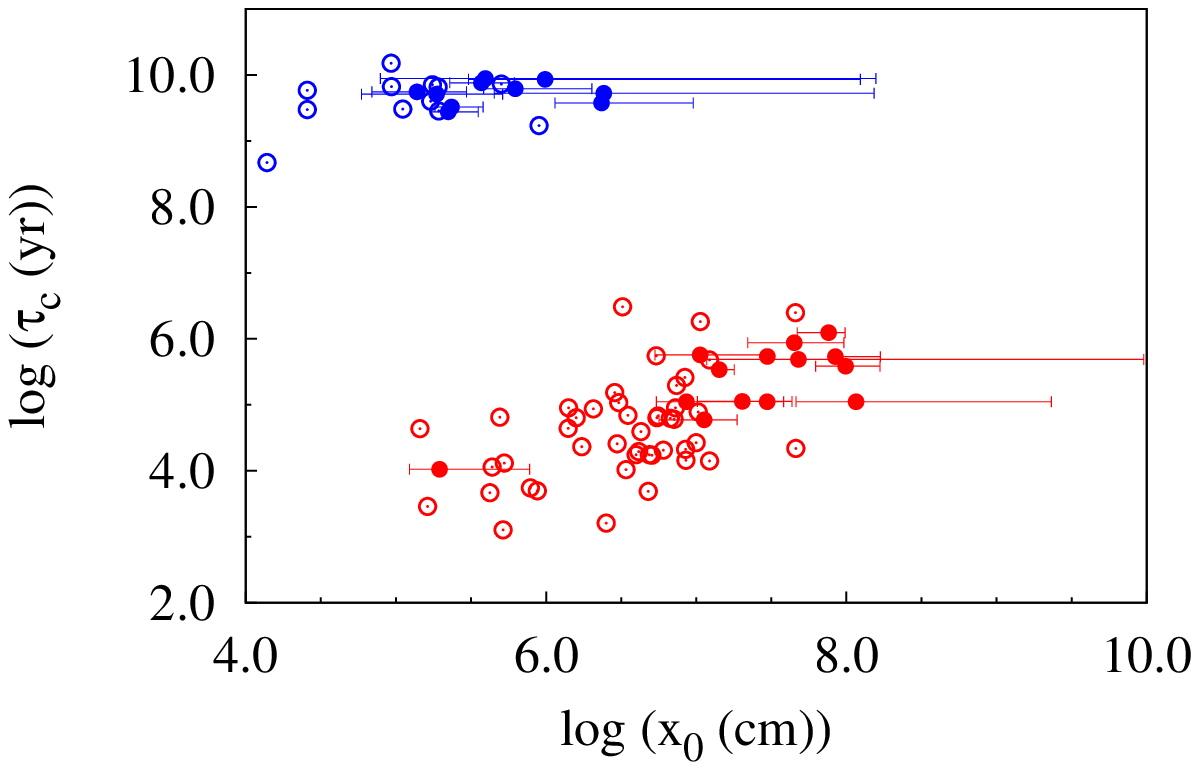}
\includegraphics[width=0.33\textwidth]{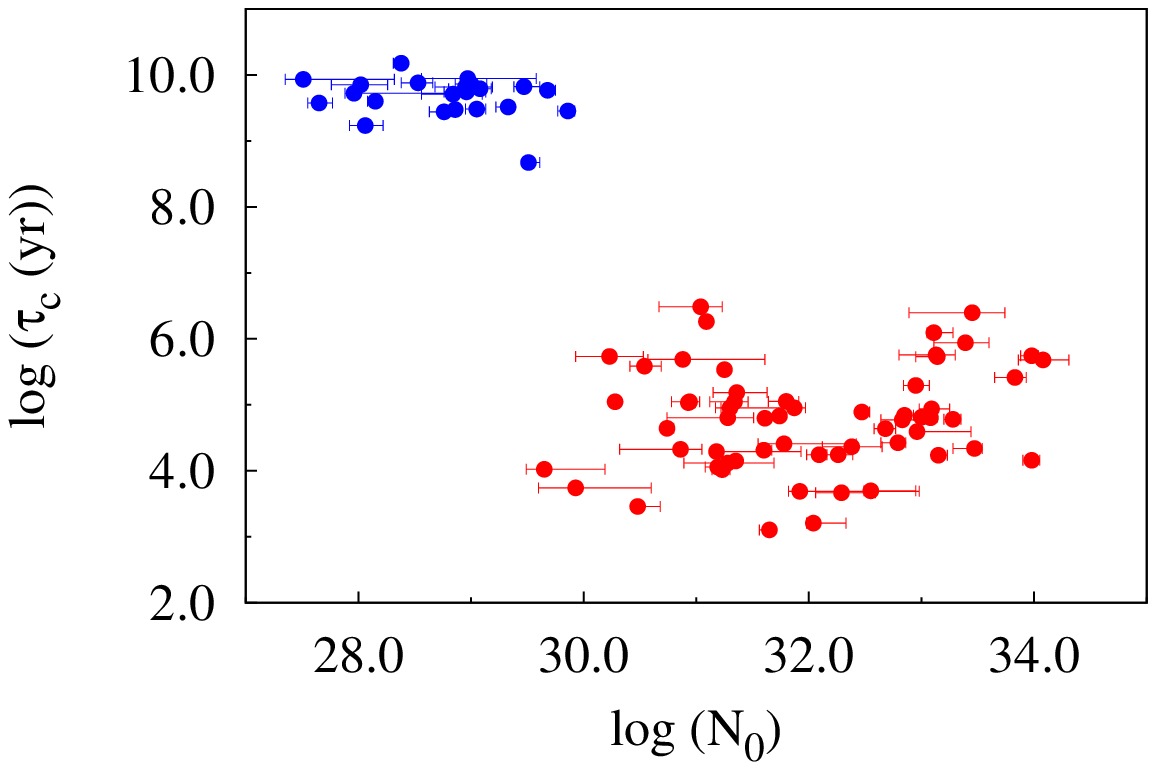}\\
\includegraphics[width=0.33\textwidth]{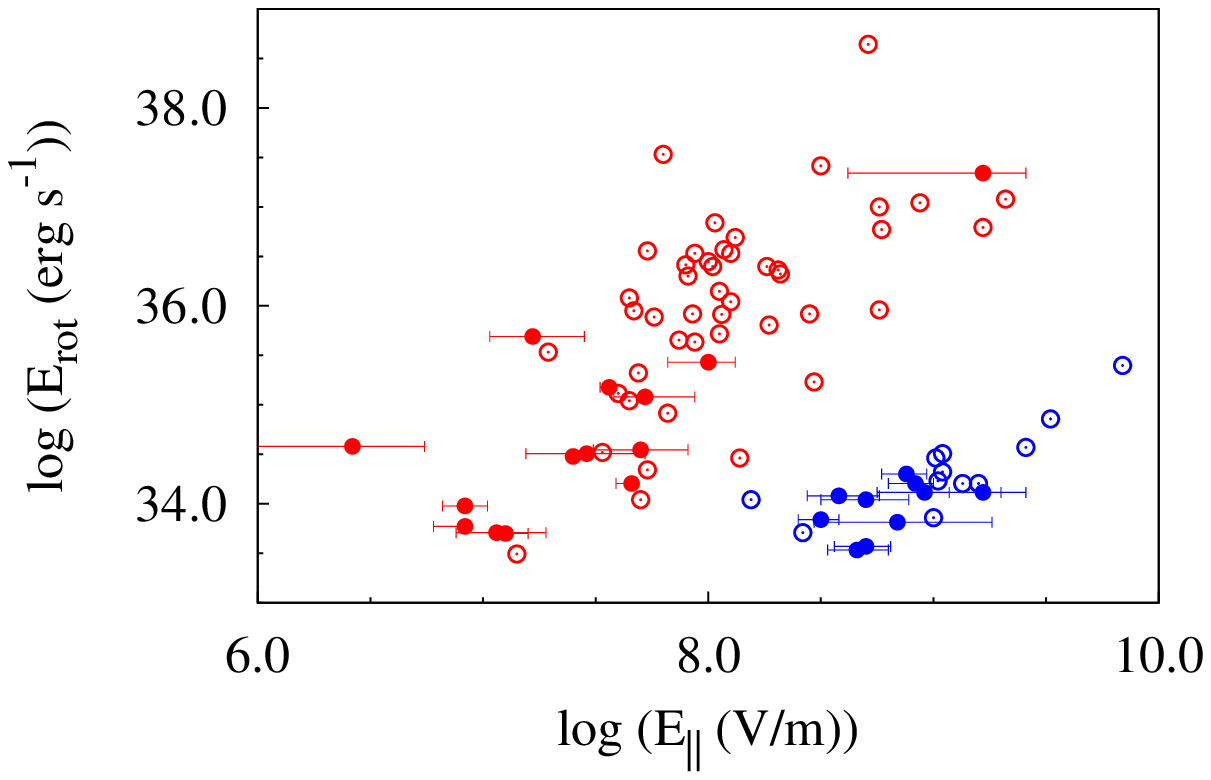}
\includegraphics[width=0.33\textwidth]{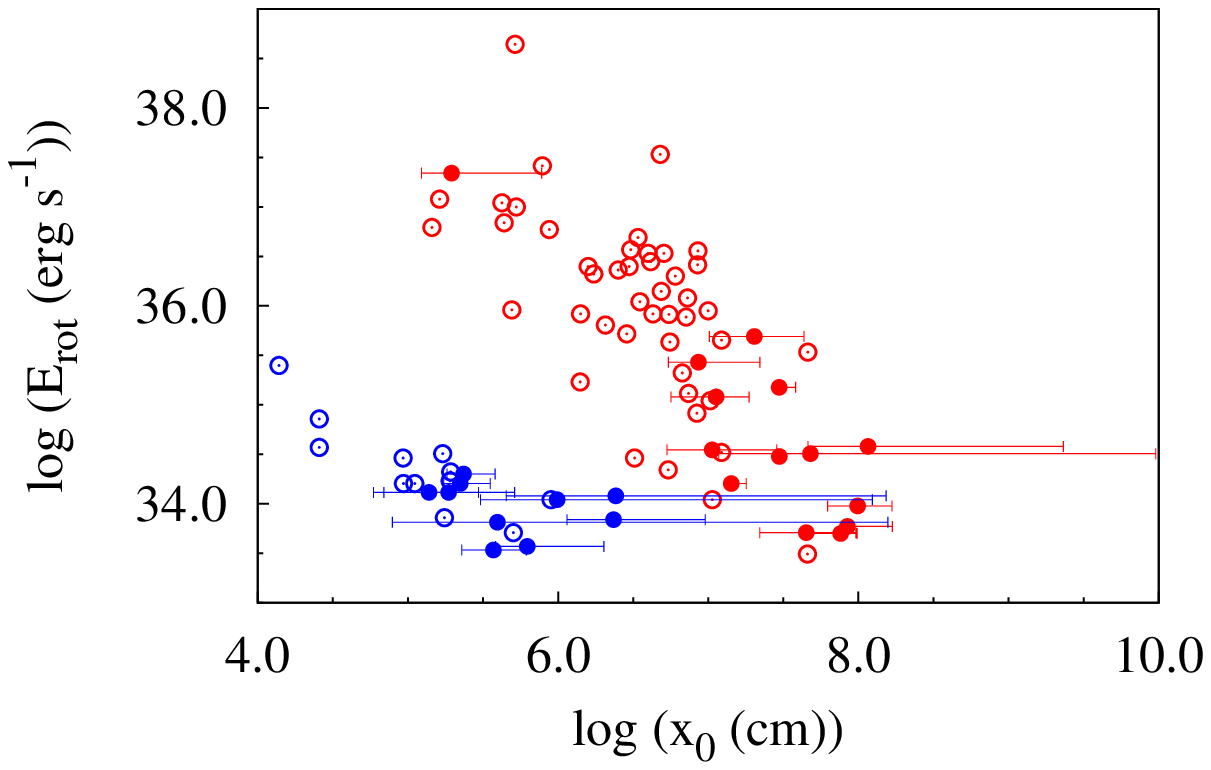}
\includegraphics[width=0.33\textwidth]{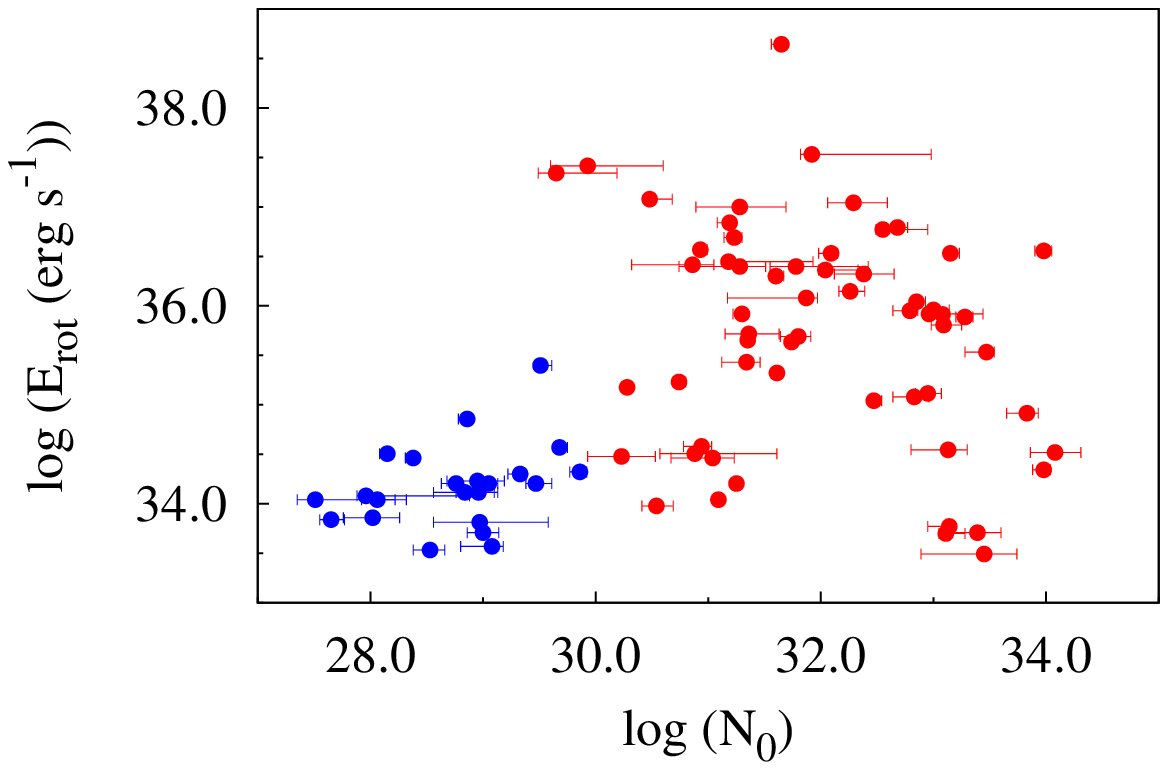}\\
\includegraphics[width=0.33\textwidth]{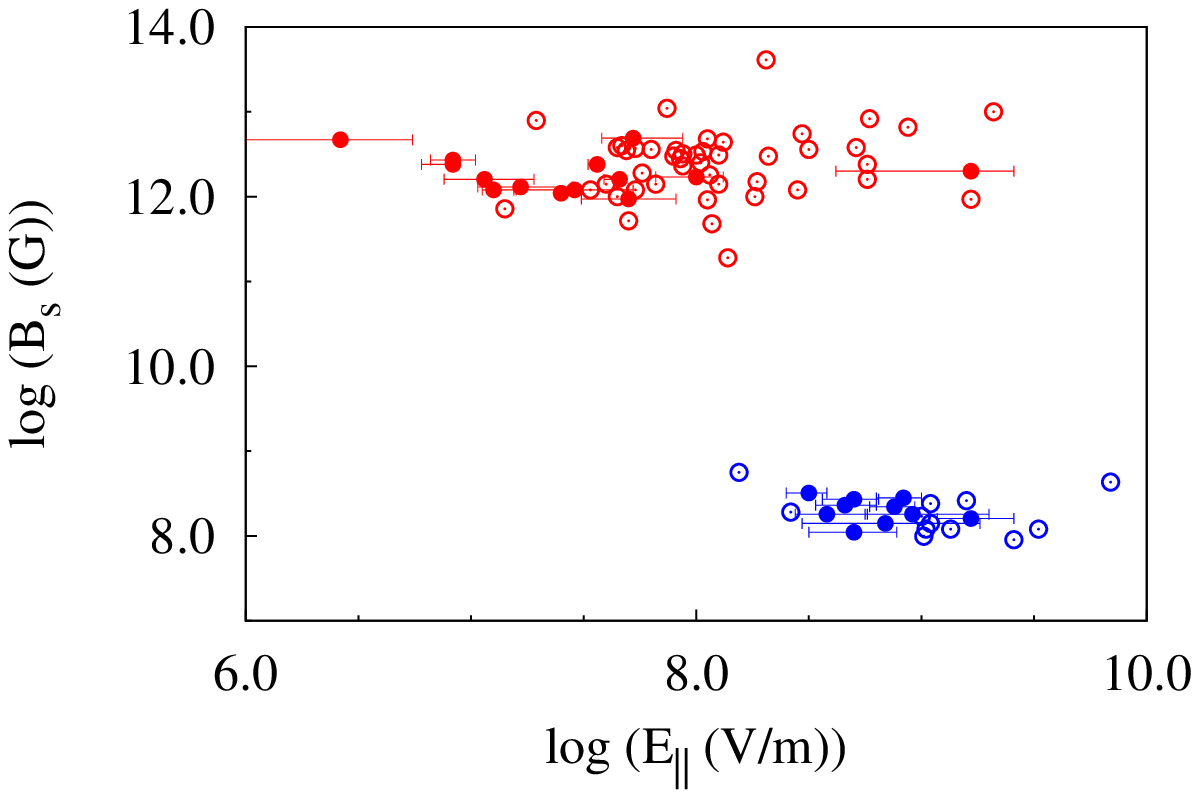}
\includegraphics[width=0.33\textwidth]{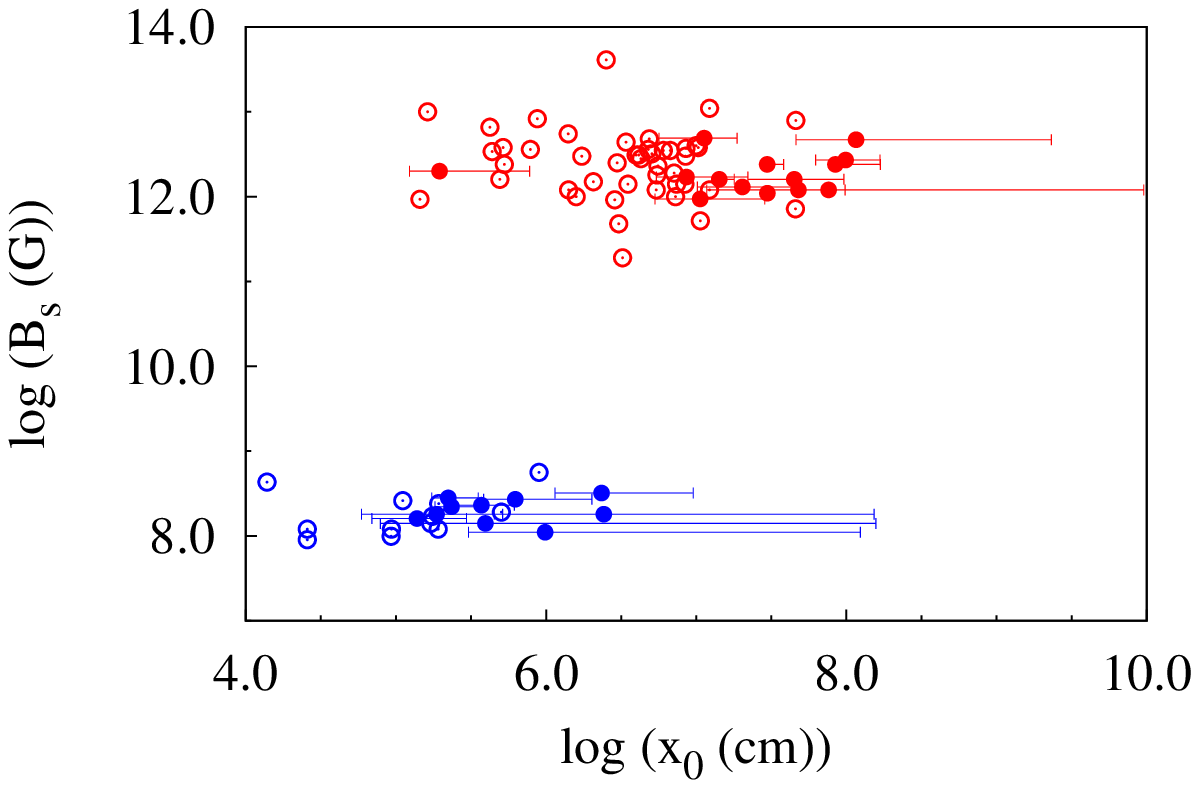}
\includegraphics[width=0.33\textwidth]{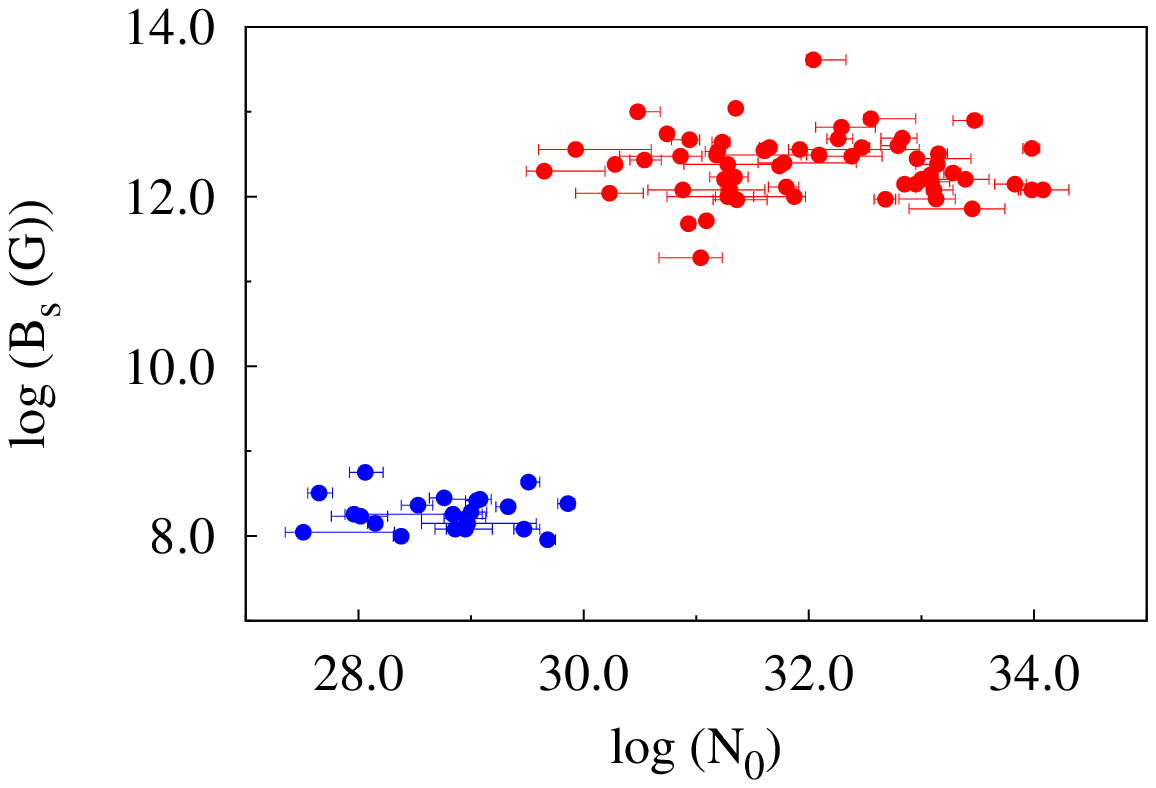}\\
\includegraphics[width=0.33\textwidth]{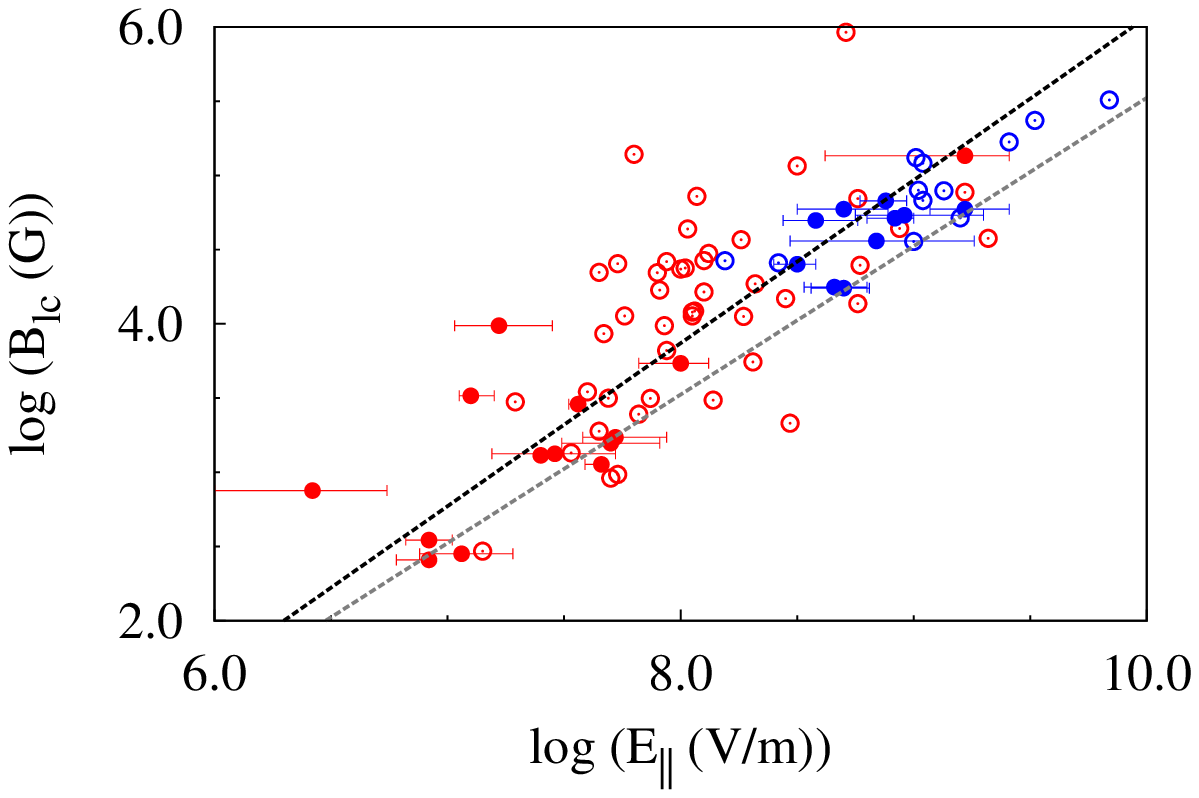}
\includegraphics[width=0.33\textwidth]{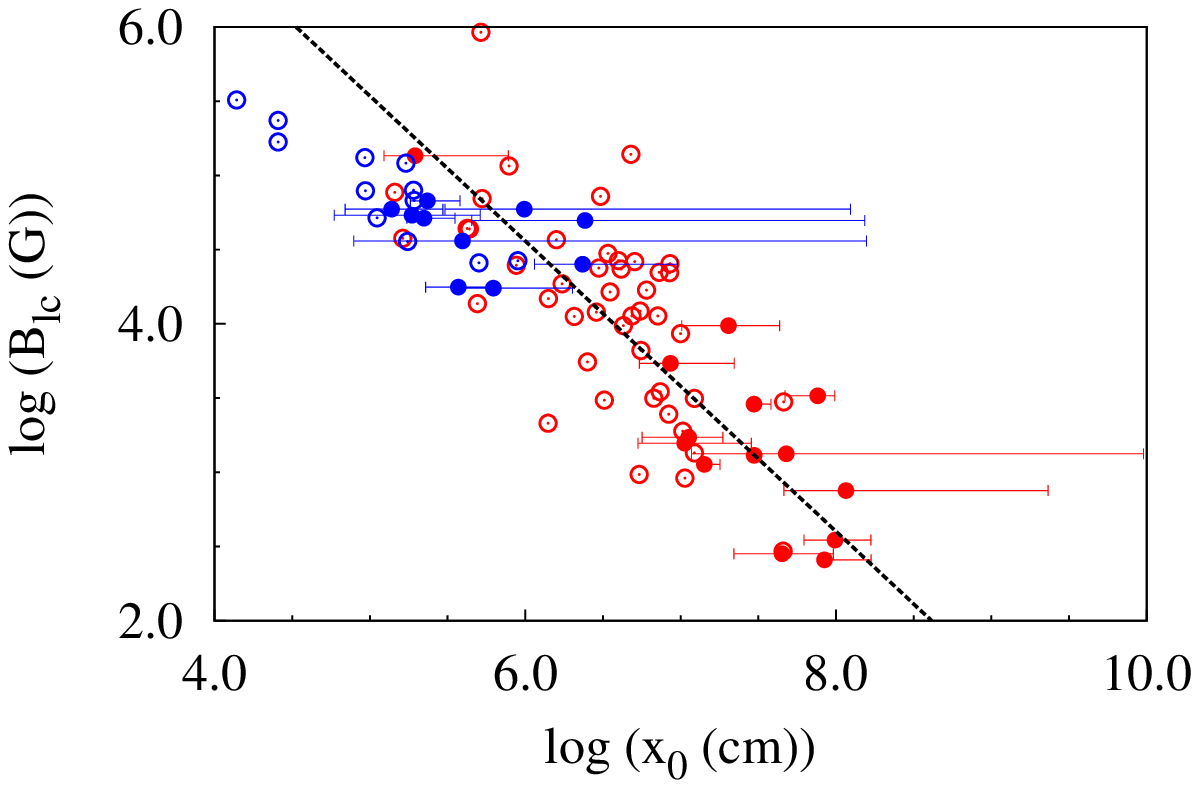}
\includegraphics[width=0.33\textwidth]{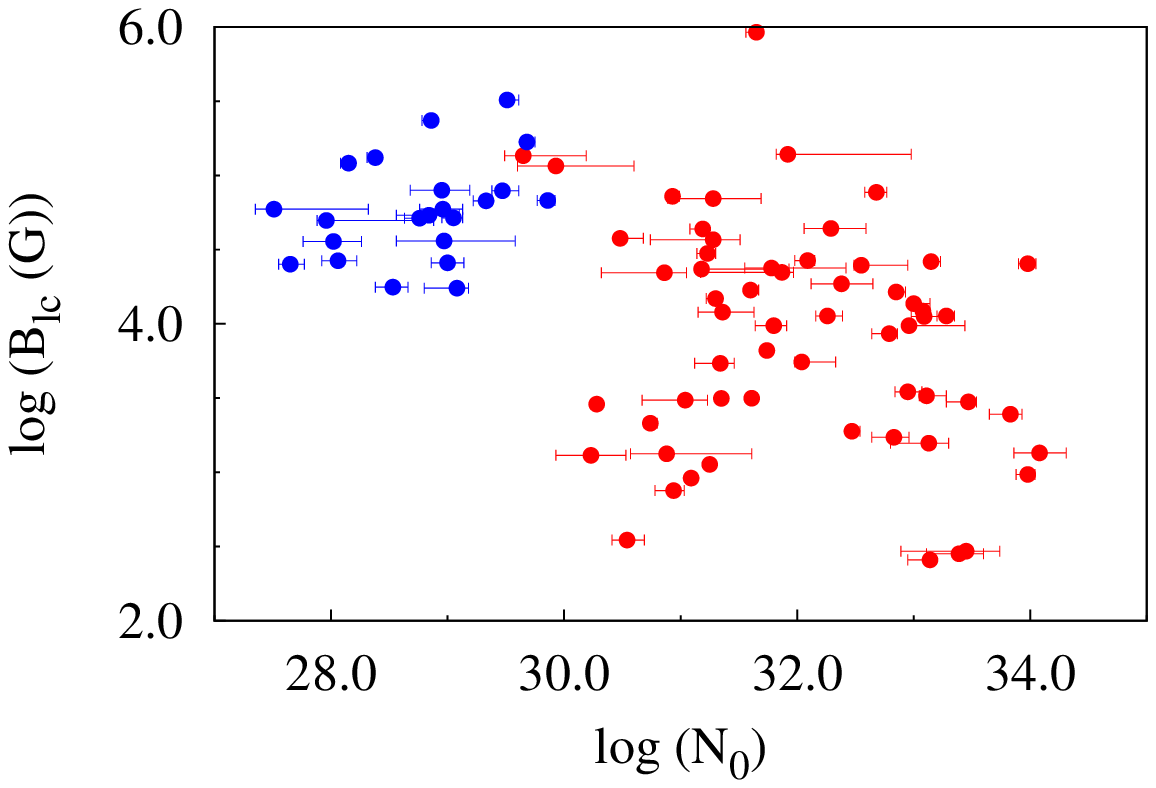}
\end{center}
\caption{Search for correlations among the timing-derived parameters of the pulsars: $\tau$, $E_{\rm rot}$, $B_s$, and $B_{\rm lc}$ with the three best-fitting SC model parameters. Color and empty/filled codings are the same as Fig.~\ref{fig:corr_parameters}. The bold dashed lines represent the fits, when the correlation is significant. In the bottom left panel, the light dashed line identifies $E_\parallel = B_{\rm lc}$. The magnetic field in the accelerating region is even larger than this, and can be computed according to its defition in Table 1. See text for a discussion.}
\label{fig:corr_timing}
\end{figure*}

\subsection{Correlations with timing-derived properties}\label{sec:correlations_timing_derived}

Since YPs and MSPs span very different ranges of values of $P$ and $\dot{P}$, it is natural to search for possible correlations between the best-fitting parameters of the SC
model and the timing properties of the pulsars. The panels of Fig.~\ref{fig:corr_timing} explore these possible correlations against the characteristic age, the rotational energy loss, the surface magnetic field, and the magnetic field at the light cylinder, defined as 
\begin{eqnarray}
 && \tau_c = \frac{P}{2 \dot P }~, \label{eq:tau_c}\\
 && \dot{E}_{\rm rot} = 3.9 \times 10^{46} P^{-3}  \dot{P} \;\;\; {\rm erg} \; {\rm s^{-1}}, \label{eq:erot} \\
 && B_s = 6.4 \times 10^{19} ~ P^{1/2}\dot{P}^{1/2} \;\;\; {\rm G}, \label{eq:bs} \\
 && B_{\rm lc} = 5.9\times 10^8 P^{-5/2} \dot{P}^{1/2} \;\;\; {\rm G}, \label{eq:blc}
\end{eqnarray}
where $P$ is implicitly expressed in seconds and $\dot P$ in s~s$^{-1}$. Magnetic field values are estimated according to the widely-used vacuum spin-down formula, and assuming standard values of moment of inertia $I=10^{45}$ g~cm$^2$, and NS radius $R_\star=10$ km. Since their actual physical values could vary, an intrinsic uncertainty by a factor of a few is expected on the inferred values $B_s$ and $B_{\rm lc}$. Below we discuss these quantities and their correlations. All the estimates above are just different combination of powers of $P$ and $\dot{P}$. Other quantities can be construed, like, e.g., the surface electric voltage $\Delta V = ({B_s 4\pi^2 R_\star^3})/({2 c P^2}) $ but this, being proportional to $P^{-3/2}\dot{P}^{1/2} $, is directly proportional to $\dot E_{\rm rot}^{0.5}$. Thus, any correlation found with $\dot{E}_{\rm rot}$ would reflect in a similar one with the voltage. When we find a correlation of note, we list in Table~\ref{tab:correlations} the parameters of a linear fitting, together with the Pearson test coefficient.

\subsubsection{Spin-down age $\tau_c$}

The range of values of $\tau_c$, Eq.~(\ref{eq:tau_c}), spans more than seven orders of magnitude, well separating the two sub-samples: $\tau_c \sim 10^3$-$10^6$ yr for the YPs, $\tau_c \sim 10^9$-$10^{10}$ yr for the MSPs. There is no visible correlation between $\tau_c$ and any of the three model parameters, in any of the sub-samples. We also recall that $\tau_c$ is, more properly said, the spin-down timescale, and it can be a fair estimate of the real age only if the spin-down torque mechanism has not suffered variations with time (e.g., evolution of the large-scale dipolar magnetic field, accretion phases...), and the present spin period is much larger than the period at birth. Note that the latter assumption could not hold for some fast-spinning YPs and MSPs, which likely underwent accretion phases with different spin-down torque mechanisms. Thus, the lack of correlation is not a surprise.

\subsubsection{Spin-down energy loss $\dot E_{\rm rot}$}

The second row in Fig.~\ref{fig:corr_timing} shows the search for correlations of 
the model parameters with the spin-down energy loss, Eq.~(\ref{eq:erot}). There appears a separation of the two sub-samples of pulsars, due to the discussed differences in $E_\parallel$, and to the fact that MSPs have lower spin-down power than most YPs. Values of $\dot E_{\rm rot}$ span more than five orders of magnitude, with Crab being by far the most energetic pulsar of the sample. Hints that for both populations, the larger is the spin-down, the larger is $E_\parallel$ might be argued, but is not at the level of having the Pearson coefficient $r>0.85$ for the well-constrained pulsars in both sub-samples.

\begin{figure*}
\begin{center}
\includegraphics[width=0.33\textwidth]{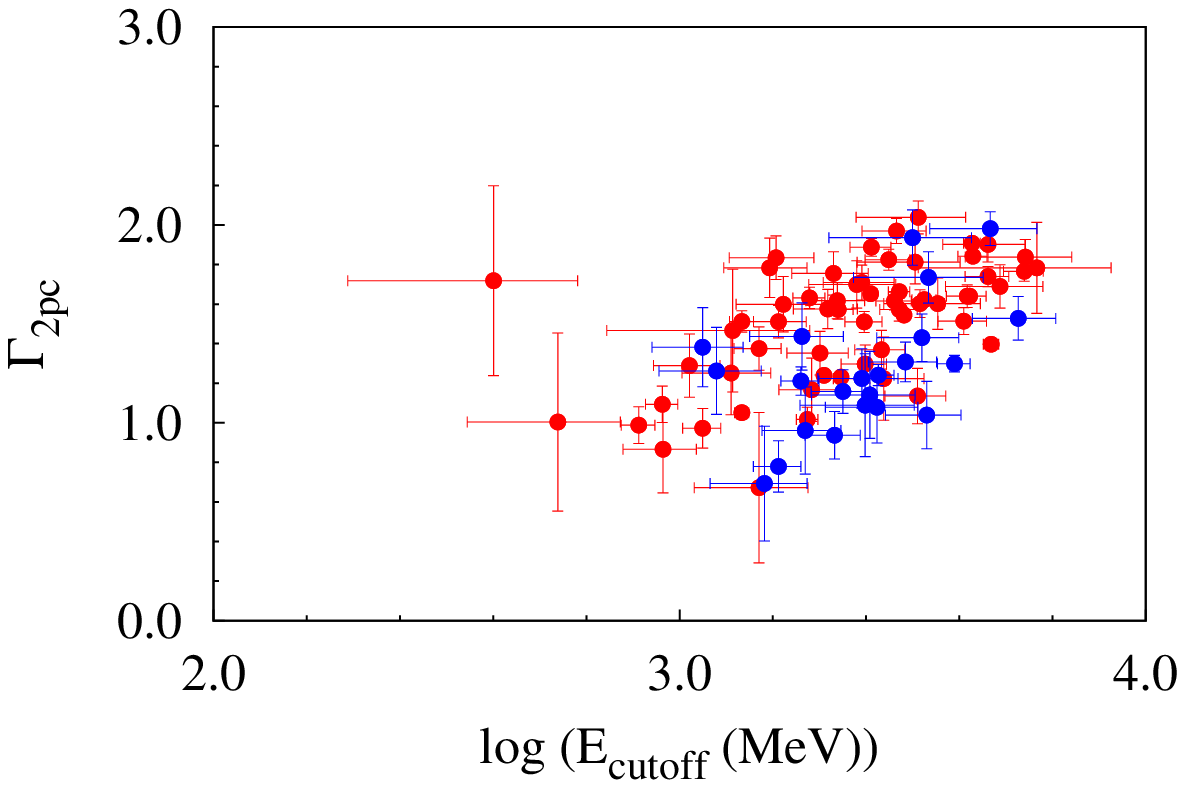}
\includegraphics[width=0.33\textwidth]{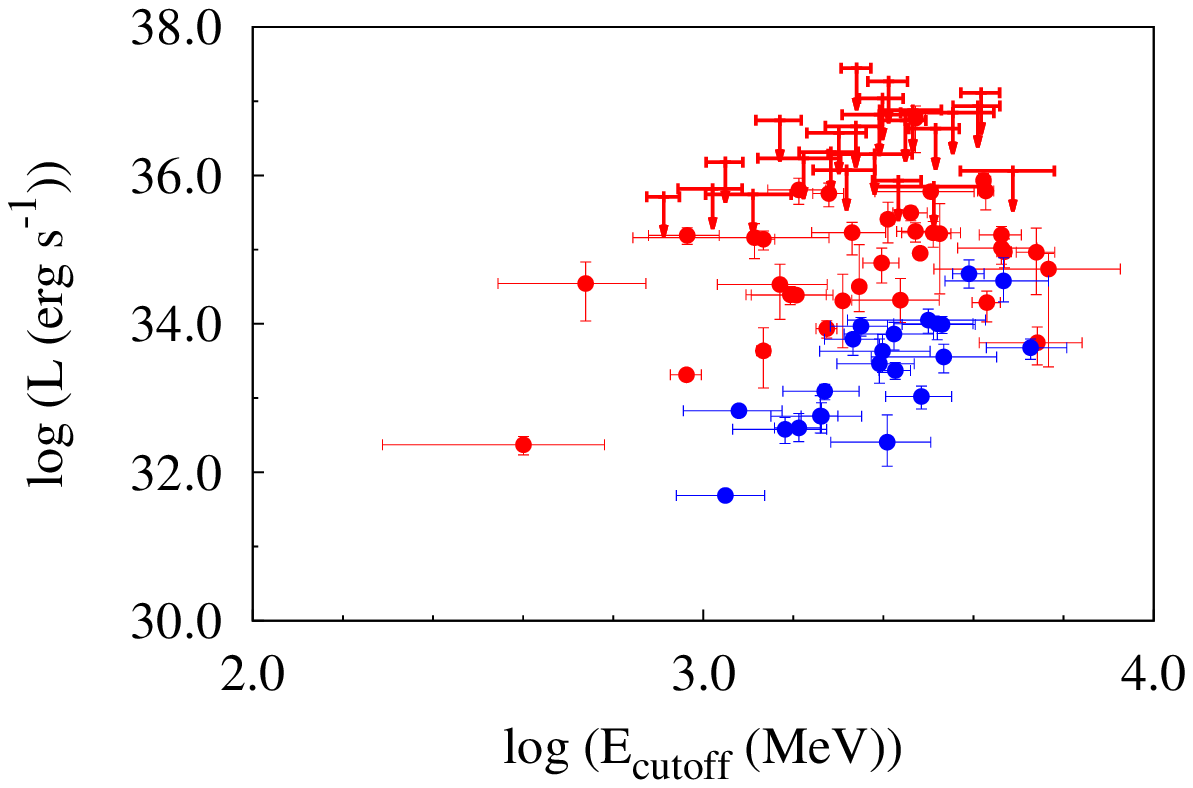}
\includegraphics[width=0.33\textwidth]{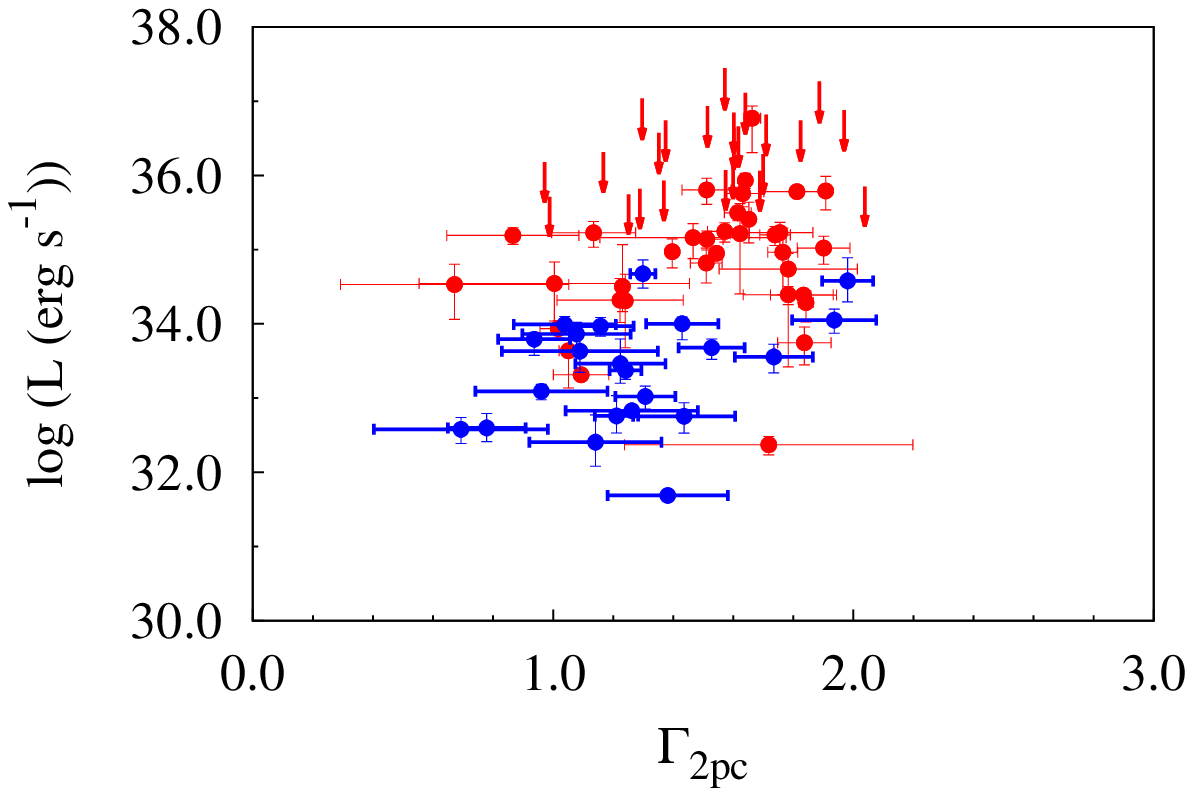}
\end{center}
\caption{Search for correlations between the three parameters of the best-fitting PLEC1 models, taken from the 2PC. Color coding is the same as Fig.~\ref{fig:corr_parameters}.}
\label{fig:corr_parameters_PLEC}
\end{figure*}

\subsubsection{Surface magnetic field $B_s$}

The range of values of $B_s$, Eq.~(\ref{eq:bs}), covers almost six orders of magnitude, again keeping the two sub-samples well separated (third row in Fig.~\ref{fig:corr_timing}). MSPs are expected to be less magnetic, since they are much older than YPs, and their magnetic fields have been strongly dissipated. We find no correlation between $B_s$ and any of the model parameters. This is expected if $\gamma$-ray emission is coming from the outer magnetosphere.

\subsubsection{Magnetic field at light cylinder $B_{\rm lc}$}

The canonical value of $B_{\rm lc}$, Eq.~(\ref{eq:blc}), defined simply by rescaling $B_s$, with $(R_\star/r)^3$, has to be taken as a rough estimate of the magnetic field in the gap. In fact, our model assumes a softer magnetic field decay (described by $B=B_s(R_\star/x)^b$, with $b=2.5$, \citealt{paper1} and Table \ref{tab:parameters}), which is qualitatively more consistent with the numerical simulations of magnetospheres of rotating NSs. Nevertheless, the classical estimate $B_{\rm lc}$ still gives a relative idea of the magnetic field far from the surface across the sample, and we checked that all trends found with $B_{\rm lc}$ would be found also assuming the scaling $(R_\star/r)^{2.5}$. Moreover, in our models the spectra are quite insensitive to the exact value of $b$. 

Despite MSPs are less magnetized (smaller $B_s$), they have on average larger values of light cylinder field ($B_{\rm lc}\sim 10^4$-$10^5$ G) as compared to YPs ($B_{\rm lc}\sim 10^2$-$10^5$ G), since their light cylinder is much closer, located at a few stellar radii. In the fourth row of Fig.~\ref{fig:corr_timing} we explore correlations between the model parameters and $B_{\rm lc}$. The uppermost outlier (in $B_{\rm lc}$) in these plots is the Crab pulsar. 

We find that the values of $B_{\rm lc}$ correlate with the accelerating field $E_\parallel$ resulting from our model fits. More interestingly, YPs and MSPs form an indistinct continuous in this plot (see Fig.~\ref{fig:corr_timing}). The inferred relation between $B_{\rm lc}$ and $ E_\parallel$ (quoted in Table~\ref{tab:correlations})  is very close to a direct proportionality, $E_\parallel \propto B_{\rm lc}$. This correlation, together with the fact that $E_\parallel$ does not correlate with $B_s$ is an evidence that the accelerating gap locates in the outer magnetosphere. However, note that since the correlation would appear also further away than the light cylinder, we cannot say whether the accelerating location is close to the light cylinder (as in the classical OG models), or farther away (as in the striped wind models). 

Looking at correlations with the other two parameters, we consistently find a correlation such that an increasing value of $x_0$ implies a decreasing value of $B_{\rm lc}$. No correlation of $B_{\rm lc}$ with $N_0$ is visible.

In Fig. \ref{fig:corr_timing}, bottom left panel, we also show the line corresponding to having $B_{\rm lc}=E_\parallel$. 
%
%
We have discussed our model compared with those of the outer gap magnetosphere in \cite{paper1} (see especially the section 3.3 and Table 2 therein), and \cite{paper2}. We discussed there in detail the many underlying assumptions of the outer gap, as well as the caveats in the computations of their main features. Among them, those applying to the formula quoted for $E_\parallel$, which is used to define a death line separating gamma-ray emitting pulsars, $E_\parallel \propto B_{\rm lc} f^2$, where $f$ is the fractional gap width, and $f<1$. We concluded there that the specific values one can obtain for $f$ from this formula has doubtful physical content. $B_{\rm lc}$ is only an estimate of the value of the local magnetic field at light cylinder, assuming a dependence of a static dipole. However, neither the real radial dependence of the field nor the exact location of the acceleration place are known. In the outer gap models, in addition, there are a number of assumptions related with the photon fields that participate in the bootstrap to sustain the gap itself, and some problems related with the computation itself (cross sections, resonant scattering, see the references quoted above for details), and all of that is carried over to the estimations of $f$. It is nevertheless interesting that the $B_{\rm lc}$ vs. $E_\parallel$ fit we provide in Table \ref{tab:correlations} and Fig. \ref{fig:corr_timing} is comparable to (although not strictly compatible with) a slope of 1. However, the fact that the same correlation works for both young and millisecond pulsars and that it does not require enhancing the magnetic field using multi-polar components near the surface of the star as was invoked in outer gap models (e.g., \citet{zhang03}) are a new outcome of this study, even assuming physical content to the outer gap computation of $f$.

\subsection{Correlations with PLEC1 parameters}\label{sec:correlations_plec}

In this section we study whether our best-fitting SC model parameters present any correlation with the PLEC1 phenomenological fits of the 2PC, as defined by
\begin{equation}\label{eq:cutoff}
 \frac{dN}{dE} = K_{\rm pc} E^{-\Gamma_{\rm 2pc}} \exp\left(-\frac{E}{E_{\rm cutoff}}\right)~.
\end{equation}
Thus, here we focus on $\Gamma_{\rm 2pc}$, $E_{\rm cutoff}$, and the 0.1--100 GeV luminosity $L$, related to the normalization $K_{\rm pc}$ and to the distance (see Appendix~\ref{app:fits} for details). We recall that the former parameters, different from the set $x_0$, $N_0$ and $E_\parallel$, come from a likelihood fit to the $\gamma$-ray data, and are not associated to the predictions of any concrete theoretical model.

\begin{figure*}
\begin{center}
\includegraphics[width=0.33\textwidth]{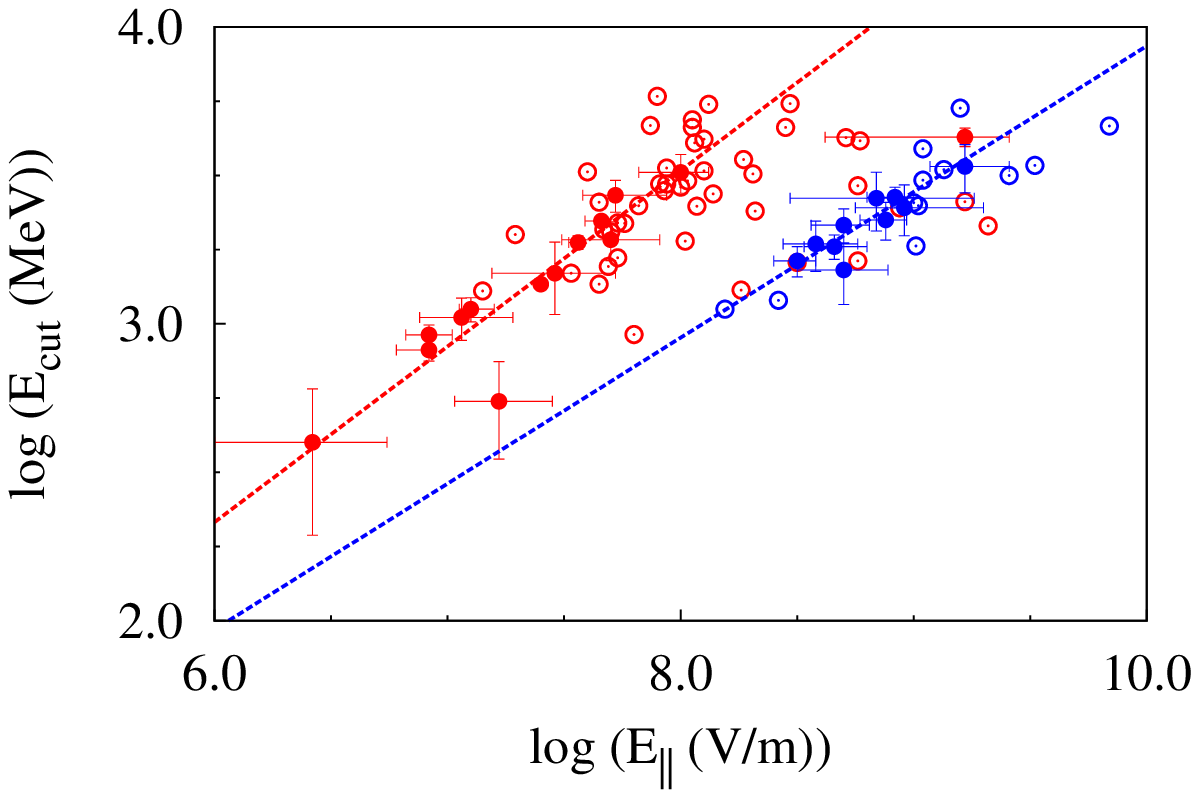}
\includegraphics[width=0.33\textwidth]{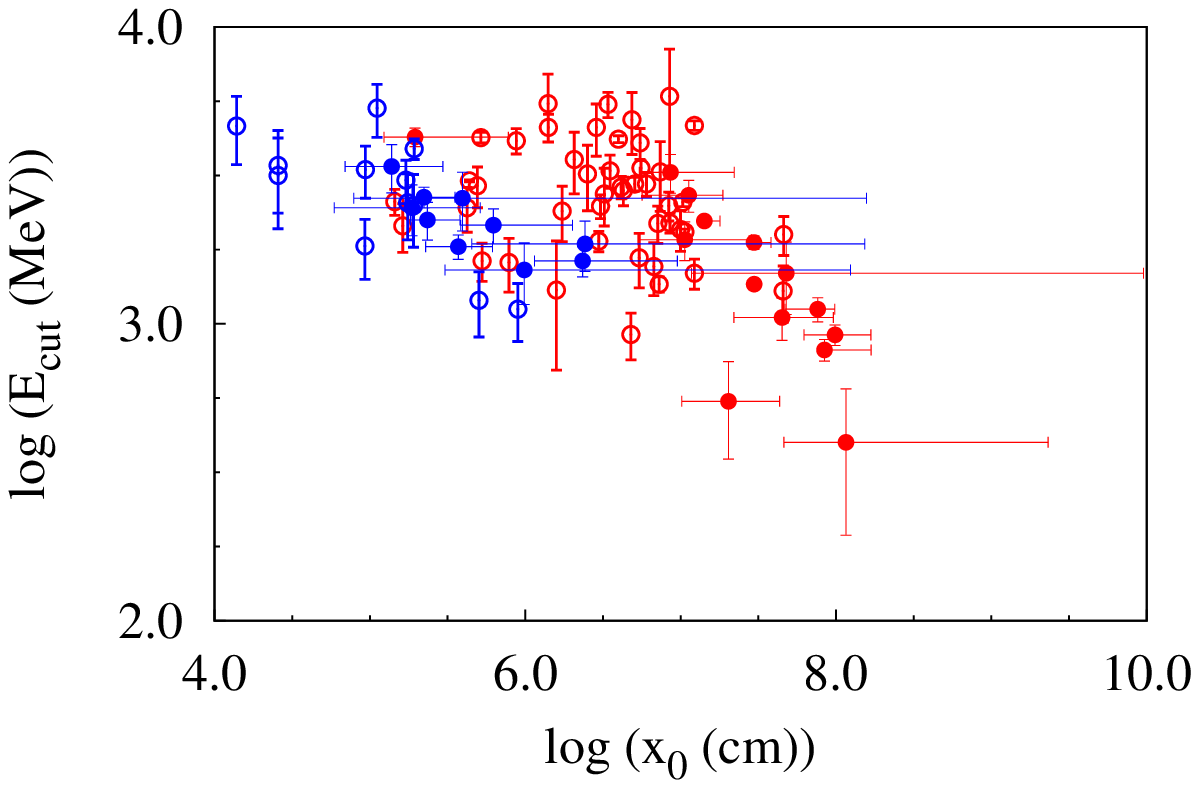}
\includegraphics[width=0.33\textwidth]{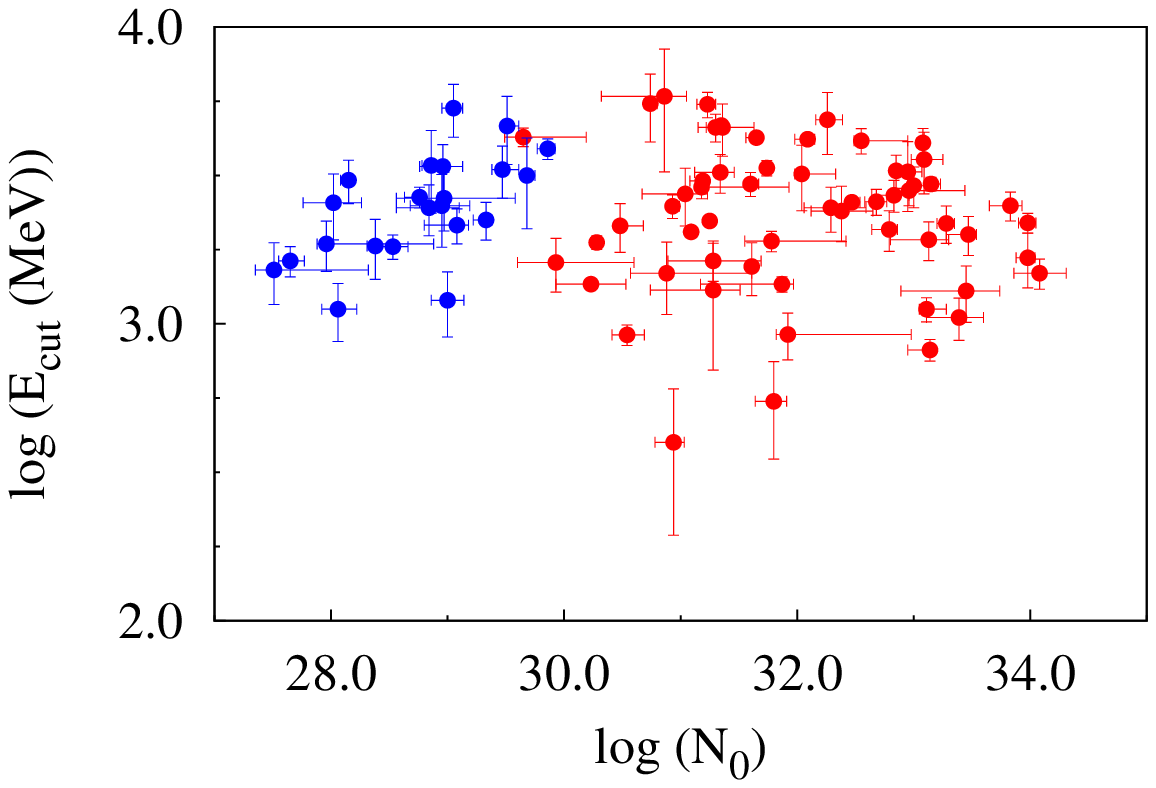}\\
\includegraphics[width=0.33\textwidth]{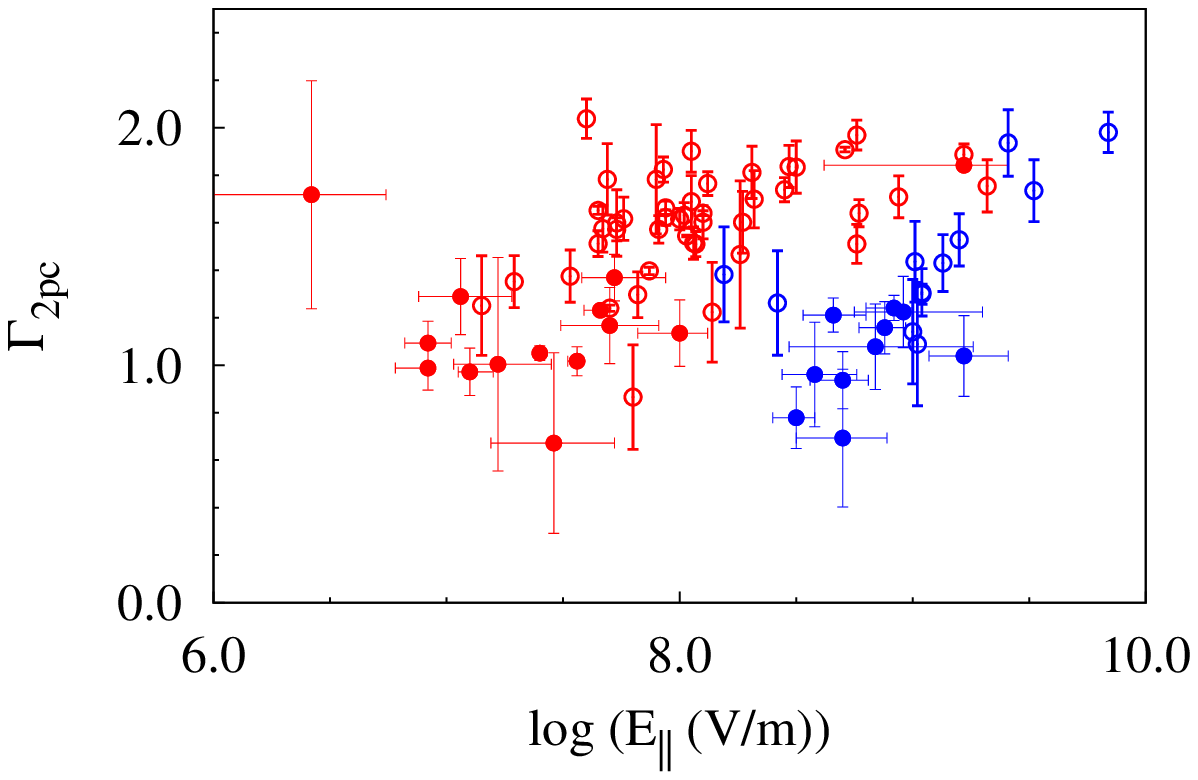}
\includegraphics[width=0.33\textwidth]{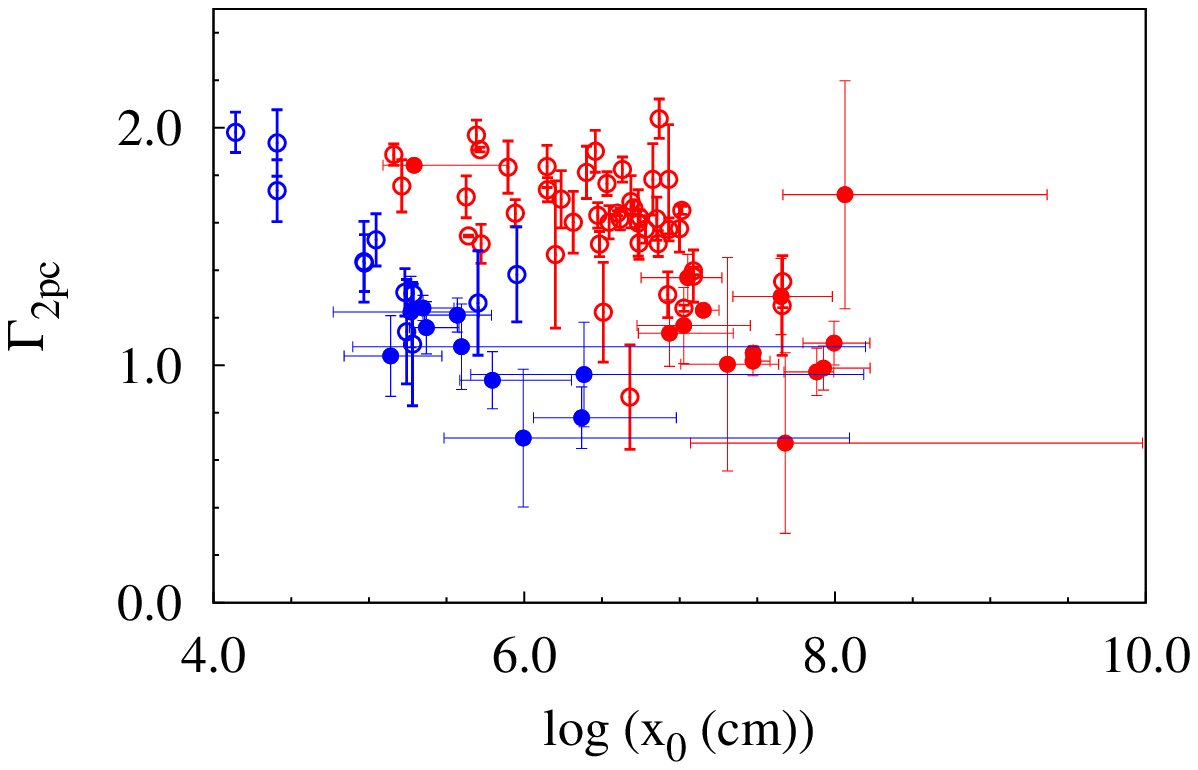}
\includegraphics[width=0.33\textwidth]{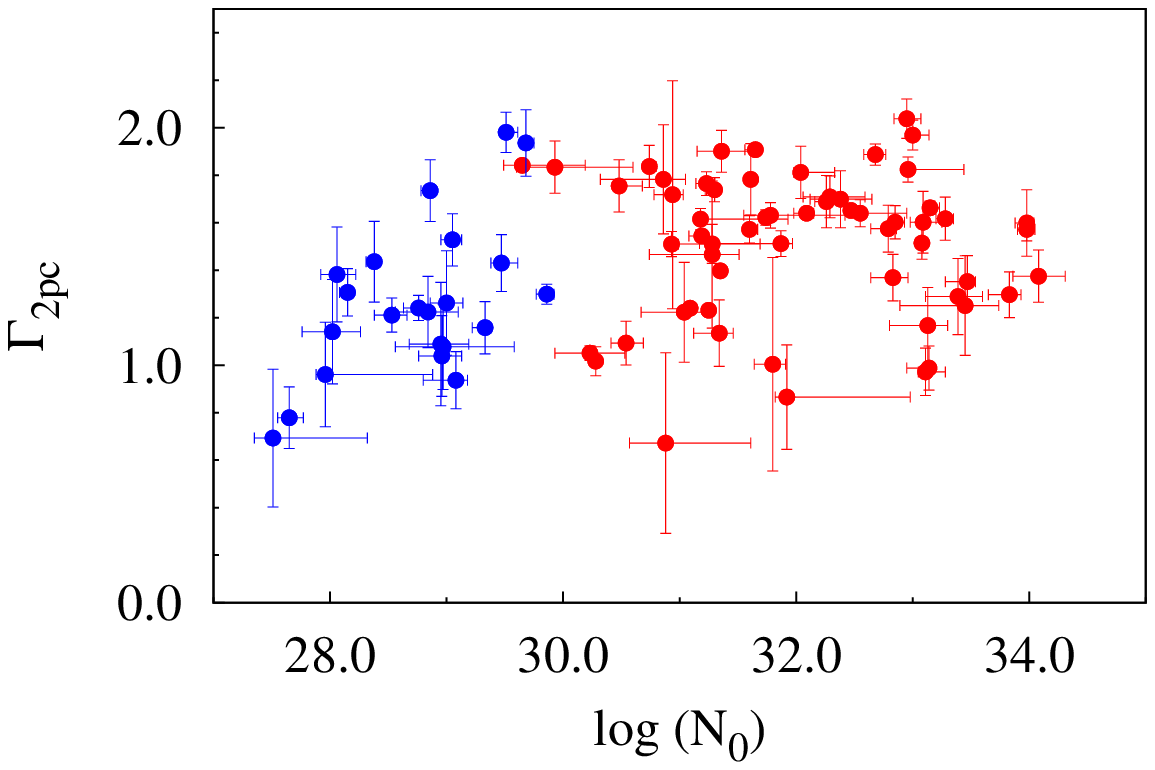}\\
\includegraphics[width=0.33\textwidth]{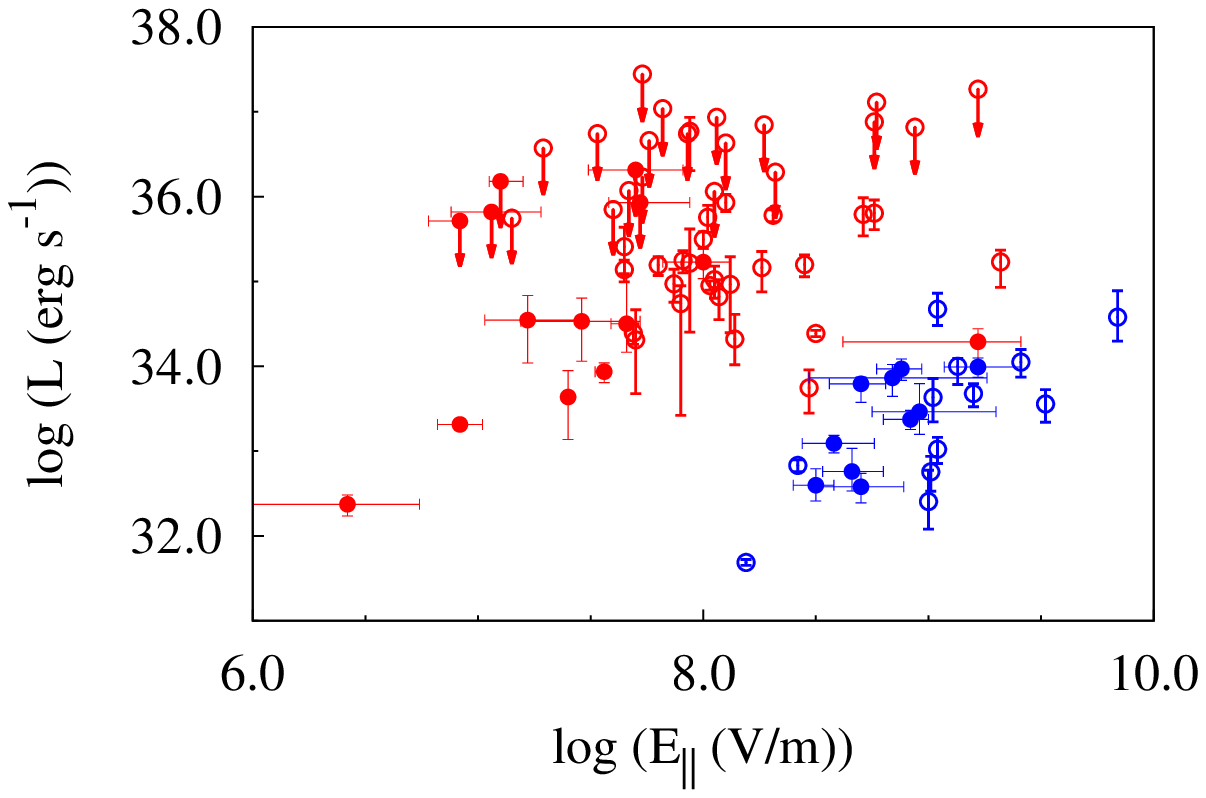}
\includegraphics[width=0.33\textwidth]{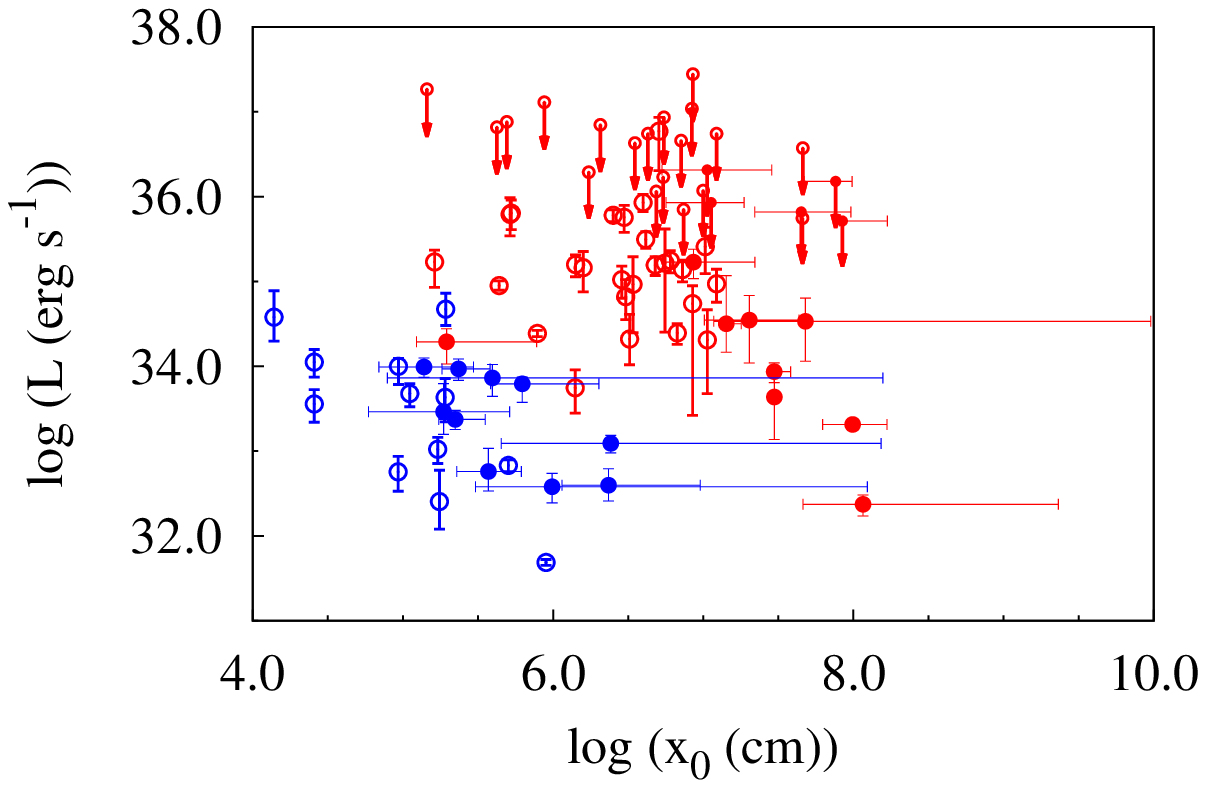}
\includegraphics[width=0.33\textwidth]{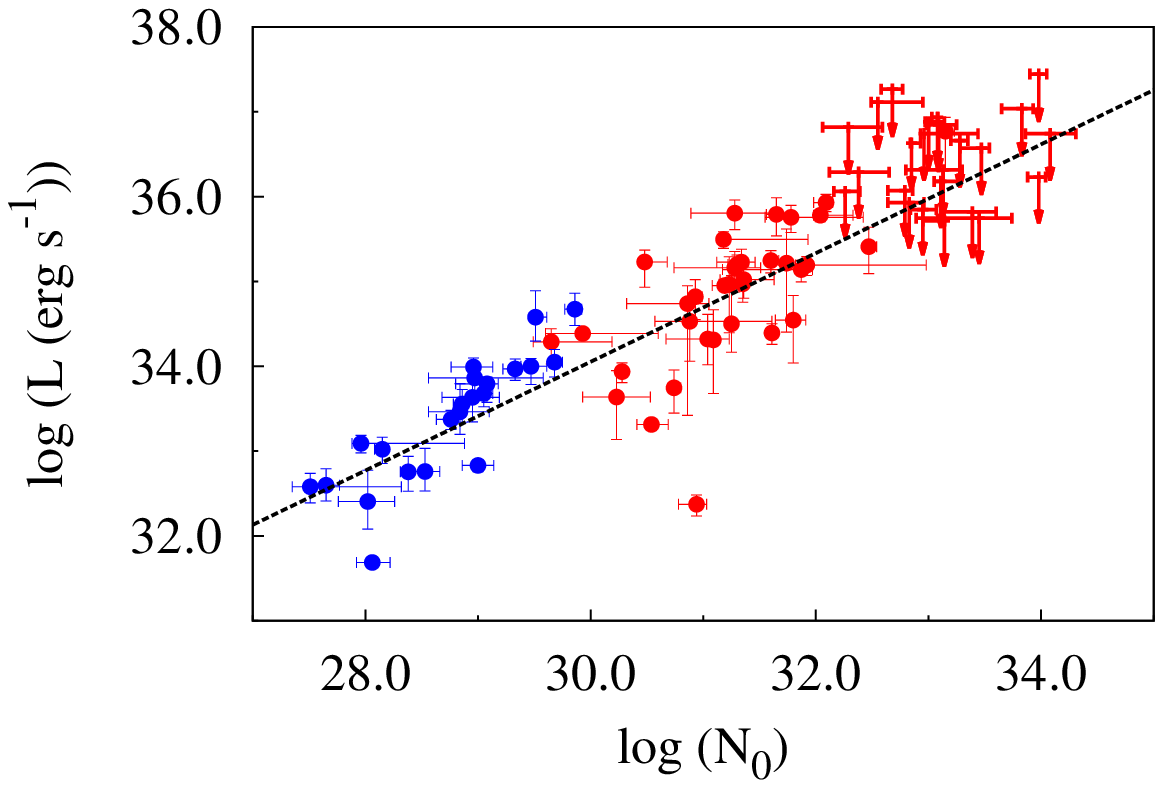}
\end{center}
\caption{Search for correlations among the phenomenological parameters of the PLEC1 model fitted to {\it Fermi}-LAT data: $E_{\rm cut}$, $\Gamma_{\rm 2pc}$, and 0.1--100 GeV luminosity $L$ with the best-fitted SC model parameters: $E_\parallel$, $x_0$, and $N_0$.  Color and empty/filled codings are the same as Fig.~\ref{fig:corr_parameters}.}
\label{fig:corr_PLEC}
\end{figure*}

To put any arising trend 
in perspective it is interesting first to use the PLEC1 fits from 2PC \citep{2fpc} to see if there is any self-correlation among them for the pulsars 
in our sample. This is explicitly shown in Fig.~\ref{fig:corr_parameters_PLEC}.
Whereas the luminosity distinguishes among YPs and MSPs, as we have commented above, 
neither $\Gamma_{\rm 2pc}$ nor the cutoff energy $E_{\rm cutoff}$ are able to clearly separate the sample. For instance, the lowest and highest values of 
$\Gamma_{2pc}$ are attained both by a YP and a MSP. The $L$--$E_{\rm cutoff}$ plane, particularly for MSPs, show an apparent positive trend, where the
highest luminosities are achieved at the largest $E_{\rm cutoff}$, what can be understood in the framework of a SC model as explained below when we discuss
the plane $L$--$E_\parallel$.

\subsubsection{Energy cutoff $E_{\rm cutoff}$}

The energy cutoff traces the radiation released by the most energetic particles, i.e., the ones that have reached the maximum values of Lorentz factor. Thus, it is not surprising to find that 
the larger it is $E_\parallel$ the larger are the $E_{\rm cutoff}$ values, for both sub-samples. What seems to be a rising relation is broken at the largest values $E_\parallel$. It is intrinsic to the curvature losses to have a limit in the cutoff energies (see the saturated Lorentz factors in Fig.\ref{fig:traj}). No other clear correlations with $N_0$ and $x_0$ are found.

\subsubsection{Power-law index $\Gamma_{\rm 2pc}$}

The plane $\Gamma_{\rm 2pc}$--$E_\parallel$ ($\Gamma_{2pc}$--$x_0$)
might also present a hint for a possible positive (negative) trend (albeit not at a level producing $r>0.85$). 
Both parameters $E_\parallel$ and $x_0$ (the degeneracy of many cases makes this obvious) 
play a role in regulating the observed $\gamma$-ray slope at sub-peak energies (and the spectrum in general).
Finally, $N_0$ does not present any obvious trend with the power-law index, which is expectable since there is also no relation between the $\gamma$-ray luminosity and the slope (see below).

\subsubsection{$\gamma$-ray luminosity $L$}

The luminosity in $\gamma$-rays is a separating parameter (as we have seen above) between the two sub-samples of pulsars. Note that for some YPs, we only have upper limits due to the unknown distance (its upper limit is taken as the edge of the galaxy in a given direction). The $\gamma$-ray luminosity relation with the normalization $N_0$ is expected by definition. Hints of a correlation between $L$ and $E_\parallel$ could be argued (but not at the level of $r>0.85$ for the well-constrained pulsars in the sample). If further data confirms it,
can be physically motivated by the fact that the largest Lorentz factors attained by particles in larger electric fields produce a more intense and more energetic photon flux. Since the radiation losses in MSPs are more efficient (closer light cylinder), particles are less energetic and the two sub-sample are divided in this plane. No clear trend is seen with $x_0$, due to the larger errors in the determination of the latter parameter.

\begin{table}
\centering
\scriptsize
\vspace{0.2cm}
\caption{Correlation fits between couples of well-constrained parameters (solid circles in the figures). For the $N_0$-parameter 
all considered pulsar fits are well-constrained.
We show the best linear fits $y=A+Bx$, with
the associated Pearson coefficients, $r$. Units of parameter values are the same used throughout the
paper, and are omitted here. Only very significant correlations are reported ($r>0.85)$. Number in parenthesis give the
pulsars considered.}
\label{tab:correlations}
\begin{tabular}{llrr  r}
\hline
\hline
x & y & $A$ & $B$ & $r$\\
\hline
(24/81)\\
\hline
$\log E_\parallel$ & $\log x_0$ & 17.2$\pm$0.5 & -1.31$\pm$0.07 & -0.974\\
$\log B_{\rm lc}$ & $\log E_\parallel$ & 4.95$\pm$0.08 & 0.78$\pm$0.02 & 0.916\\
$\log B_{\rm lc}$ & $\log x_0$ & 10.8$\pm$0.2 & -1.07$\pm$0.05 & -0.906\\
\hline
(81)\\
\hline
$\log N_0$ & $\log L$ & 15$\pm$2 & 0.64$\pm$0.06 & 0.852\\
\hline
YPs (14/59)\\
\hline
$\log E_{cut}$ & $\log E_\parallel$ & -1.2$\pm$0.4 & 0.59$\pm$0.06 & -0.881\\
$\log E_\parallel$ & $\log(x_0/R_{lc})$ & 5$\pm$1 & -0.9$\pm$0.1 & -0.921\\
\hline
MSPs (10/22)\\
\hline
$\log E_{cut}$ & $\log E_\parallel$ & -1$\pm$1 & 0.5$\pm$0.1 & -0.892\\
\hline
\hline
\end{tabular}
\end{table}

\subsection{Lightcurve parameters}\label{sec:correlations_lc}


We have also tested the possible appearance of correlations among the three SC model parameters with lightcurve features, with the latter being as described in the 2PC. In particular, we considered the number of peaks in the lightcurve, the radio lag, and the $\gamma$-ray peak separation. 
In principle, one could expect some correlation of some of these parameters with $x_0$ or $x_0/R_{\rm lc}$, if the latter values were effectively prompted by some geometrical effect. 
However, we find no significant trend in any of the planes.

\section{Conclusions}\label{sec:conclusions}

This work represents the first systematic spectral fitting of $\gamma$-ray pulsar spectra using a physically-based model including particle dynamics along a magnetic field line and its associated SC radiation. Our models are effectively described by only three parameters. In order to determine their value we have extensively explored thousands of models for each pulsar,  calculating the evolution of the particle energetics and the locally emitted radiation in each. 

In a few cases on which we have individually commented above, residuals are seen above $\sim 10$ GeV. This can be due to a relatively important contribution of inverse Compton scattering (e.g., as in the Crab and Vela pulsars, see \citealt{abdo10a,abdo10b,lyutikov12,paper3} for discussions), or been spuriously generated due to the superposition of quantitatively different phase-resolved spectra, all compatible with SC radiation (like in Geminga, \citealt{paper3}). 

For most of the pulsars studied, the best-fitting models are satisfactorily reproducing the phase-averaged spectra in the whole {\em Fermi}-LAT range. These models are able to systematically explain the variety of both the energy peak and the flat slope of the spectra below $\sim$ 1 GeV. The latter property is often observed, and is not compatible with models where the spectra is generated only from the most energetic particles, i.e., when they have already lost their perpendicular momenta. Our results stand for the SC process as the main generator of $\gamma$-ray pulsed emission, from most objects detected. We stress that pulsars can be fit by our models with only three free parameters, two of which are highly correlated.

This study allow us to infer physically-motivated parameters, i.e., the parallel electric field in the gap $E_\parallel$, the lengthscale over which most of the observed photons are produced $x_0$, and the number of particles responsible for such detected emission $N_0$. Although our model does not include geometry, these parameters give an idea of the effective values of these quantities in the region of the magnetosphere where the radiation is produced. Other parameters in the model have been neglected, being fixed to one and the same fiducial value for all pulsars, since they have a much minor impact on the trajectories and spectra \citep{paper3}. This is the reason why we are not able to place any strong constraint on the location of the acceleration region, being able only to exclude the near-surface region.

Some  qualitative features are discovered by this systematic approach. 
The accelerating electric field is typically of the order of $10^{8}$ V/m for YPs, and $10^9$ V/m for MSPs. There is a strong anti-correlation between $E_\parallel$ and $x_0$, which is continuous across pulsar types. The values of $x_0$ range from a few to hundreds of km, which represent a small fraction of the corresponding  magnetospheric sizes ($x_0/ R_{\rm lc} \sim 10^{-4}$--$10^{-1}$).

There is a strong correlation linking $E_\parallel$ with the magnetic field at light cylinder, $B_{\rm lc}$, which is to be considered as a proxy to the magnetic field in the gap. This, together with the absence of any correlation with the surface magnetic field $B_s$, confirms that the place of high-energy emission is located in the outer magnetosphere of pulsars. Such 
$E_\parallel$--$B_{\rm lc}$
correlation quantitatively unifies under the same trend the two sub-classes, YPs and MSPs, 
which are otherwise (except in the plane $E_\parallel$--$x_0$, as commented above) seen as disjoint populations. In this paradigm, $E_\parallel$ is larger in MSPs because their light cylinders are much closer to the surface than they are for YPs.

In order to produce gamma-ray spectra up to GeV, MSPs need an accelerating electric field that looks large in comparison to their relatively low surface magnetization. In the usual OG model, and in order to generate such particle acceleration, \cite{zhang03} invoked multipolar magnetic fields existing close to the surface, about 1000 times larger in magnitude than the dipolar surface value. Apart from the many assumptions that caveat this gap model as a whole (see \citet{paper1} for a detailed discussion), the solution found here is simpler: $E_\parallel$ correlates with the local value of $B$ in the outer regions of the magnetosphere, i.e., close to or beyond the light cylinder. Whereas in previous studies, $E_\parallel$ was inferred by adopting an over-simplified estimation involving $P$ and $B_s$, here $E_\parallel$ is inferred by considering the physical process of electric acceleration, radiative losses, and the subsequently radiated photons. The assumptions (and these latter processes themselves) make no qualitative difference between particles in the surrounding of MSPs and YPs.

The systematic finding of low values of $x_0$ in comparison with $R_{lc}$ for both classes of pulsars may point to something fundamental related with the distribution of the particles. Although our modelling relies on an effective approach, the results indicate the need for a dominant number of particles with relatively low values of Lorentz factor, one or two orders of magnitude lower than the value of saturation $\Gamma \sim 10^7$--$10^8$.  On one side, one could appeal to the cascading process as the mechanism able to replenish a great number of particles which quickly lose their perpendicular momentum. Second, the beaming effect, which scales with $1/\Gamma$ favours the detection of low-energy photons. These effects, possibly responsible for the low-energy flatness of energy spectra in many pulsars, is, in our model, simplified by adopting a larger weight to the initial part of the trajectories, i.e., small $x_0$. Numerical simulations involving cascading and geometry are needed to physically reinforce our interpretation, and to allow a comparison with pulse profiles. 


We also note that, both in radio and in $\gamma$-rays, the gross spectral properties, like the luminosity, or the energy peak, show no or weak correlations with timing properties. Thus, although correlations between rotational energy and luminosity have been proposed many times, we remind that the observations show a huge dispersion, which cannot be neglected. This strengthens the importance of the correlations we find, particularly because they unify YPs and MSPs. 

An important prediction is that, if we extrapolate the correlation $E_\parallel(B_{\rm lc})$ to the magnetar regime ($B_{\rm lc}\sim 10^{-1}$--$10^2$~G due to their long periods, $P\sim 2$--$10$ s), it is clear that $E_\parallel$ would be very weak, $E_\parallel \lesssim 10^6$ V/m, and unable to provide particles energetic enough as to emit $\gamma$-rays. Thus, we expect magnetars not to be visible as pulsed $\gamma$-ray emitters (unless a very different mechanism from SC radiation is at play).

Finally, we note that with the forthcoming publication of the third {\em Fermi}-LAT pulsar catalog, with a longer time baseline and the new Pass 8 analysis, a larger and better quality sample could be analyzed. With such sample, for which more than 1/3  will consist of pulsars not contained in our study, we shall look deeper and in an independent way into the trends here illustrated and their physical implications.

\section*{Acknowledgements}

This research was supported by the grants AYA2012-39303 and SGR2014-1073. DV thanks the many colleagues with which he has collaborated and shared good time during these years of research in astrophysics.

\bibliography{og}

\appendix

\section{Details of the fitting model}\label{app:formulae}

\subsection{Formulae of SC radiation}

The single-particle SC radiation power is \citep{cheng96,paper0}
\begin{equation}
\label{eq:sed_synchrocurv}
 \frac{dP_{\rm sc}}{dE} = \frac{\sqrt{3} e^2 \Gamma y}{4\pi \hbar r_{\rm eff} } [ (1 + z) F(y) - (1 - z) K_{2/3}(y)]~,
\end{equation}
where
\begin{eqnarray}
 && z= (Q_2 r_{\rm eff})^{-2} ~, \label{eq:z}\\
 && F(y) = \int_y^\infty K_{5/3}(y') dy'~,\label{eq:f_y}\\
 && y=\frac{E}{E_c} ~, \\
 && E_c = \frac{3}{2}\hbar cQ_2\Gamma^3~,\label{eq:echar}\\
 && r_{\rm gyr} = \frac{mc^2\Gamma\sin\alpha}{eB}~,\\
 && Q_2 = \frac{\cos^2\alpha}{r_c}\sqrt{1 + 3\xi  + \xi^2 + \frac{r_{\rm gyr}}{r_c}} \label{eq:q2}~, \\
 && \xi = \frac{r_c}{r_{\rm gyr}}\frac{\sin^2\alpha}{\cos^2\alpha}~, \label{eq:xi}\\
 && r_{\rm eff} = \frac{r_c}{\cos^2\alpha}\left(1 + \xi+ \frac{r_{\rm gyr}}{r_c}  \right)^{-1}~,\\
 && g_r =  \frac{r_c^2}{r_{\rm eff}^2}\frac{[1 + 7(r_{\rm eff}Q_2)^{-2}]}{8 (Q_2r_{\rm eff})^{-1}}~.\label{eq:gr}
\end{eqnarray}
Here, $m$ and $\Gamma$ are the rest mass and the Lorentz factor of the particle, $\alpha$, $r_{\rm gyr}$ and $r_c$ are the pitch angle, the Larmor radius, and the radius of curvature of its trajectory, respectively, $e$ the elementary charge, $B$ the local strength of the magnetic field, $\hbar$ the reduced Planck constant, $c$ is the speed of light, $K_n$ are the modified Bessel functions of the second kind of index $n$, $E$ is the photon energy, $E_c$ is the characteristic energy of the emitted radiation. In the limits of high or vanishing perpendicular momentum, Eq.~(\ref{eq:sed_synchrocurv}) reduces to purely synchrotron ($\xi\gg1$) or curvature radiation ($\xi\ll 1$) formulae, respectively. In any case, the peak of the spectrum is located close to $E_c$ and, for energies $E\ll E_c$, the dominant term in Eq.~(\ref{eq:sed_synchrocurv}), $F(y)$, provides $dP_{\rm sc}/dE \sim E^{0.25}$. 

\subsection{Particle dynamics}

For each model, we simulate the motion of charged particles by evolving the parallel and perpendicular momenta of particles according to the equations of motion \citep{paper0}
\begin{eqnarray}
 && \frac{d(p\sin\alpha)}{d t} = - \frac{P_{\rm sc}\sin\alpha}{v}~, \label{eq:motion_perp} \\
 && \frac{d(p\cos\alpha)}{d t} = eE_\parallel - \frac{P_{\rm sc}\cos\alpha}{v}~, \label{eq:motion_par}
\end{eqnarray}
where $p=\Gamma m v$ is the momentum of the particle, $v$ its spatial velocity, $E_\parallel$ is the parallel component of the electric field, and $P_{\rm sc}$ is the single-particle SC power, obtained by integrating Eq.~(\ref{eq:sed_synchrocurv}) in energy:
\begin{equation}\label{eq:power_synchrocurv}
 P_{\rm sc} = \frac{2e^2 \Gamma^4 c}{3 r_c^2} g_r.
\end{equation}
At each position in the gap we use the corresponding local values of $B(x)=B_s(R_\star/x)^b$ and $r_c(x)=R_{\rm lc}(x/R_{\rm lc})^\eta$ ($b$ and $\eta$ are model parameters, here fixed to 2.5 and 0.5, respectively, \citealt{paper2,paper3}), to compute Eqs.~(\ref{eq:motion_perp})-(\ref{eq:motion_par}) consistently.

In Fig.~\ref{fig:traj}, and the associated text, we describe the main outcome of our simulated trajectories. Here we derive why the correlation $E_\parallel \sim 1/x_0$, discussed in the text and shown in many cases in Figs.~\ref{fig:contours1}-\ref{fig:contours5}, is expected to appear. When the electric acceleration term dominates over the SC power, $eE_\parallel\gg P_{\rm sc}\cos\alpha/v$, one can simplifies Eq.~(\ref{eq:motion_par}), considering also that $\cos\alpha\rightarrow 1$ very fast, as seen in the numerically computed trajectories \citep{paper0}: $d\Gamma/dt \sim eE_\parallel/mv$. Since $t = x/v$, then, in the accelerating regime, $E_\parallel \sim 1/x$. In other words, the linear rise of the trajectory ($\Gamma(x)$, left panels of Fig.~\ref{fig:traj}) can be displaced to lower $x$ by rising the value of $E_\parallel$. In many pulsars with relatively flat slopes, this part of the trajectory is dominating the spectrum, and the best-fitting models are endowed with a value of $x_0$ smaller than the distance at which saturated values of $\Gamma$ are reached. Therefore, the best-fitting solutions of these particular pulsars will have an intrinsic degeneracy: larger $E_\parallel$ can be compensated by re-scaling $x_0$ with $1/E_\parallel$. This quantitatively explains the correlation. 

As discussed in depth in \S\ref{sec:self-correlations}, such analytical argument explains the individual degeneracy appearing in some of the Figs.~\ref{fig:contours1}-\ref{fig:contours5}, but not the similar global trend found when comparing the best-fitting parameters for all pulsars, i.e., again, $E_\parallel \sim 1/x_0$ (top panels of Fig.~\ref{fig:corr_parameters} and first row of Table \ref{tab:correlations}).

\section{Fits of the sample}\label{app:fits}

Figures of this Appendix show individual best-fitting models compared with pulsar data, and contours of $\chi^2/\chi^2_{\rm min}$ in the plane $\log E_\parallel-\log (x_0/R_{lc})$. In each of the contour plots, we mark with a white cross the location of the best-fitting model. The extreme values of $x_0$, $E_\parallel$ and $N_0$ within the region having $\chi^2/\chi^2_{\rm min}<1.3$ are taken as the errors of model parameters and are quoted in Tables~\ref{tab:psr} and \ref{tab:msp} (see also \ref{sec:errors}). Due to the finite steps of the explored parameter values, the resolution of the contours can be visually insufficient at times, especially when data suffer very small relative errors (like in the Vela pulsar). Red and blue lines or labels indicate YPs and MSPs, respectively, following the color coding used in the main text.

Observational spectra are taken from the publicly available data of the 2PC.
Data consist of a number of $N_{\rm bin}\sim 5-10$ energy bins (with extremes $E_1$ and $E_2$, and weighted central energy $E_{\rm cent}$), each one with its associated photon flux, $F^{\rm bin}_{\rm obs}$ (in units photons~cm$^{-2}$s$^{-1}$), and its associated statistical error, $\delta F^{\rm bin}_{\rm obs}$. We consider the isotropic luminosity, 
\begin{equation} 
L^{\rm bin}_{\rm obs}=4\pi d^2 F^{\rm bin}_{\rm obs}, 
\end{equation}
where $d$ is the distance of the pulsar to the Earth, and we neglect possible beaming effects which would reduce the inferred luminosity. We plot the binned functions $E^2dN/dE \equiv E_{\rm cent}^2 L^{\rm bin}$, in units erg~s$^{-1}$, both for the theoretical models and data.

To explore the space of parameters, we span a grid of 101x51 equi-spaced values of $[\log E_\parallel,\log x_0]$, covering two and five orders of magnitude, respectively. The range of $E_\parallel$ varies case by case, while $x_0$ always spans $[0.0001-10] R_lc$. We also evaluate the case of uniform effective distribution of particle. For each pair of values $(E_\parallel,x_0)$, we evaluate the expected spectrum over a grid of hundreds of points in the 0.1--100 GeV range. Then, we renormalize it by the best-fitting value of $N_0$, found by means of scanning a grid having progressively finer steps, up to the numerical convergence, defined by $\delta N_0/N_0 < 10^{-3}$. Therefore, we integrate the SC photon spectrum (related to the SC energy spectrum, Eq.~\ref{eq:sed_x}, by $dN_{\rm gap}/dE=(1/E)dP_{\rm gap}/dE$) in each bin (and normalized by the bin width), to obtain the binned SC photon spectrum (number of photons per unit energy):
\begin{equation}
L^{\rm bin}_{\rm gap} = \frac{1}{E_2-E_1} \int_{E_1}^{E_2} \frac{1}{E}\frac{dP_{\rm gap}}{dE} {\rm d}E~.
\end{equation}
With this we calculate the goodness-of-fit indicator:
\begin{equation}\label{eq:chi2_app}
  {\chi}^2 =  \sum_{\rm bin} \frac{(L^{\rm bin}_{\rm obs} - L^{\rm bin}_{\rm gap})^2}{(\delta L^{\rm bin}_{\rm obs})^2}~,
\end{equation}
where $\delta L^{\rm bin}_{\rm obs} = 4\pi d^2 \delta F^{\rm bin}_{\rm obs}$ is the luminosity error for each bin. Such error neglects the (likely large) uncertainties on distance and beaming factor, but this would only change the luminosity, thus providing a different (but accordingly) inferred normalization $N_0$, i.e., without changing the spectral shape, and, leaving the constrains on $E_\parallel$ and $x_0/R_{\rm lc}$ --as well as all trends-- unscathed. When distance is unknown, we use its upper limit (the edge of the galaxy in that particular direction), and the related upper limit on $L$, to constrain $N_0$ (which, therefore, has to be taken as an upper limit).

\begin{figure*}
\begin{center}
\begin{overpic}[width=0.32\textwidth]{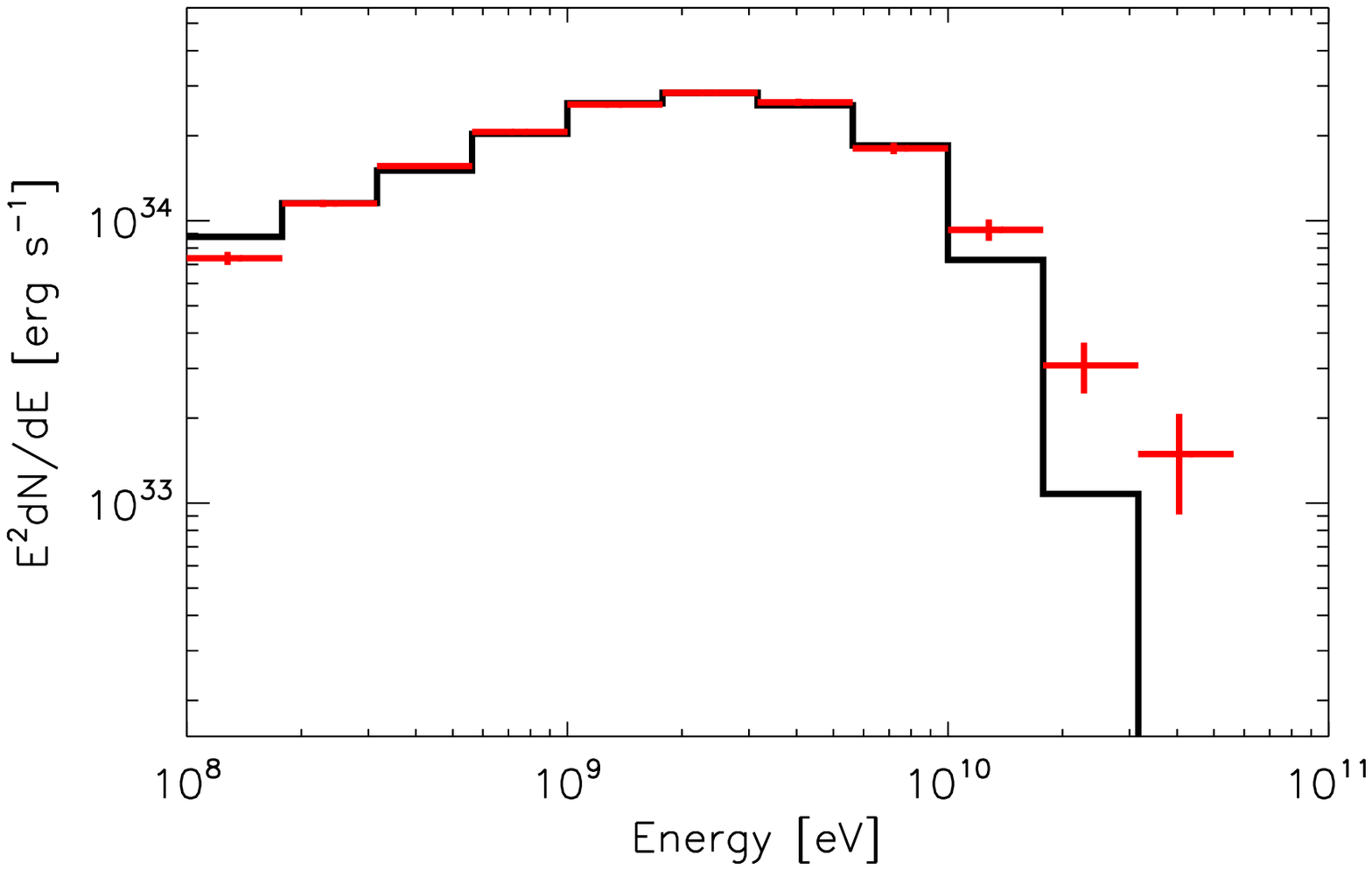}
\put(20,15){\scriptsize J0007+7303 (CTA1)}
\end{overpic}
\begin{overpic}[width=0.32\textwidth]{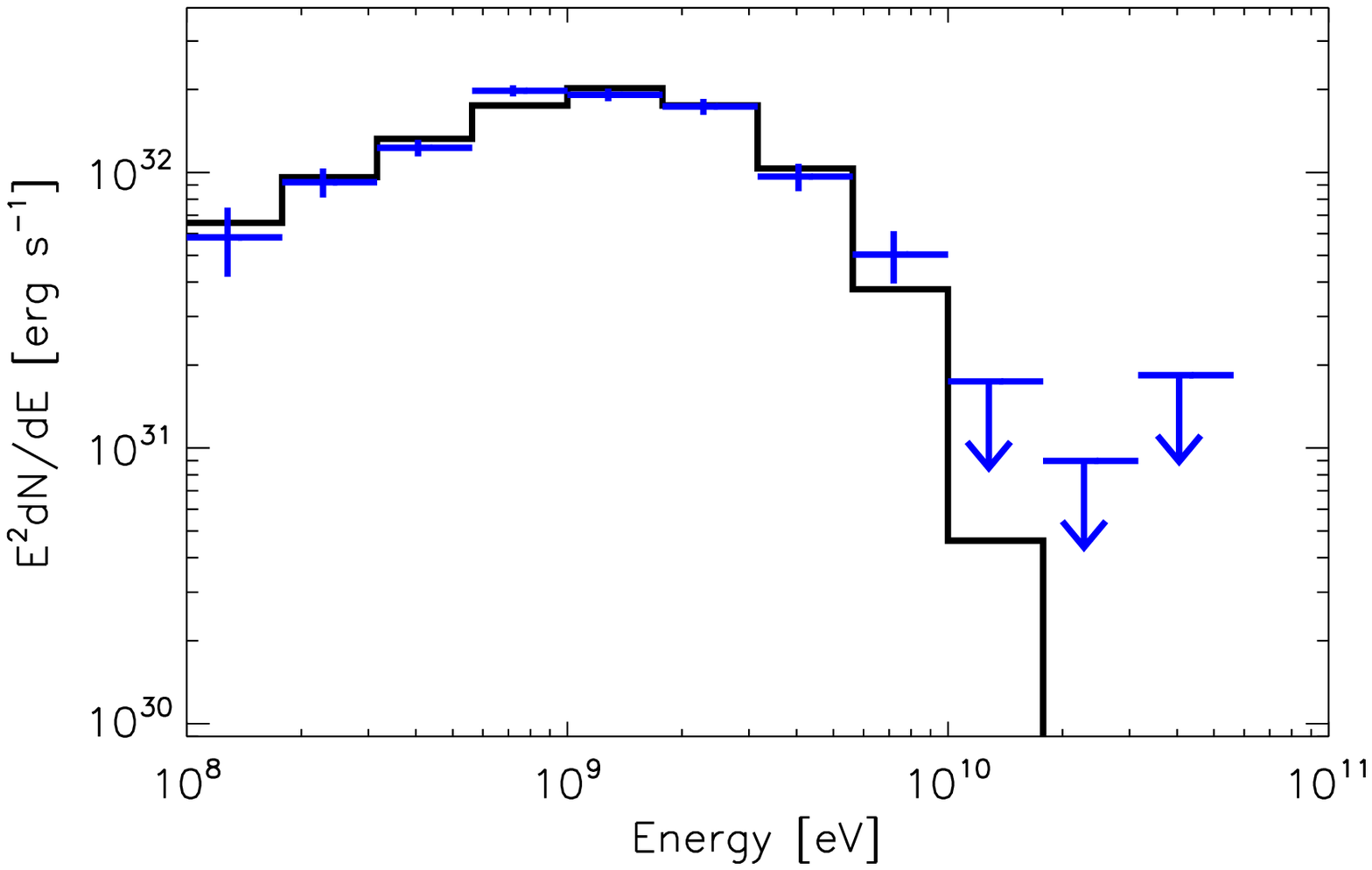}
\put(20,15){\scriptsize J0030+0451}
\end{overpic}
\begin{overpic}[width=0.32\textwidth]{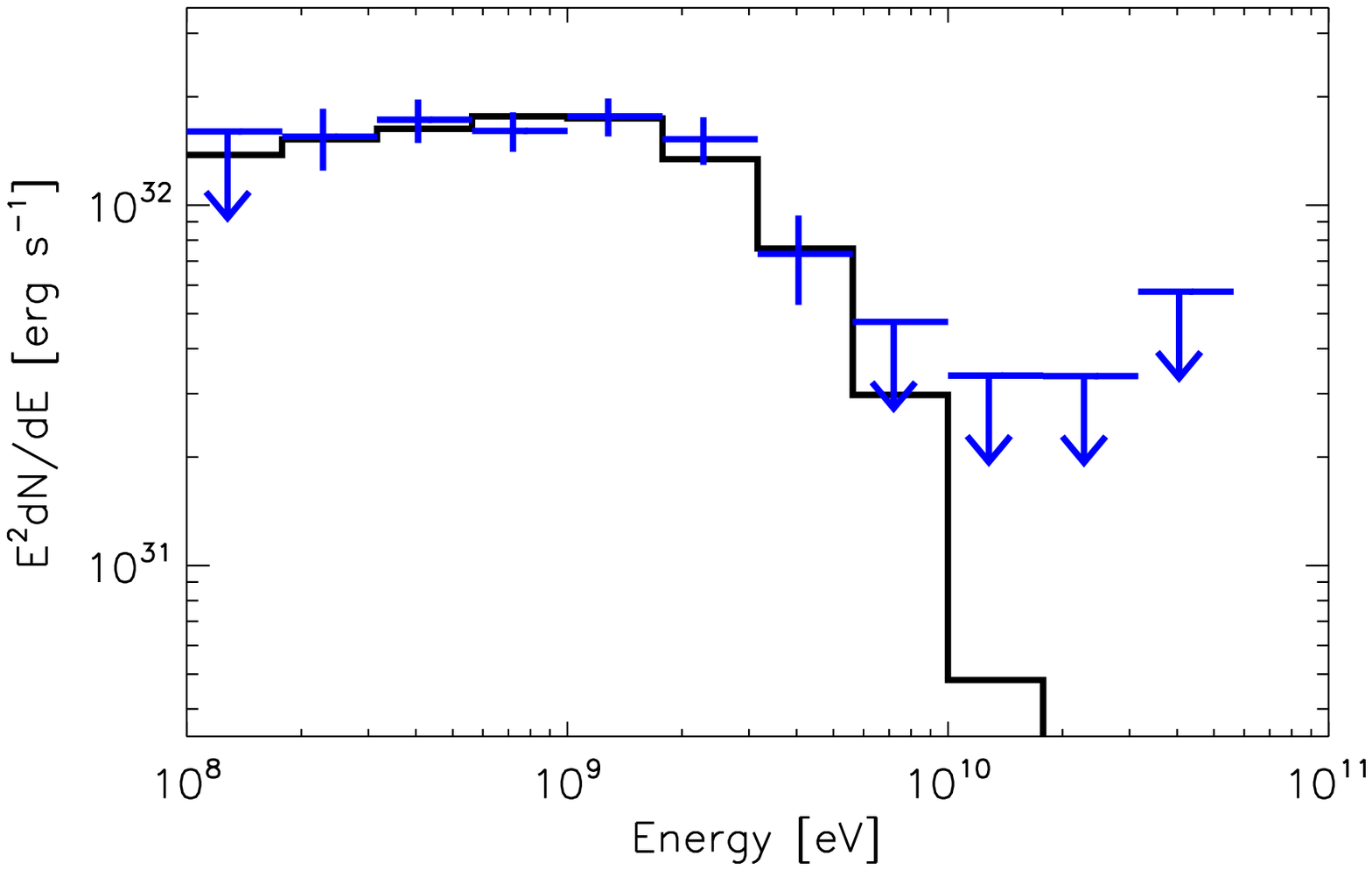}
\put(20,15){\scriptsize J0034-0534}
\end{overpic}
\begin{overpic}[width=0.32\textwidth]{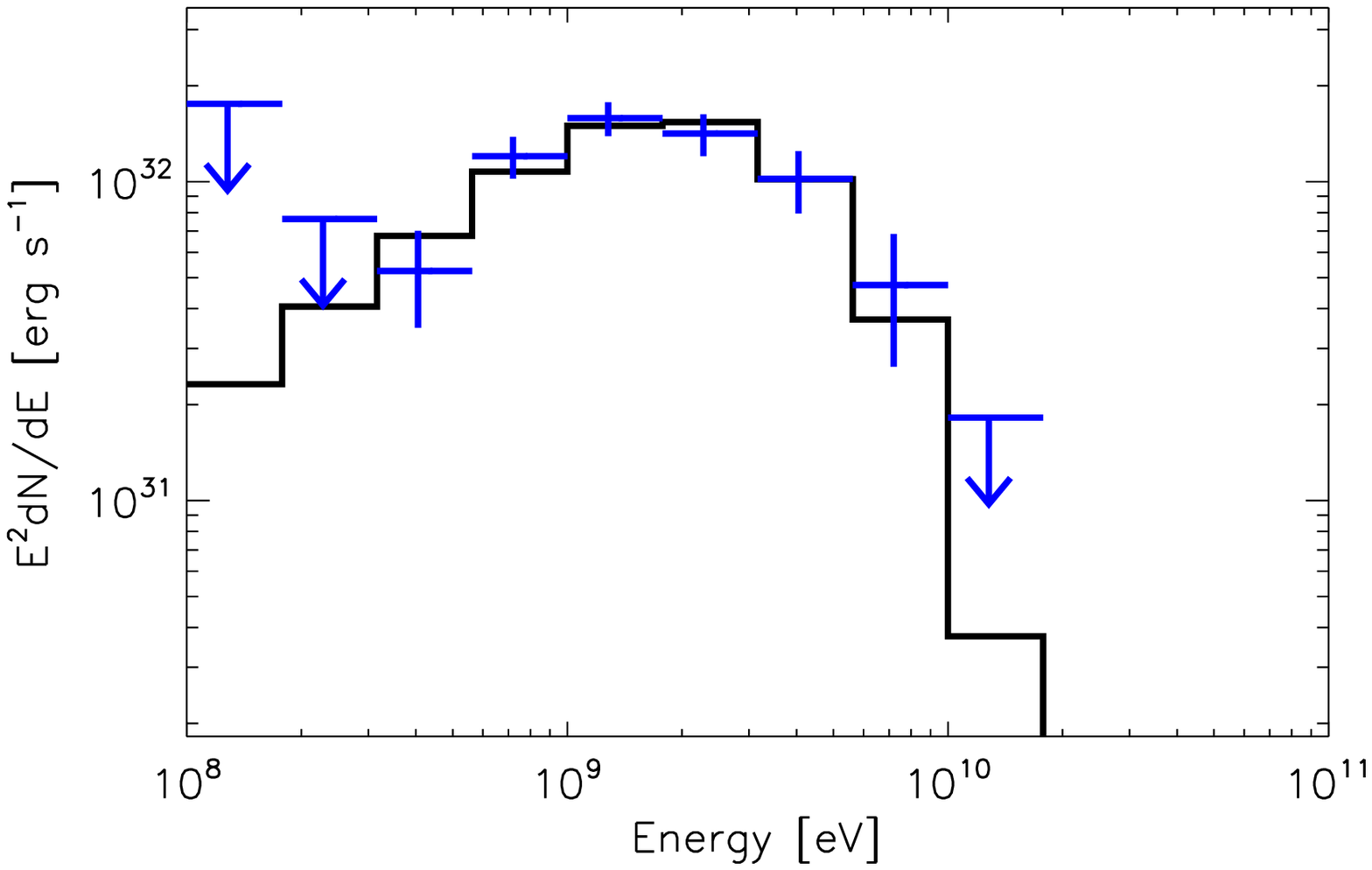}
\put(20,15){\scriptsize J0101-6422}
\end{overpic}
\begin{overpic}[width=0.32\textwidth]{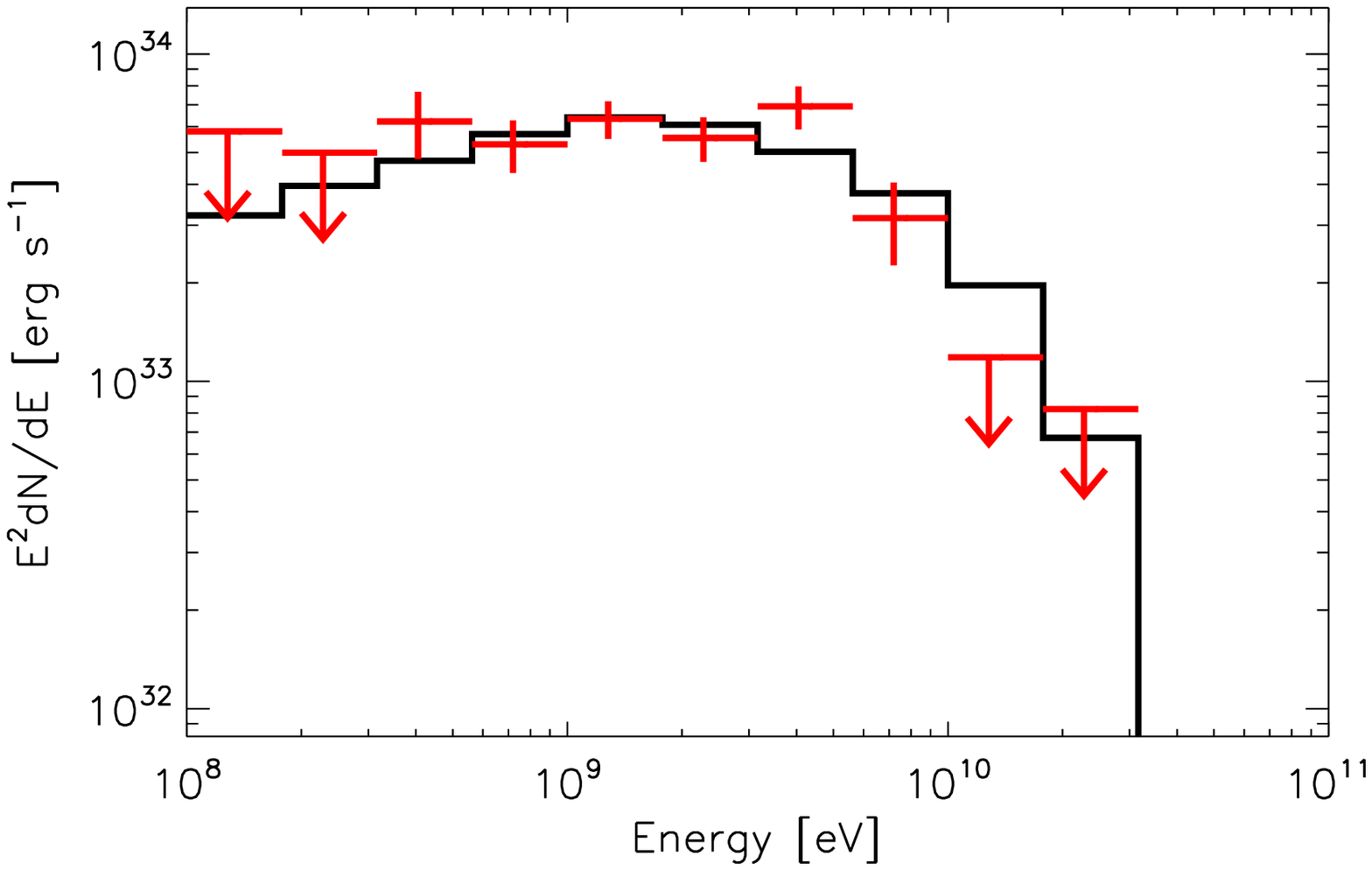}
\put(20,15){\scriptsize J0106+4855}
\end{overpic}
\begin{overpic}[width=0.32\textwidth]{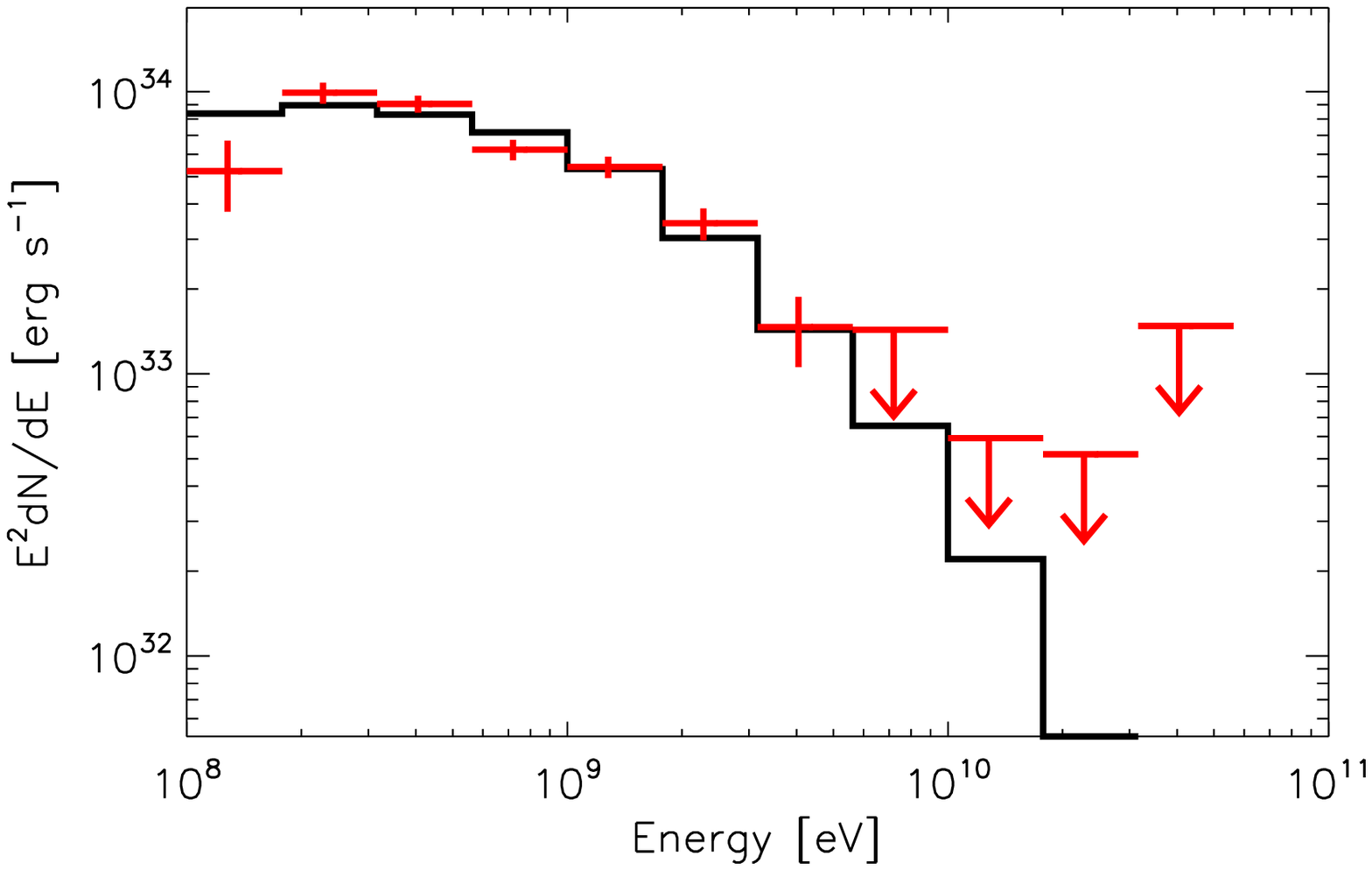}
\put(20,15){\scriptsize J0205+6449}
\end{overpic}
\begin{overpic}[width=0.32\textwidth]{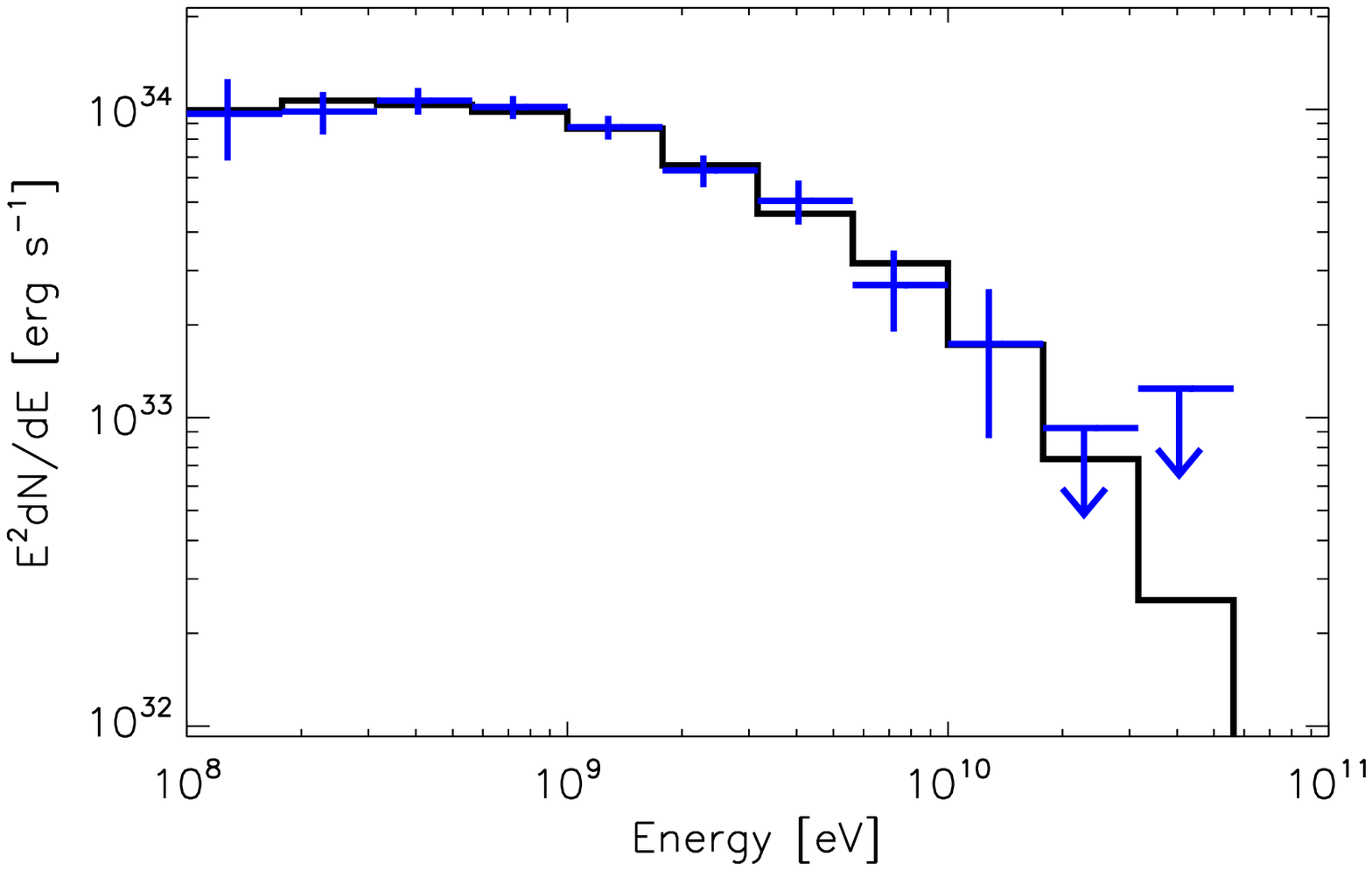}
\put(20,15){\scriptsize J0218+4232}
\end{overpic}
\begin{overpic}[width=0.32\textwidth]{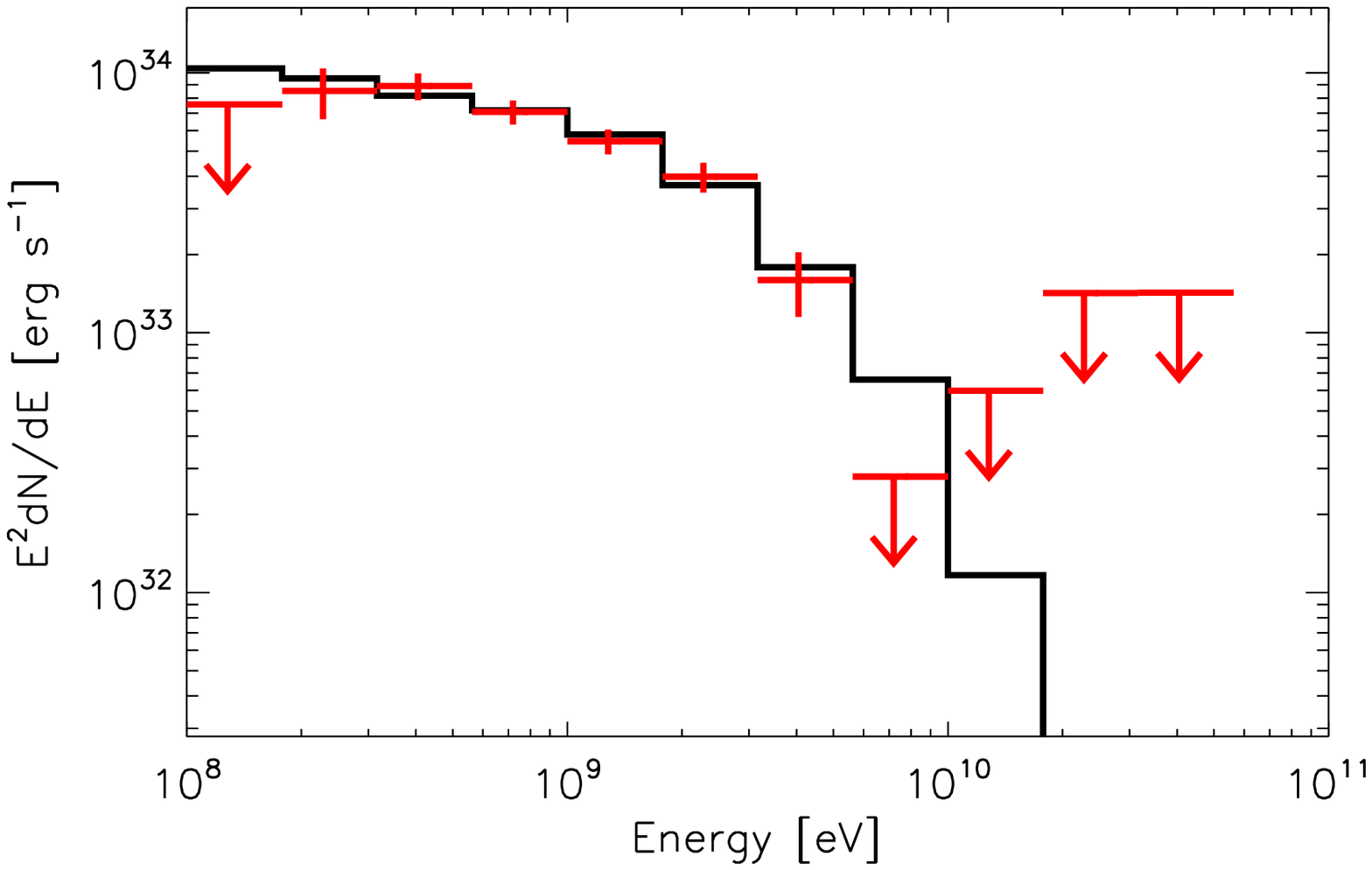}
\put(20,15){\scriptsize J0248+6021}
\end{overpic}
\begin{overpic}[width=0.32\textwidth]{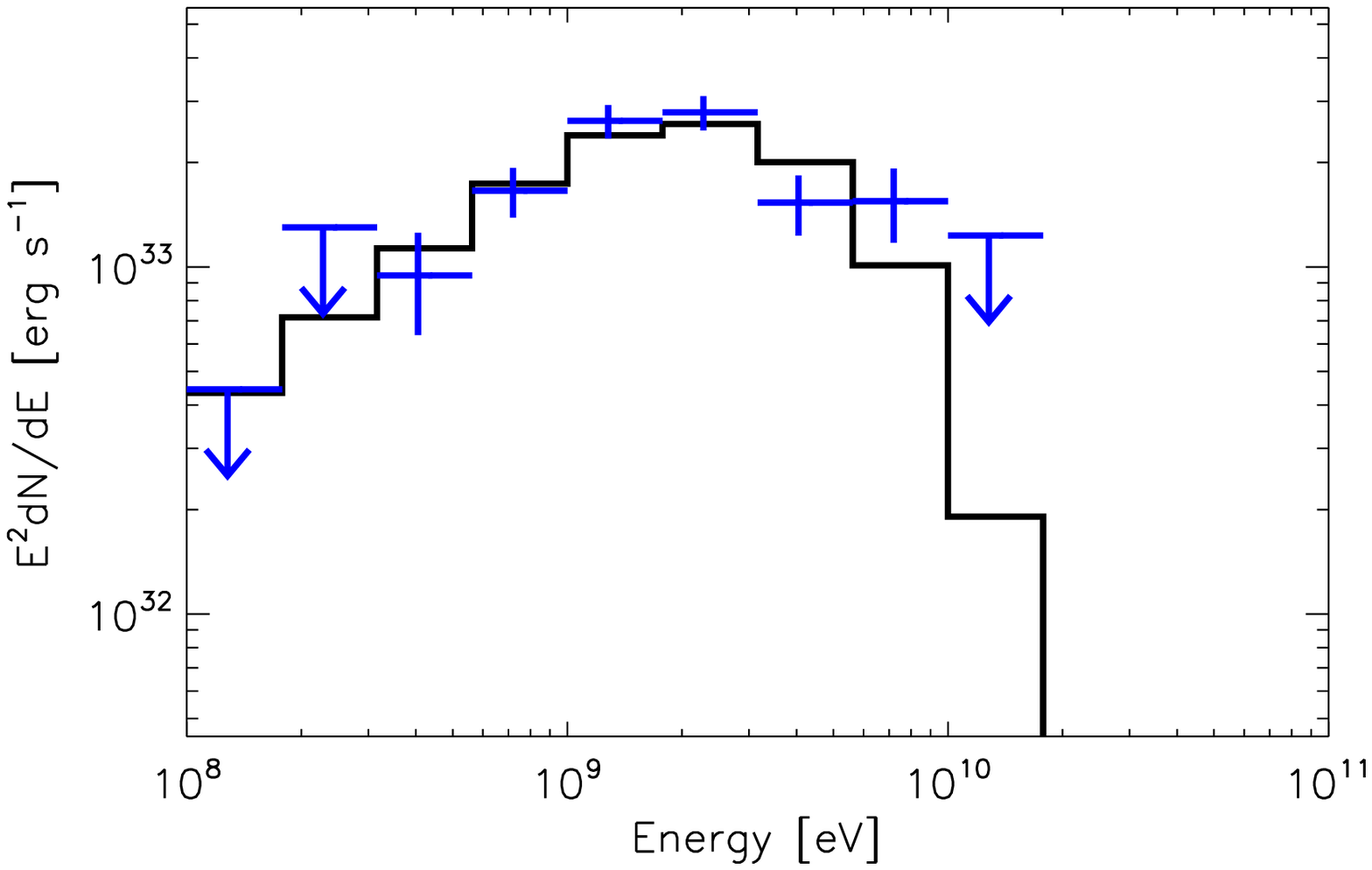}
\put(20,15){\scriptsize J0340+4130}
\end{overpic}
\begin{overpic}[width=0.32\textwidth]{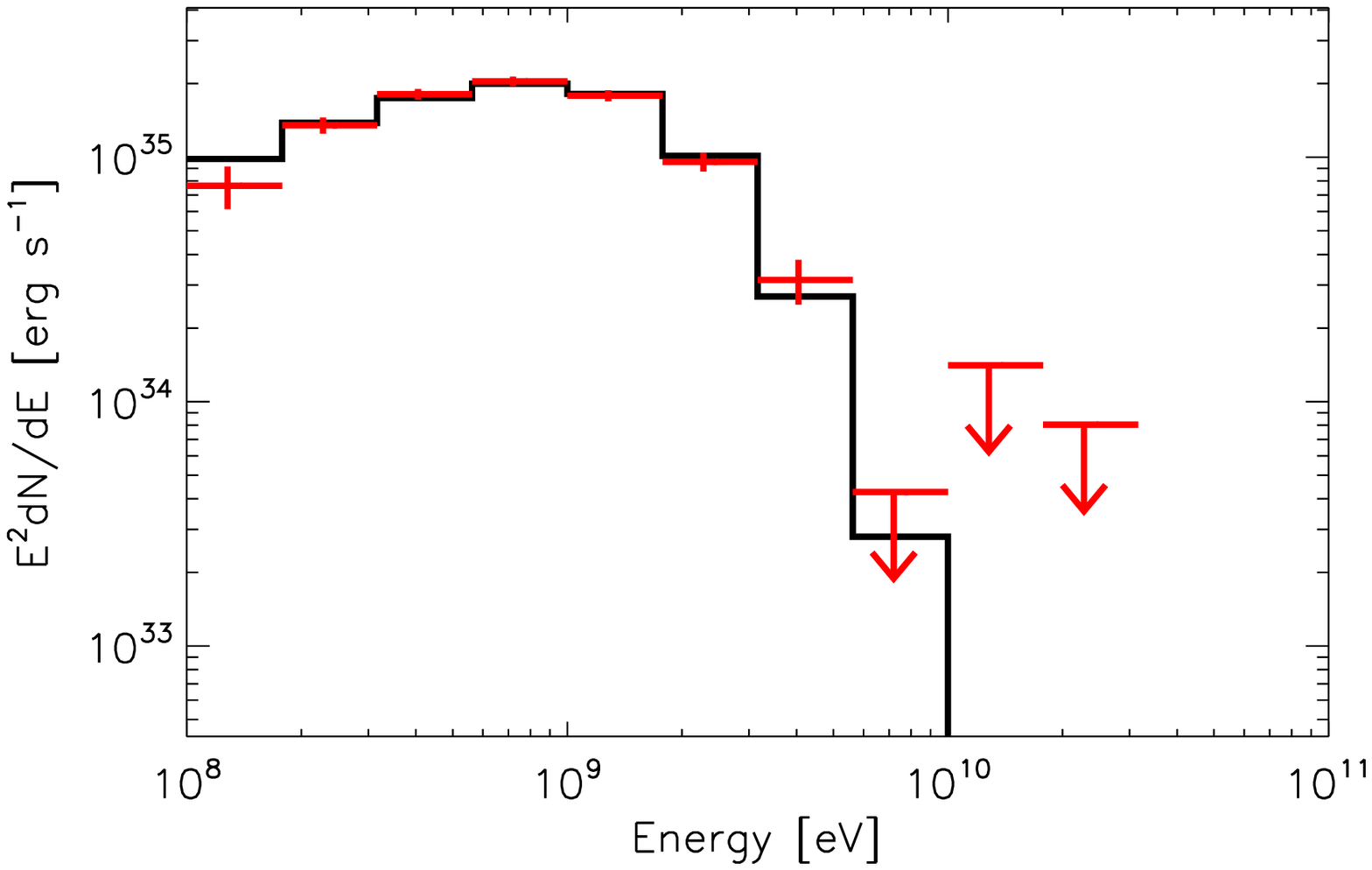}
\put(20,15){\scriptsize J0357+3205}
\end{overpic}
\begin{overpic}[width=0.32\textwidth]{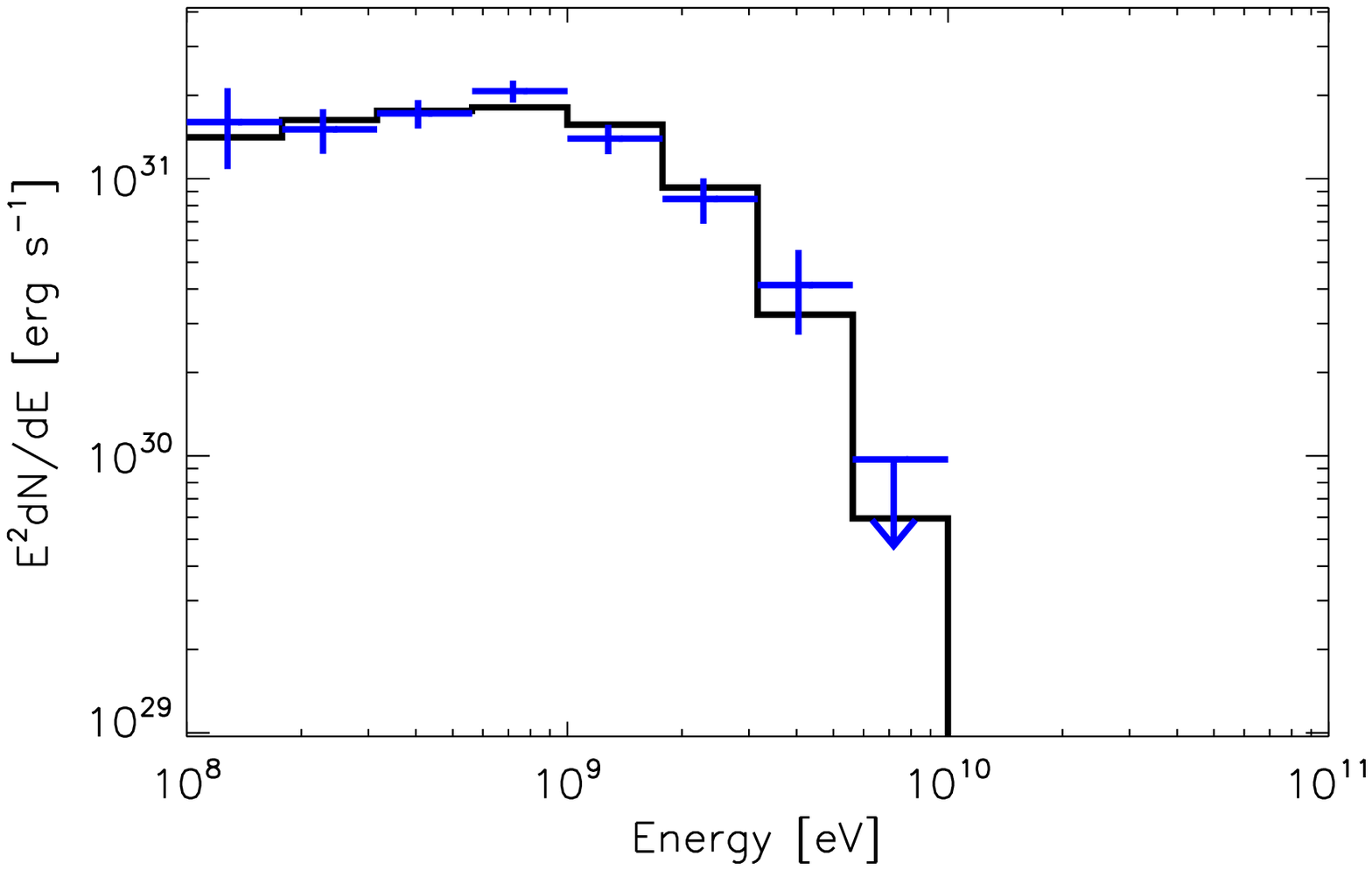}
\put(20,15){\scriptsize J0437-4715}
\end{overpic}
\begin{overpic}[width=0.32\textwidth]{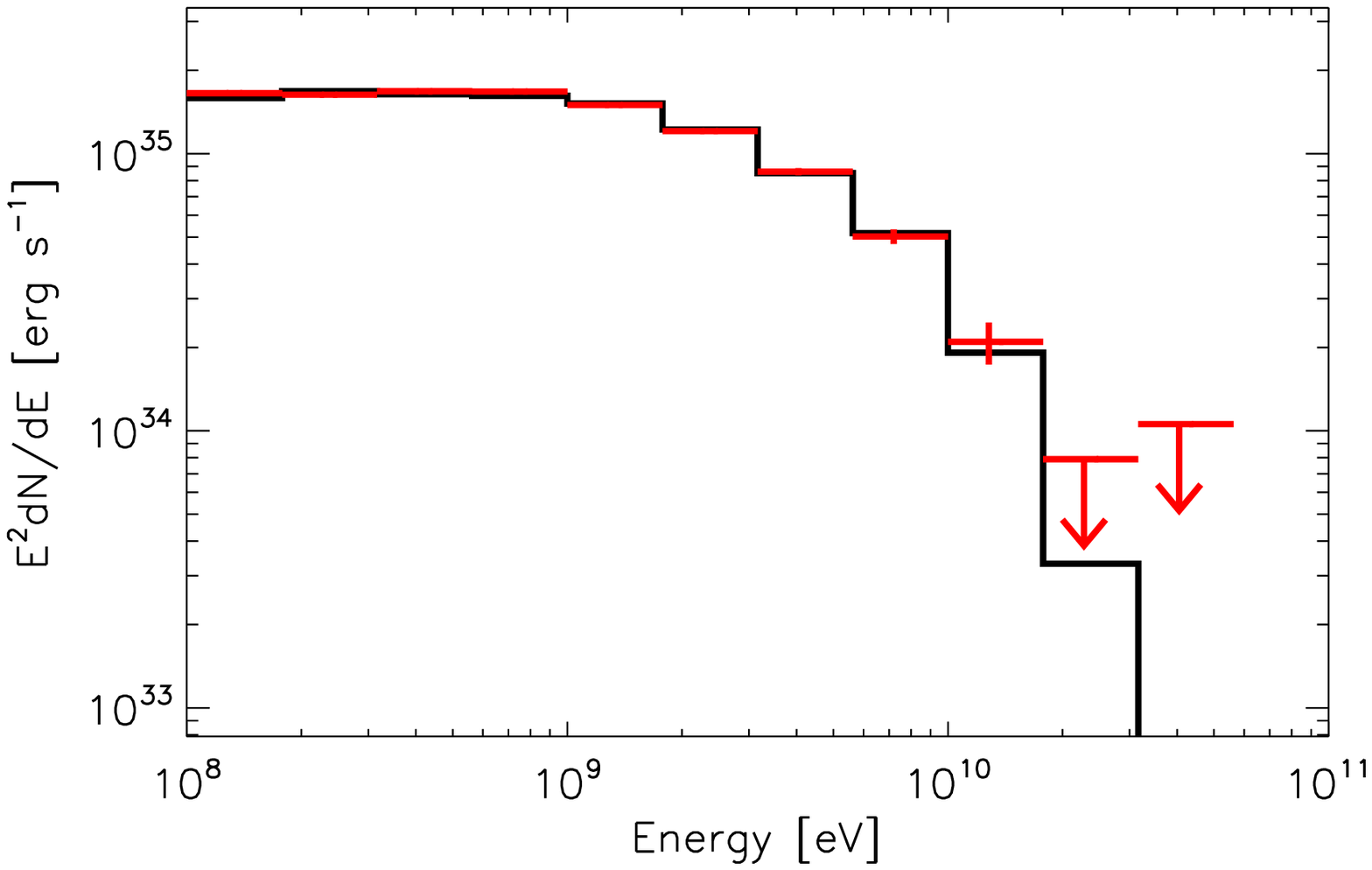}
\put(20,15){\scriptsize J0534+2200 (Crab)}
\end{overpic}
\begin{overpic}[width=0.32\textwidth]{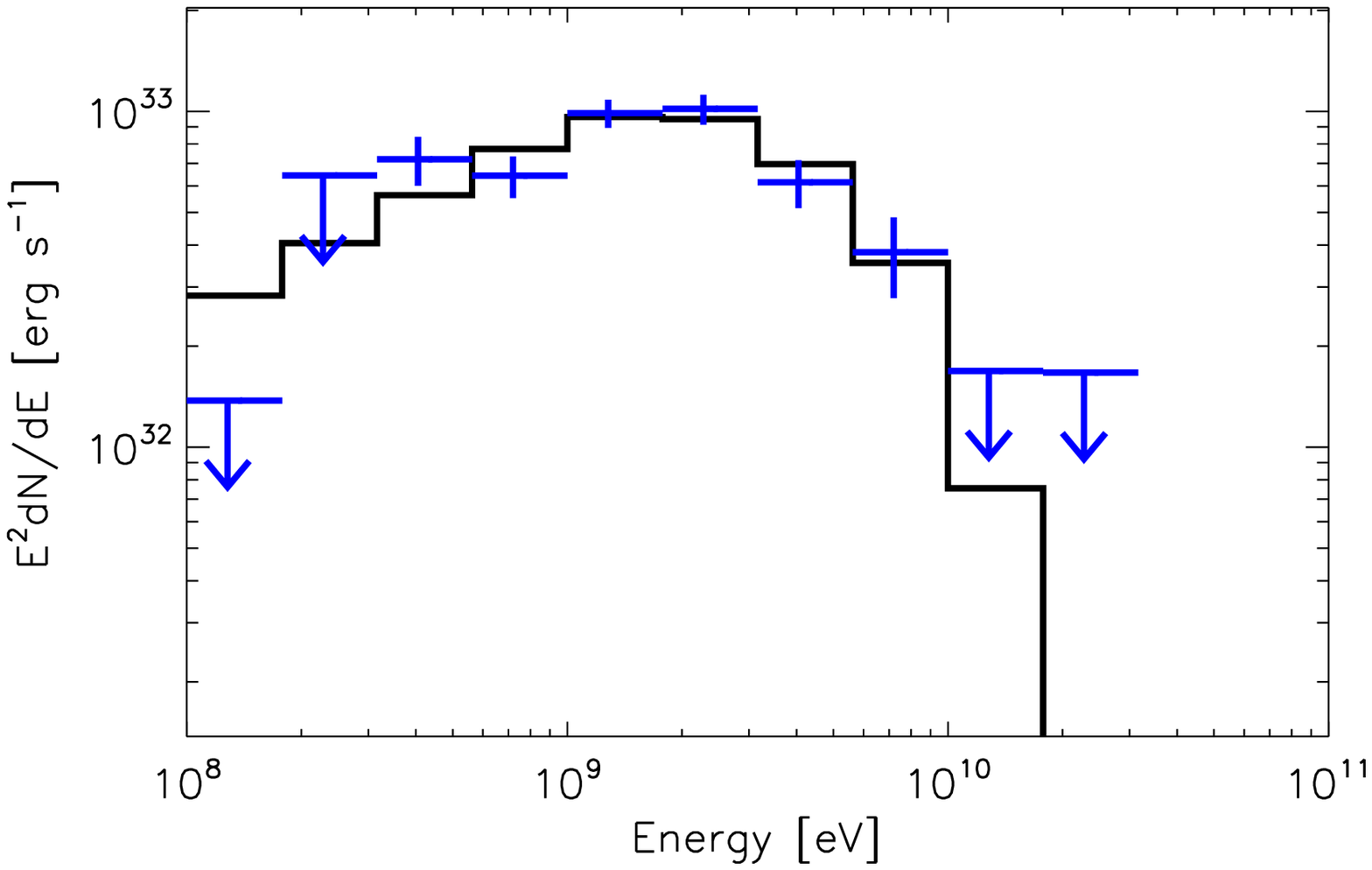}
\put(20,15){\scriptsize J0613-0200}
\end{overpic}
\begin{overpic}[width=0.32\textwidth]{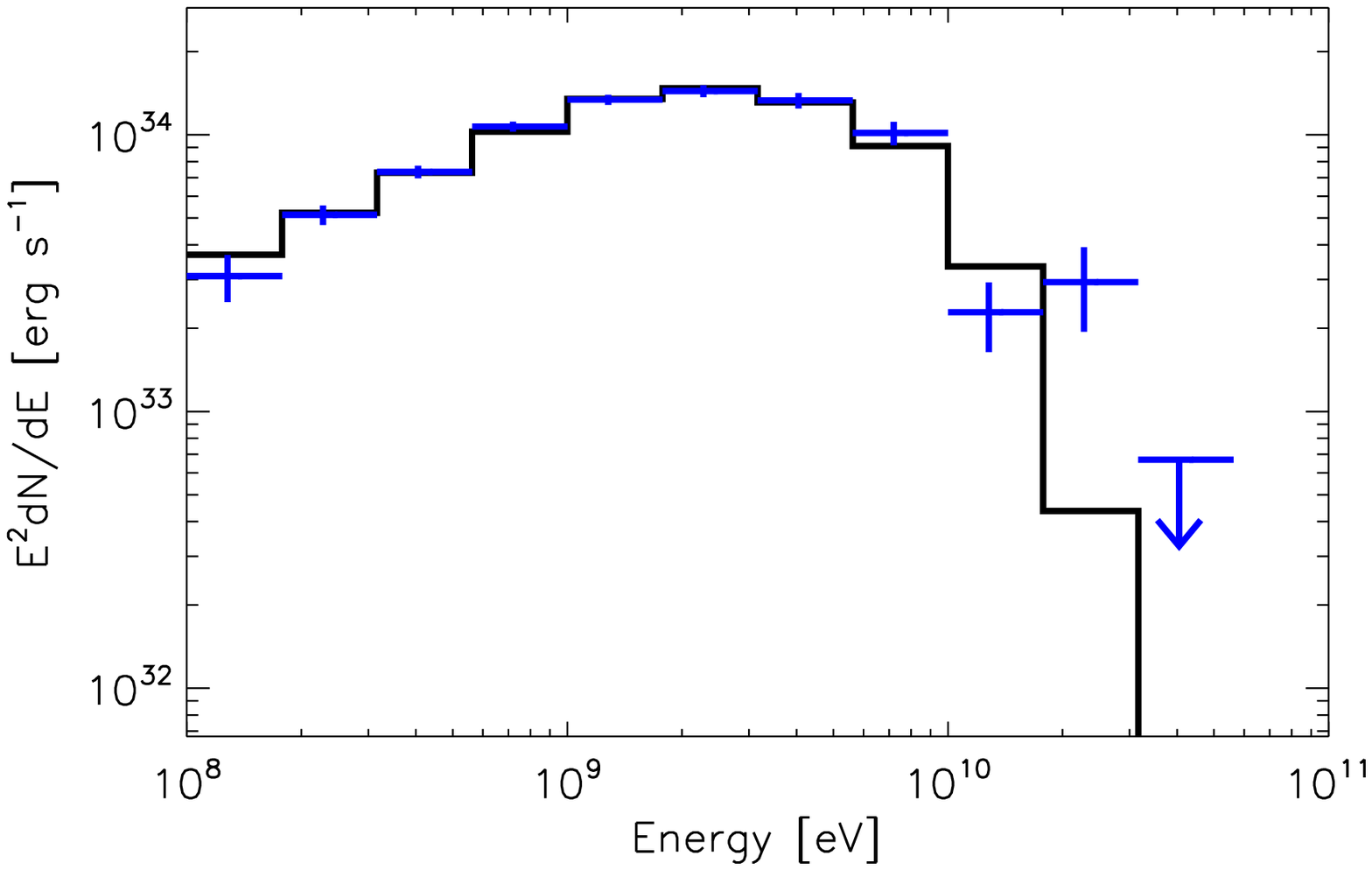}
\put(20,15){\scriptsize J0614-3329}
\end{overpic}
\begin{overpic}[width=0.32\textwidth]{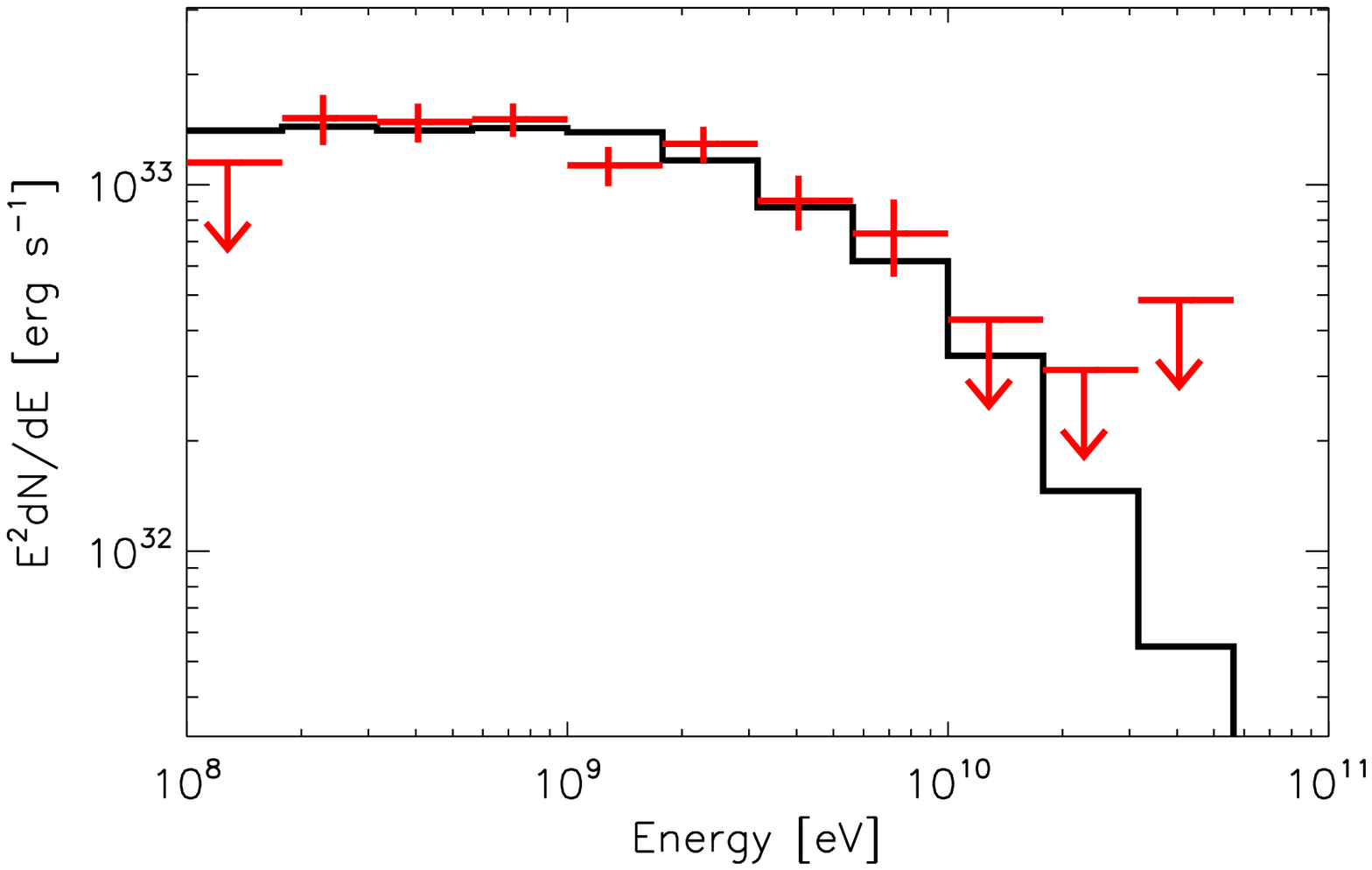}
\put(20,15){\scriptsize J0631+1036}
\end{overpic}
\begin{overpic}[width=0.32\textwidth]{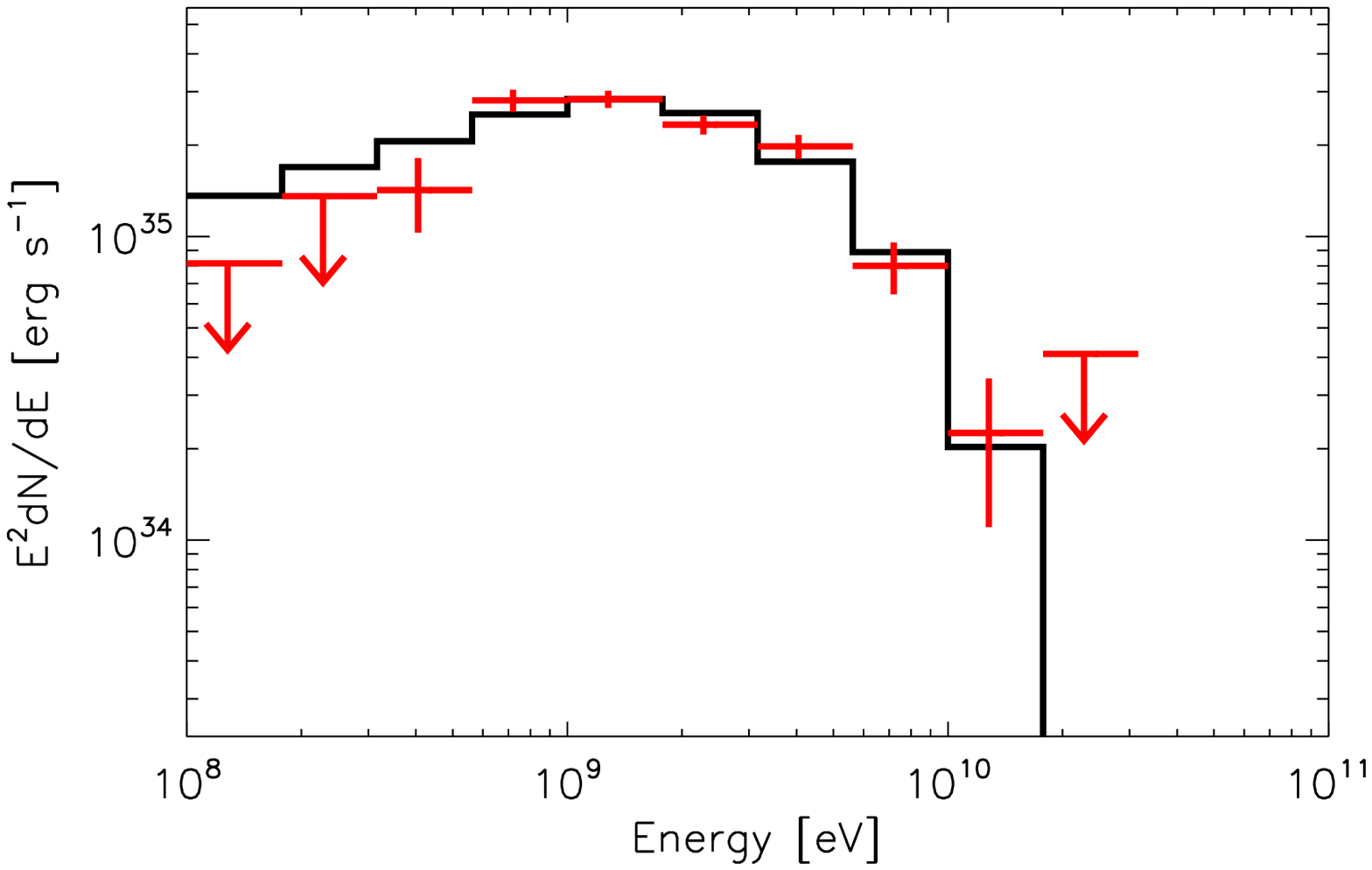}
\put(20,15){\scriptsize J0633+0632}
\end{overpic}
\begin{overpic}[width=0.32\textwidth]{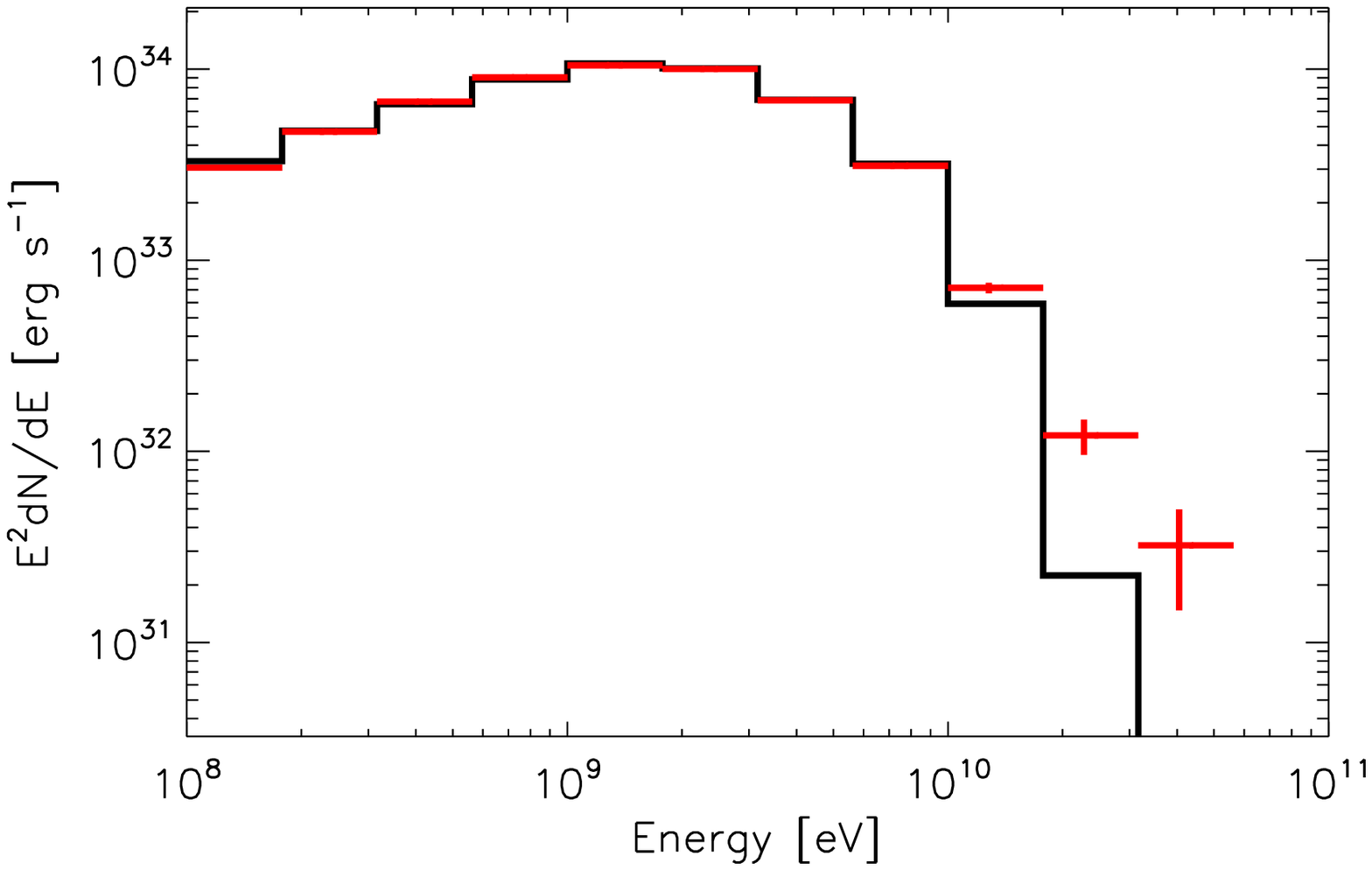}
\put(20,15){\scriptsize J0633+1746 (Geminga)}
\end{overpic}
\begin{overpic}[width=0.32\textwidth]{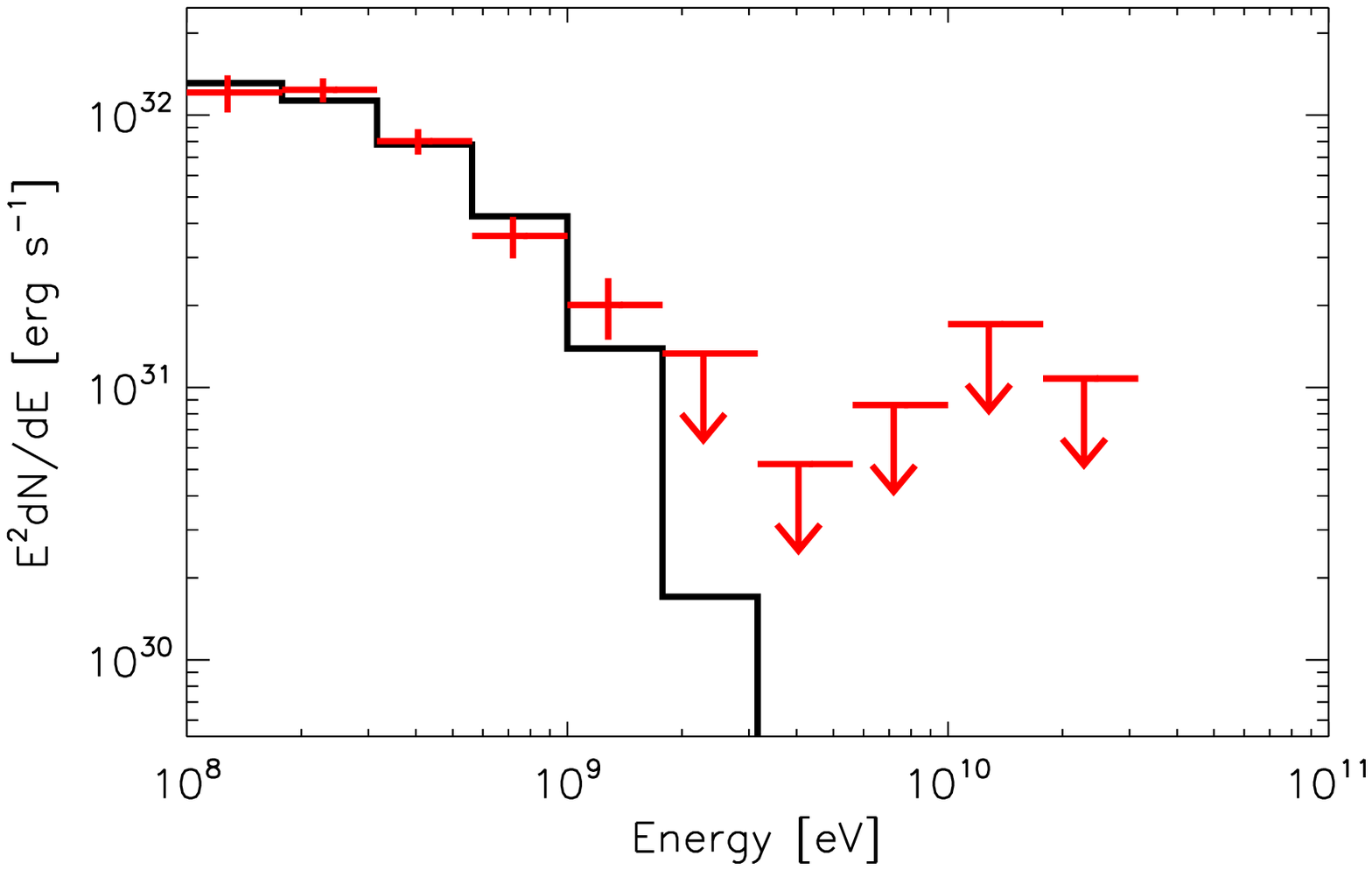}
\put(20,15){\scriptsize J0659+1414}
\end{overpic}
\end{center}
\caption{Fits of the {\it Fermi}-LAT spectral data and models for the pulsars considered in the sample (I). We plot the binned functions $E^2dN/dE \equiv E_{\rm cent}^2 L^{\rm bin}$, in units erg~s$^{-1}$, both for the theoretical models and data. Red and blue lines indicate YPs and MSPs, respectively.}
\label{fig:best_fit1}
\end{figure*}

\begin{figure*}
\begin{center}
\begin{overpic}[width=0.32\textwidth]{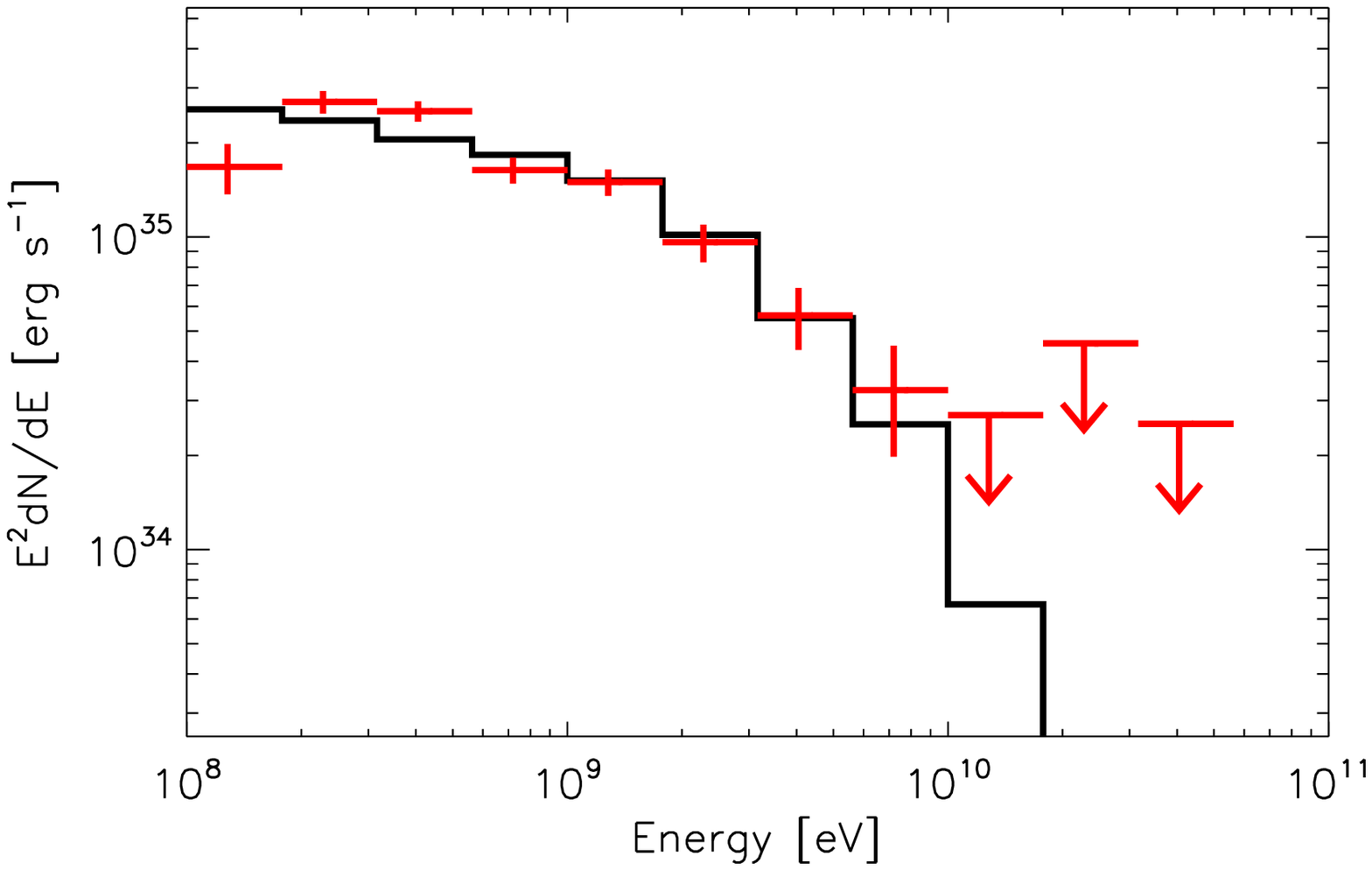}
\put(20,15){\scriptsize J0734-1559}
\end{overpic}
\begin{overpic}[width=0.32\textwidth]{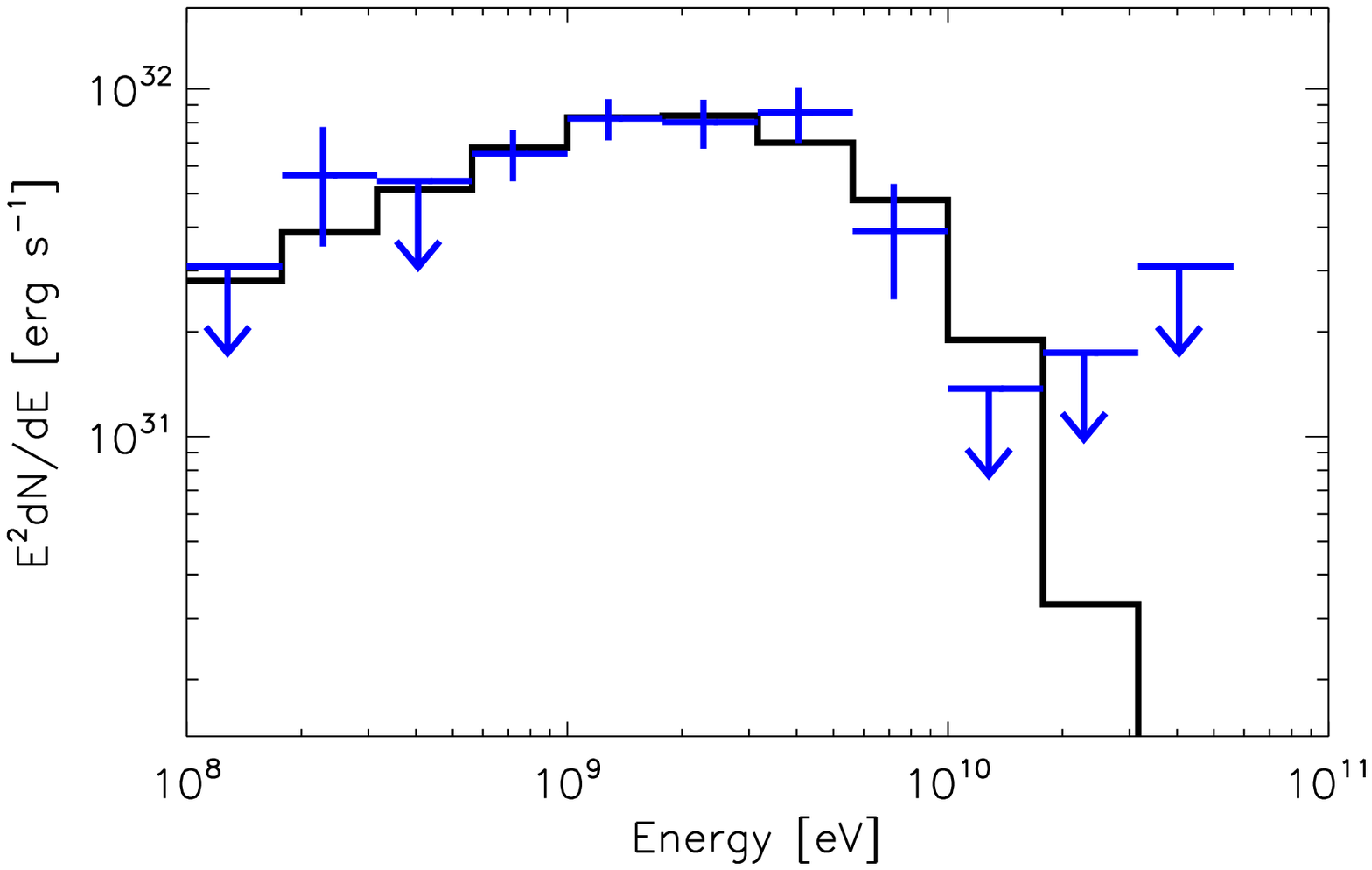}
\put(20,15){\scriptsize J0751+1807}
\end{overpic}
\begin{overpic}[width=0.32\textwidth]{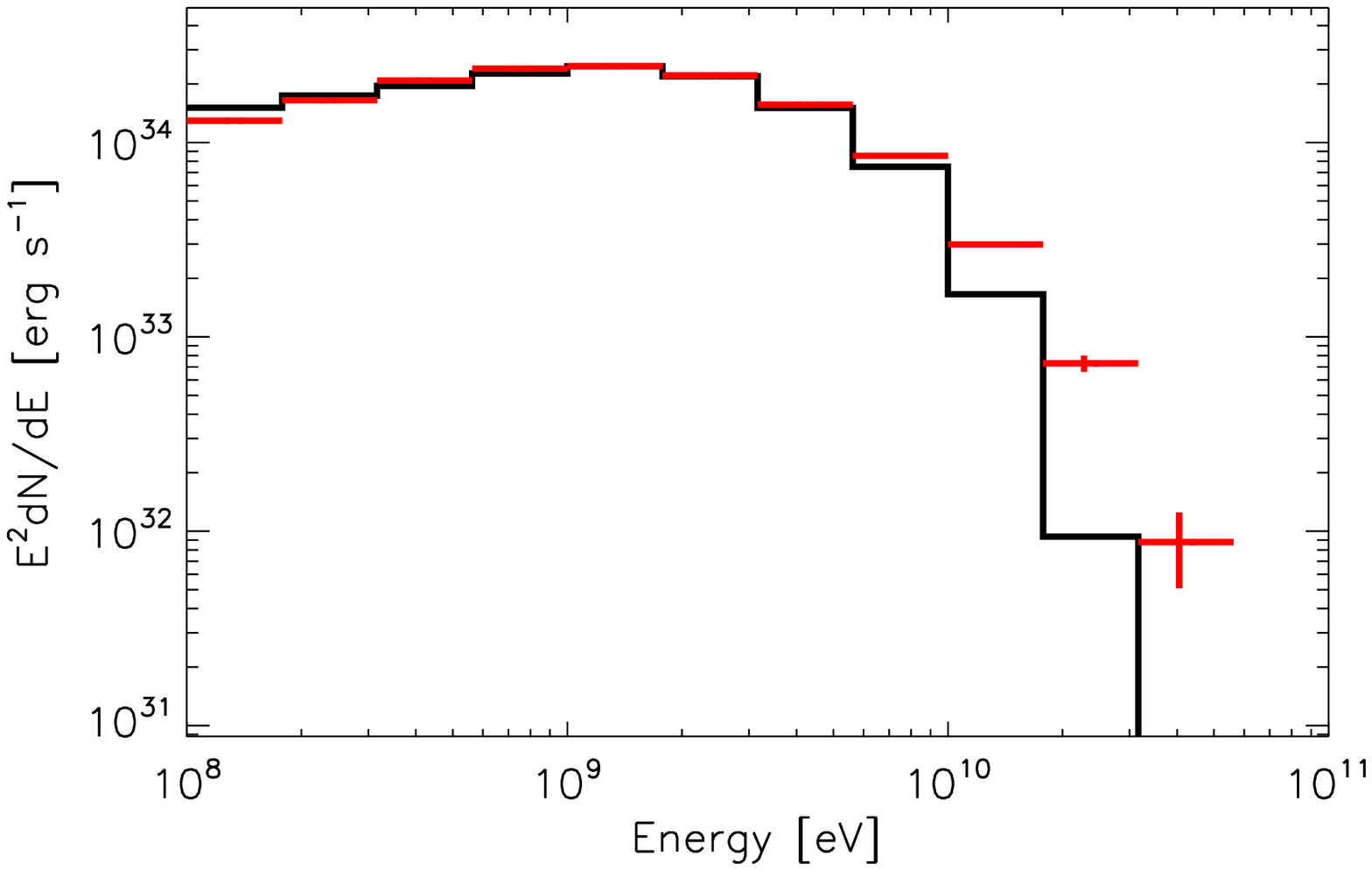}
\put(20,15){\scriptsize J0835-4510 (Vela)}
\end{overpic}
\begin{overpic}[width=0.32\textwidth]{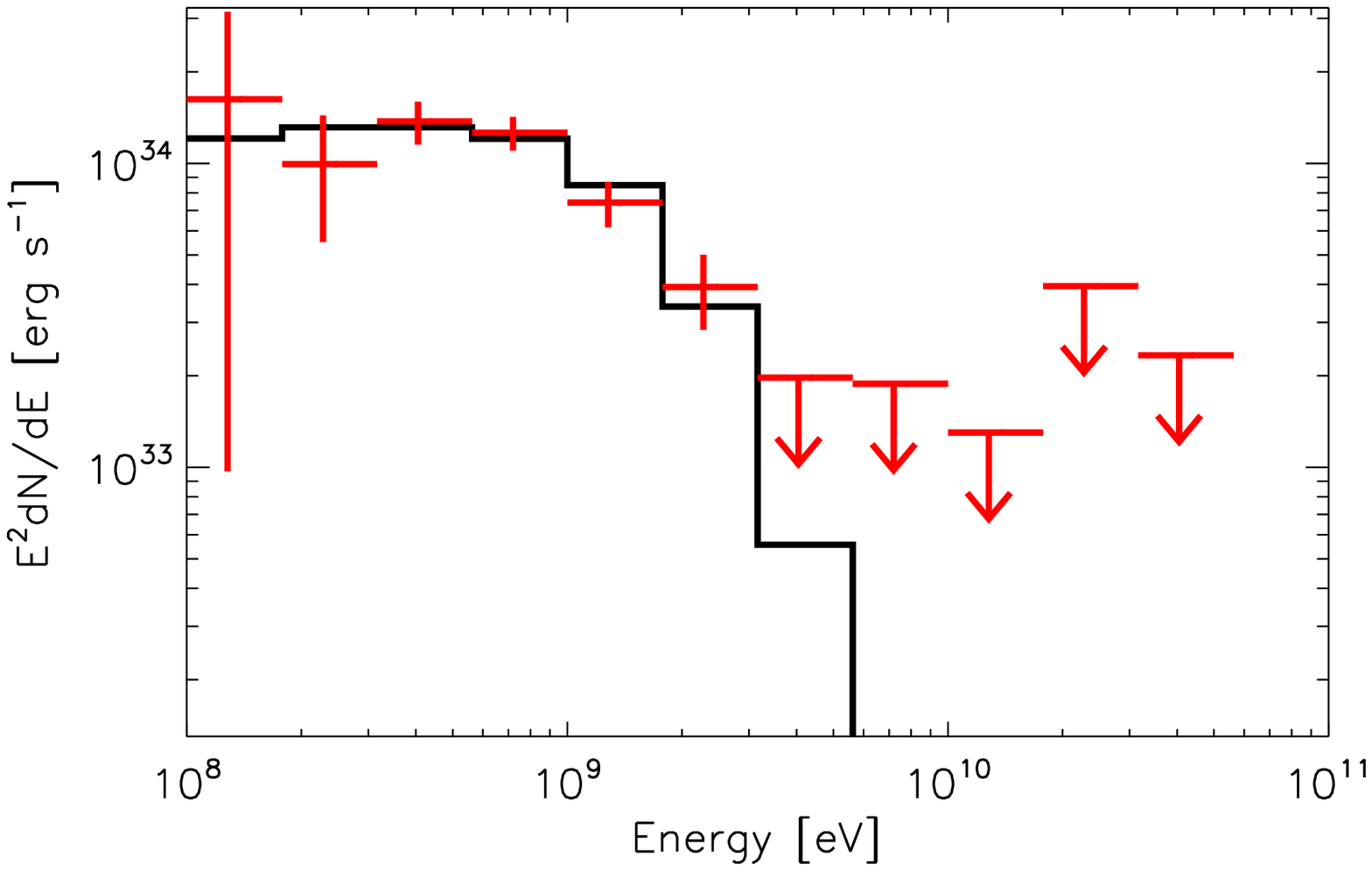}
\put(20,15){\scriptsize J0908-4913}
\end{overpic}
\begin{overpic}[width=0.32\textwidth]{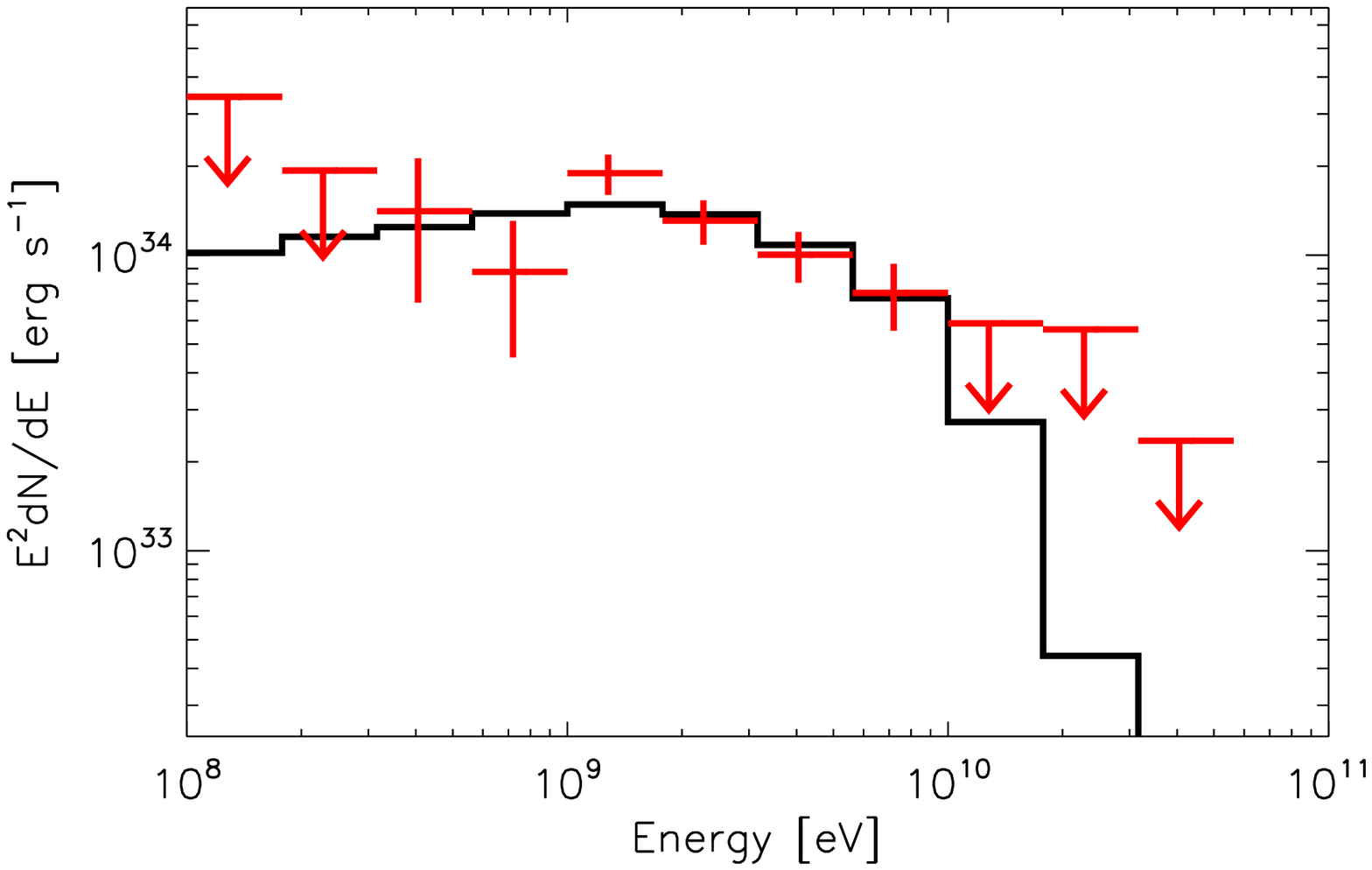}
\put(20,15){\scriptsize J1016-5857}
\end{overpic}
\begin{overpic}[width=0.32\textwidth]{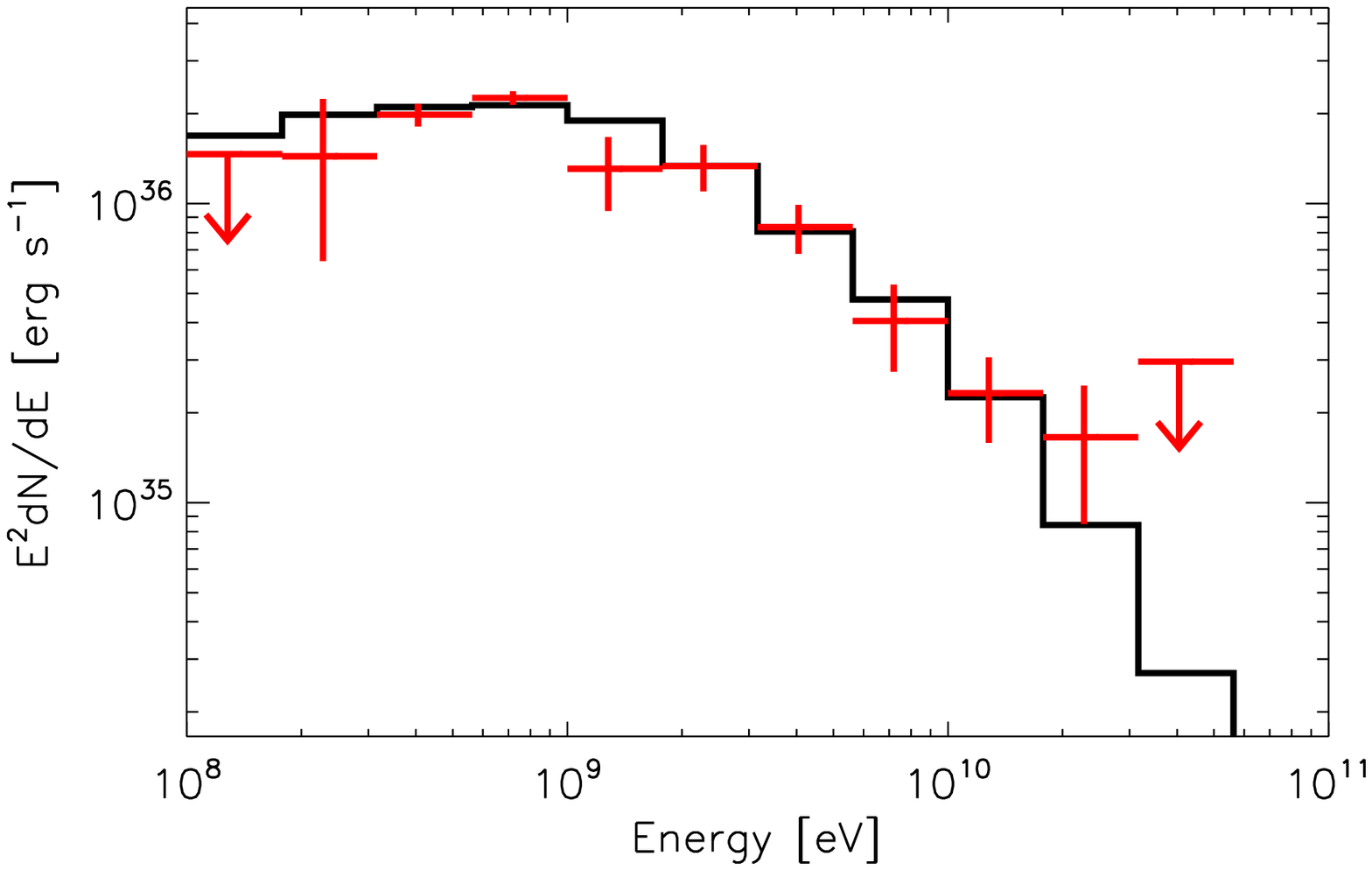}
\put(20,15){\scriptsize J1023-5746}
\end{overpic}
\begin{overpic}[width=0.32\textwidth]{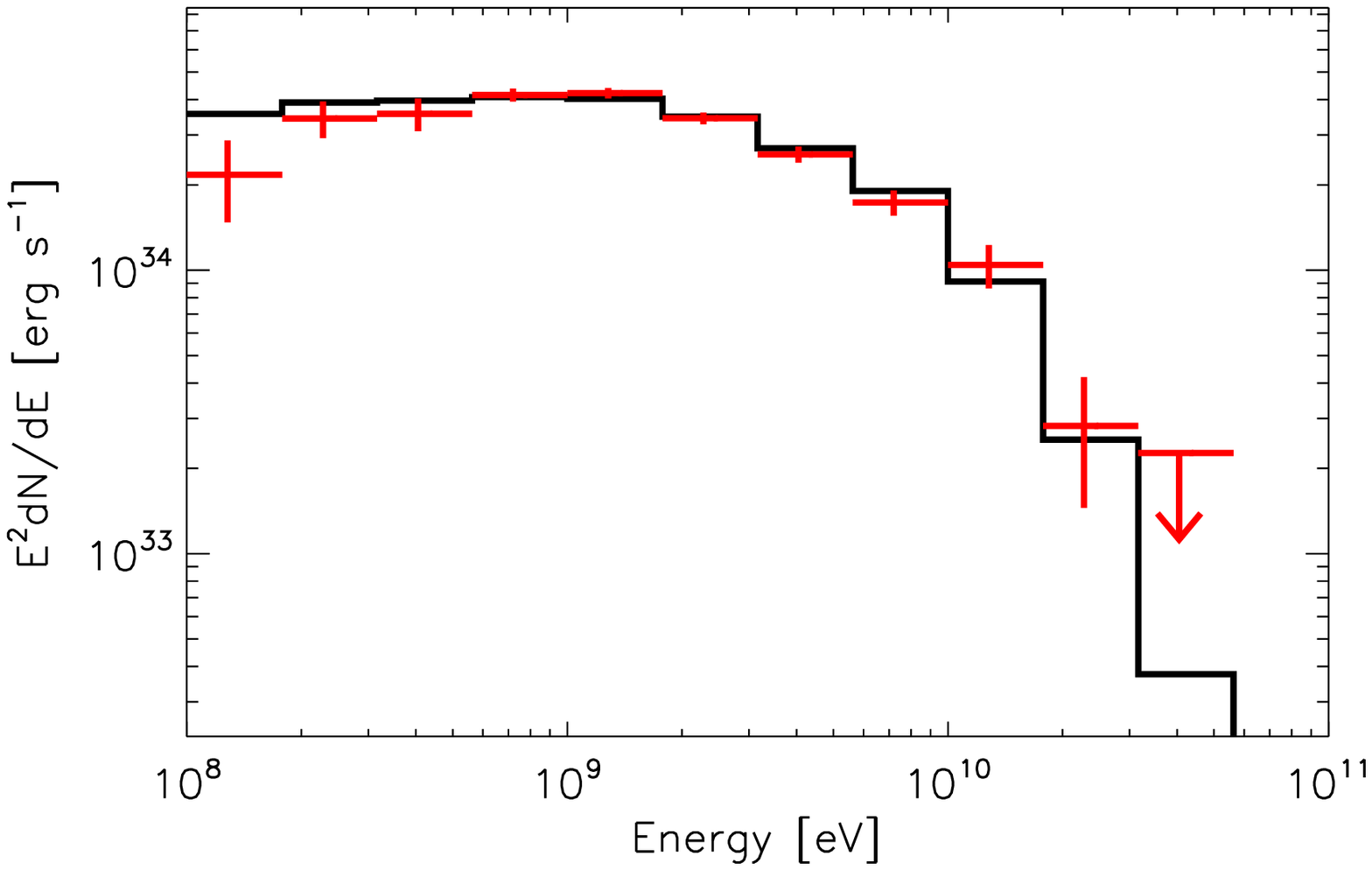}
\put(20,15){\scriptsize J1028-5819}
\end{overpic}
\begin{overpic}[width=0.32\textwidth]{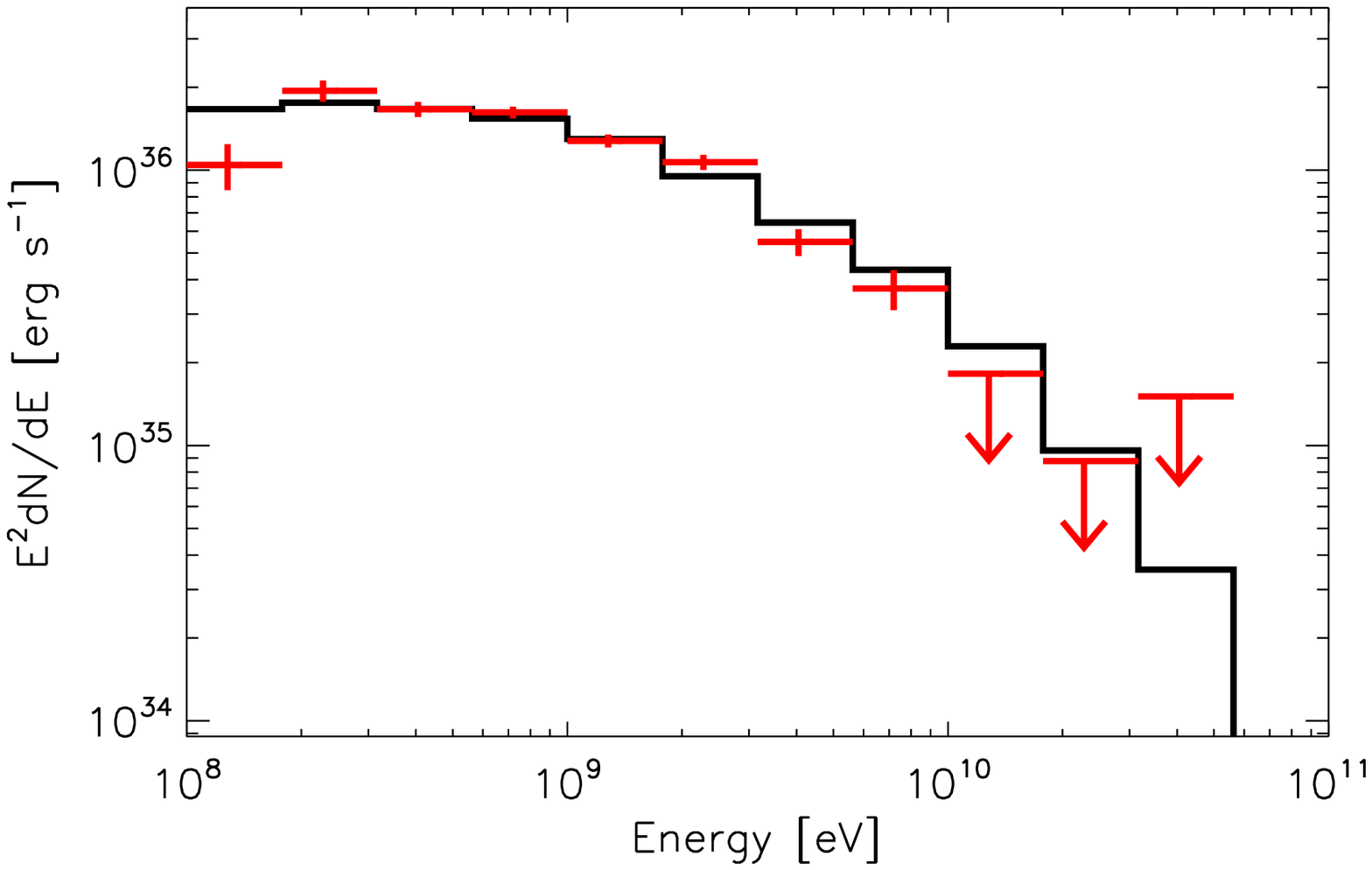}
\put(20,15){\scriptsize J1044-5737}
\end{overpic}
\begin{overpic}[width=0.32\textwidth]{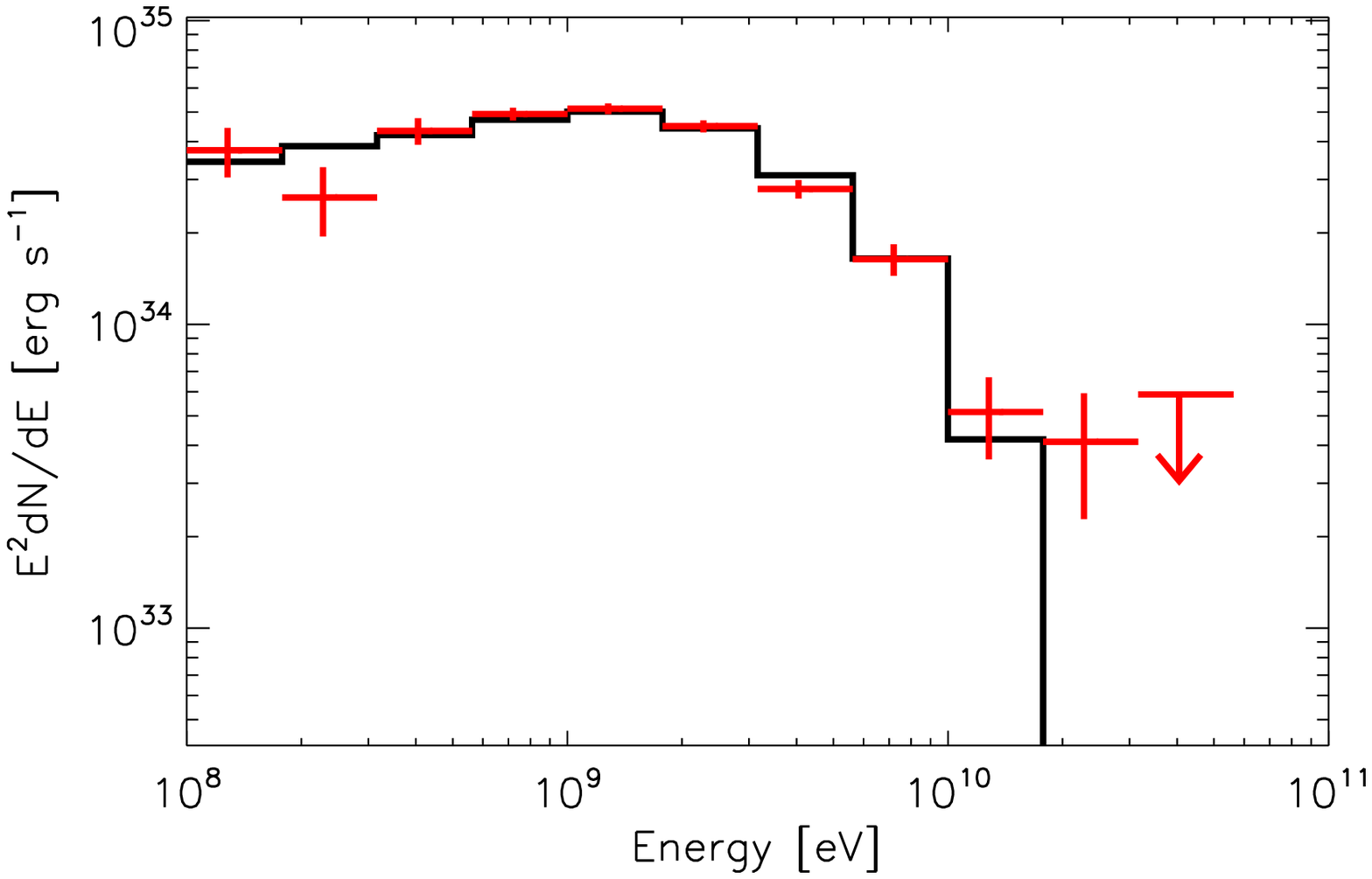}
\put(20,15){\scriptsize J1048-5832}
\end{overpic}
\begin{overpic}[width=0.32\textwidth]{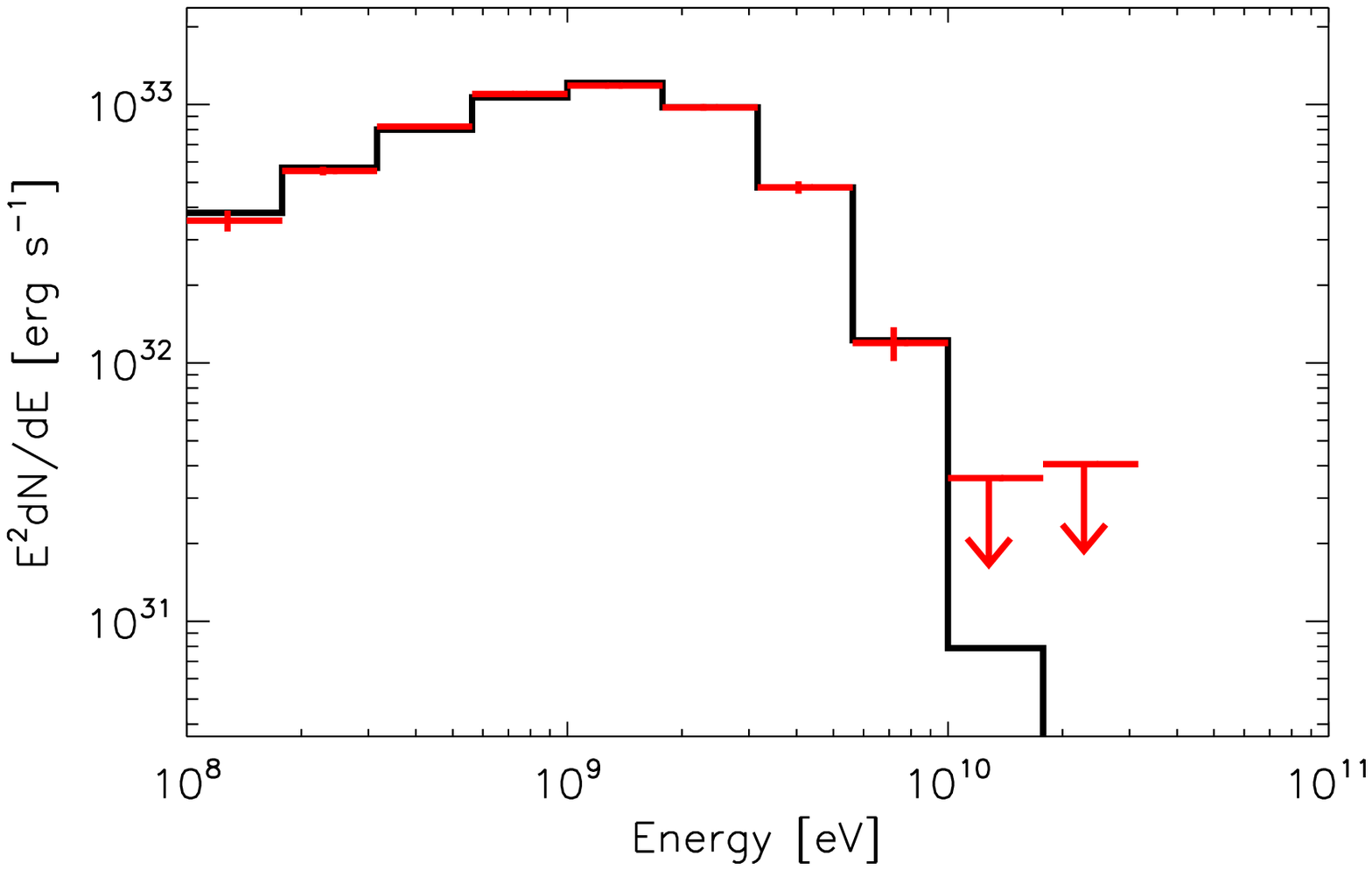}
\put(20,15){\scriptsize J1057-5226}
\end{overpic}
\begin{overpic}[width=0.32\textwidth]{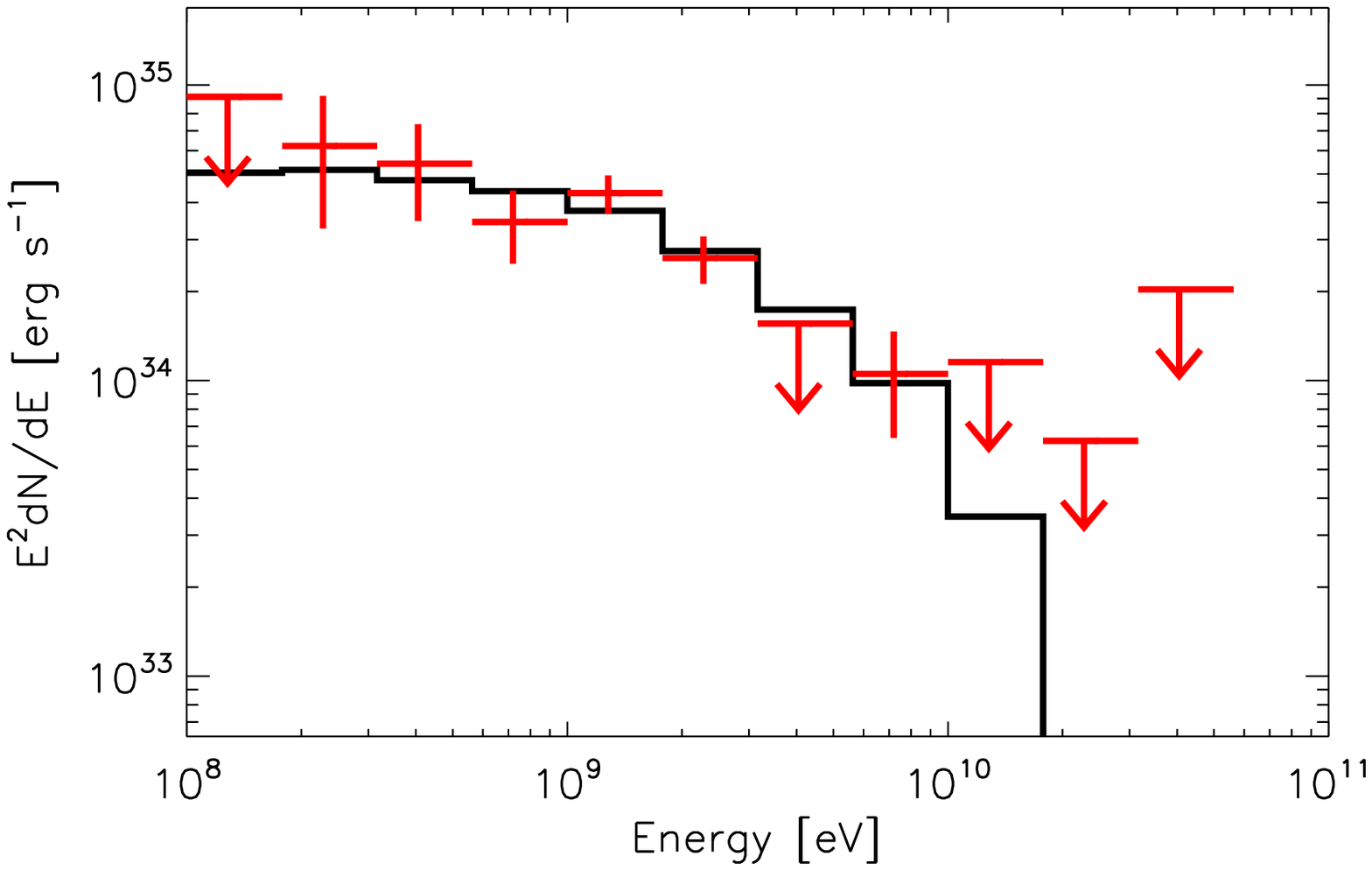}
\put(20,15){\scriptsize J1105-6107}
\end{overpic}
\begin{overpic}[width=0.32\textwidth]{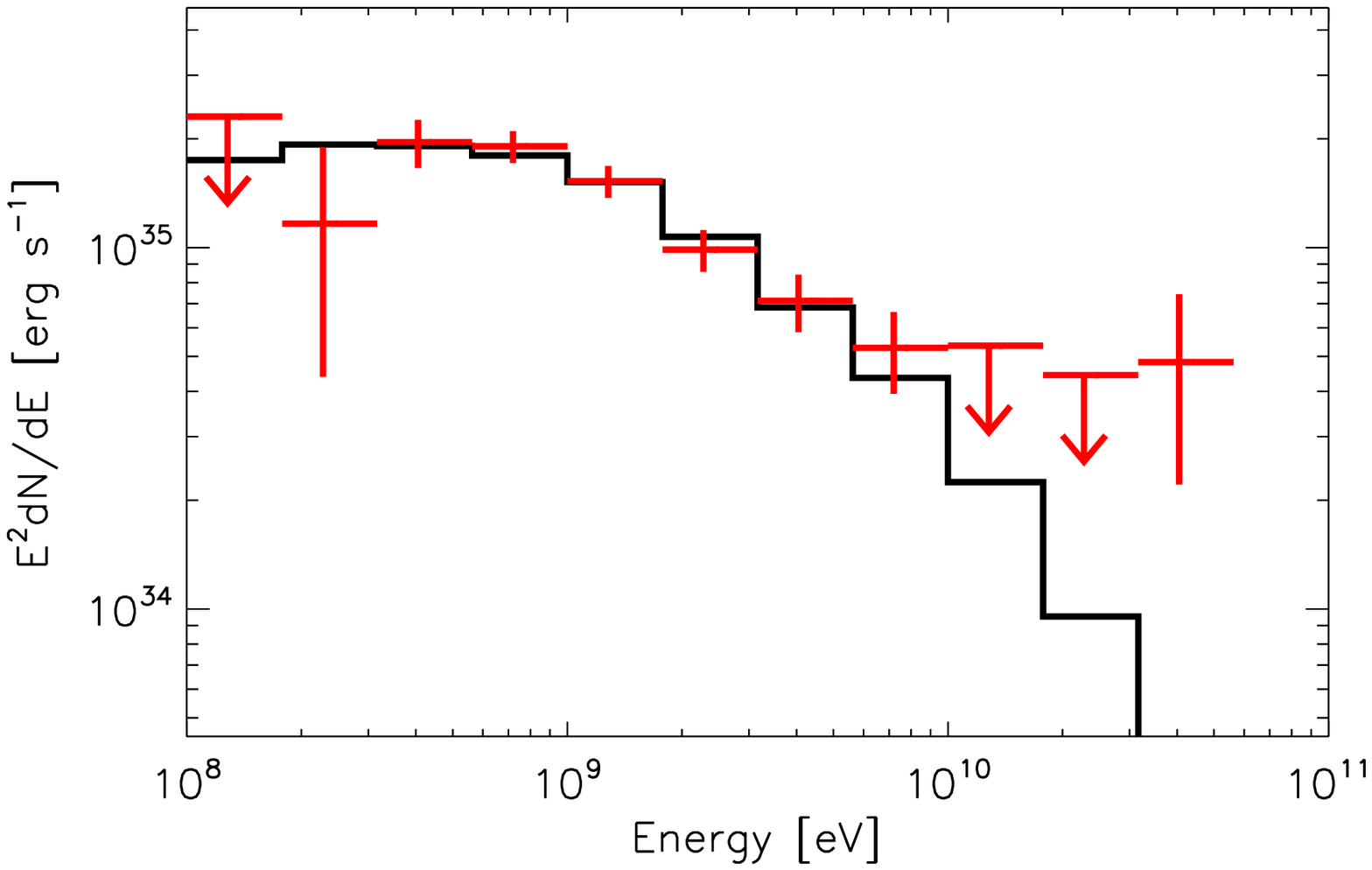}
\put(20,15){\scriptsize J1119-6127}
\end{overpic}
\begin{overpic}[width=0.32\textwidth]{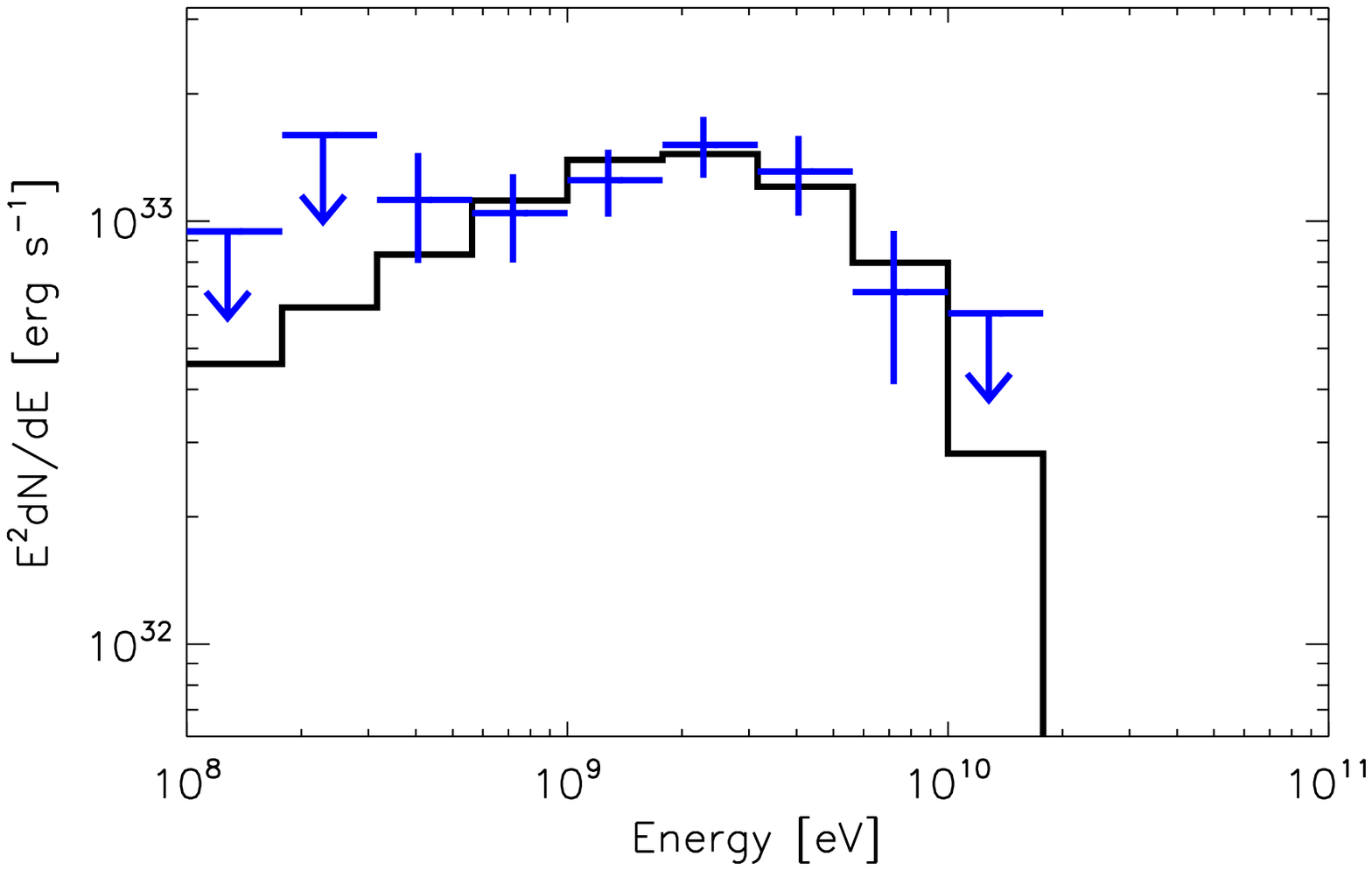}
\put(20,15){\scriptsize J1124-3653}
\end{overpic}
\begin{overpic}[width=0.32\textwidth]{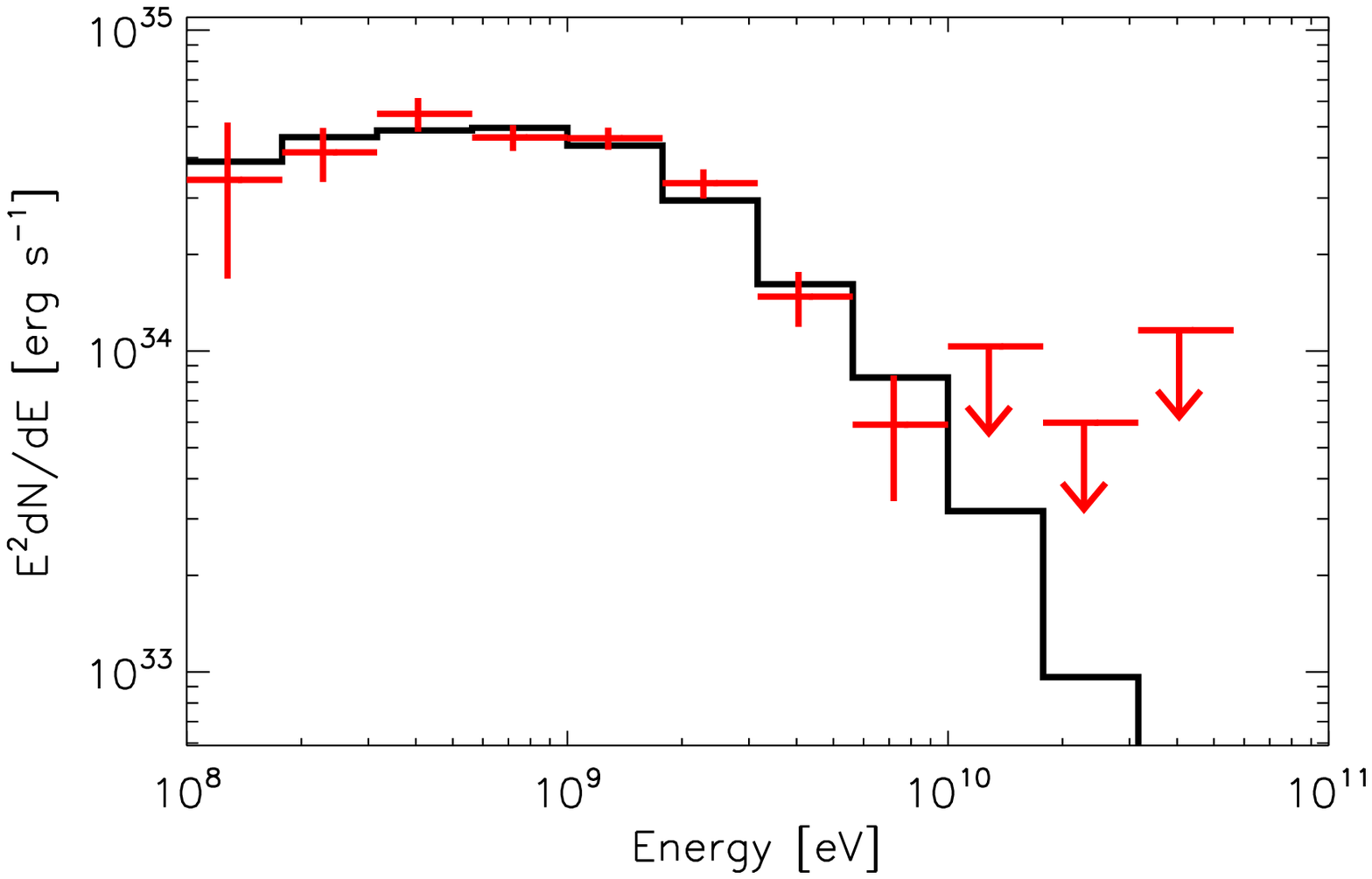}
\put(20,15){\scriptsize J1124-5916}
\end{overpic}
\begin{overpic}[width=0.32\textwidth]{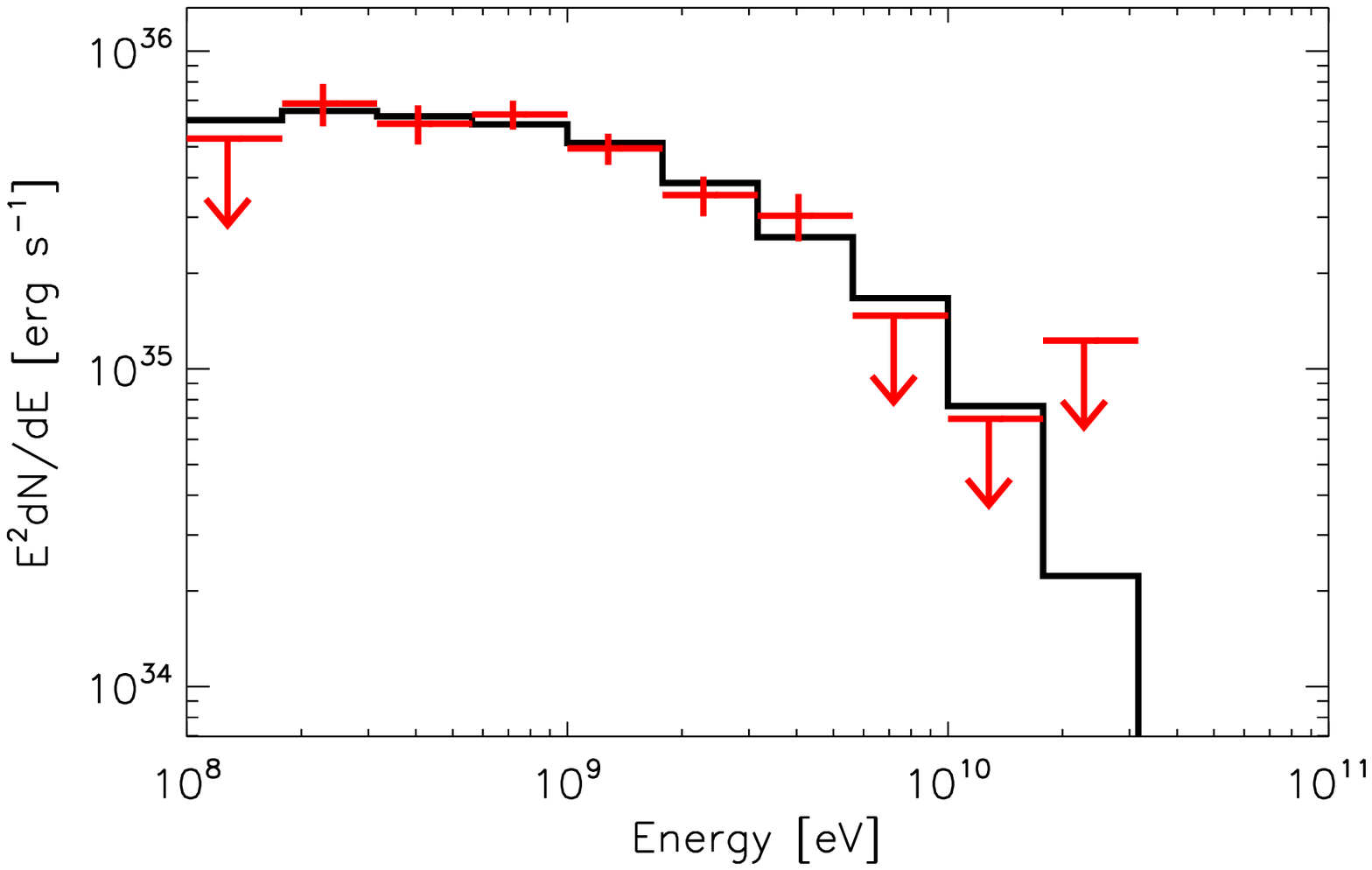}
\put(20,15){\scriptsize J1135-6055}
\end{overpic}
\begin{overpic}[width=0.32\textwidth]{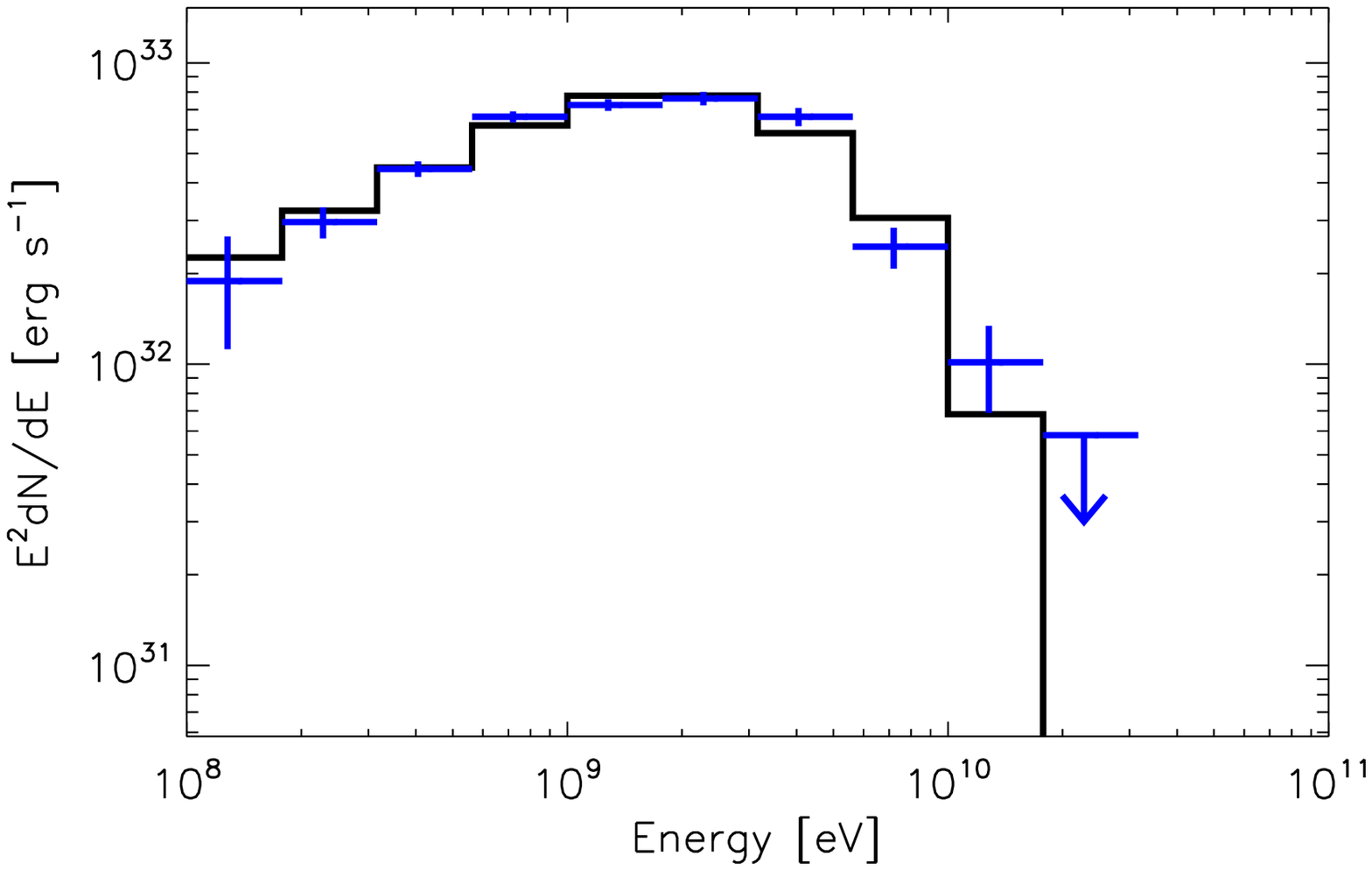}
\put(20,15){\scriptsize J1231-1411}
\end{overpic}
\begin{overpic}[width=0.32\textwidth]{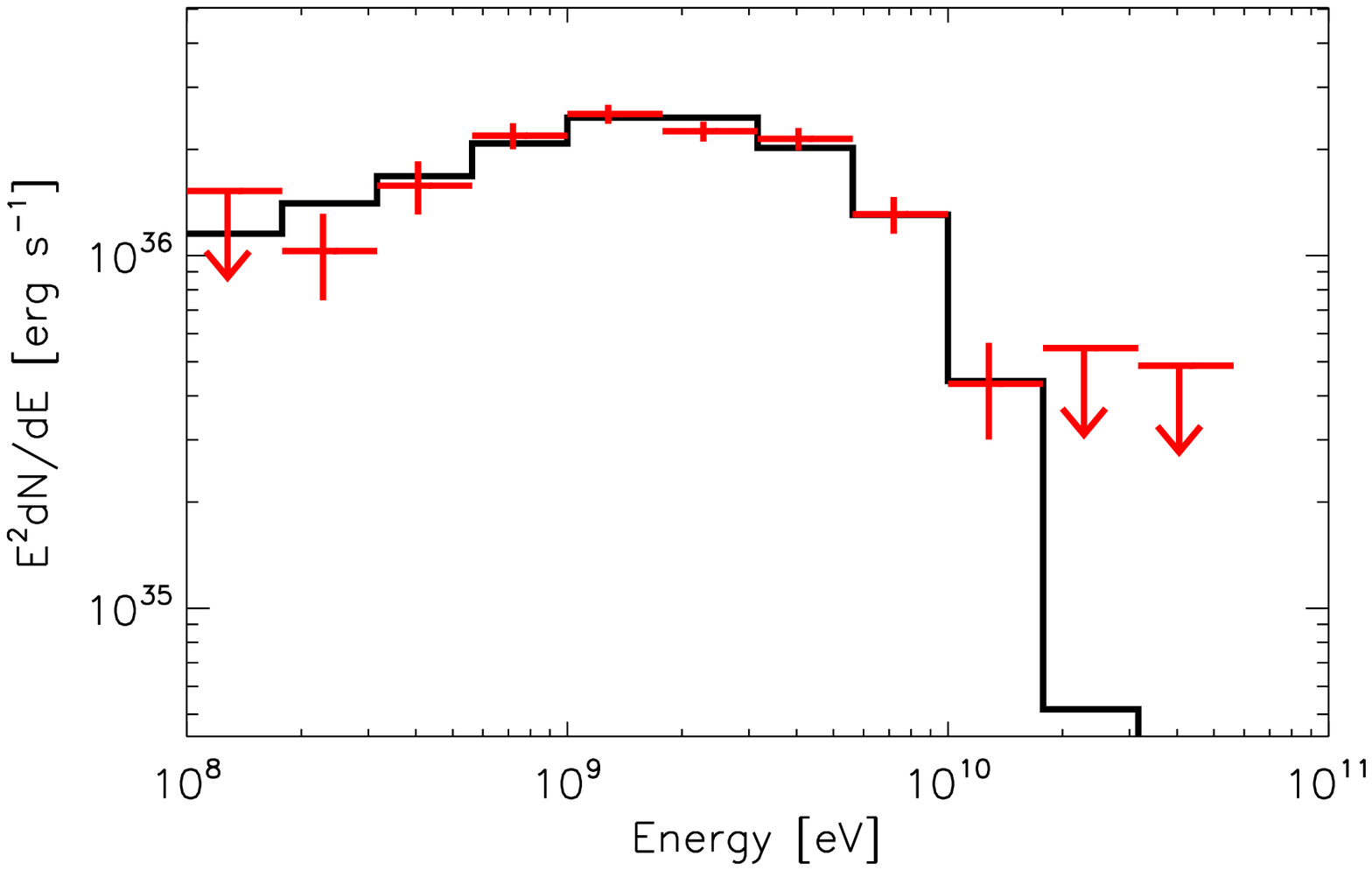}
\put(20,15){\scriptsize J1413-6205}
\end{overpic}
\begin{overpic}[width=0.32\textwidth]{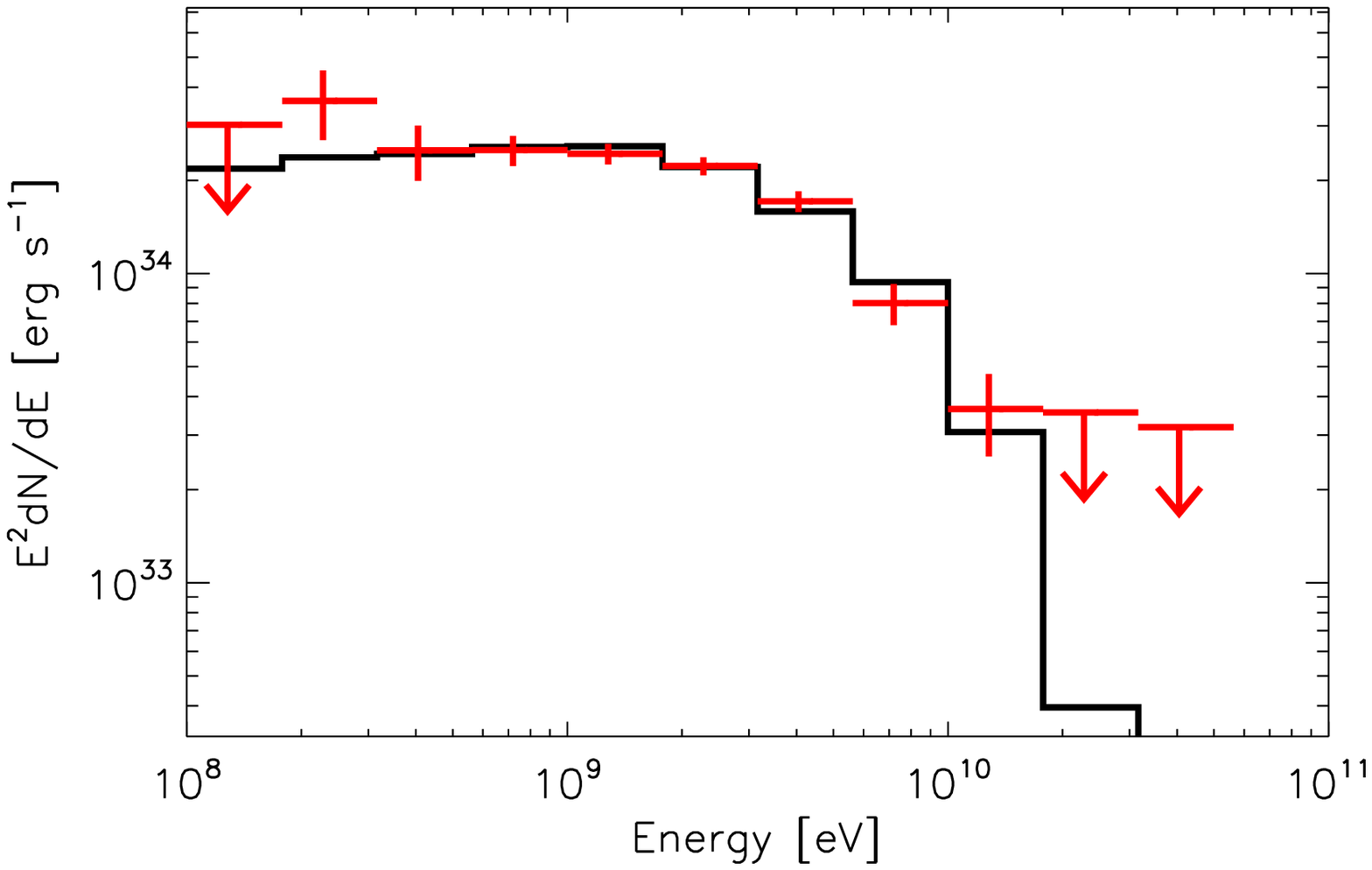}
\put(20,15){\scriptsize J1418-6058}
\end{overpic}
\end{center}
\caption{Fits of the {\it Fermi}-LAT spectral data and models for the pulsars considered in the sample (II).}
\label{fig:best_fit2}
\end{figure*}

\begin{figure*}
\begin{center}
\begin{overpic}[width=0.32\textwidth]{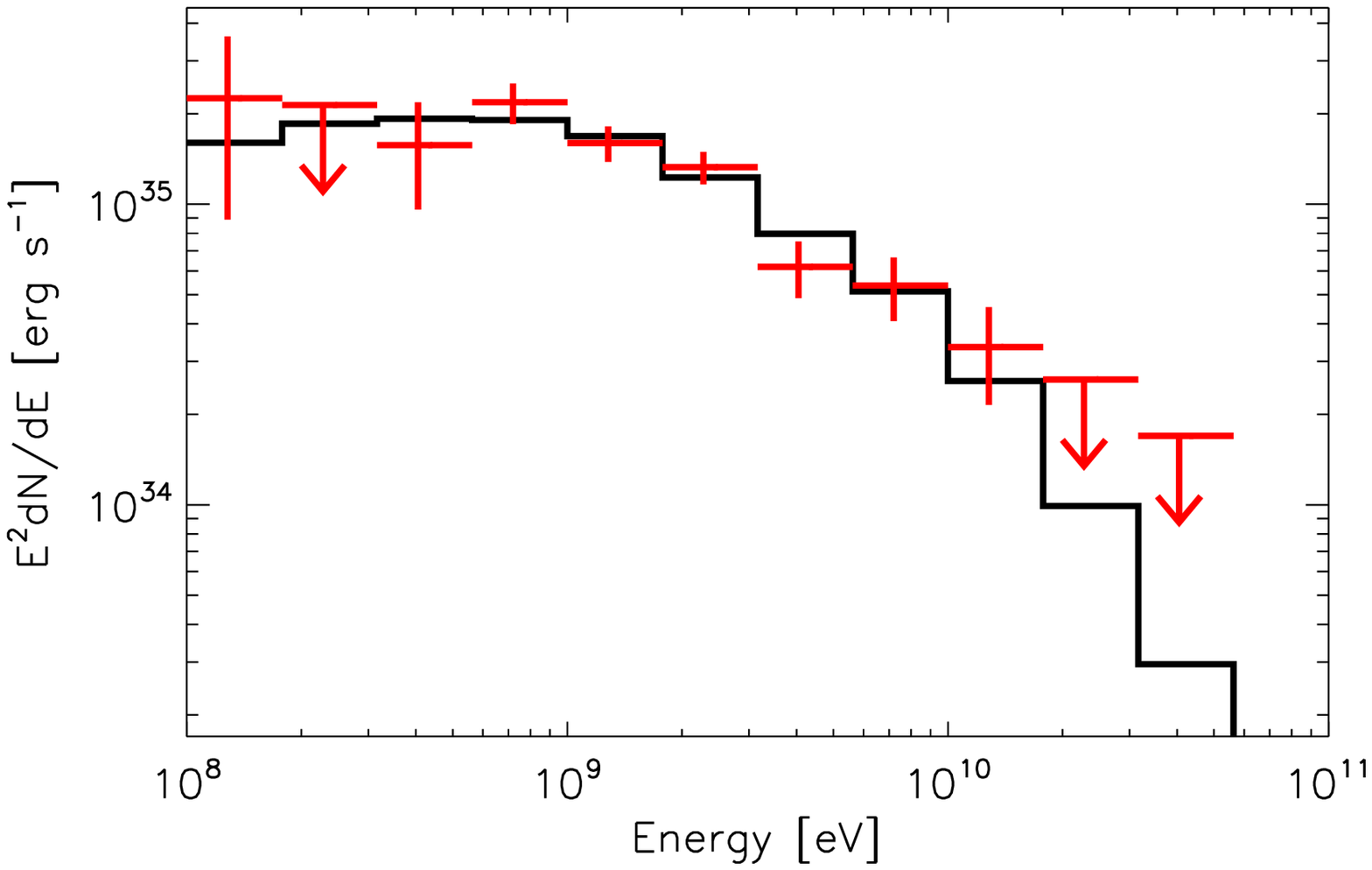}
\put(20,15){\scriptsize J1420-6048}
\end{overpic}
\begin{overpic}[width=0.32\textwidth]{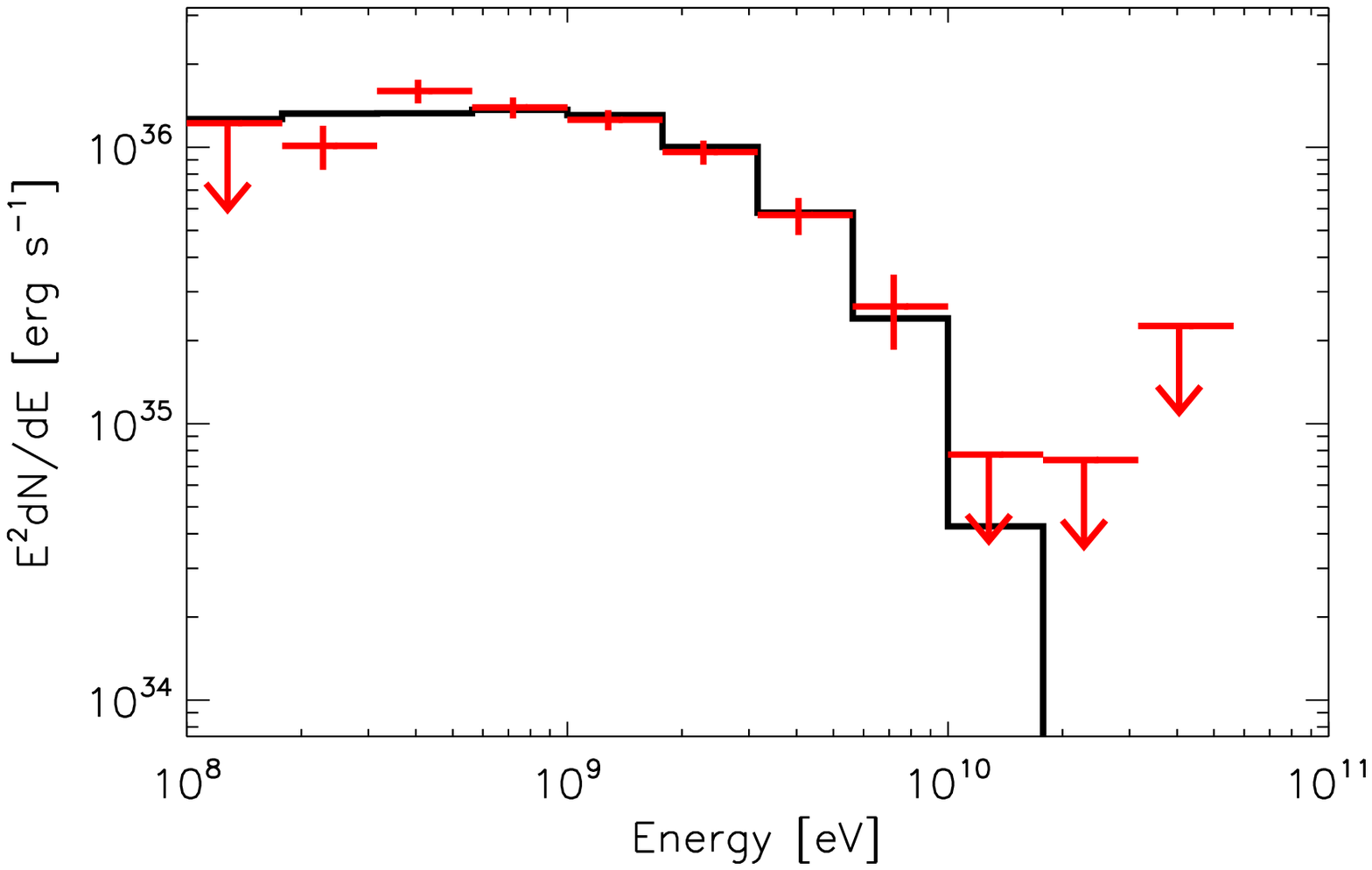}
\put(20,15){\scriptsize J1429-5911}
\end{overpic}
\begin{overpic}[width=0.32\textwidth]{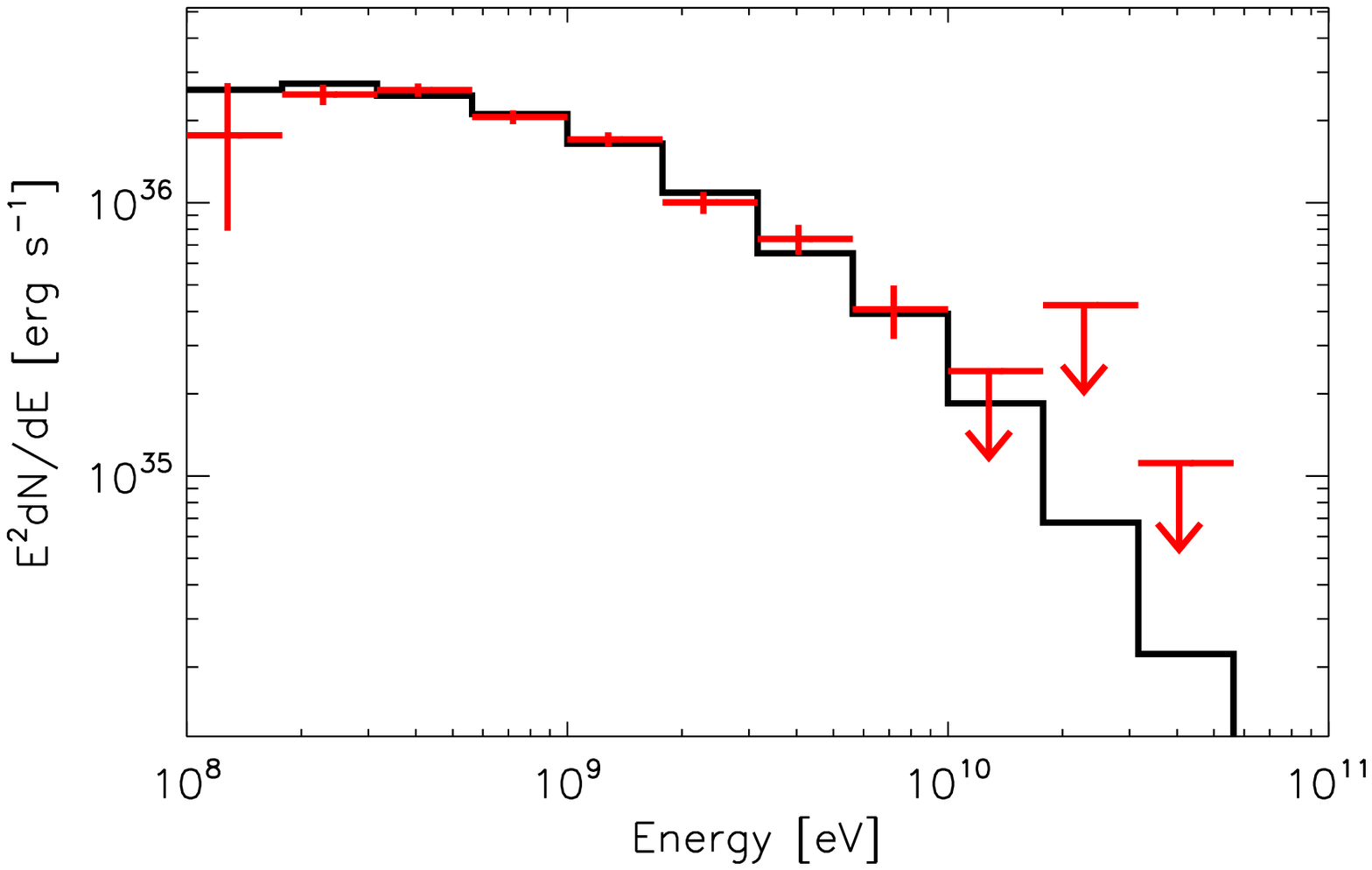}
\put(20,15){\scriptsize J1459-6053}
\end{overpic}
\begin{overpic}[width=0.32\textwidth]{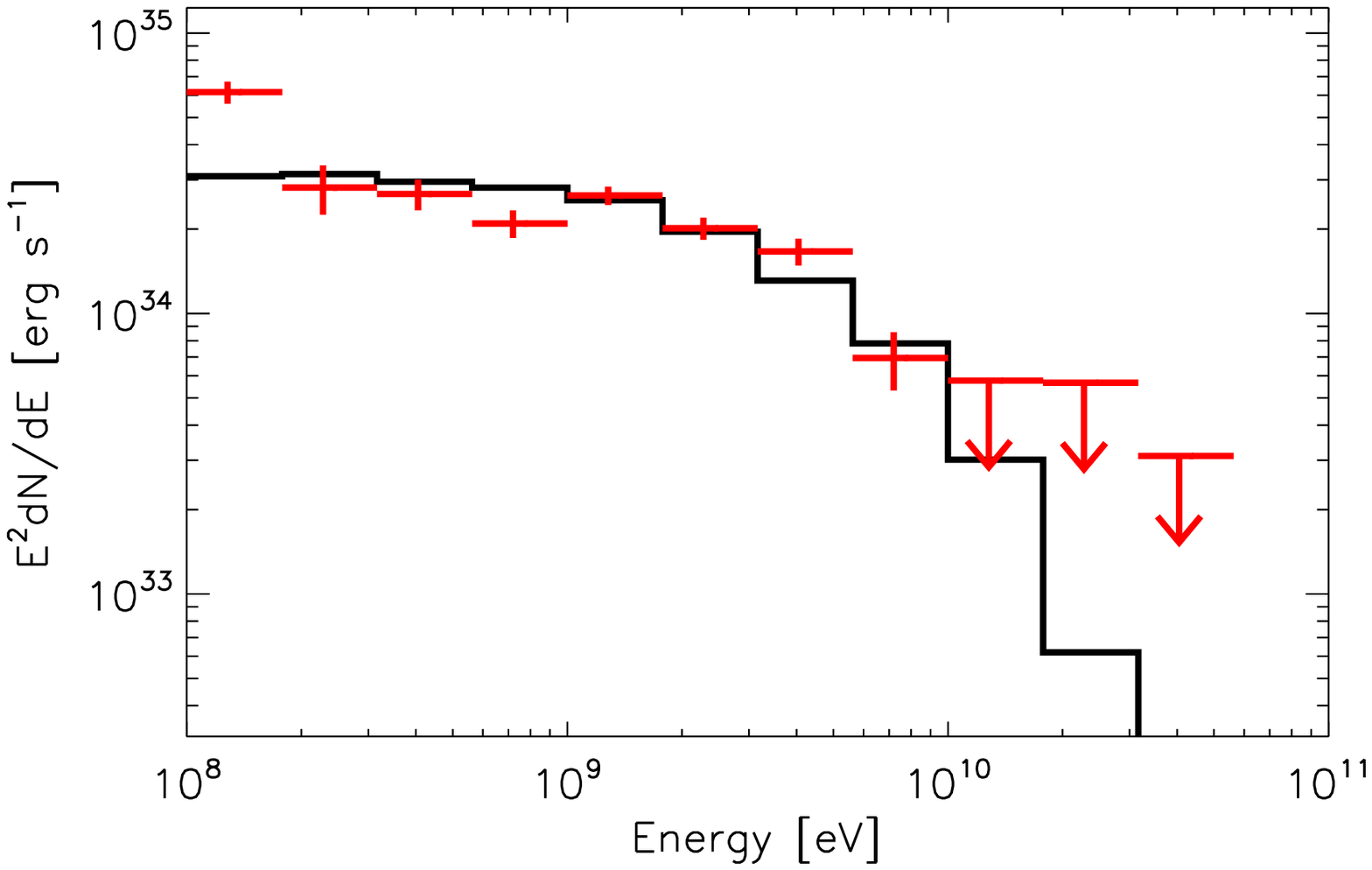}
\put(20,15){\scriptsize J1509-5850}
\end{overpic}
\begin{overpic}[width=0.32\textwidth]{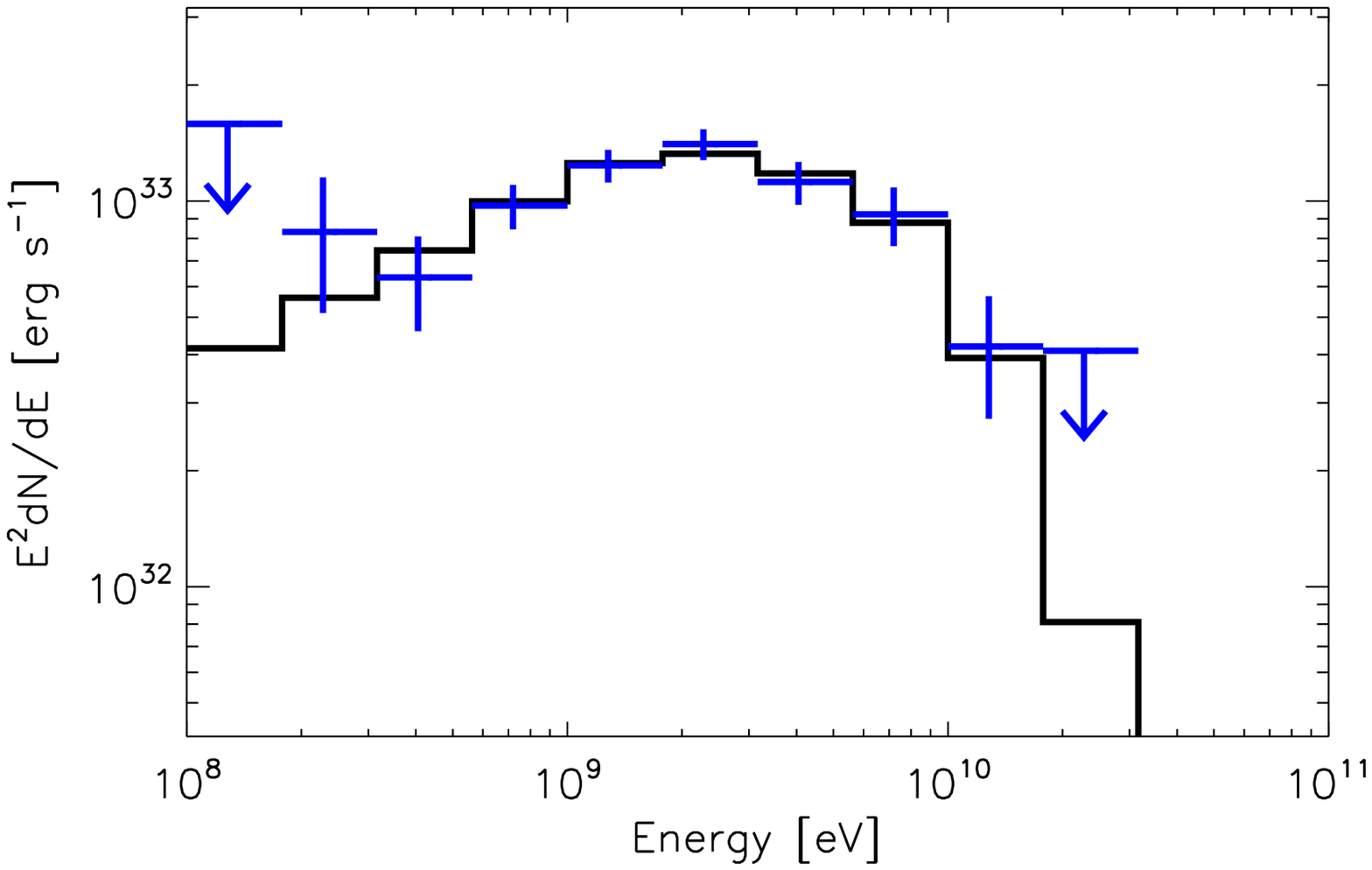}
\put(20,15){\scriptsize J1514-4946}
\end{overpic}
\begin{overpic}[width=0.32\textwidth]{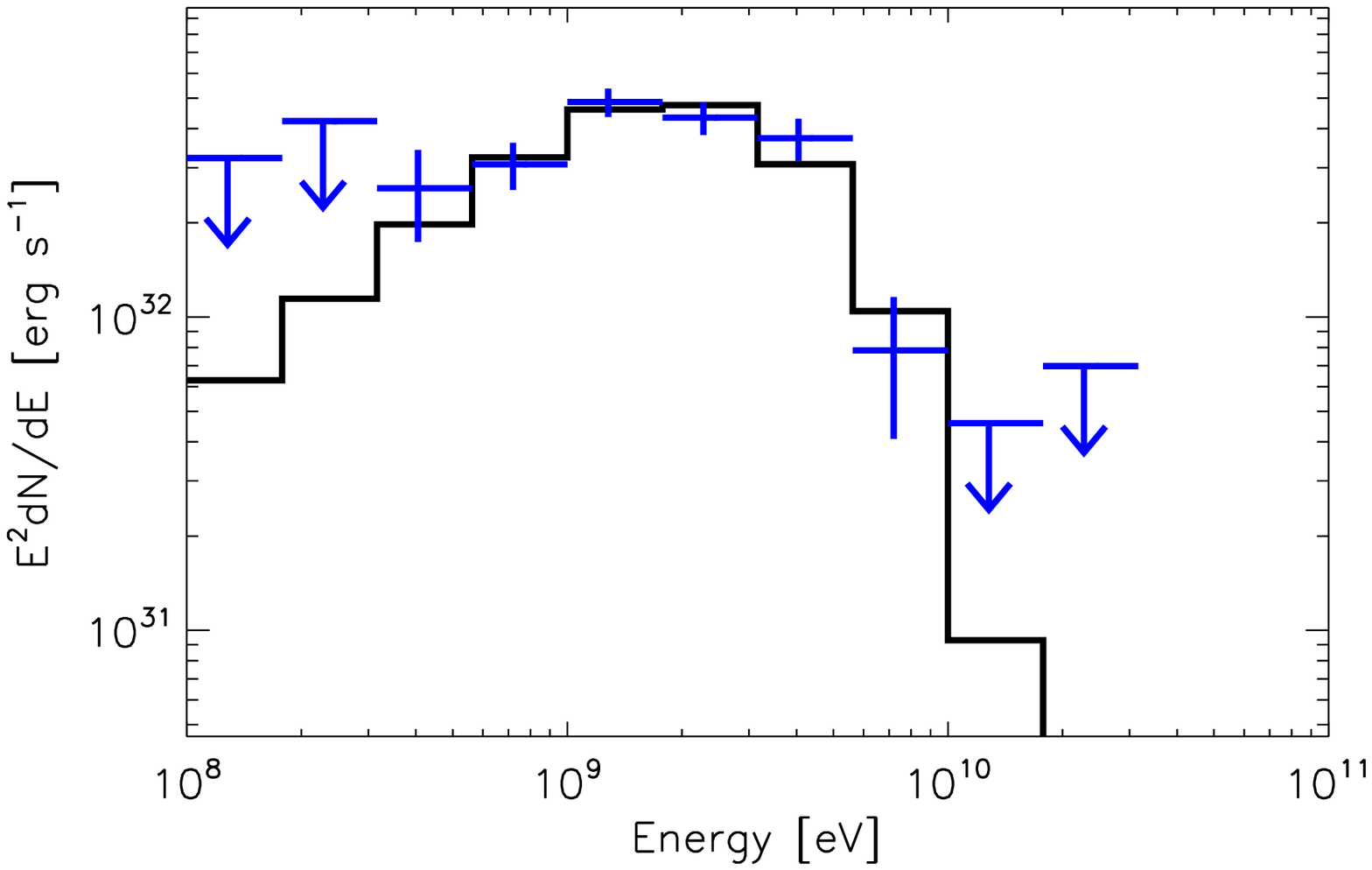}
\put(20,15){\scriptsize J1614-2230}
\end{overpic}
\begin{overpic}[width=0.32\textwidth]{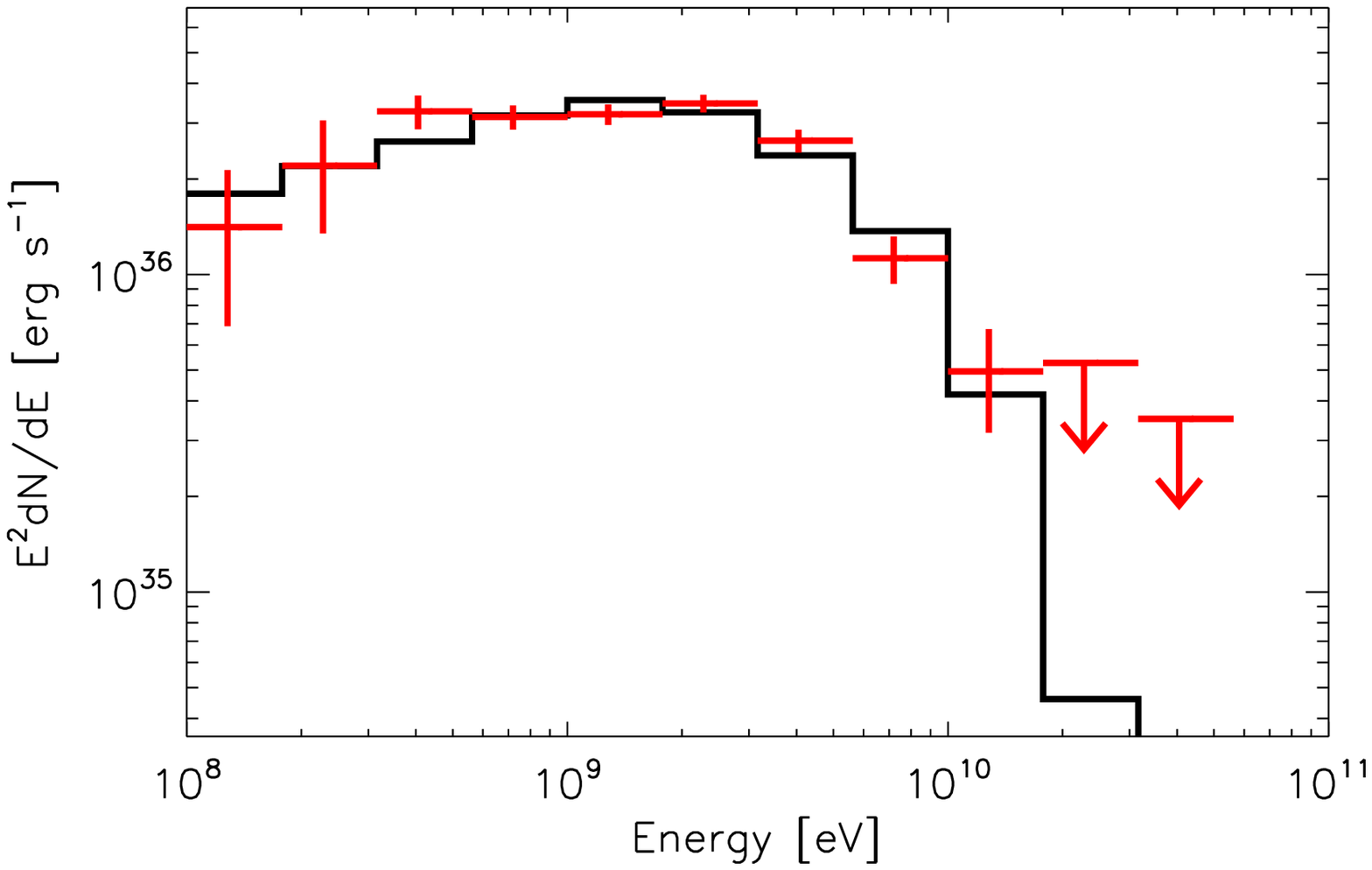}
\put(20,15){\scriptsize J1620-4927}
\end{overpic}
\begin{overpic}[width=0.32\textwidth]{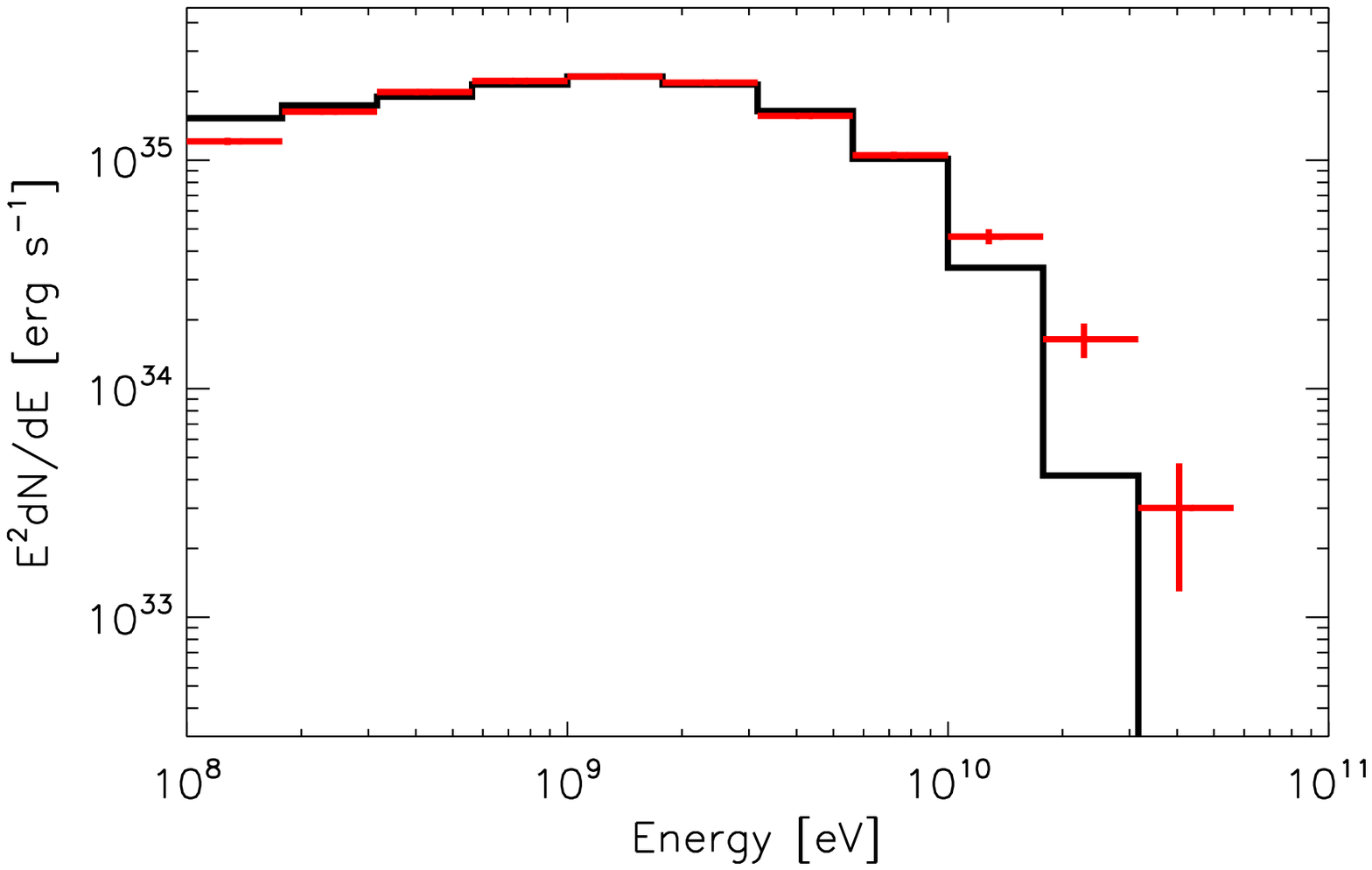}
\put(20,15){\scriptsize J1709-4429}
\end{overpic}
\begin{overpic}[width=0.32\textwidth]{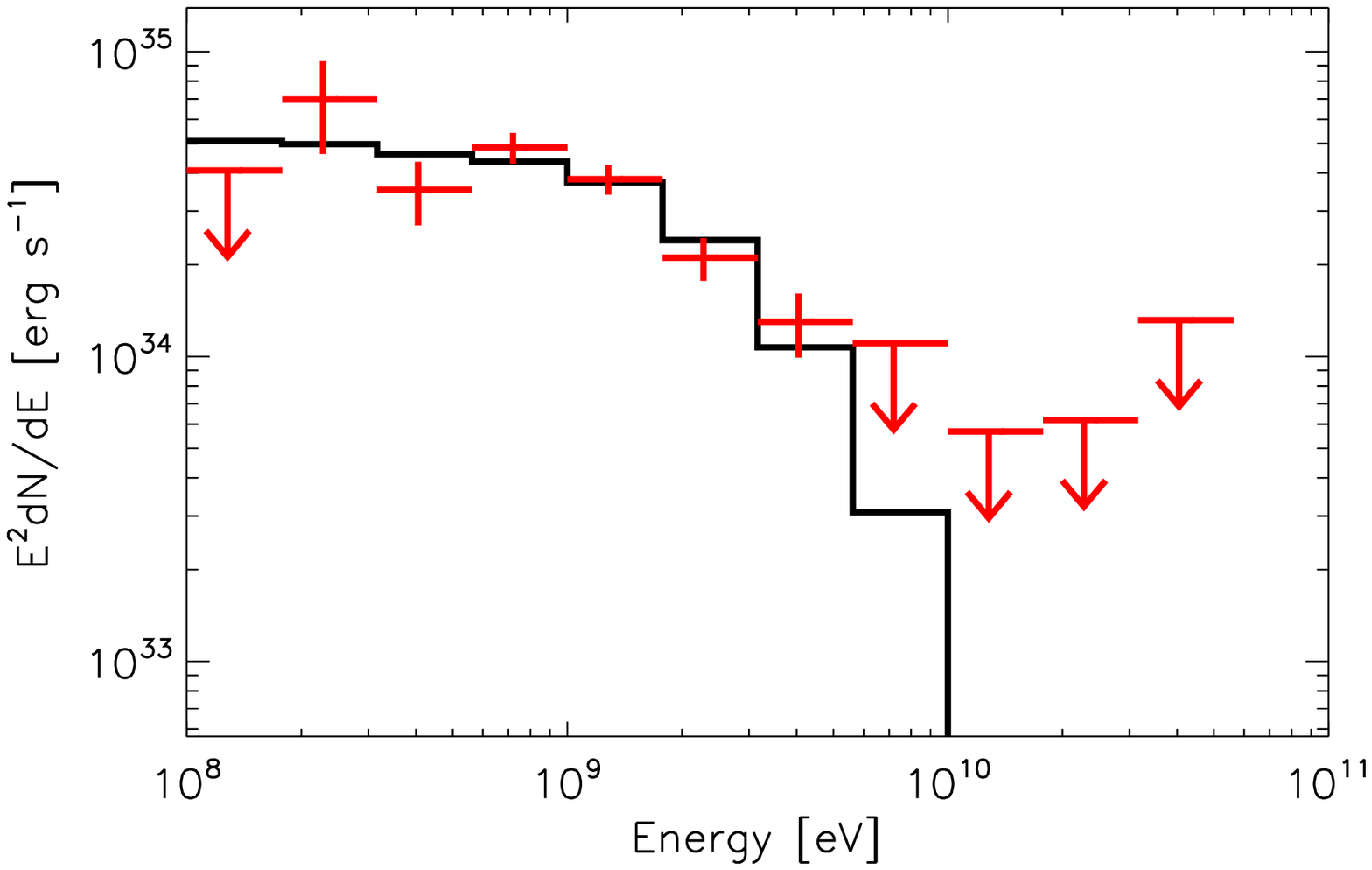}
\put(20,15){\scriptsize J1718-3825}
\end{overpic}
\begin{overpic}[width=0.32\textwidth]{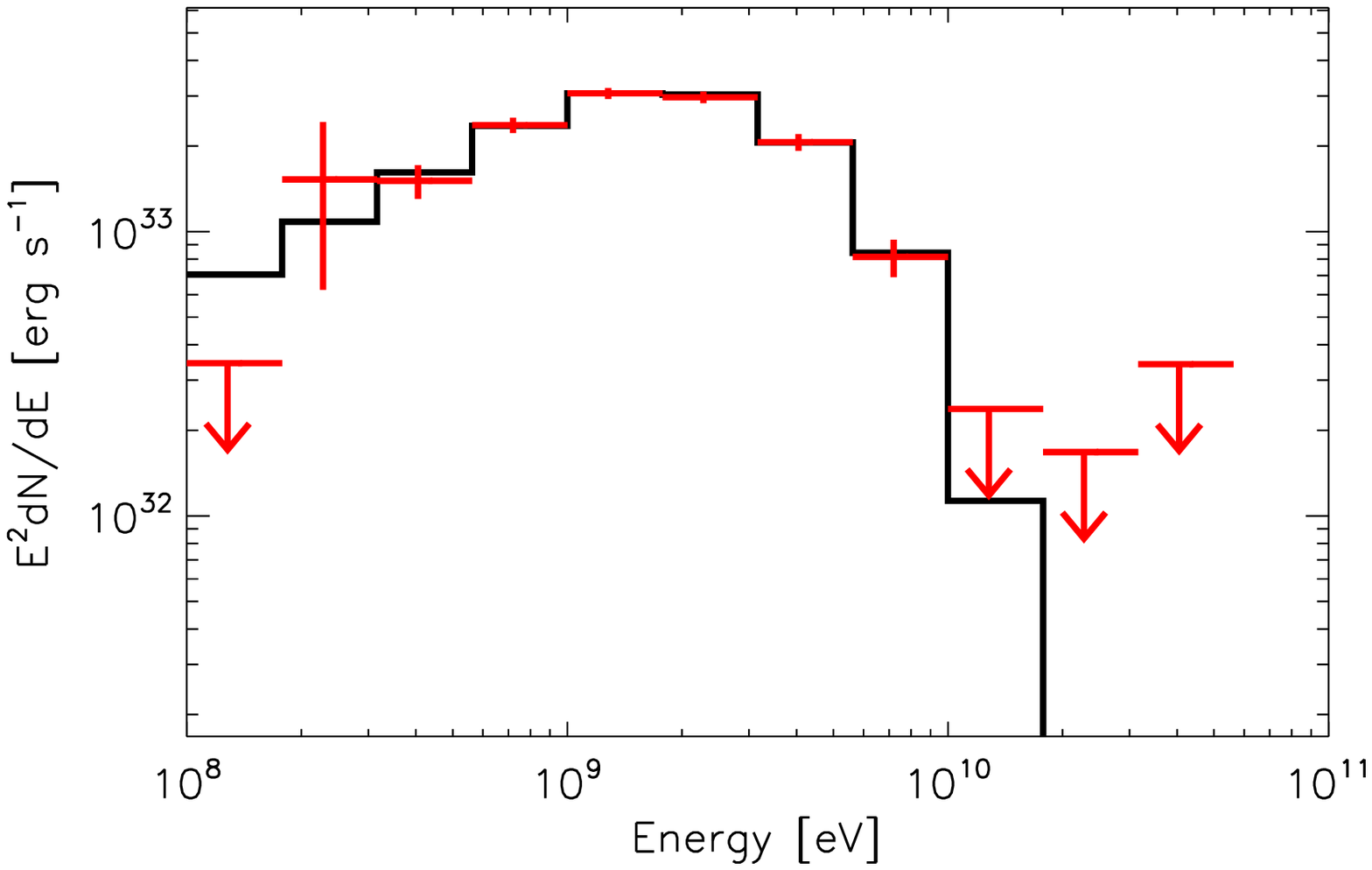}
\put(20,15){\scriptsize J1732-3131}
\end{overpic}
\begin{overpic}[width=0.32\textwidth]{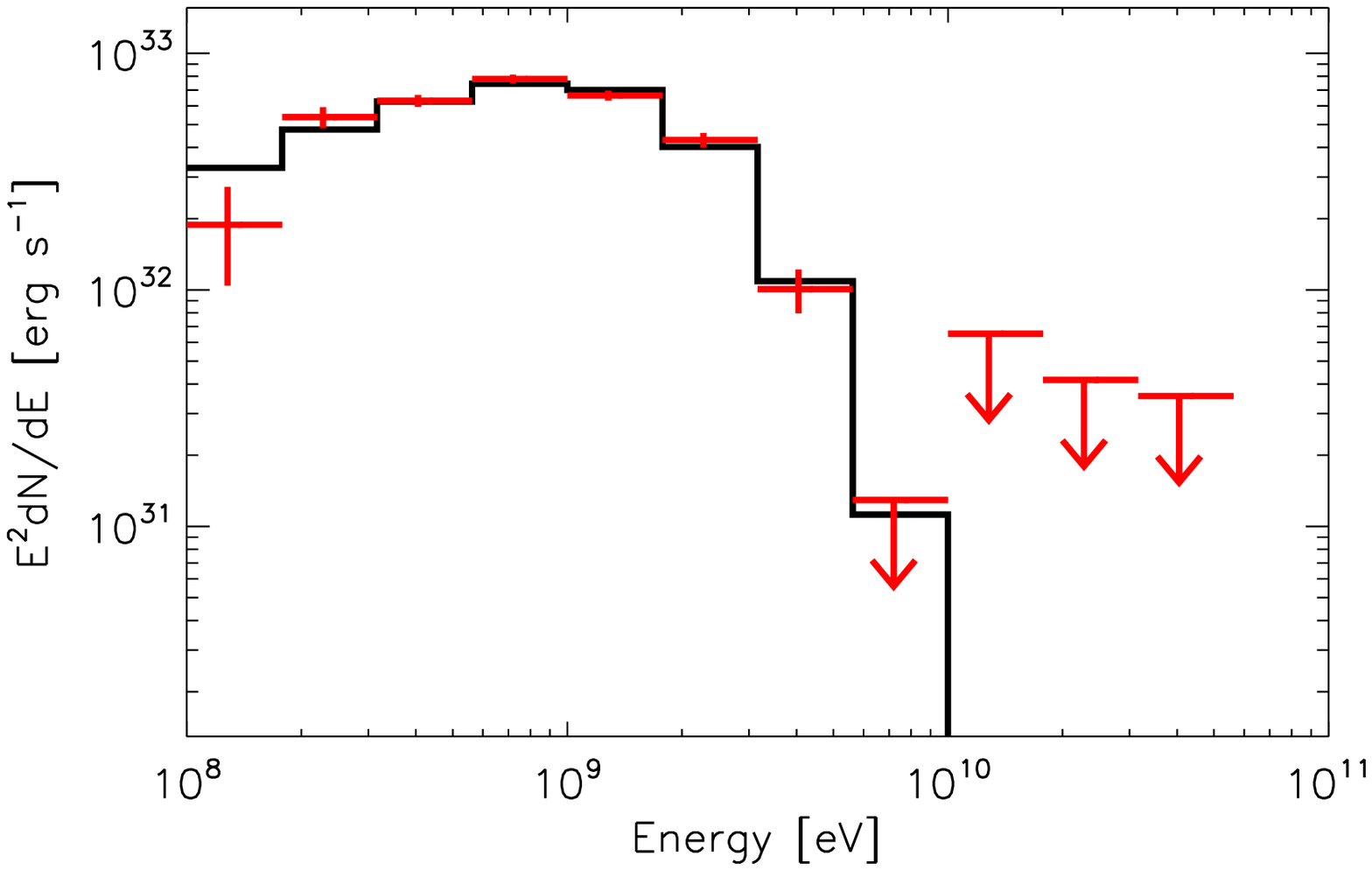}
\put(20,15){\scriptsize J1741-2054}
\end{overpic}
\begin{overpic}[width=0.32\textwidth]{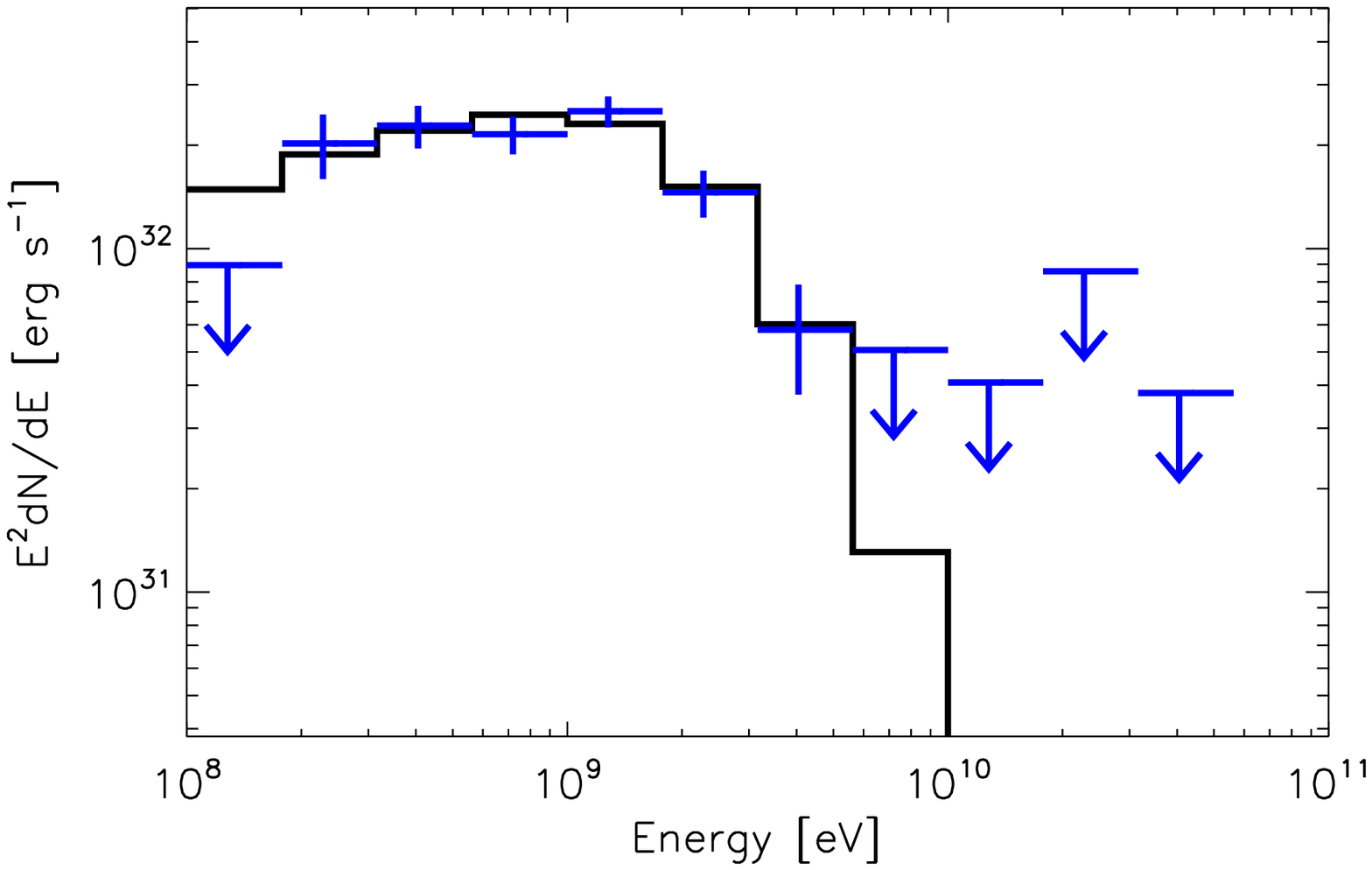}
\put(20,15){\scriptsize J1744-1134}
\end{overpic}
\begin{overpic}[width=0.32\textwidth]{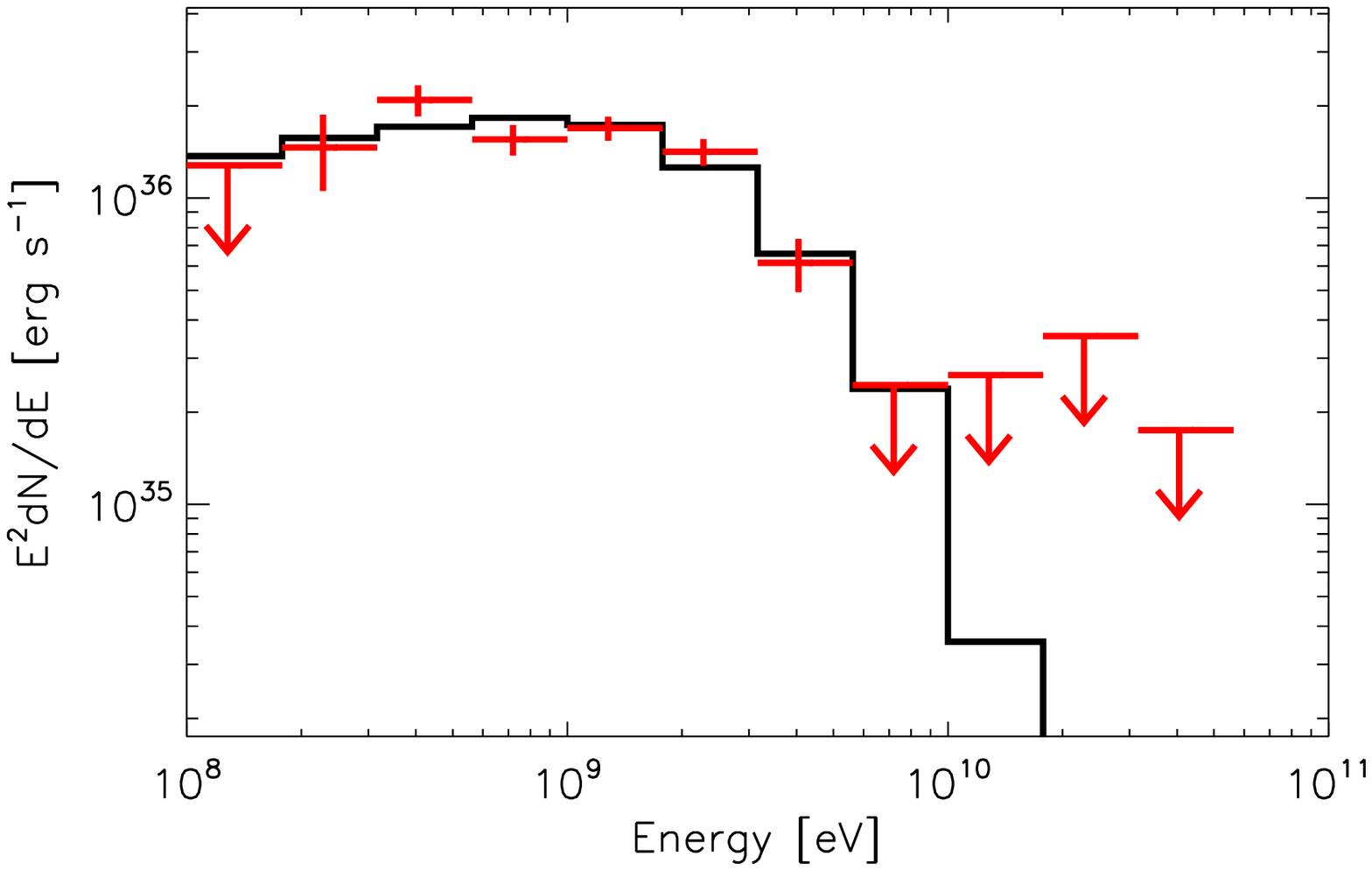}
\put(20,15){\scriptsize J1746-3239}
\end{overpic}
\begin{overpic}[width=0.32\textwidth]{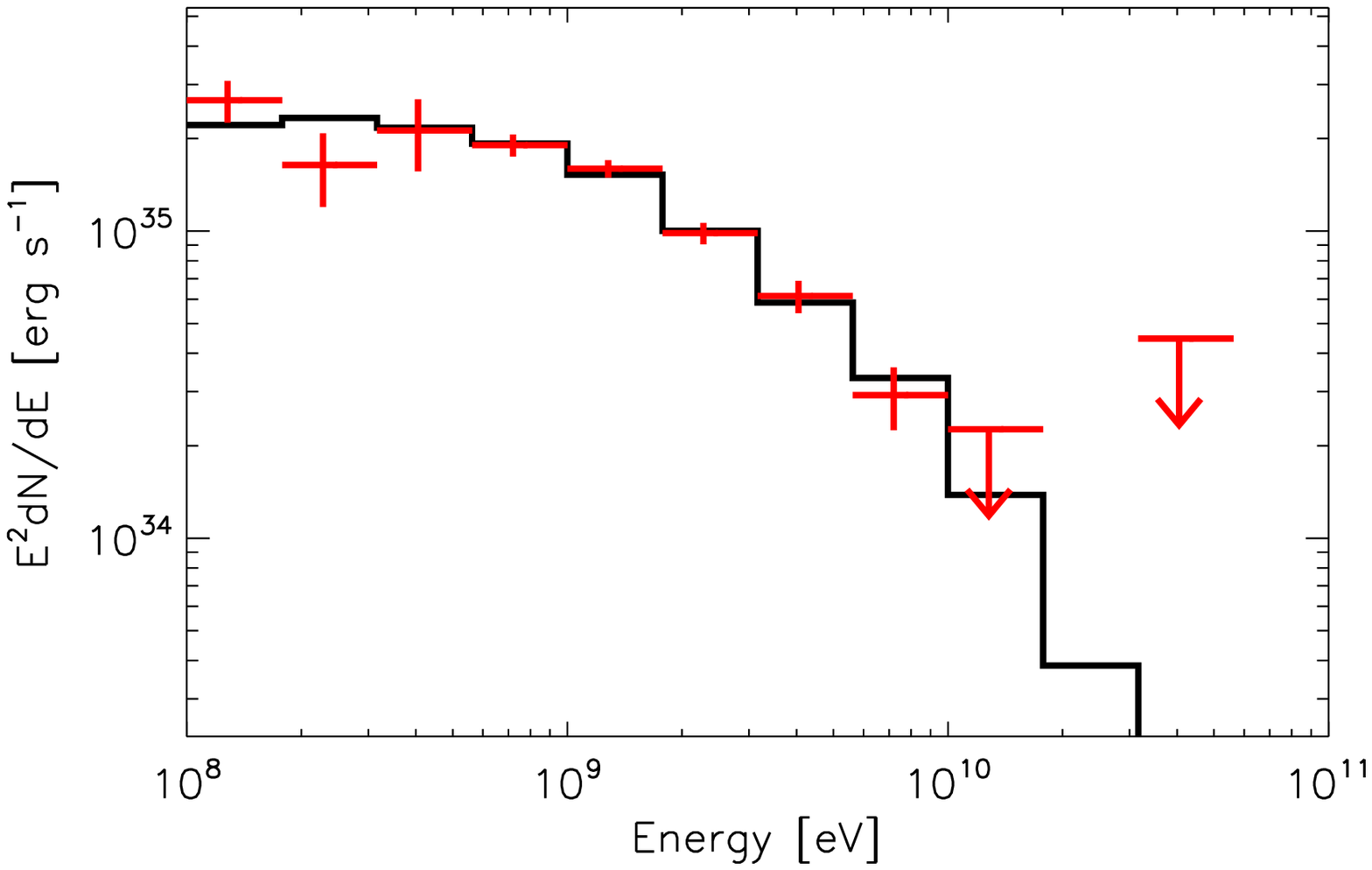}
\put(20,15){\scriptsize J1747-2958}
\end{overpic}
\begin{overpic}[width=0.32\textwidth]{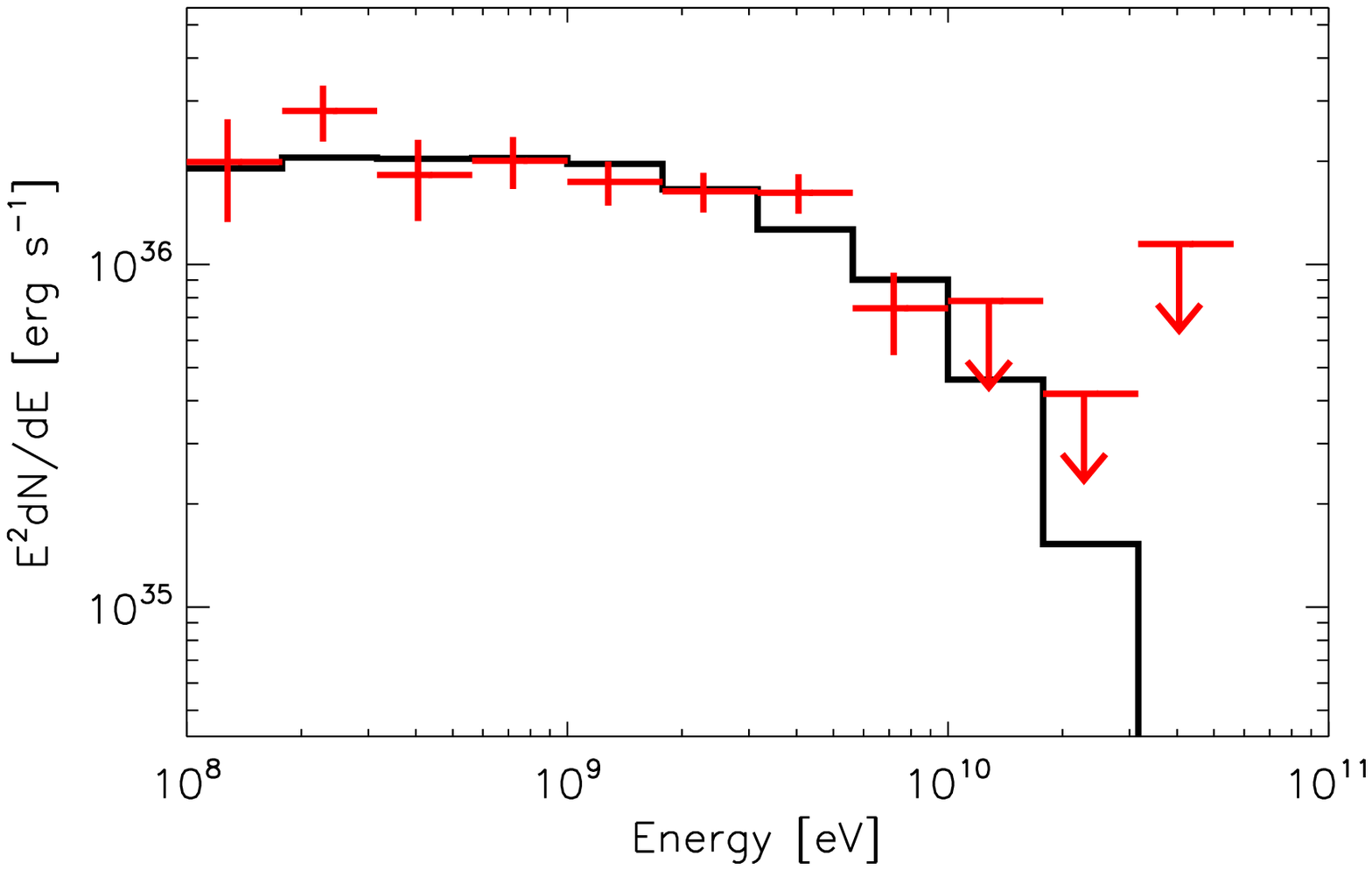}
\put(20,15){\scriptsize J1803-2149}
\end{overpic}
\begin{overpic}[width=0.32\textwidth]{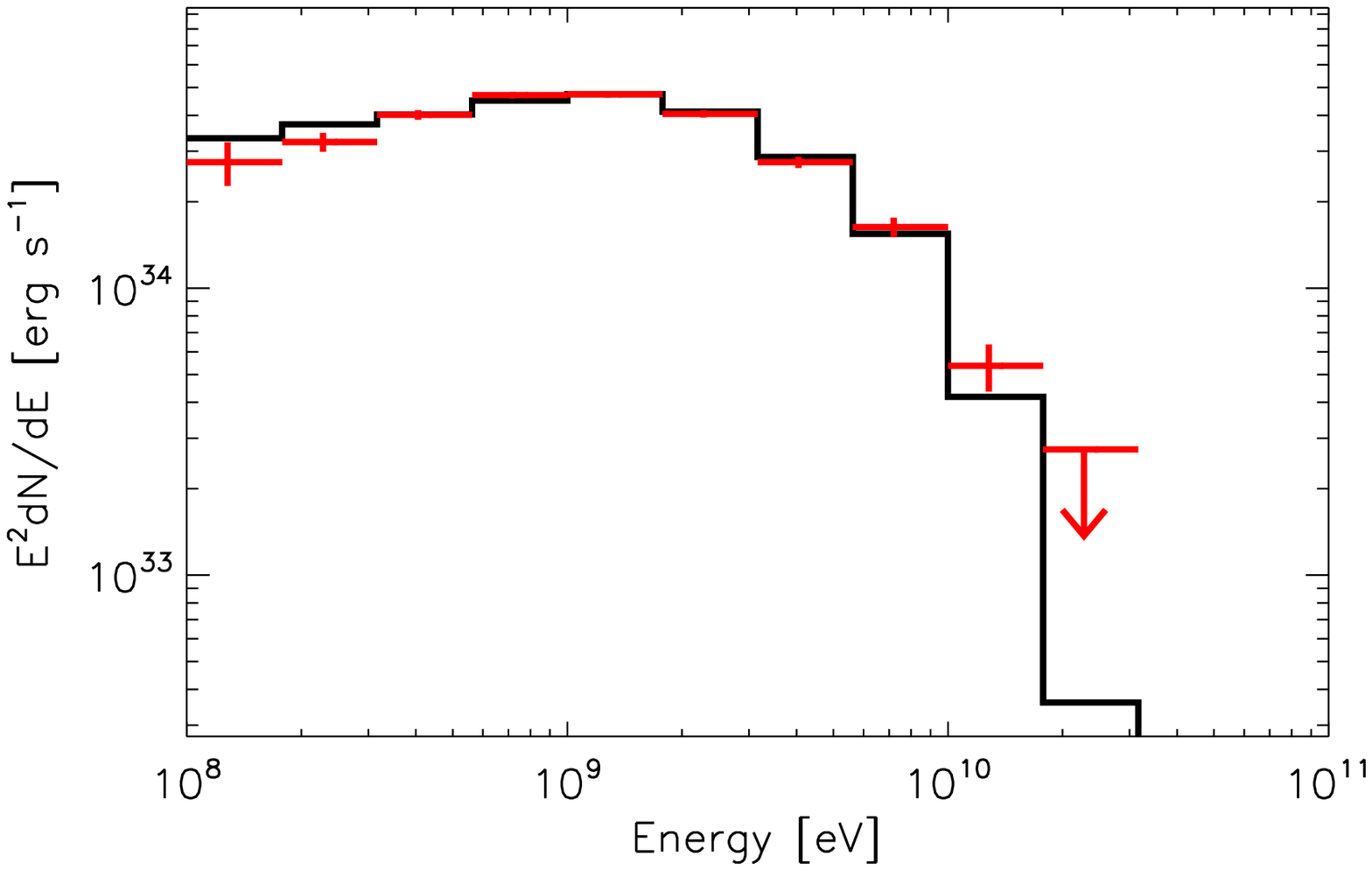}
\put(20,15){\scriptsize J1809-2332}
\end{overpic}
\begin{overpic}[width=0.32\textwidth]{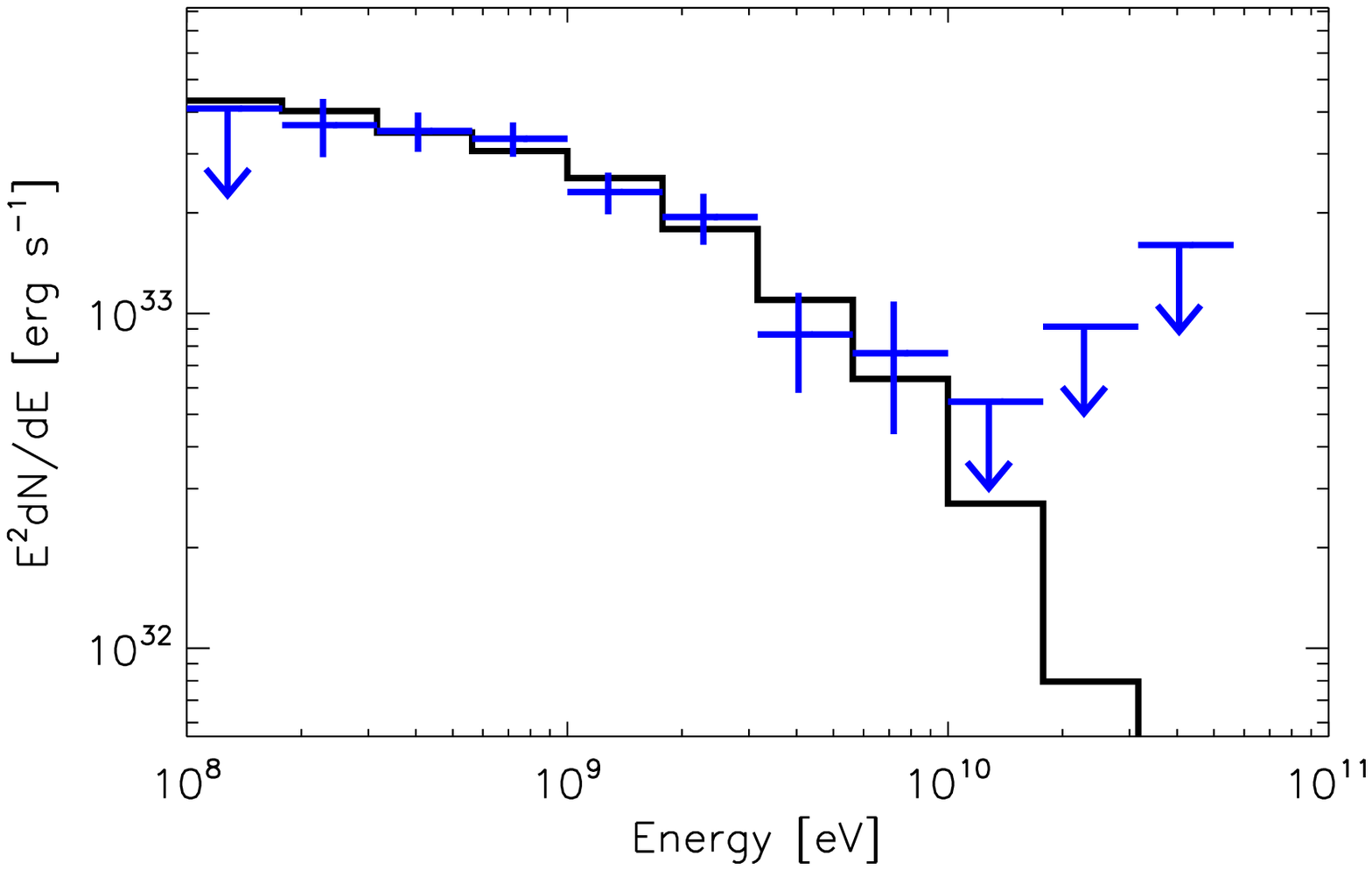}
\put(20,15){\scriptsize J1810+1744}
\end{overpic}
\begin{overpic}[width=0.32\textwidth]{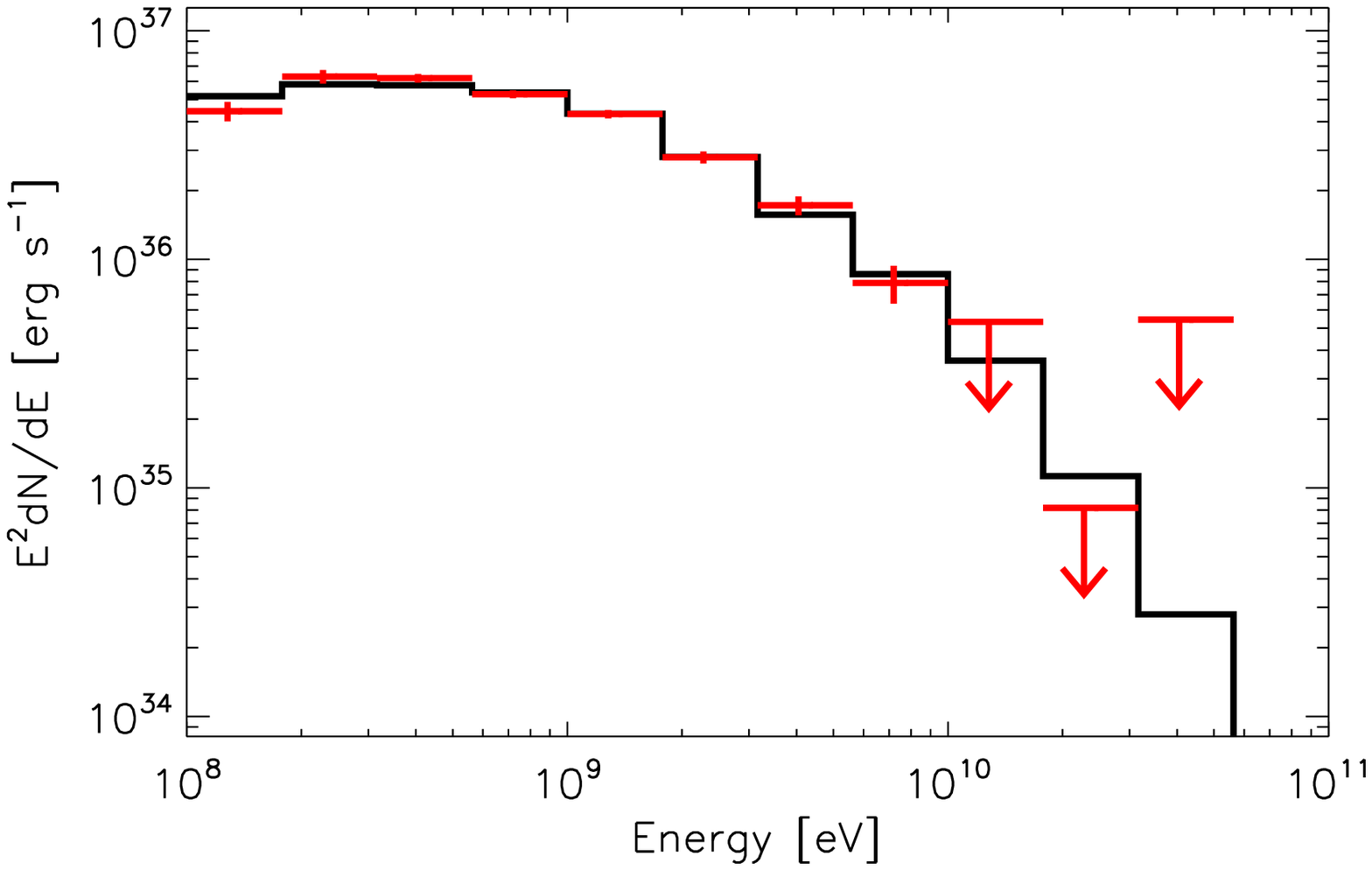}
\put(20,15){\scriptsize J1813-1246}
\end{overpic}
\end{center}
\caption{Fits of the {\it Fermi}-LAT spectral data and models for the pulsars considered in the sample (III).}
\label{fig:best_fit3}
\end{figure*}

\begin{figure*}
\begin{center}
\begin{overpic}[width=0.32\textwidth]{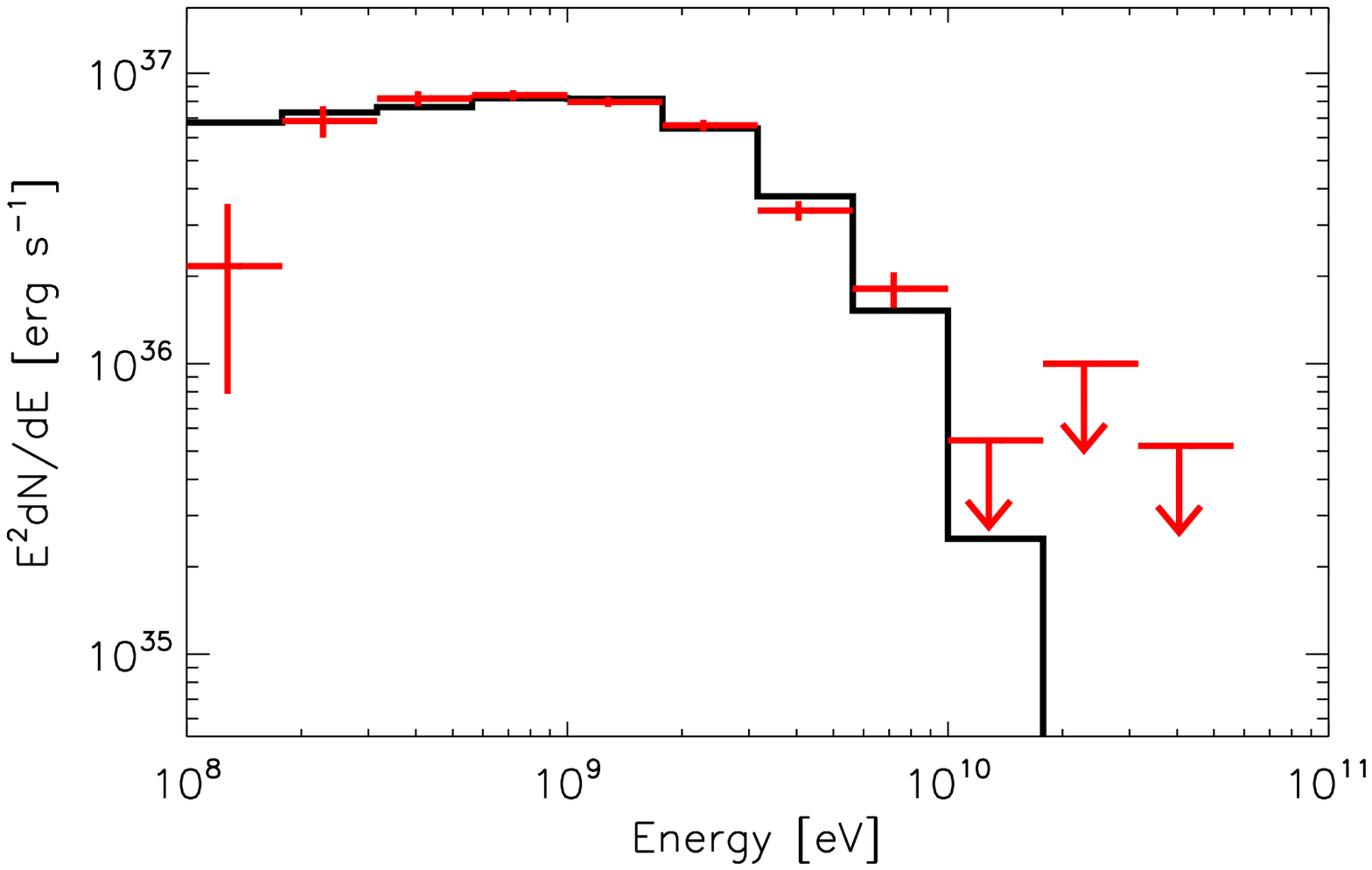}
\put(20,15){\scriptsize J1826-1256}
\end{overpic}
\begin{overpic}[width=0.32\textwidth]{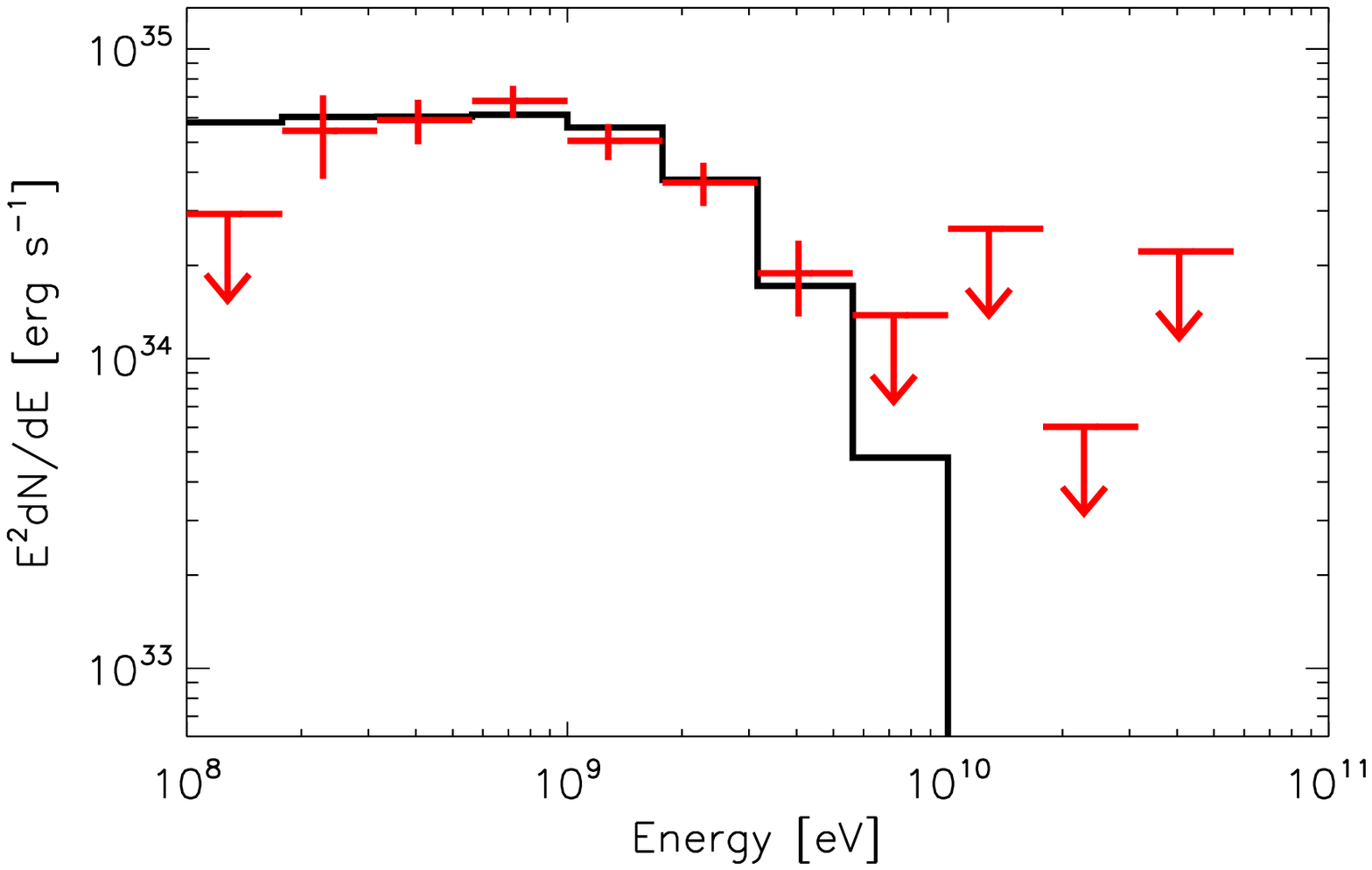}
\put(20,15){\scriptsize J1833-1034}
\end{overpic}
\begin{overpic}[width=0.32\textwidth]{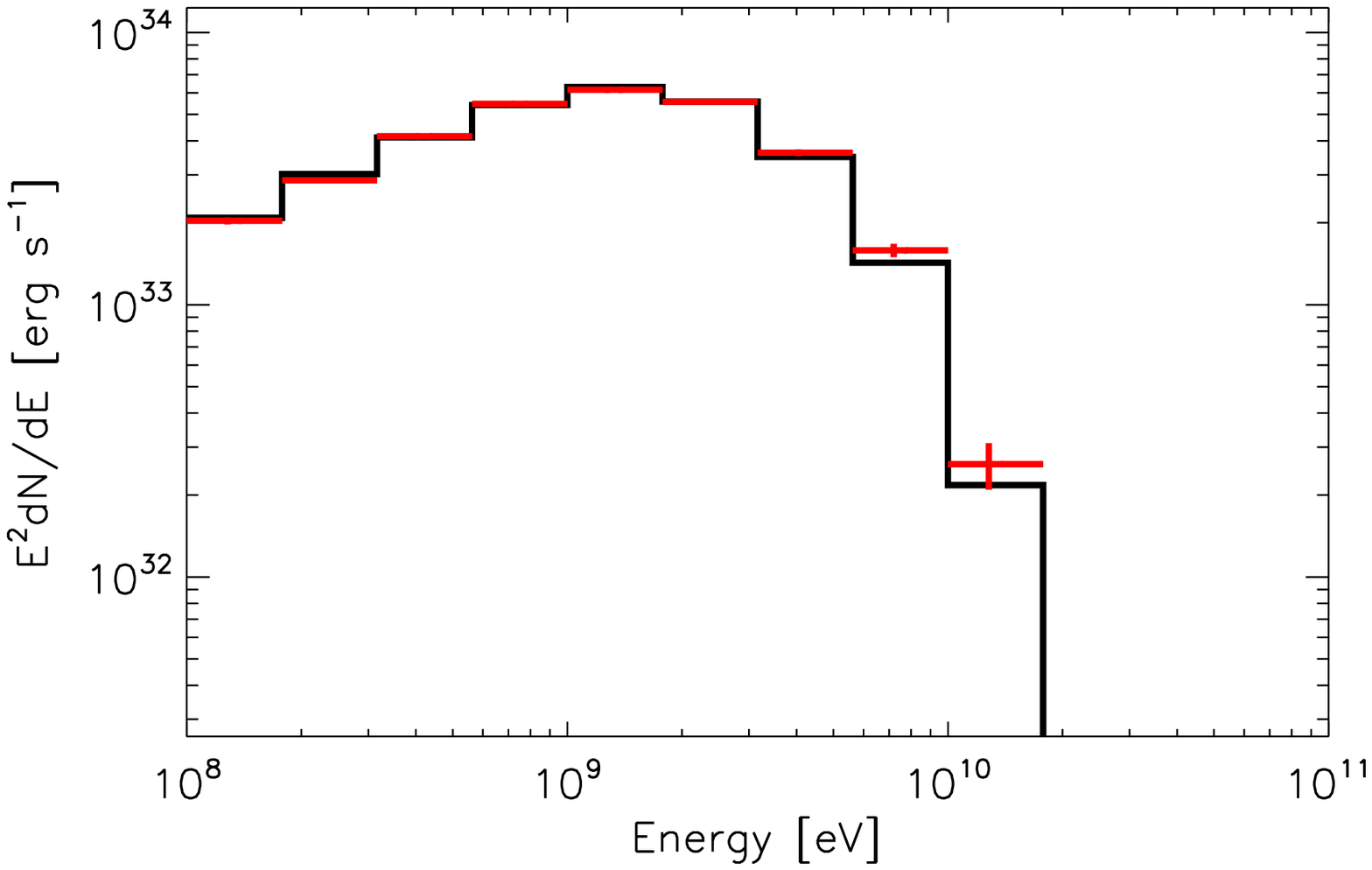}
\put(20,15){\scriptsize J1836+5925}
\end{overpic}
\begin{overpic}[width=0.32\textwidth]{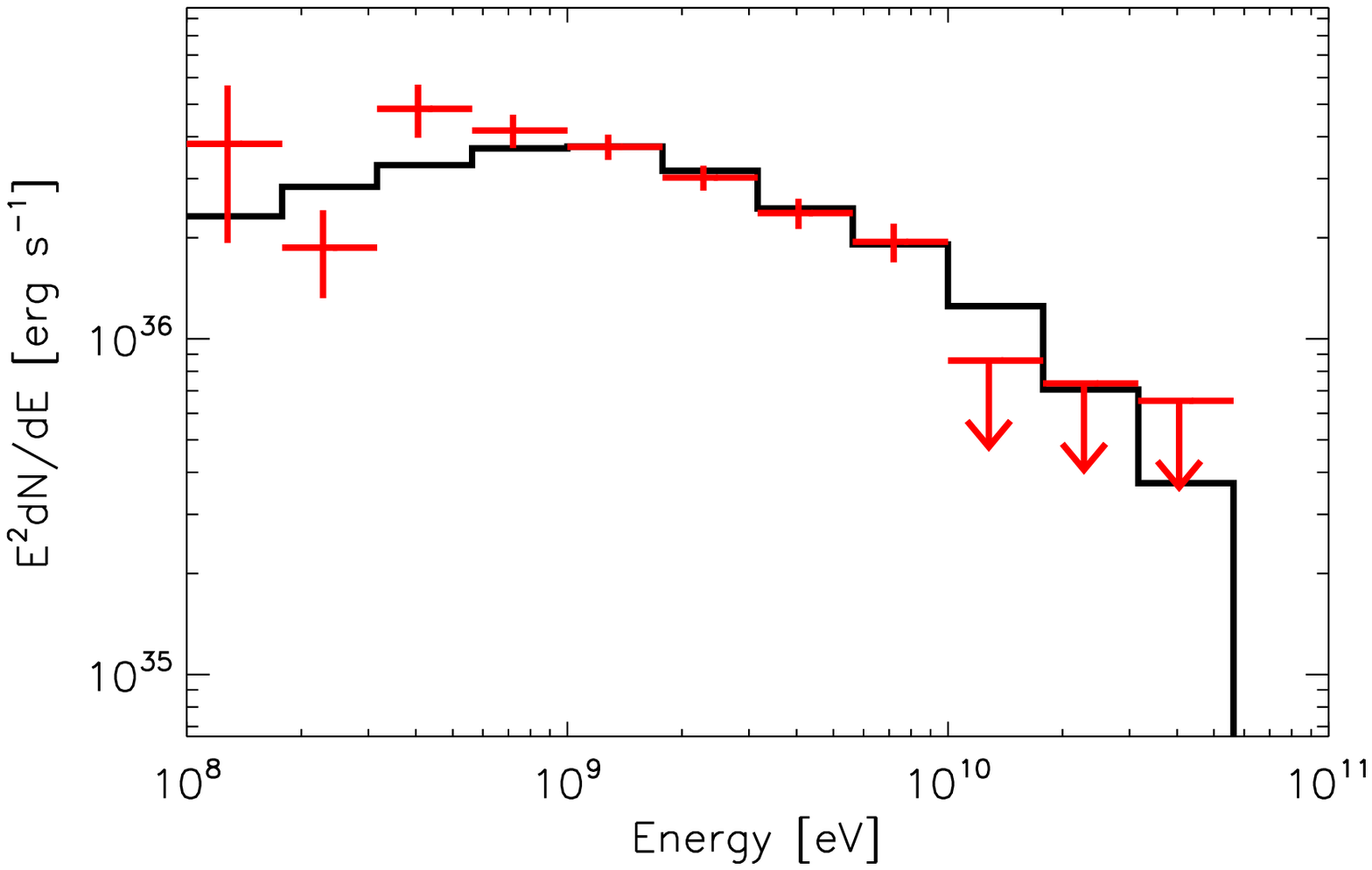}
\put(20,15){\scriptsize J1838-0537}
\end{overpic}
\begin{overpic}[width=0.32\textwidth]{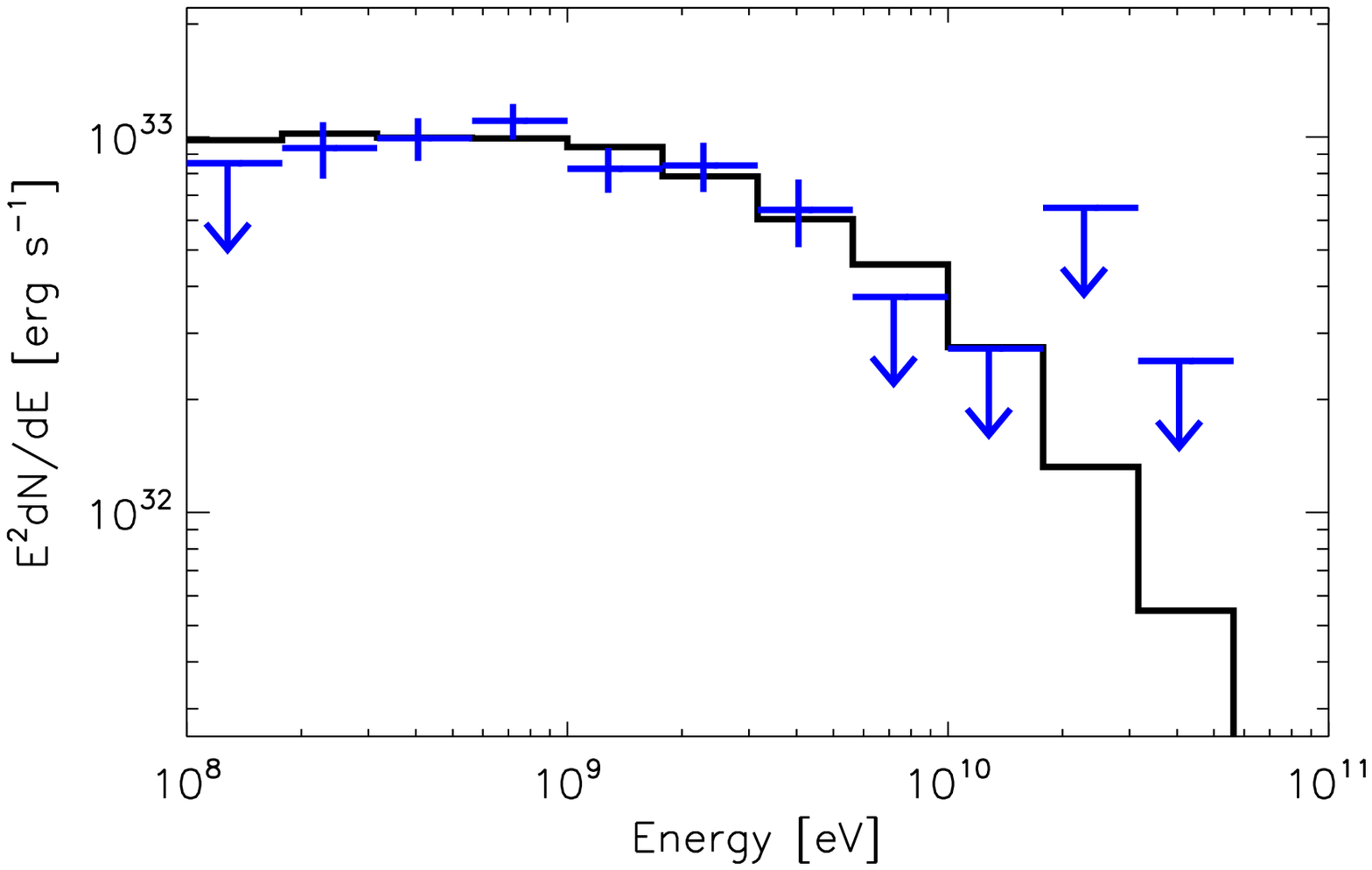}
\put(20,15){\scriptsize J1902-5105}
\end{overpic}
\begin{overpic}[width=0.32\textwidth]{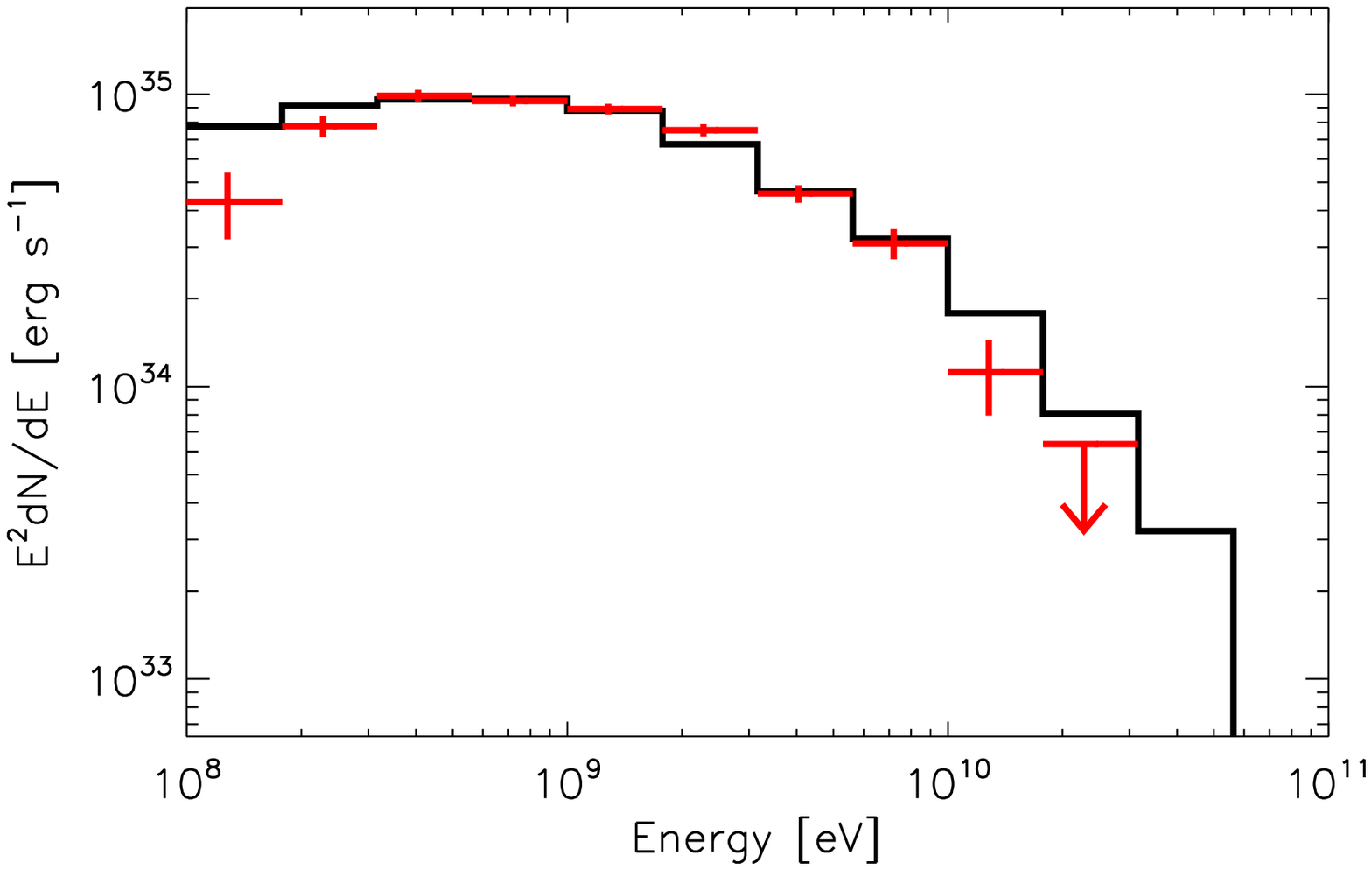}
\put(20,15){\scriptsize J1907+0602}
\end{overpic}
\begin{overpic}[width=0.32\textwidth]{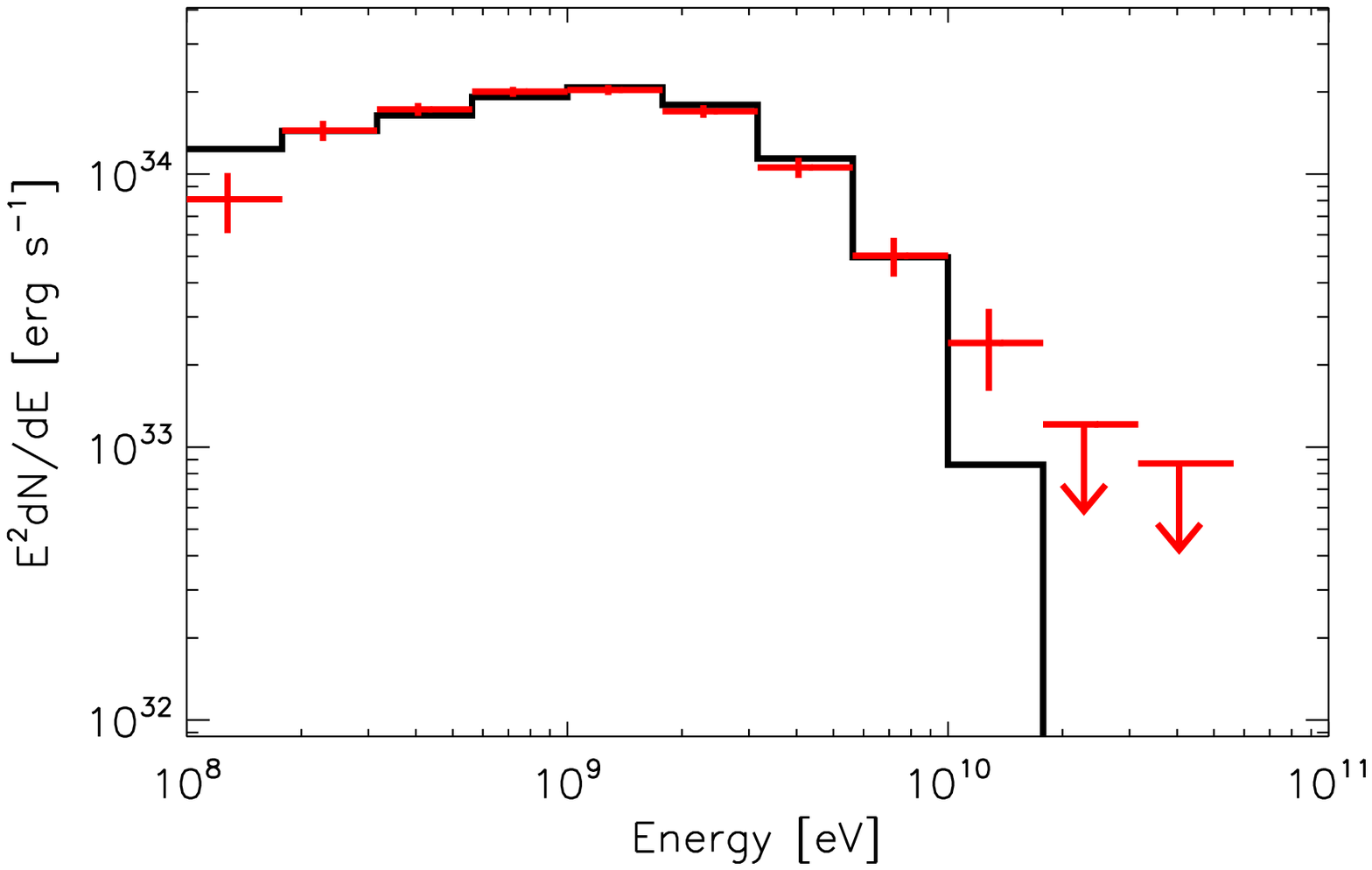}
\put(20,15){\scriptsize J1952+3252}
\end{overpic}
\begin{overpic}[width=0.32\textwidth]{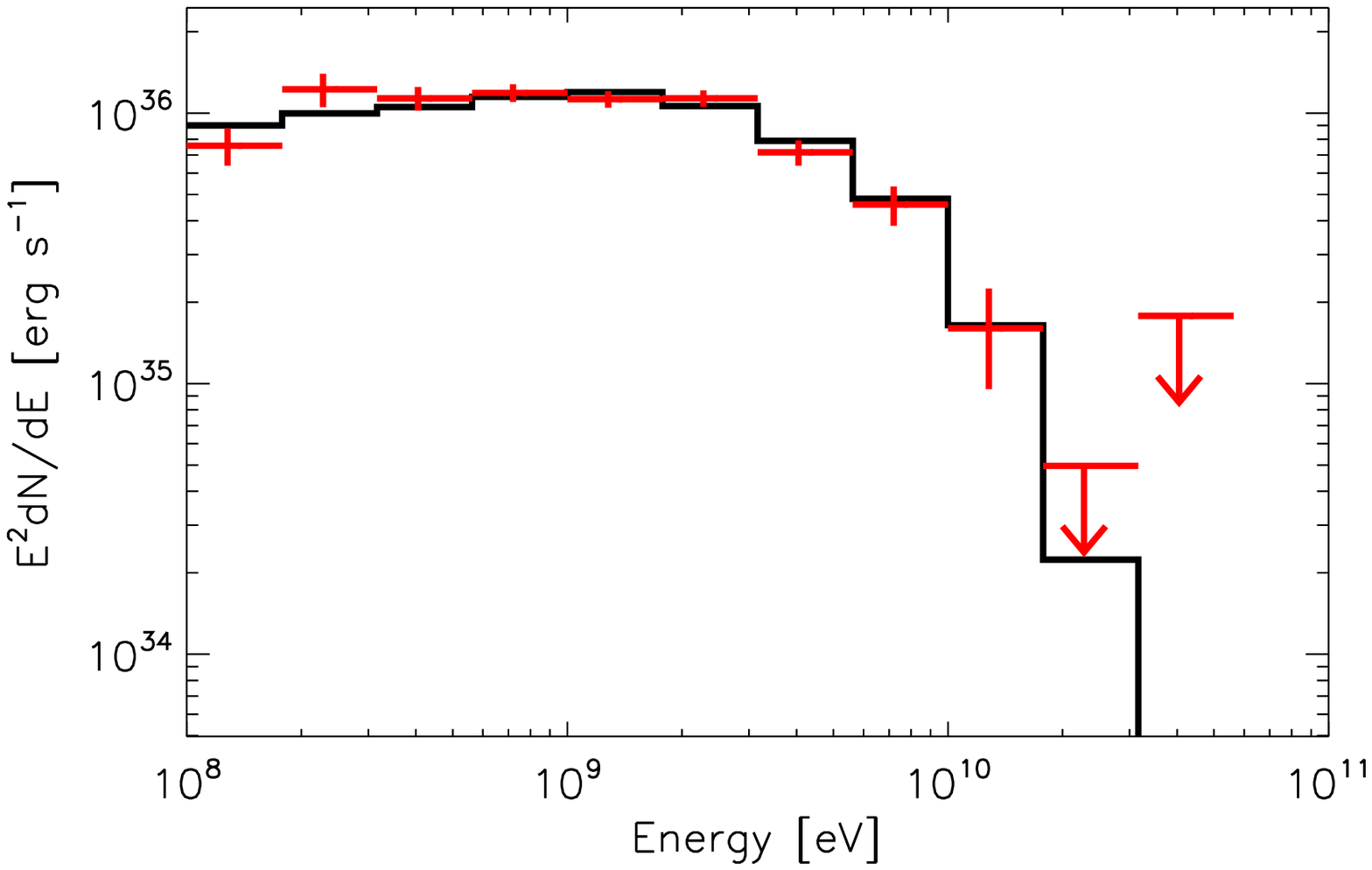}
\put(20,15){\scriptsize J1954+2836}
\end{overpic}
\begin{overpic}[width=0.32\textwidth]{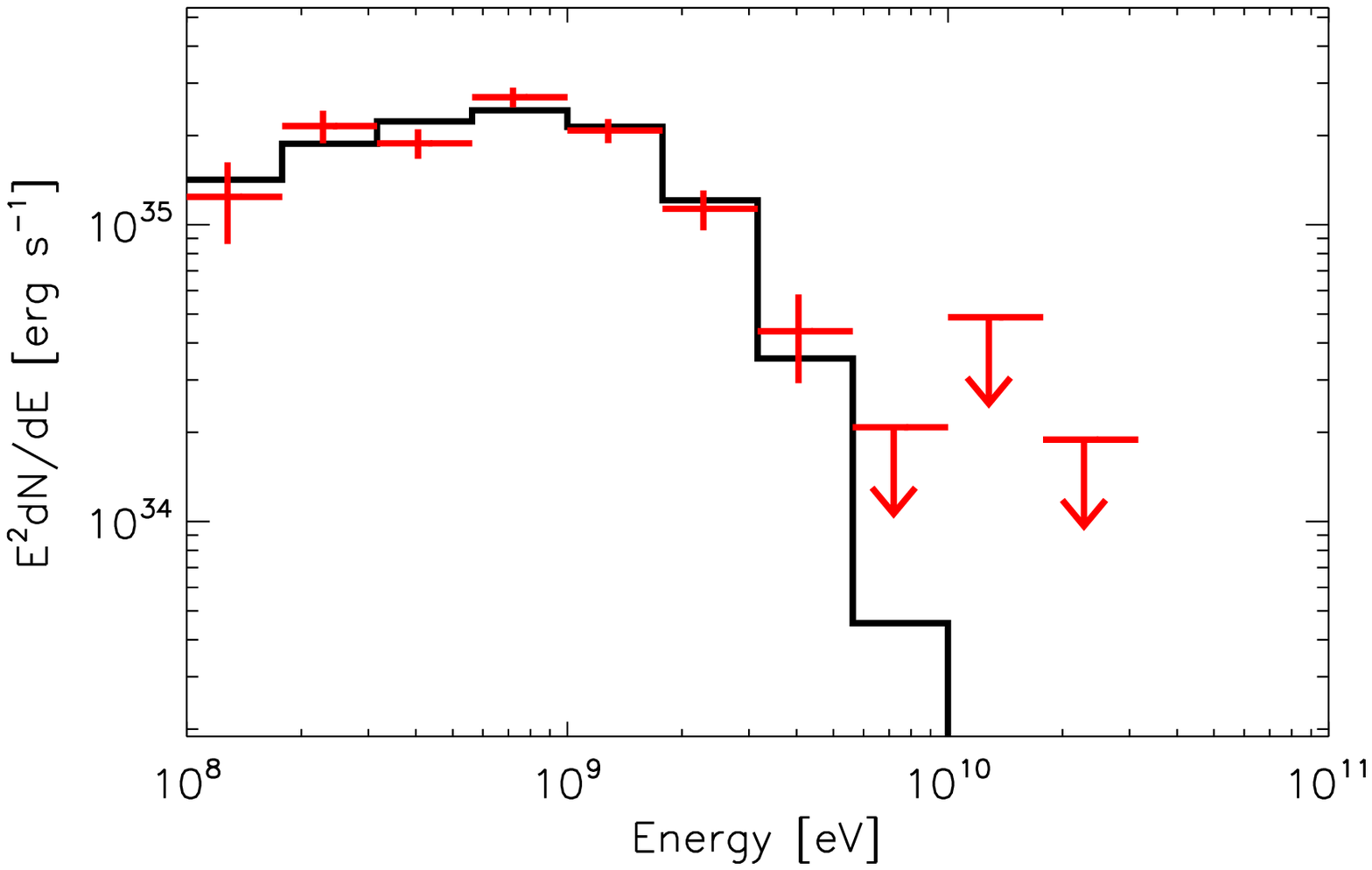}
\put(20,15){\scriptsize J1957+5033}
\end{overpic}
\begin{overpic}[width=0.32\textwidth]{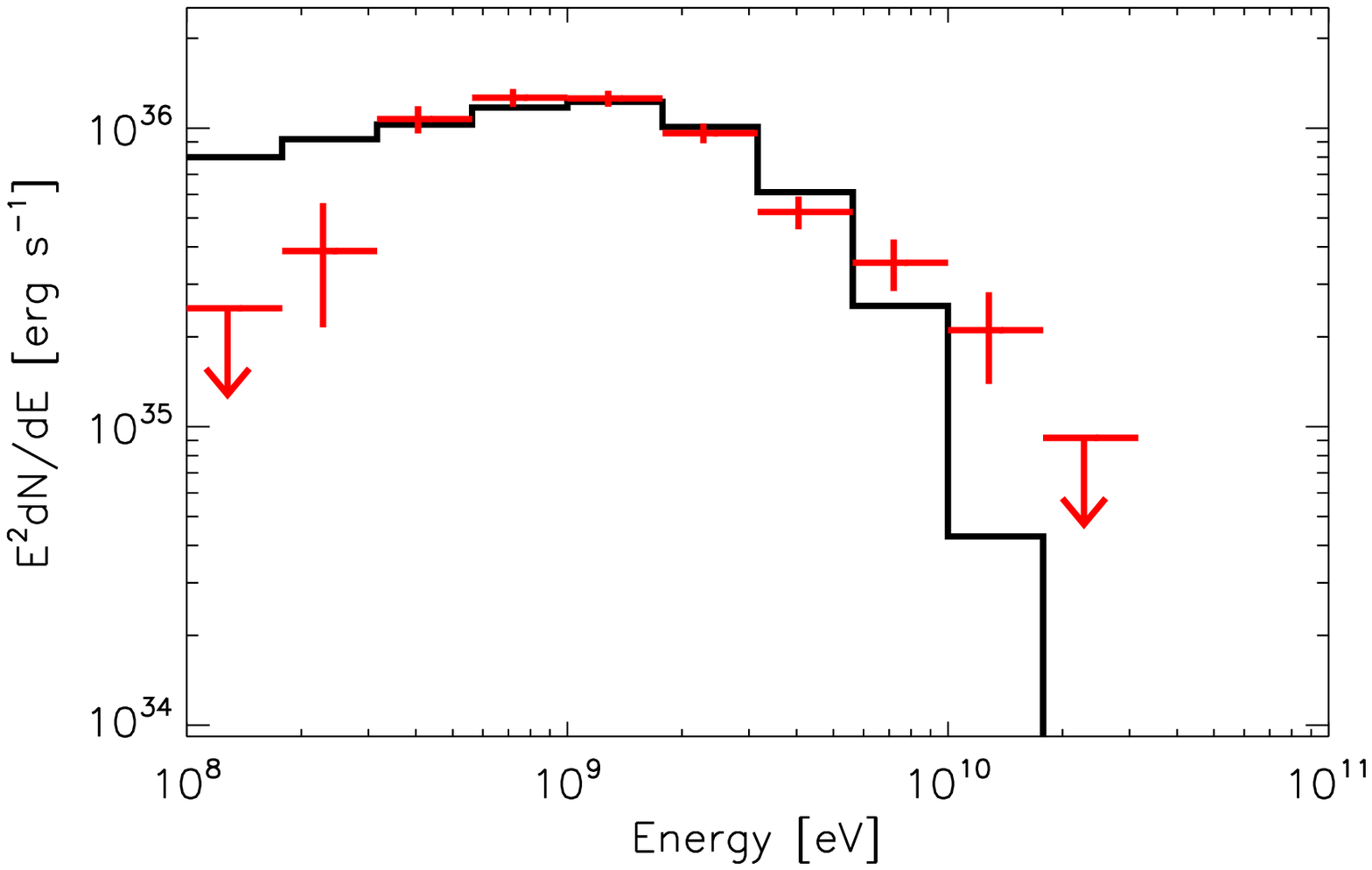}
\put(20,15){\scriptsize J1958+2846}
\end{overpic}
\begin{overpic}[width=0.32\textwidth]{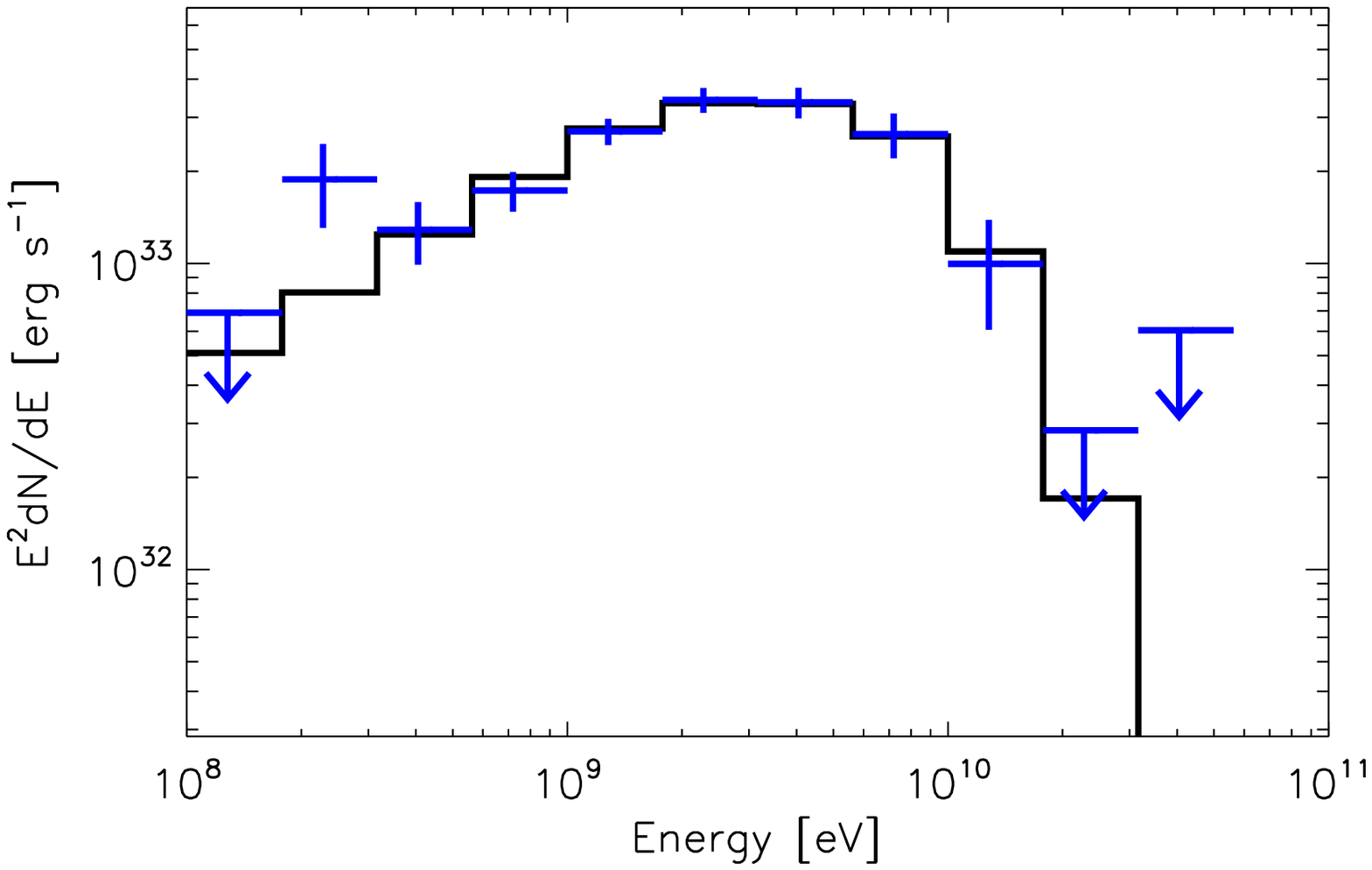}
\put(20,15){\scriptsize J2017+0603}
\end{overpic}
\begin{overpic}[width=0.32\textwidth]{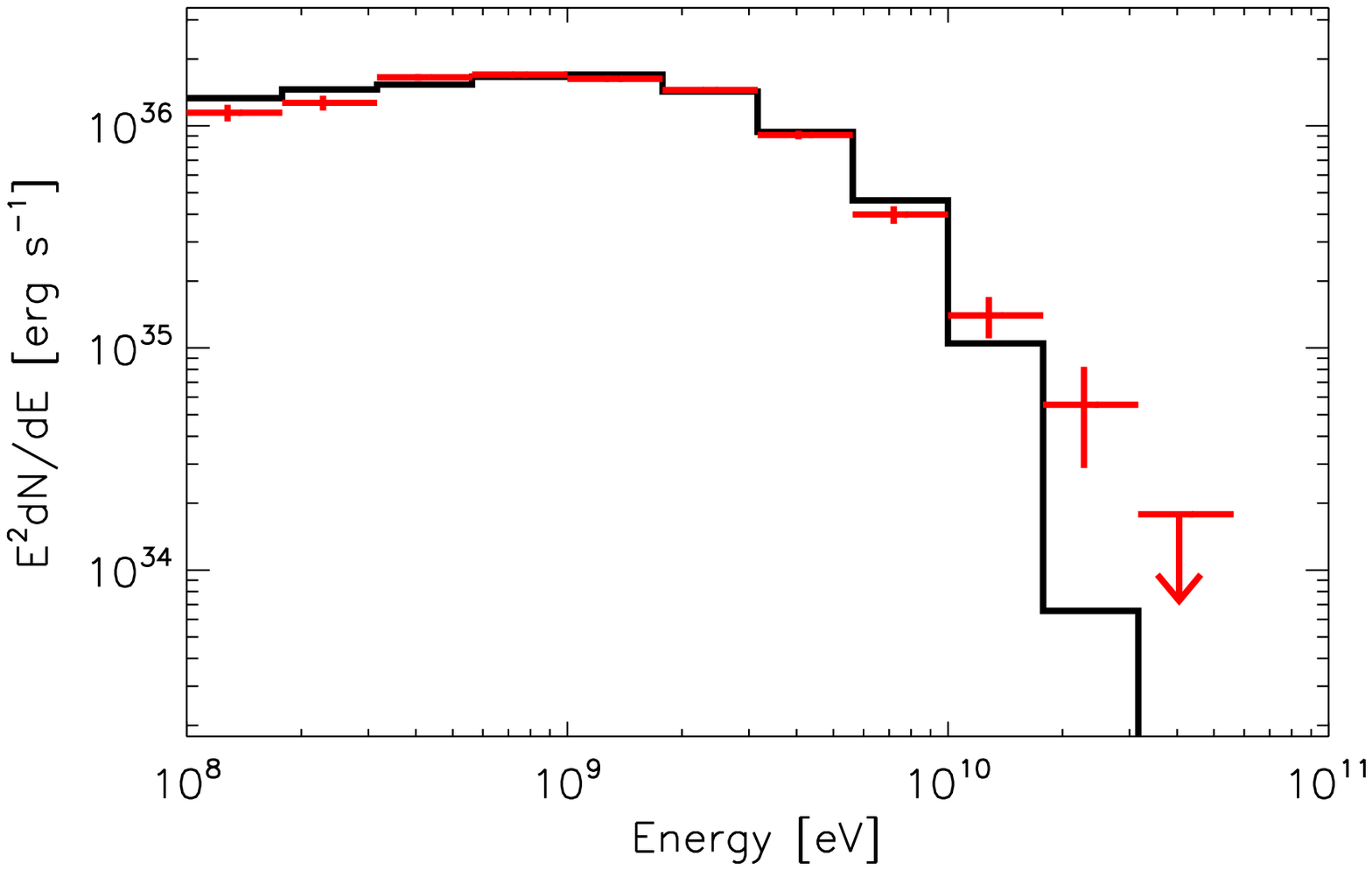}
\put(20,15){\scriptsize J2021+3651}
\end{overpic}
\begin{overpic}[width=0.32\textwidth]{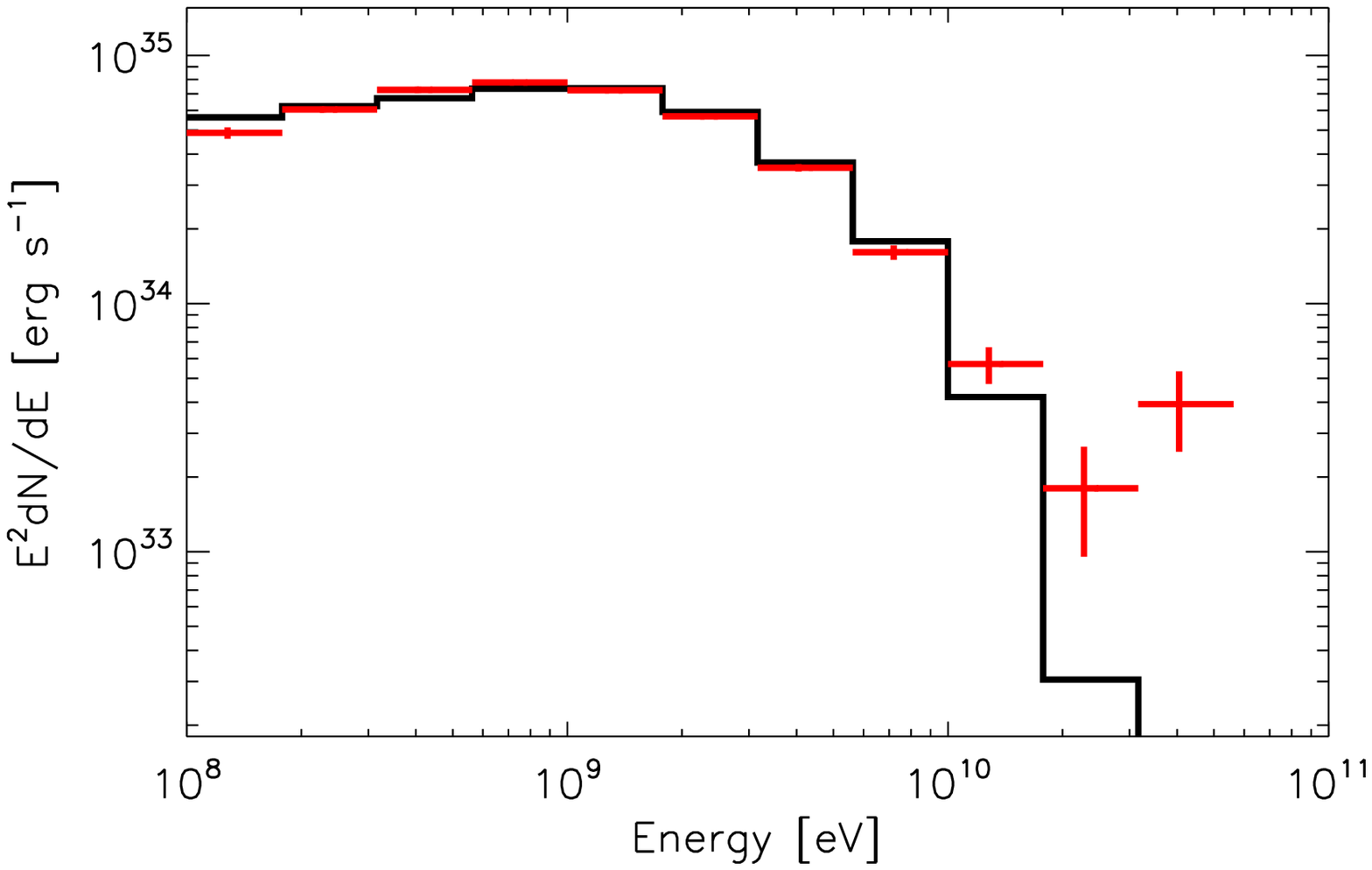}
\put(20,15){\scriptsize J2021+4026}
\end{overpic}
\begin{overpic}[width=0.32\textwidth]{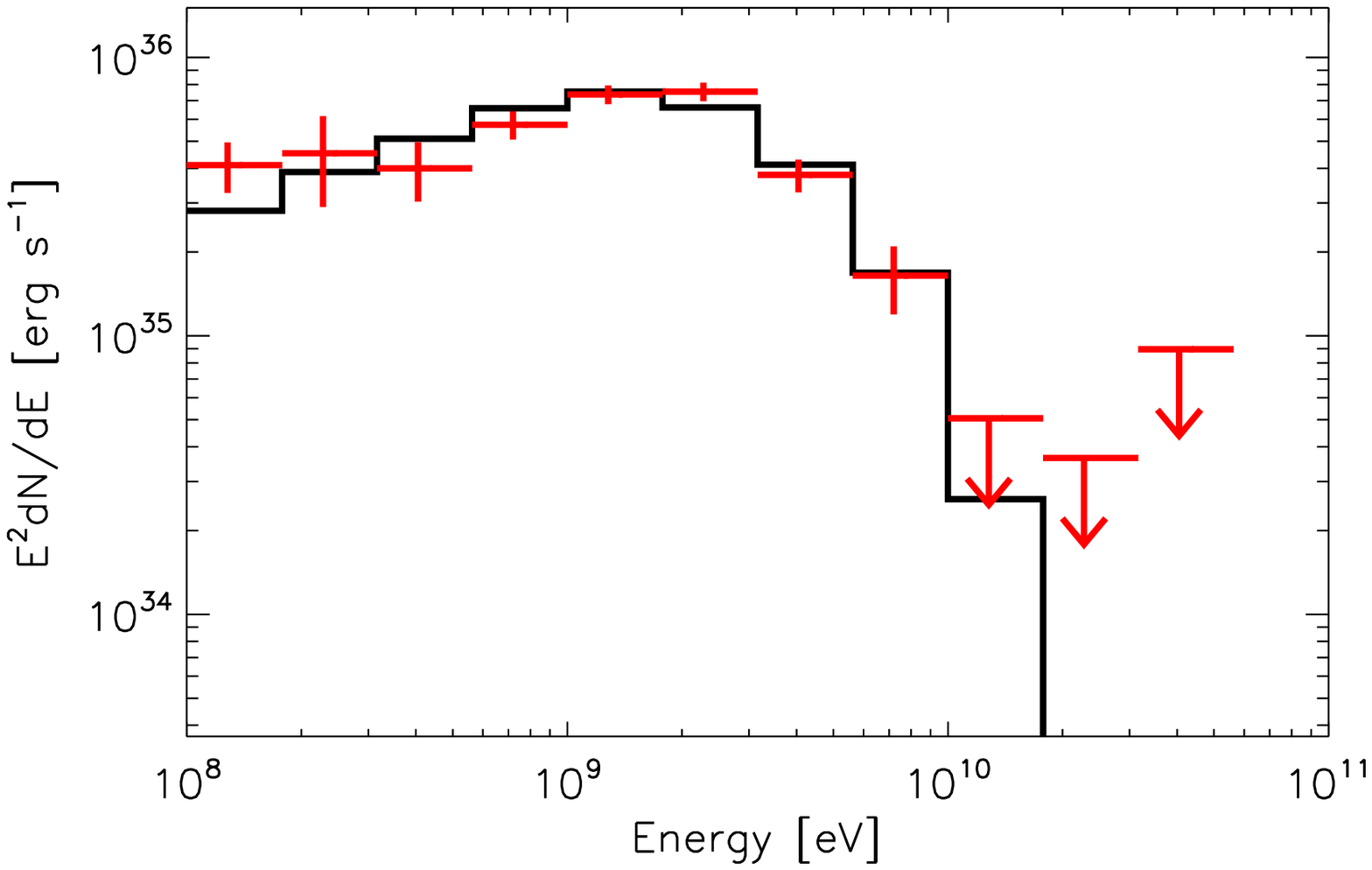}
\put(20,15){\scriptsize J2028+3332}
\end{overpic}
\begin{overpic}[width=0.32\textwidth]{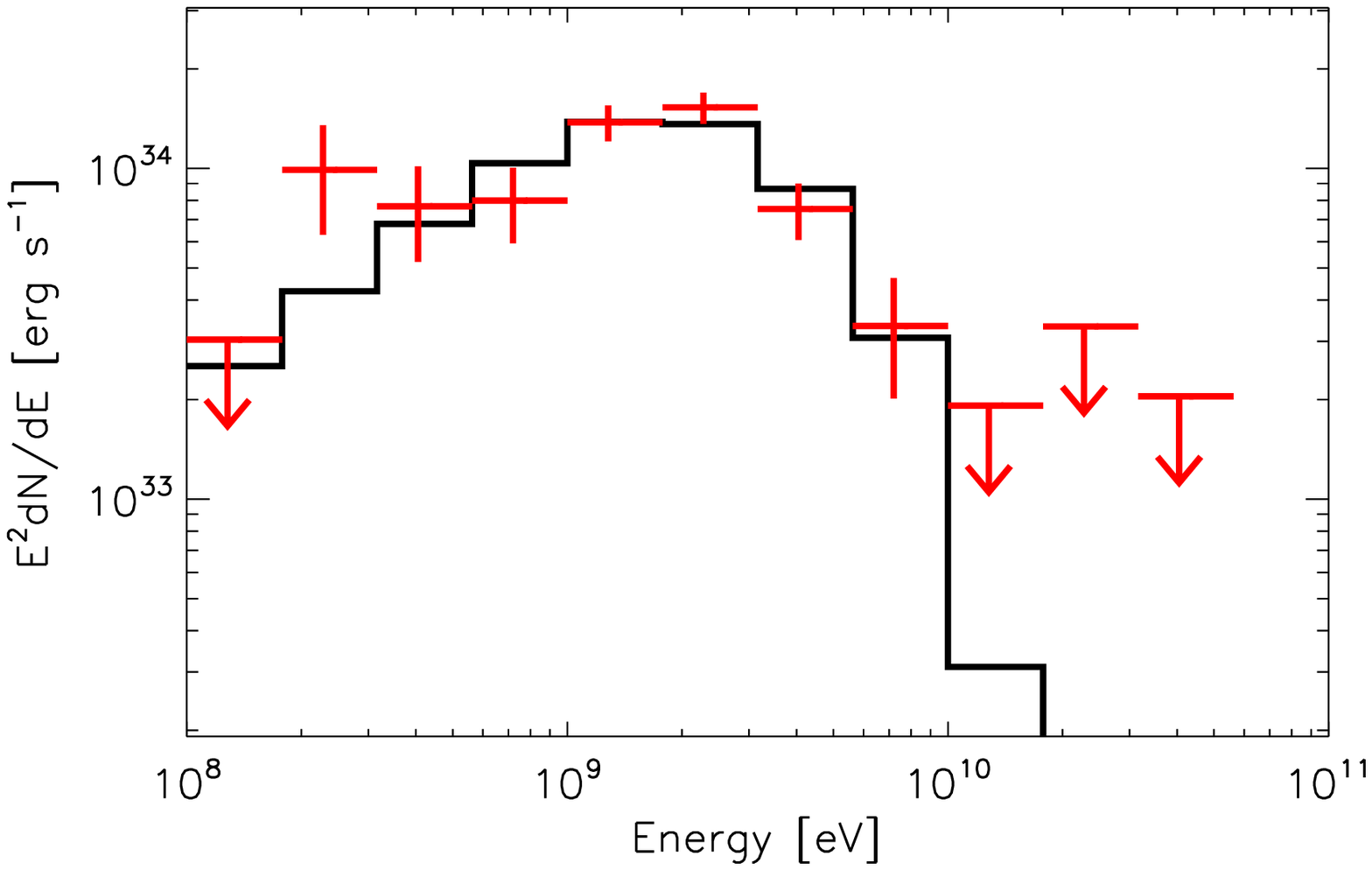}
\put(20,15){\scriptsize J2030+3641}
\end{overpic}
\begin{overpic}[width=0.32\textwidth]{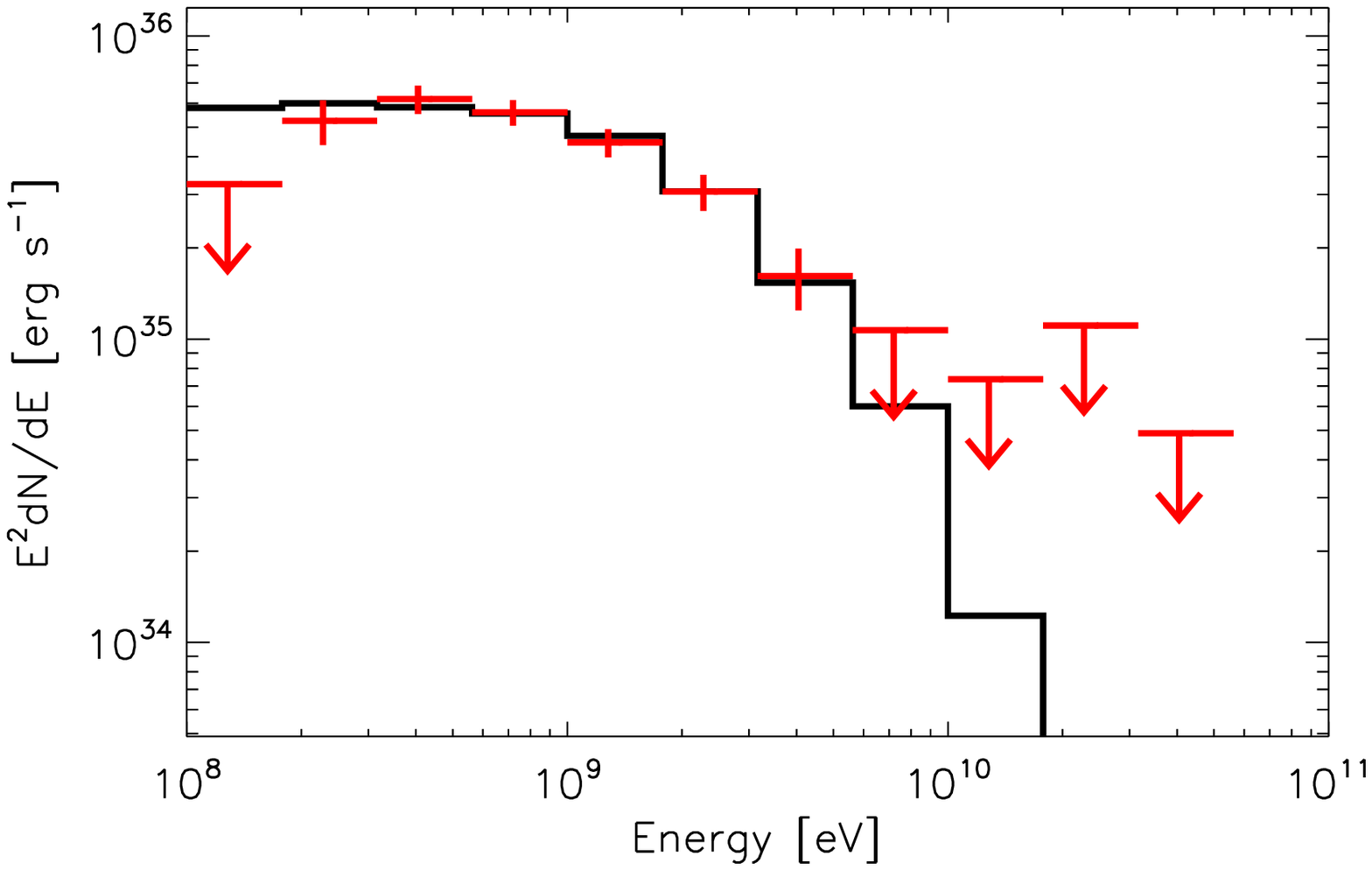}
\put(20,15){\scriptsize J2030+4415}
\end{overpic}
\begin{overpic}[width=0.32\textwidth]{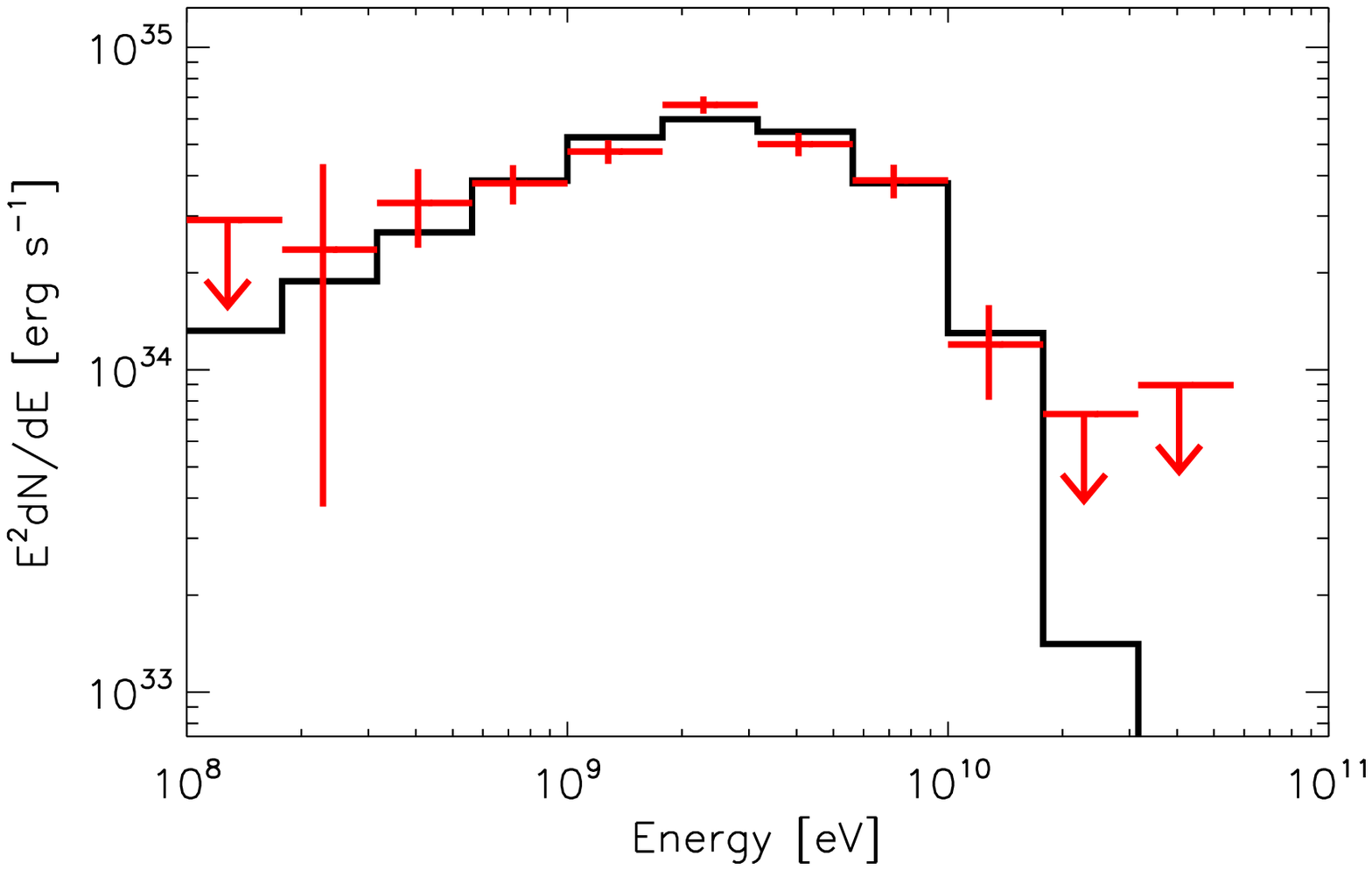}
\put(20,15){\scriptsize J2032+4127}
\end{overpic}
\begin{overpic}[width=0.32\textwidth]{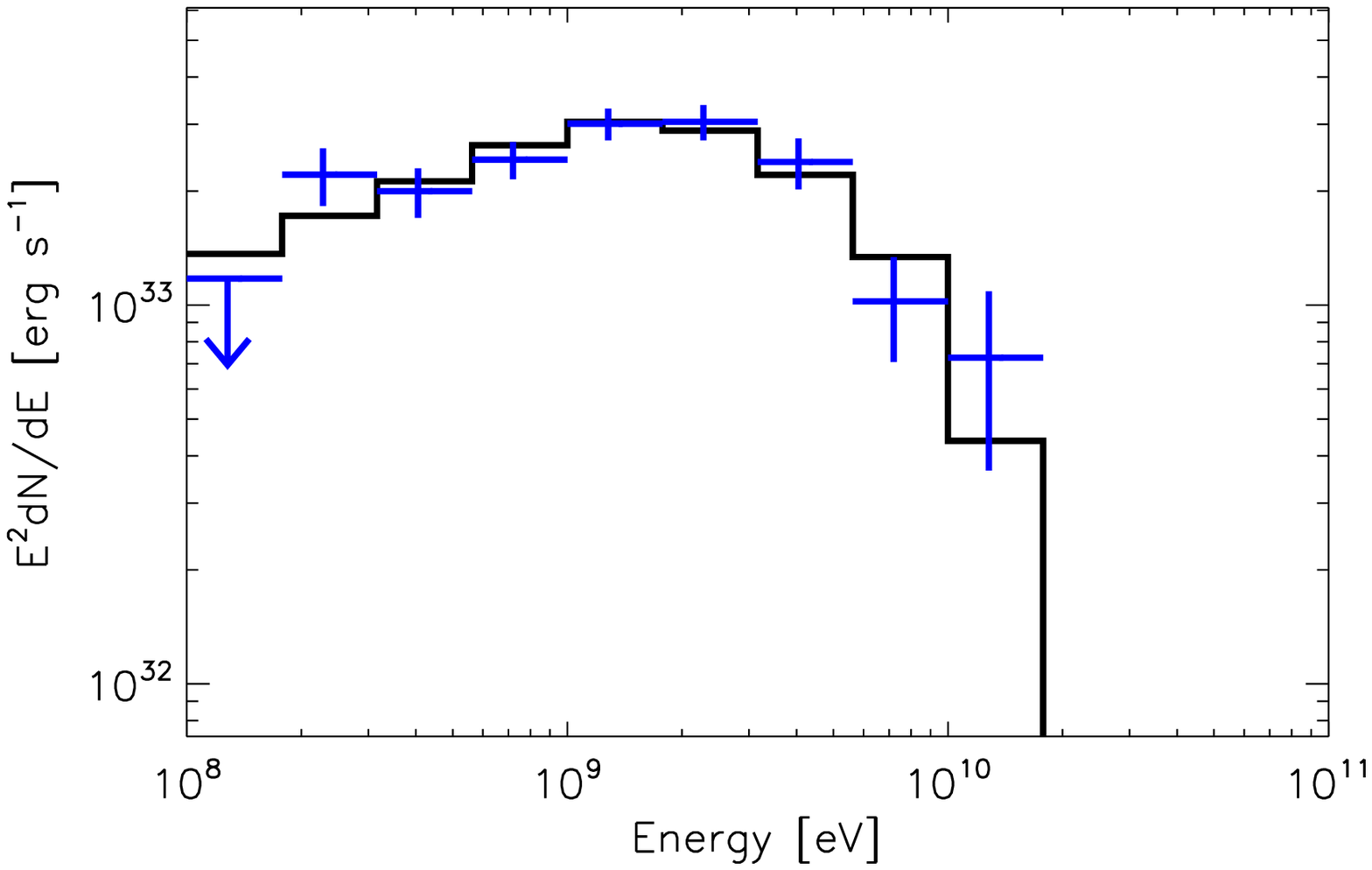}
\put(20,15){\scriptsize J2043+1711}
\end{overpic}
\end{center}
\caption{Fits of the {\it Fermi}-LAT spectral data and models for the pulsars considered in the sample (IV).}
\label{fig:best_fit4}
\end{figure*}

\begin{figure*}
\begin{center}
\begin{overpic}[width=0.32\textwidth]{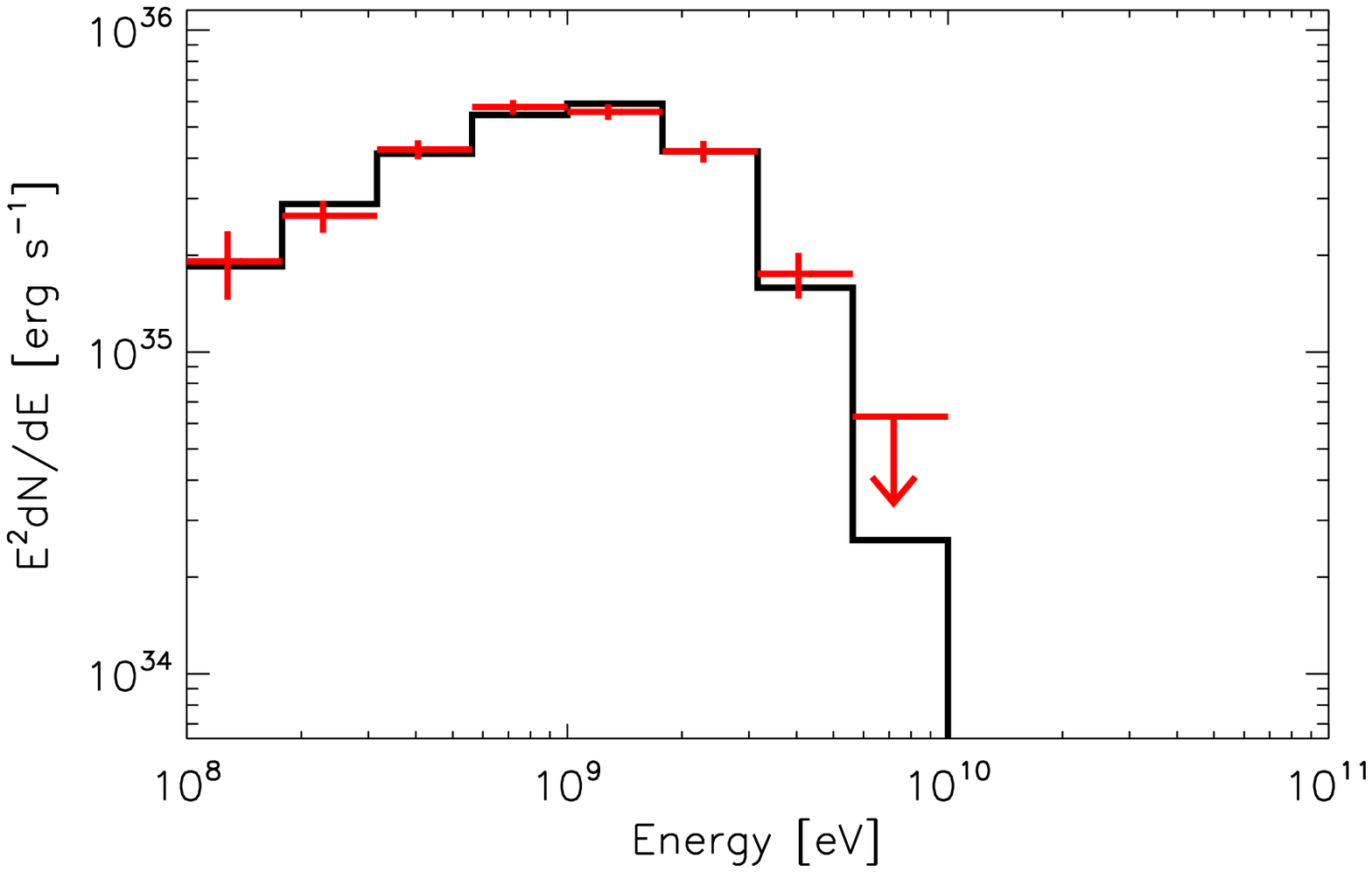}
\put(20,15){\scriptsize J2055+2539}
\end{overpic}
\begin{overpic}[width=0.32\textwidth]{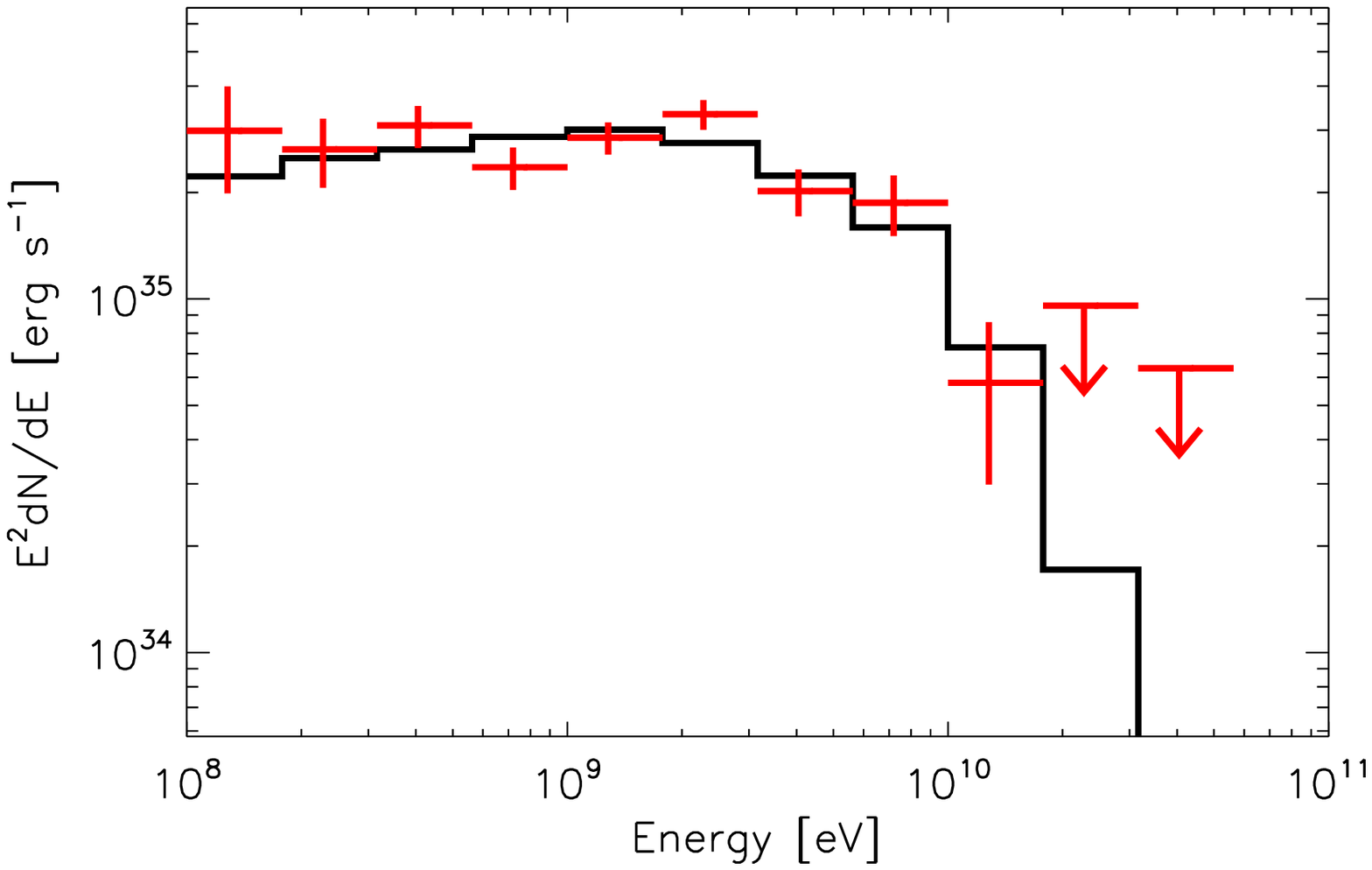}
\put(20,15){\scriptsize J2111+4606}
\end{overpic}
\begin{overpic}[width=0.32\textwidth]{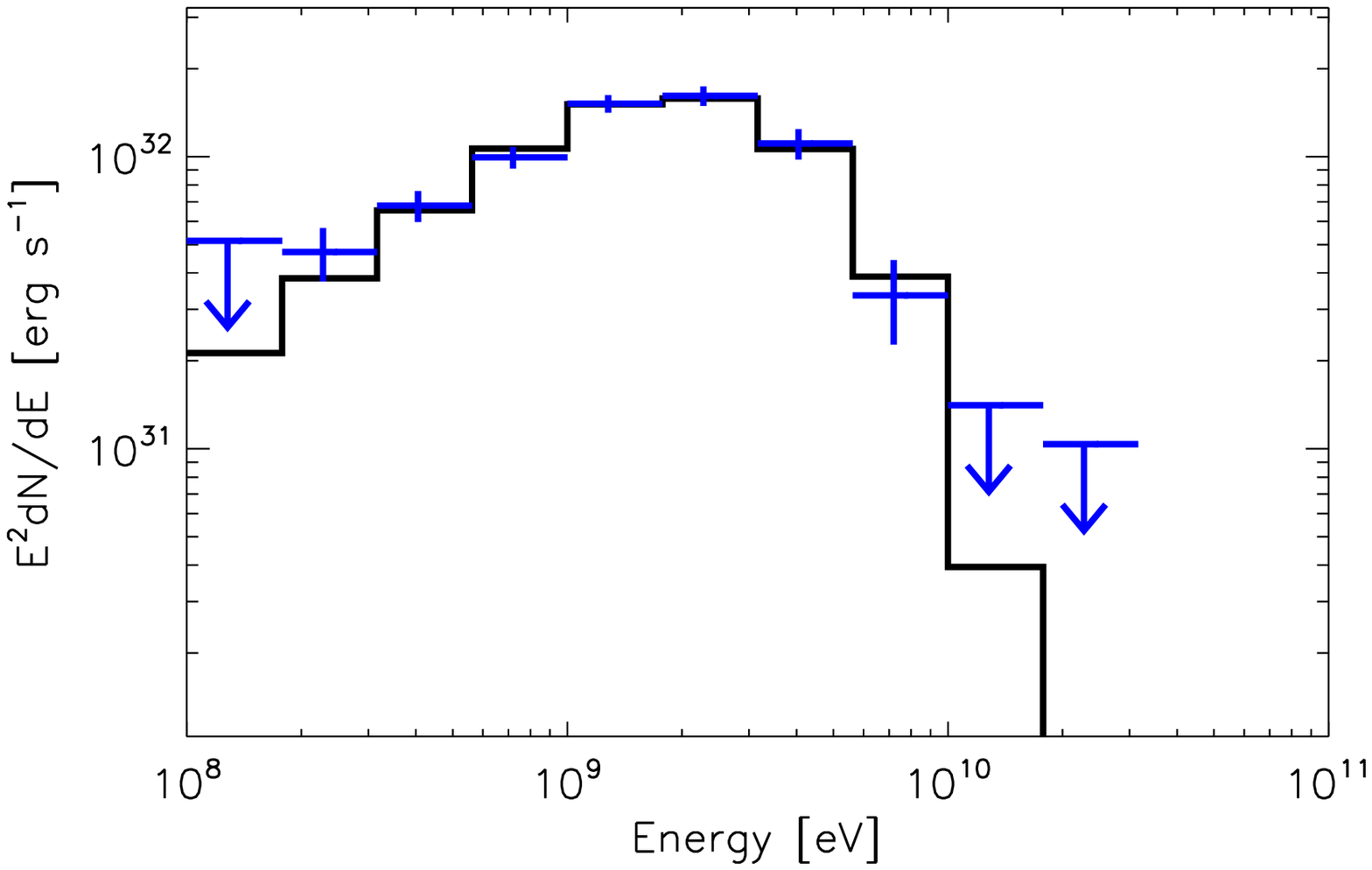}
\put(20,15){\scriptsize J2124-3358}
\end{overpic}
\begin{overpic}[width=0.32\textwidth]{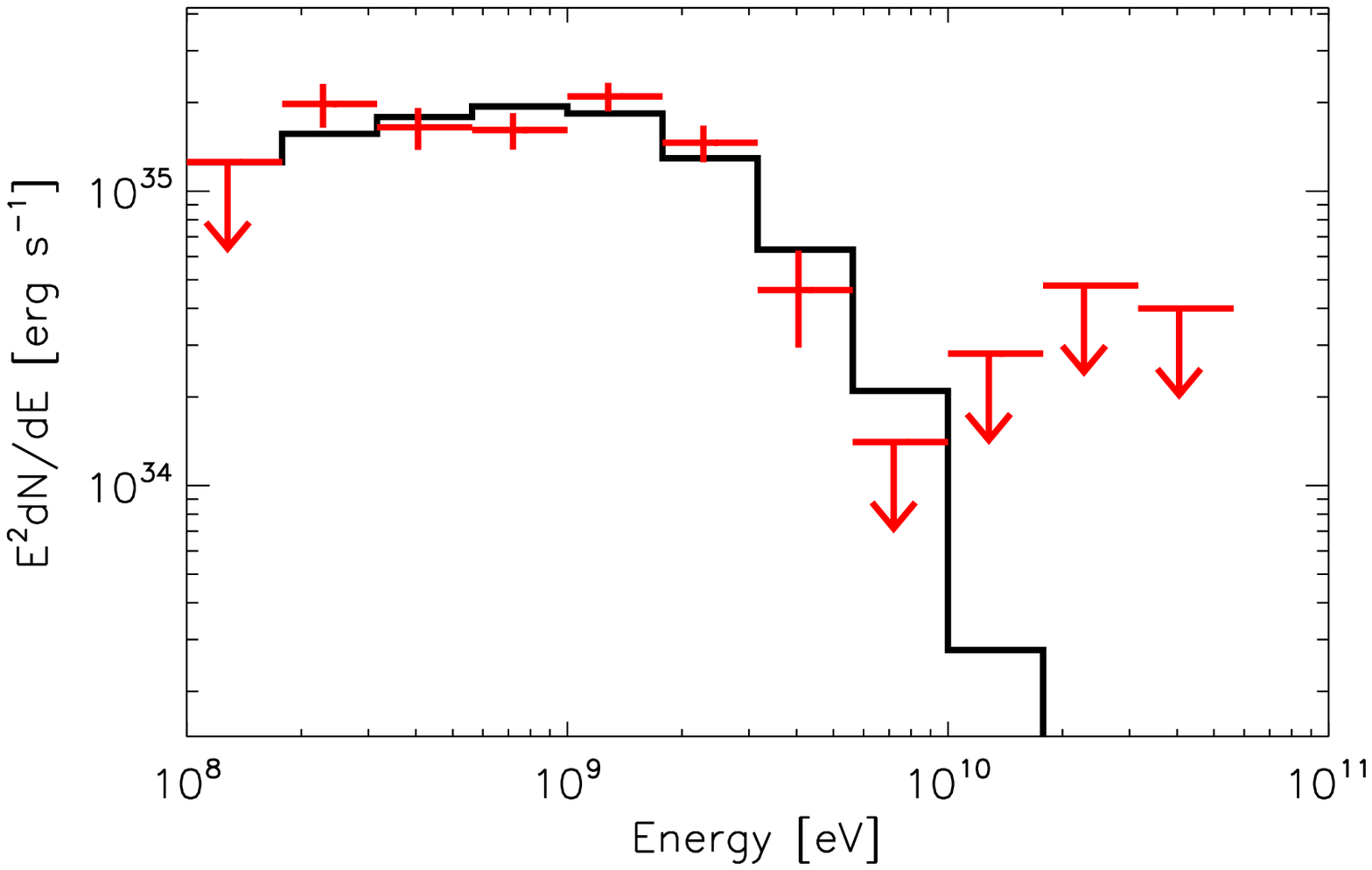}
\put(20,15){\scriptsize J2139+4716}
\end{overpic}
\begin{overpic}[width=0.32\textwidth]{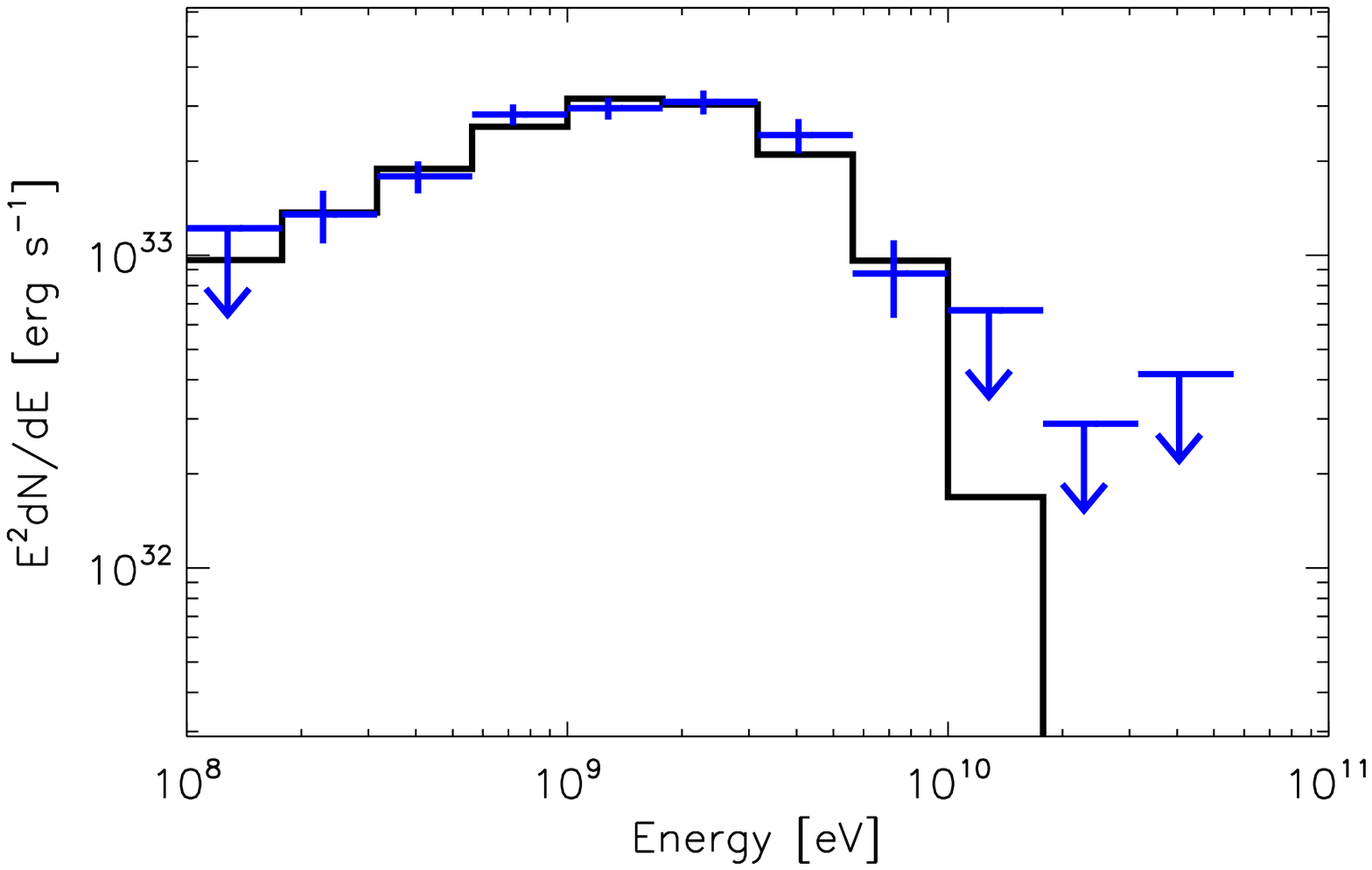}
\put(20,15){\scriptsize J2214+3000}
\end{overpic}
\begin{overpic}[width=0.32\textwidth]{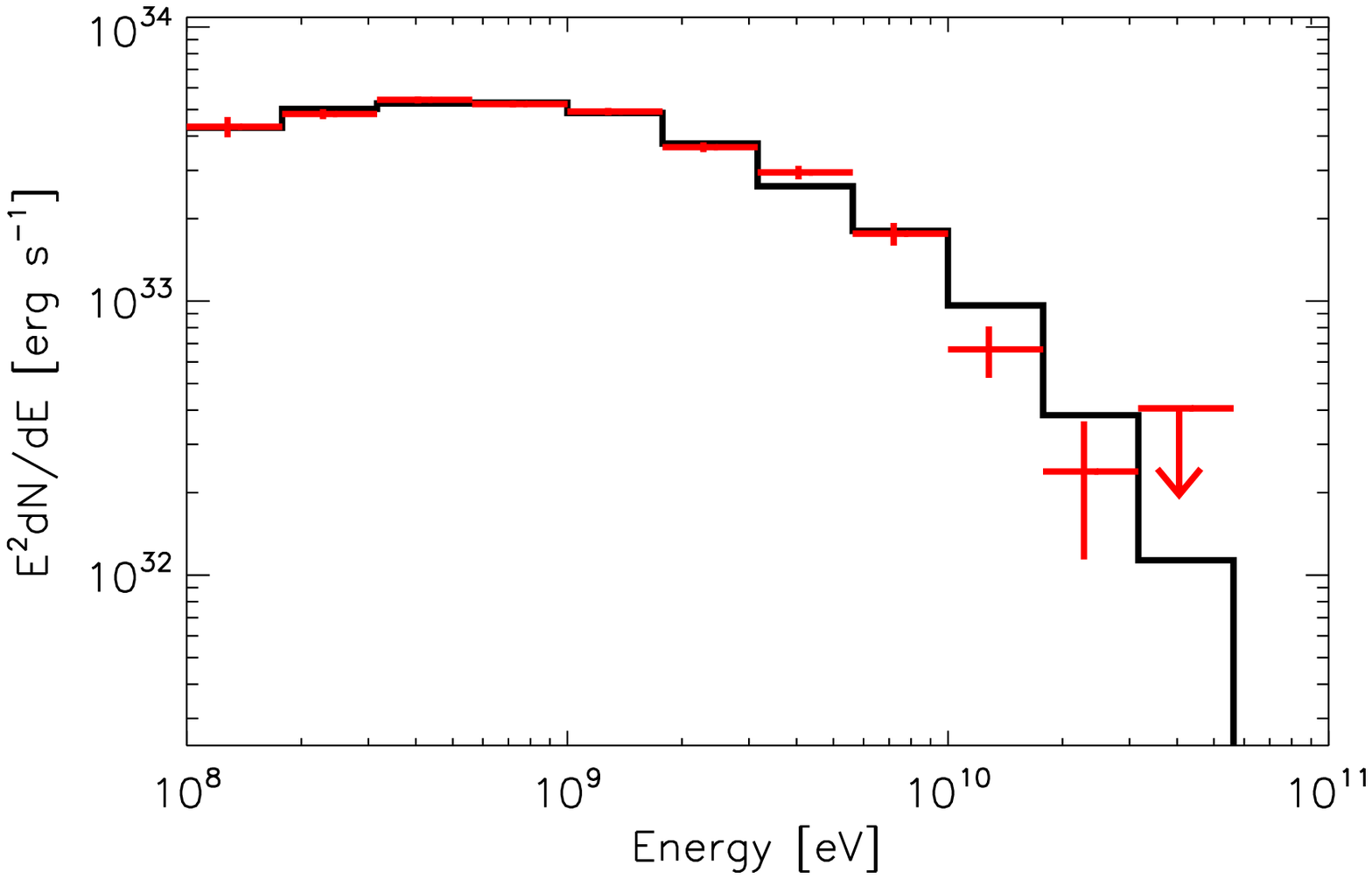}
\put(20,15){\scriptsize J2229+6114}
\end{overpic}
\begin{overpic}[width=0.32\textwidth]{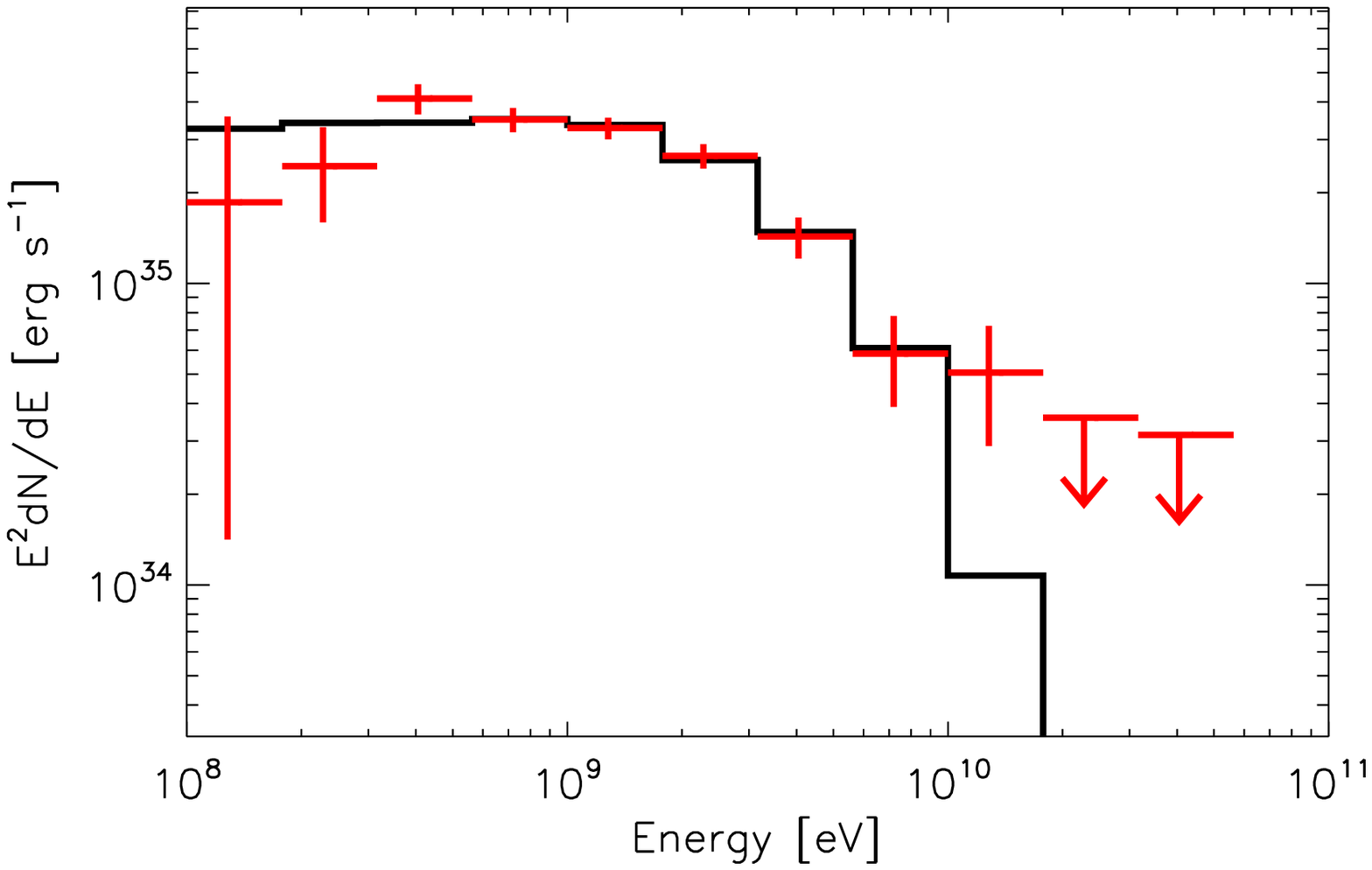}
\put(20,15){\scriptsize J2238+5903}
\end{overpic}
\begin{overpic}[width=0.32\textwidth]{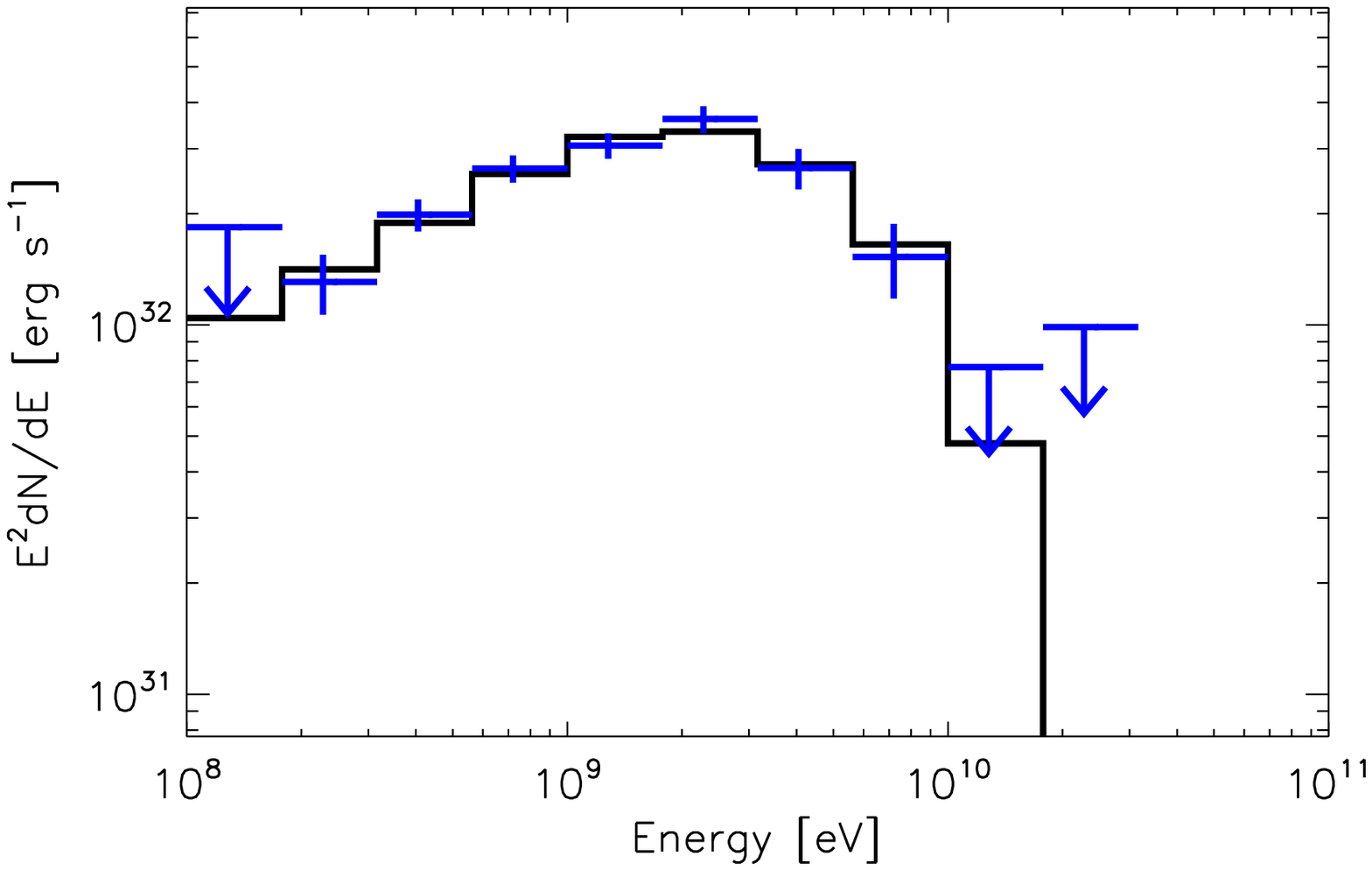}
\put(20,15){\scriptsize J2241-5236}
\end{overpic}
\begin{overpic}[width=0.32\textwidth]{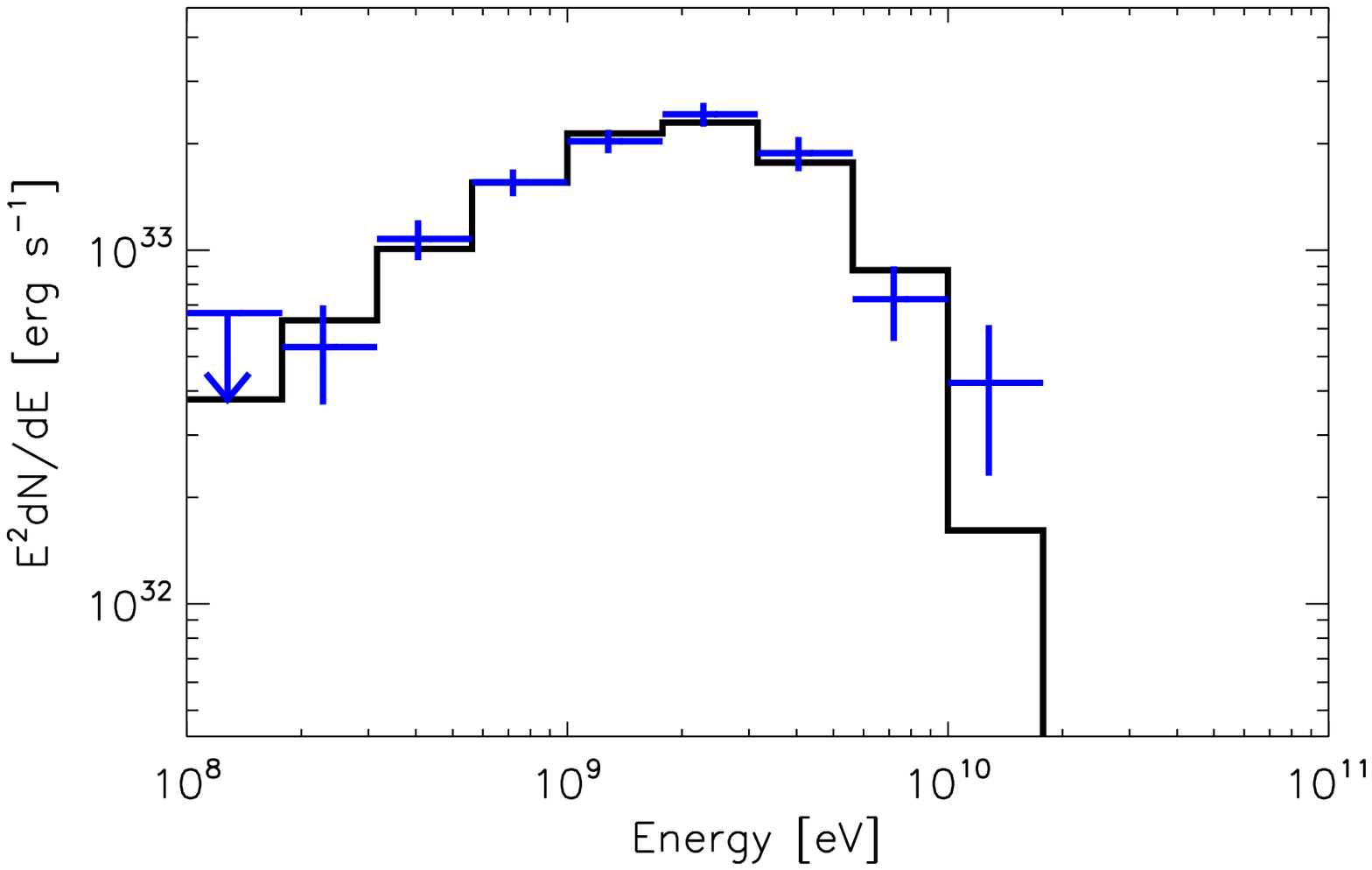}
\put(20,15){\scriptsize J2302+4442}
\end{overpic}
\end{center}
\caption{Fits of the {\it Fermi}-LAT spectral data and models for the pulsars considered in the sample (V).}
\label{fig:best_fit5}
\end{figure*}

\begin{figure*}
\begin{center}
\begin{overpic}[width=0.32\textwidth]{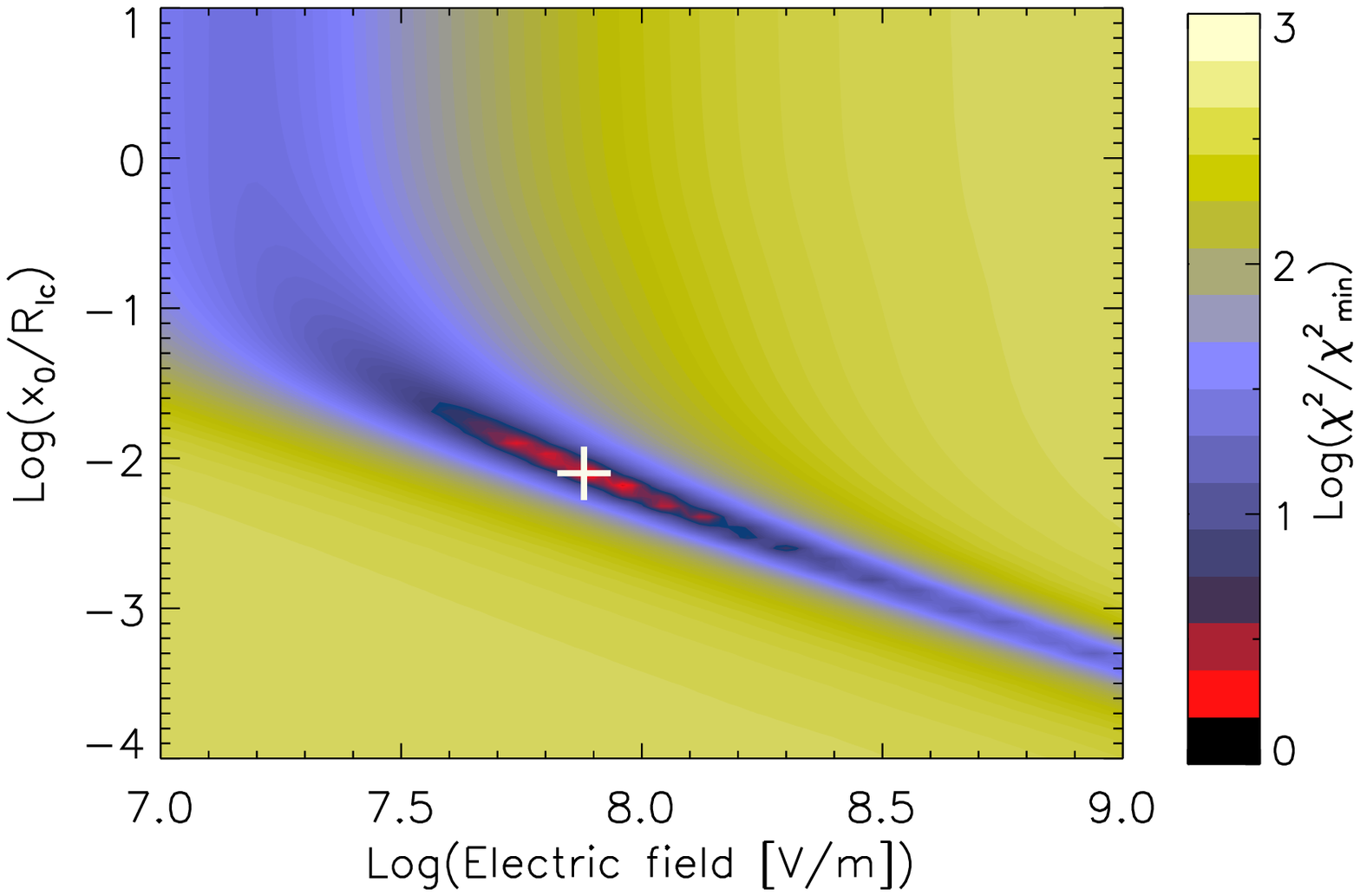}
\put(20,15){\scriptsize \red {J0007+7303 (CTA1)}}
\end{overpic}
\begin{overpic}[width=0.32\textwidth]{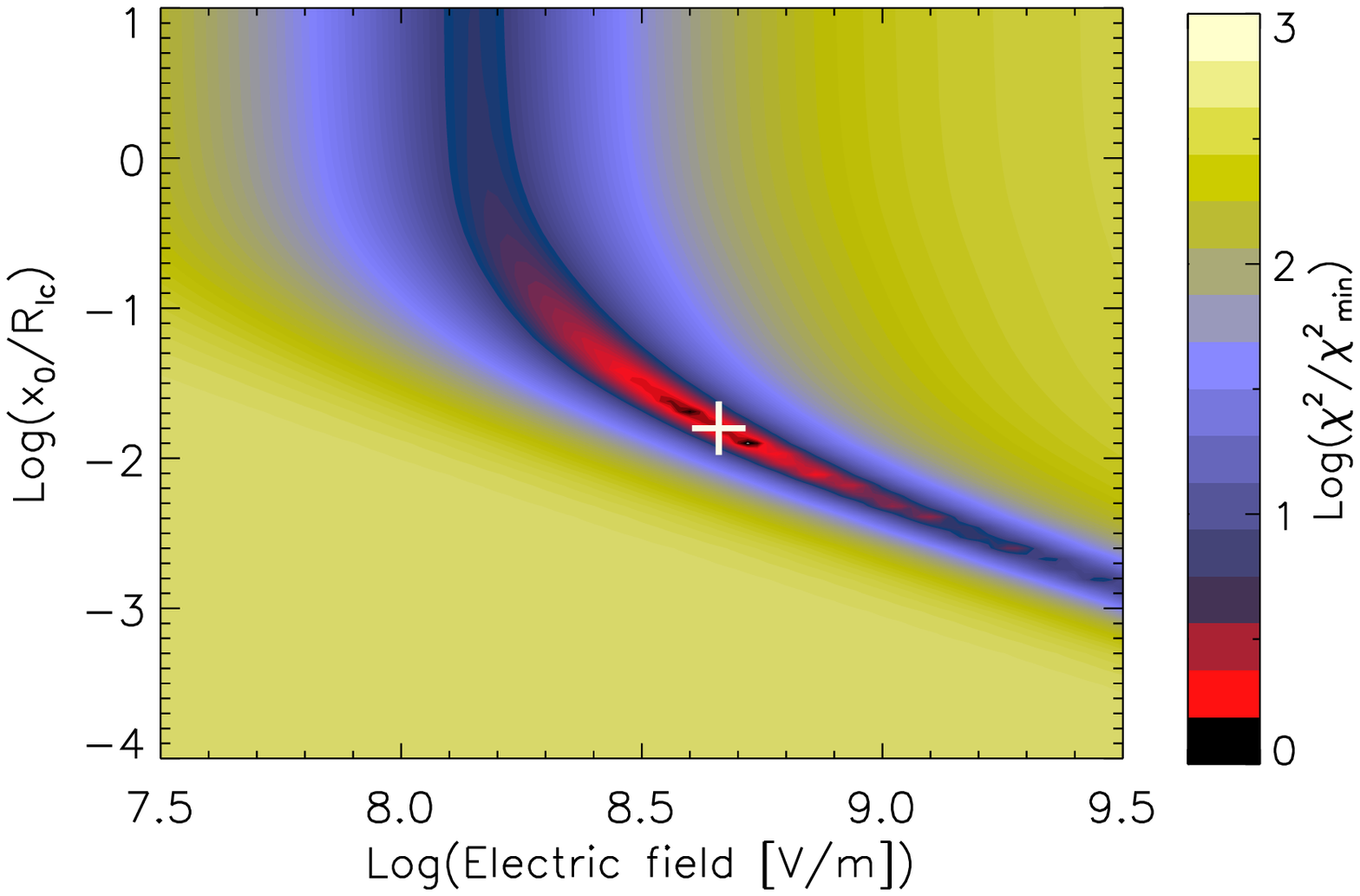}
\put(20,15){\scriptsize \blue {J0030+0451}}
\end{overpic}
\begin{overpic}[width=0.32\textwidth]{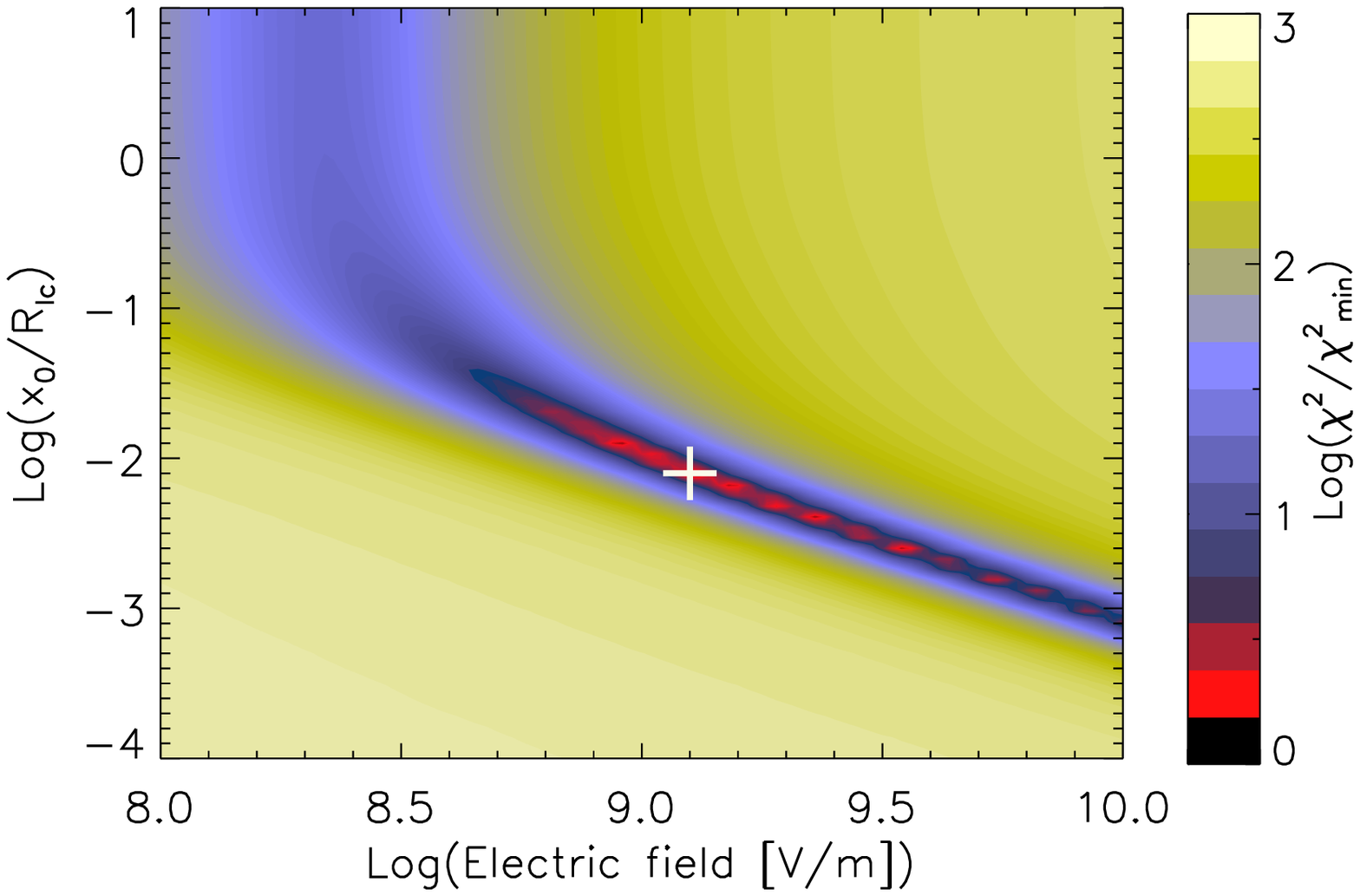}
\put(20,15){\scriptsize \blue {J0034-0534}}
\end{overpic}
\begin{overpic}[width=0.32\textwidth]{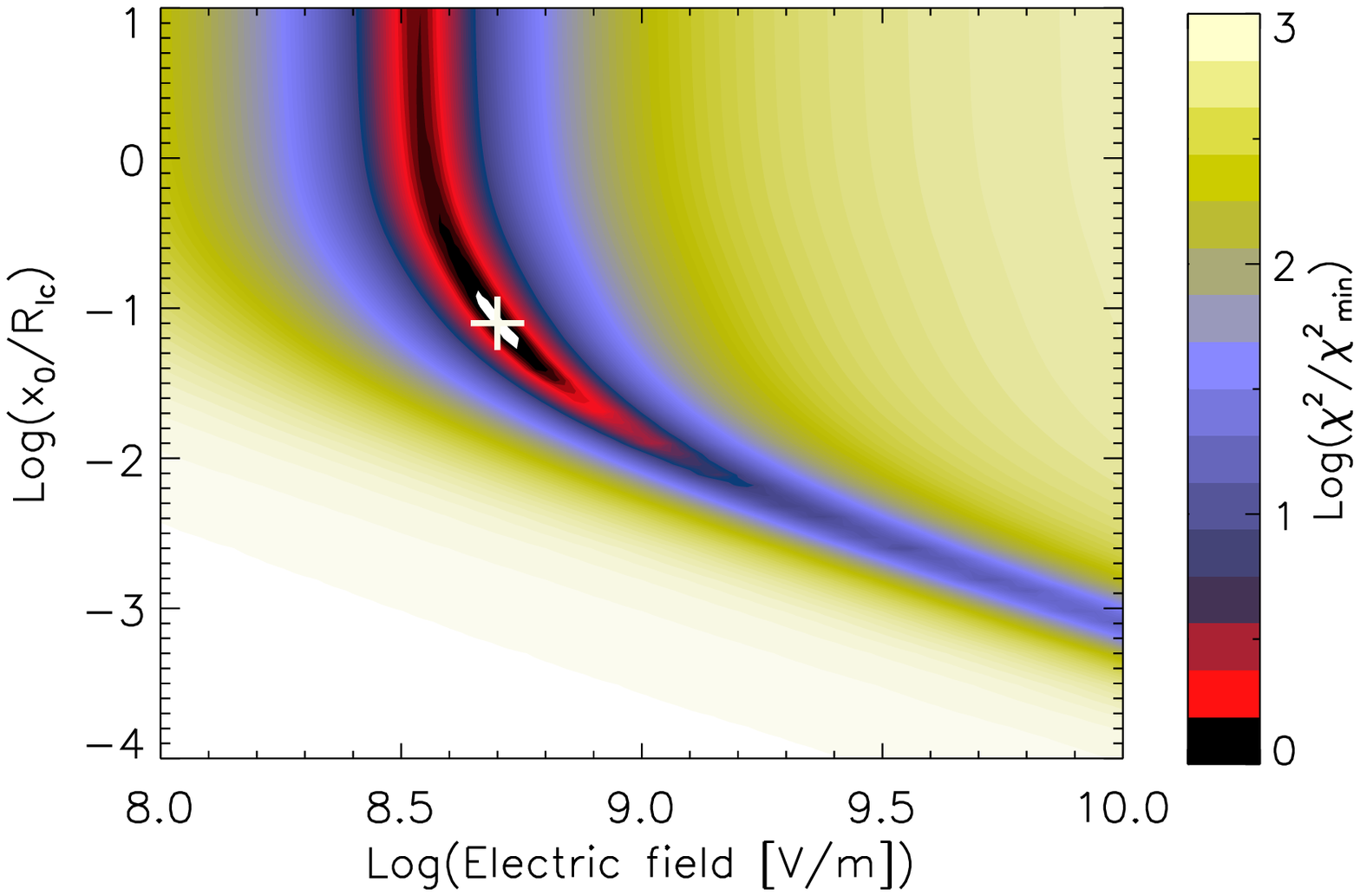}
\put(20,15){\scriptsize \blue {J0101-6422}}
\end{overpic}
\begin{overpic}[width=0.32\textwidth]{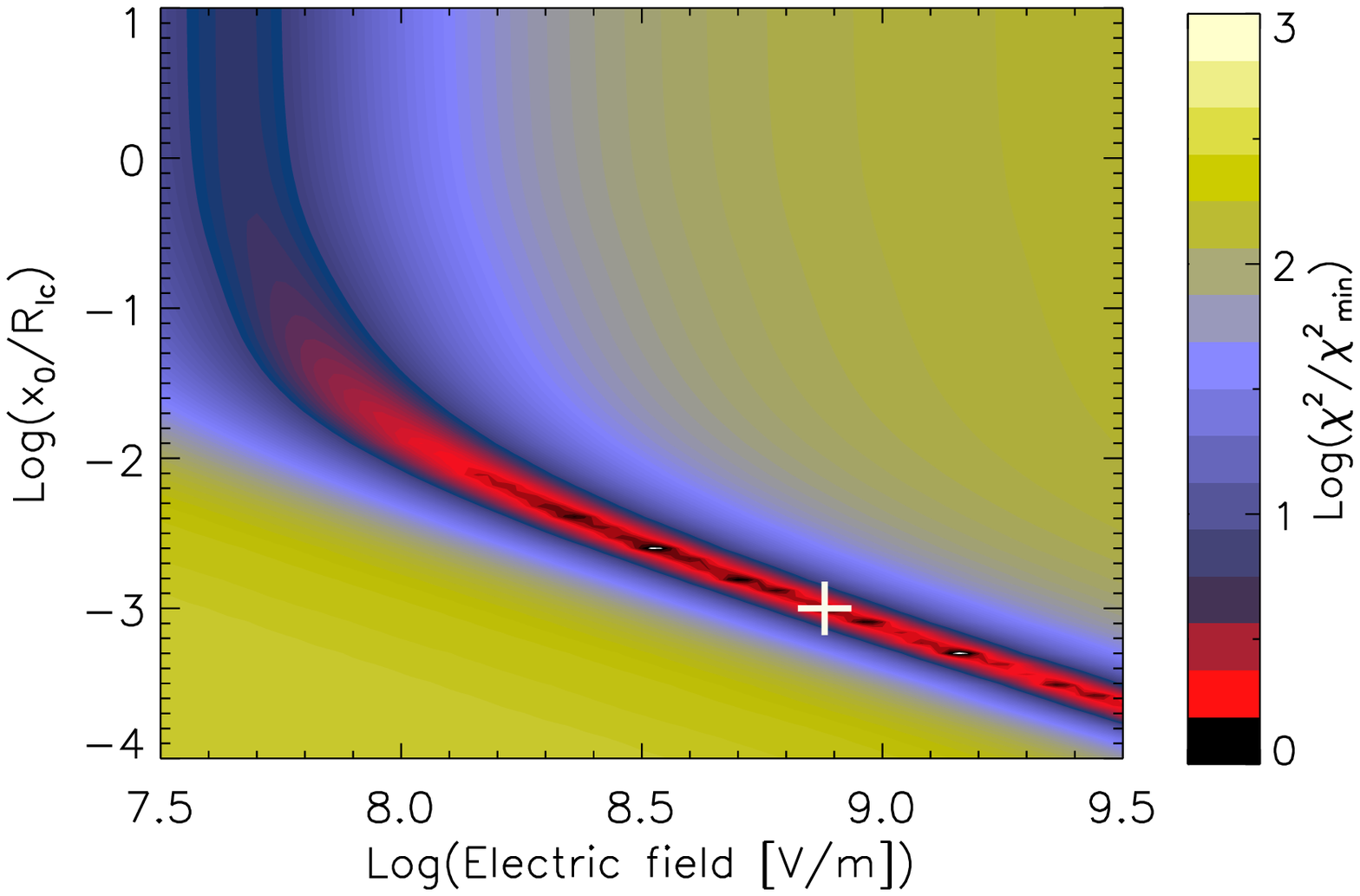}
\put(20,15){\scriptsize  \red {J0106+4855}}
\end{overpic}
\begin{overpic}[width=0.32\textwidth]{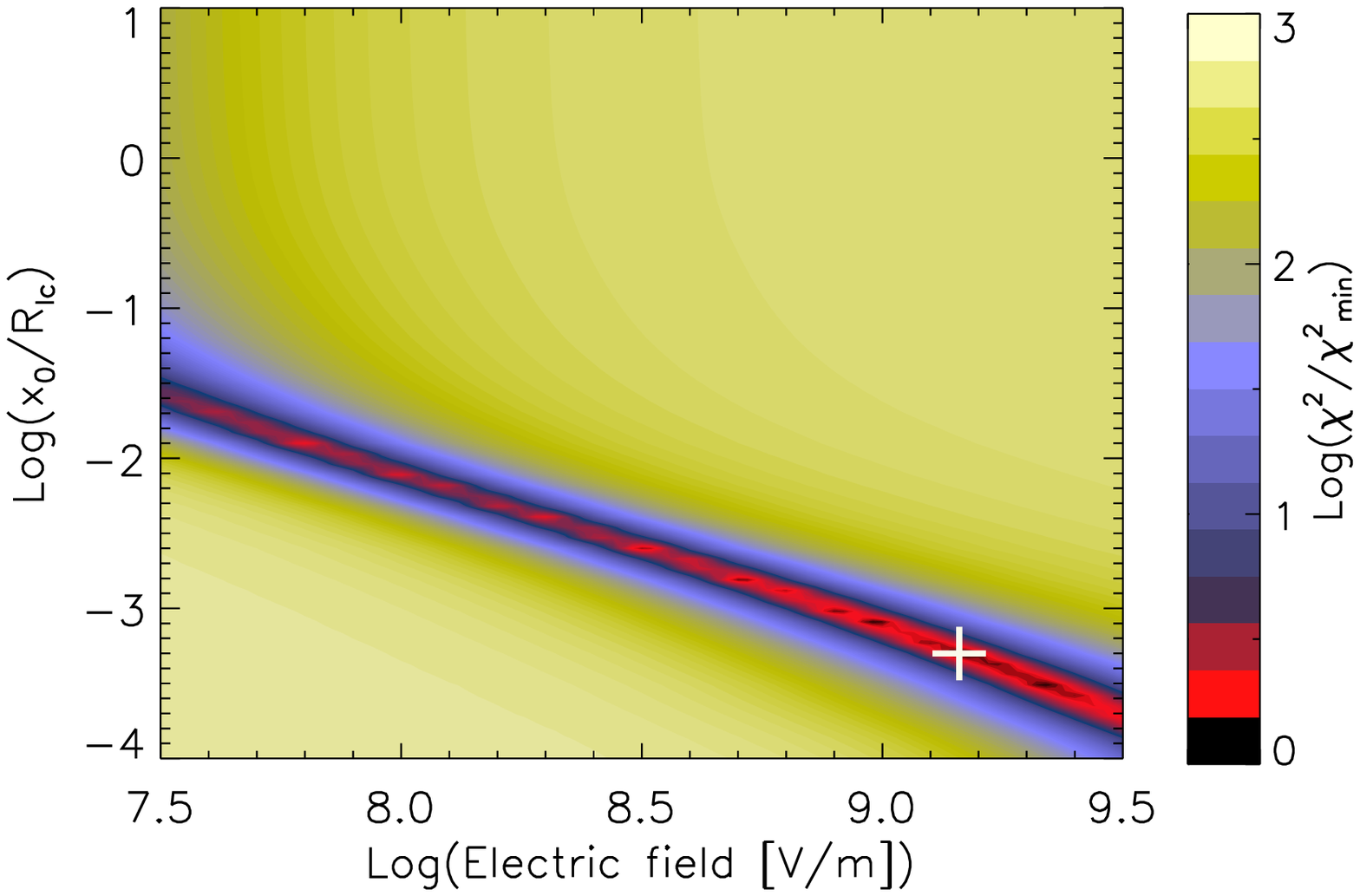}
\put(20,15){\scriptsize  \red {J0205+6449}}
\end{overpic}
\begin{overpic}[width=0.32\textwidth]{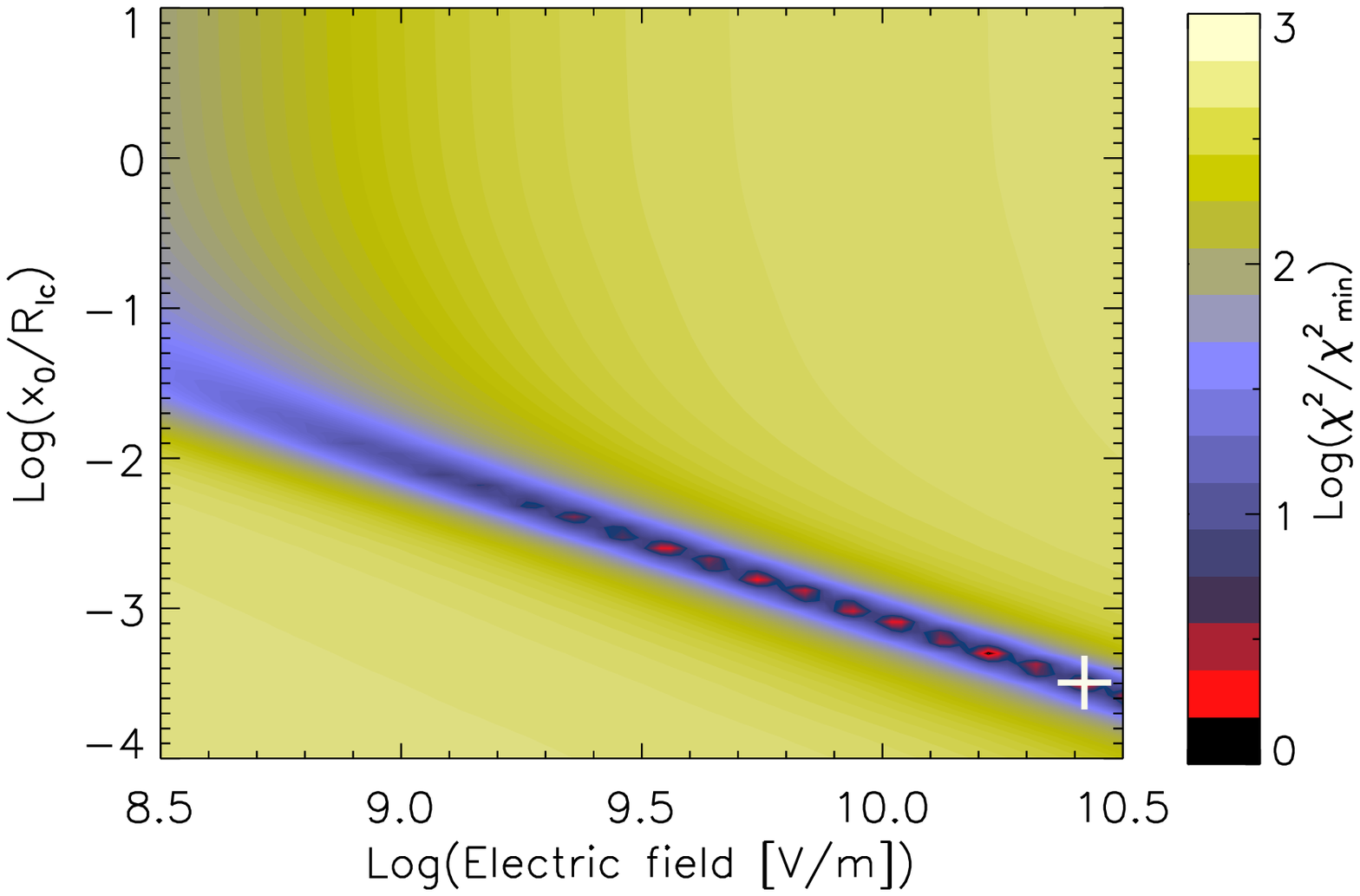}
\put(20,15){\scriptsize \blue{J0218+4232}}
\end{overpic}
\begin{overpic}[width=0.32\textwidth]{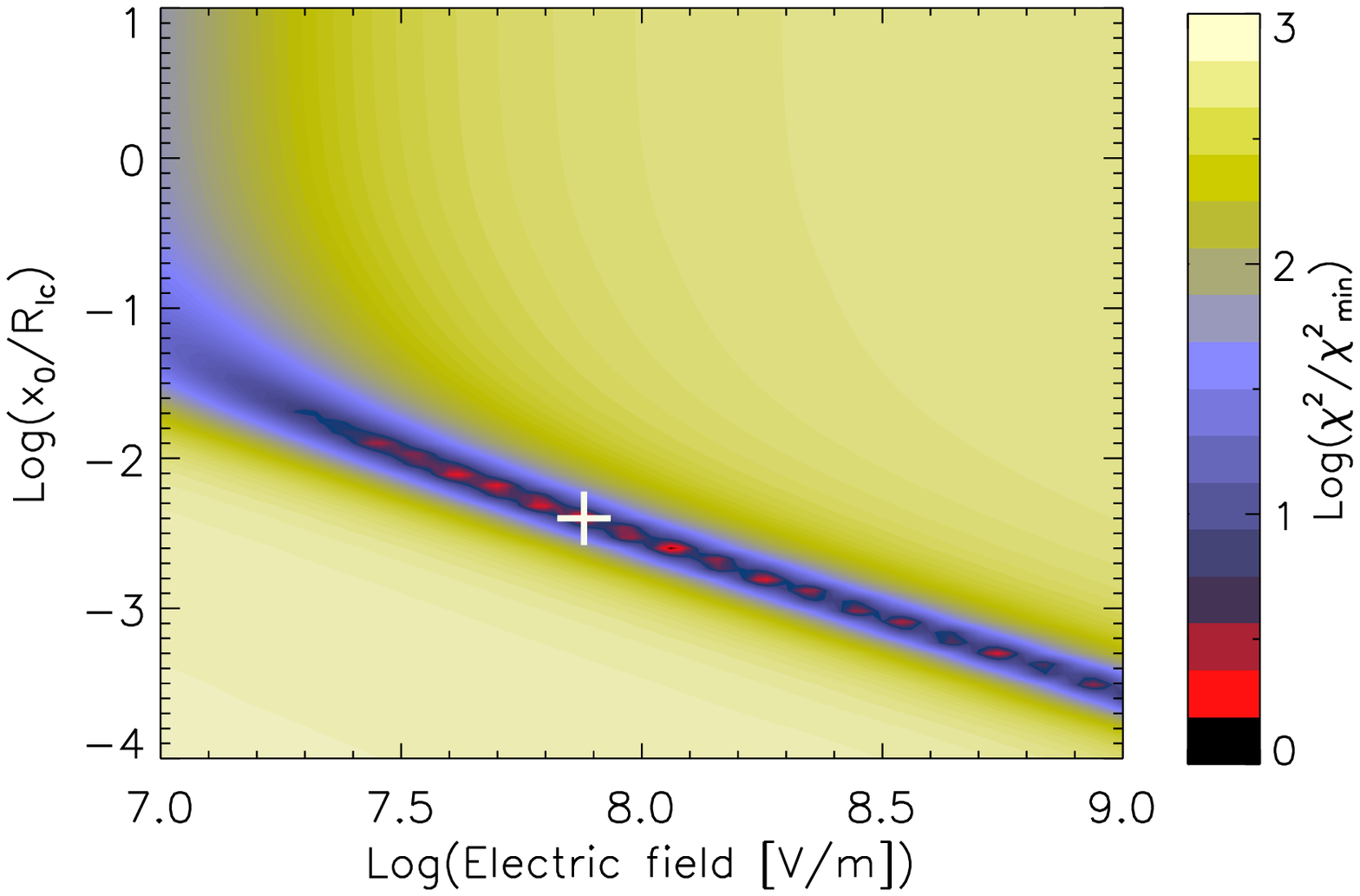}
\put(20,15){\scriptsize  \red {J0248+6021}}
\end{overpic}
\begin{overpic}[width=0.32\textwidth]{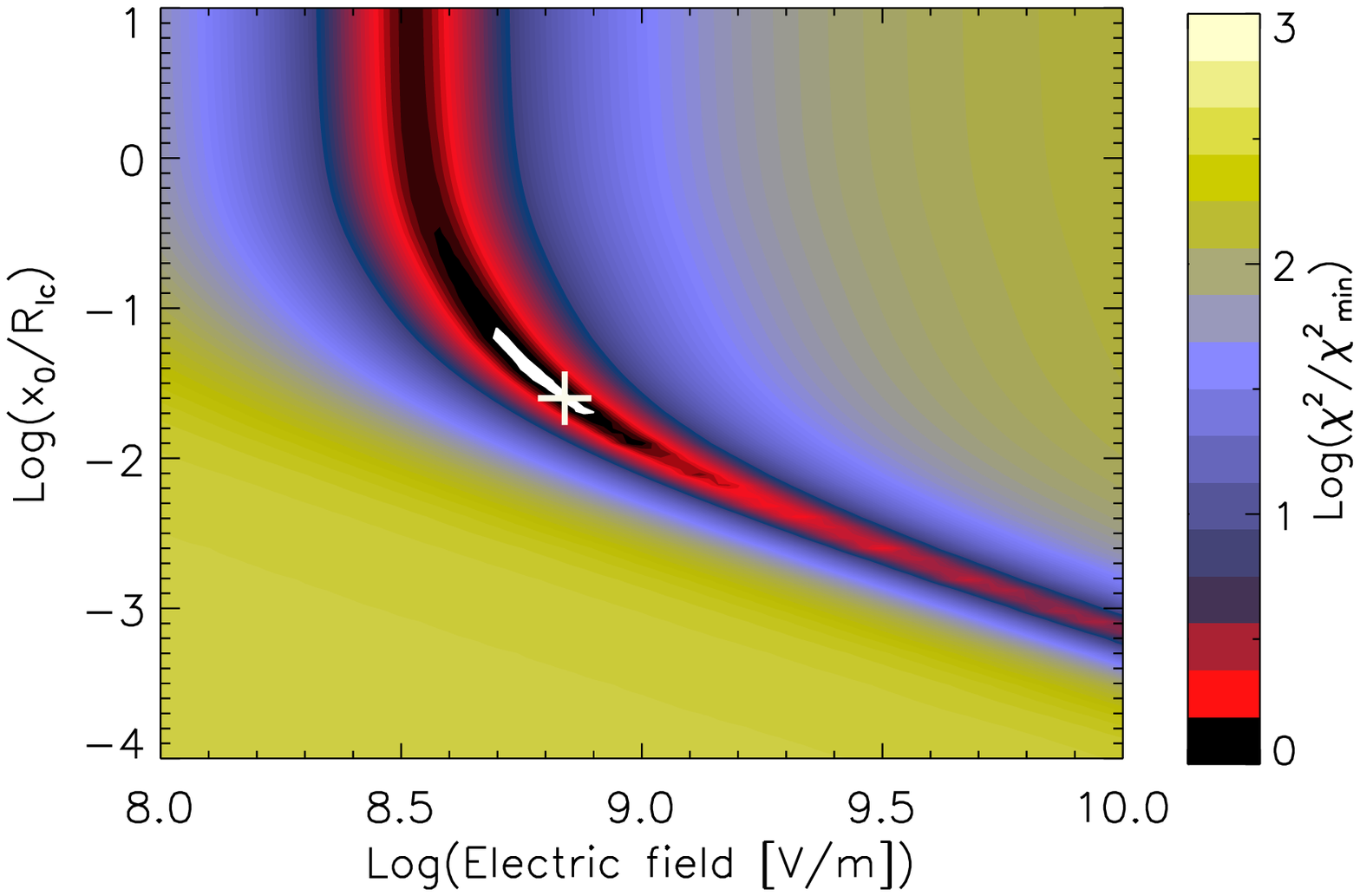}
\put(20,15){\scriptsize \blue {J0340+4130}}
\end{overpic}
\begin{overpic}[width=0.32\textwidth]{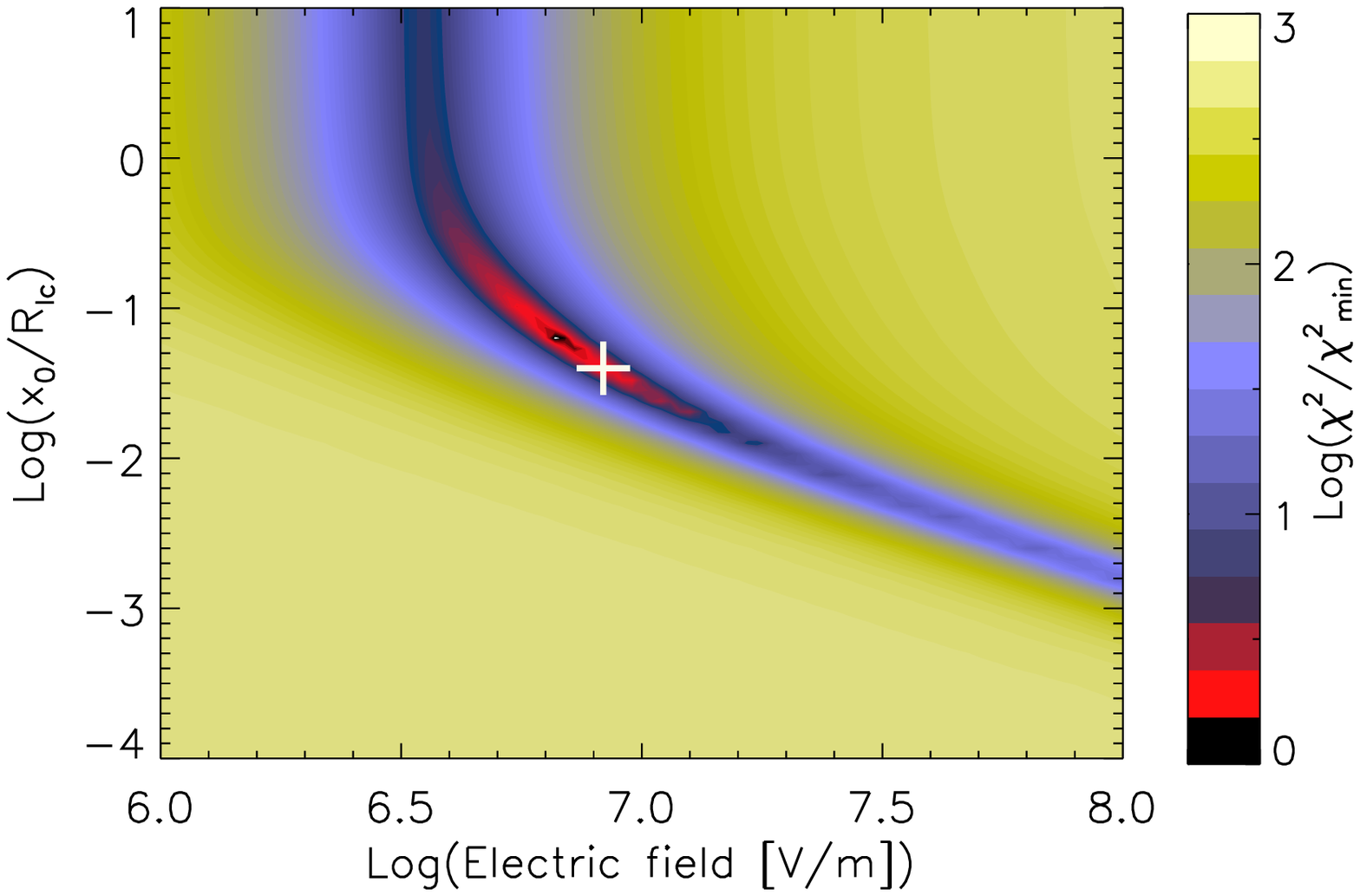}
\put(20,15){\scriptsize  \red {J0357+3205}}
\end{overpic}
\begin{overpic}[width=0.32\textwidth]{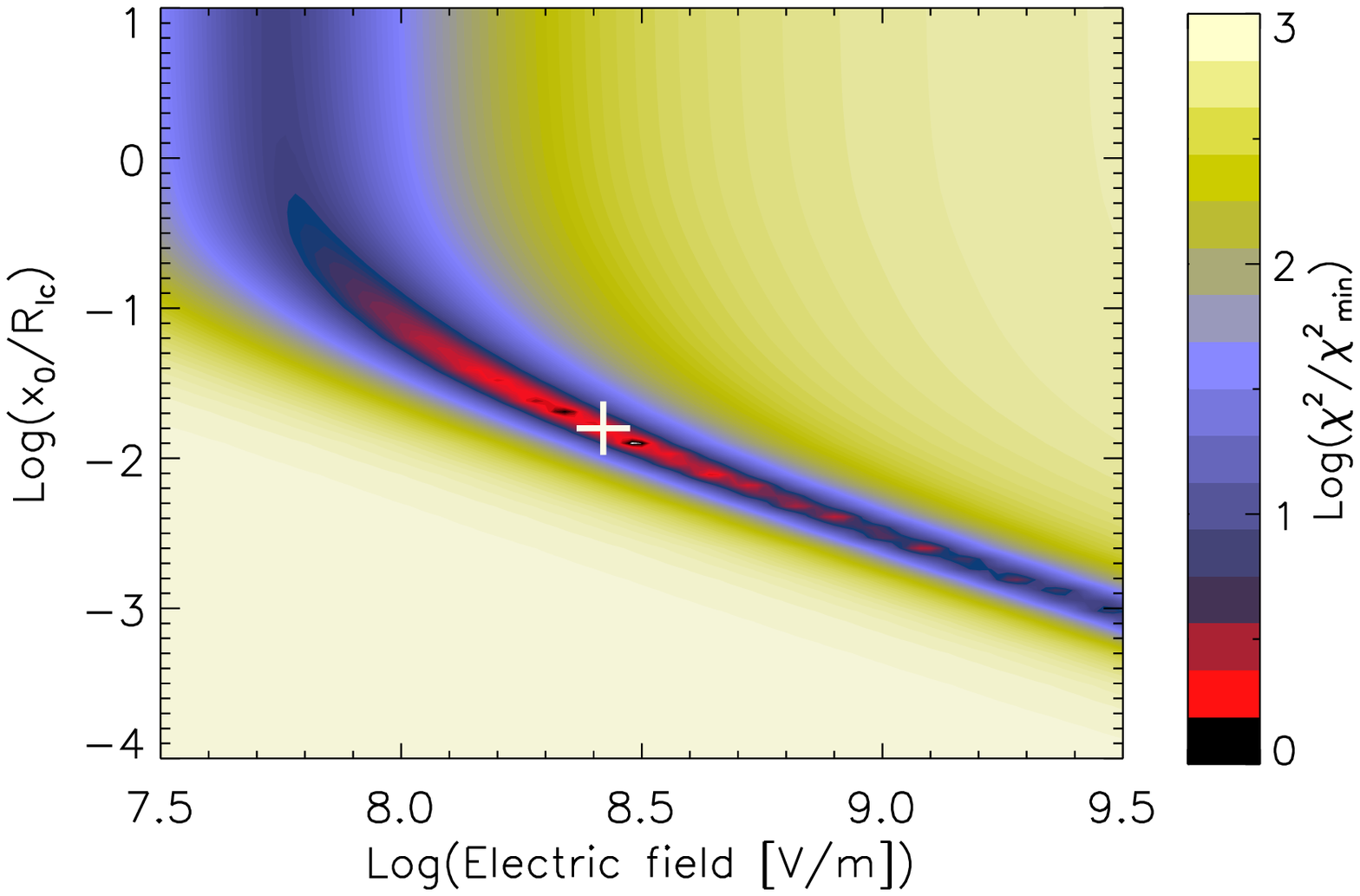}
\put(20,15){\scriptsize \blue{J0437-4715}}
\end{overpic}
\begin{overpic}[width=0.32\textwidth]{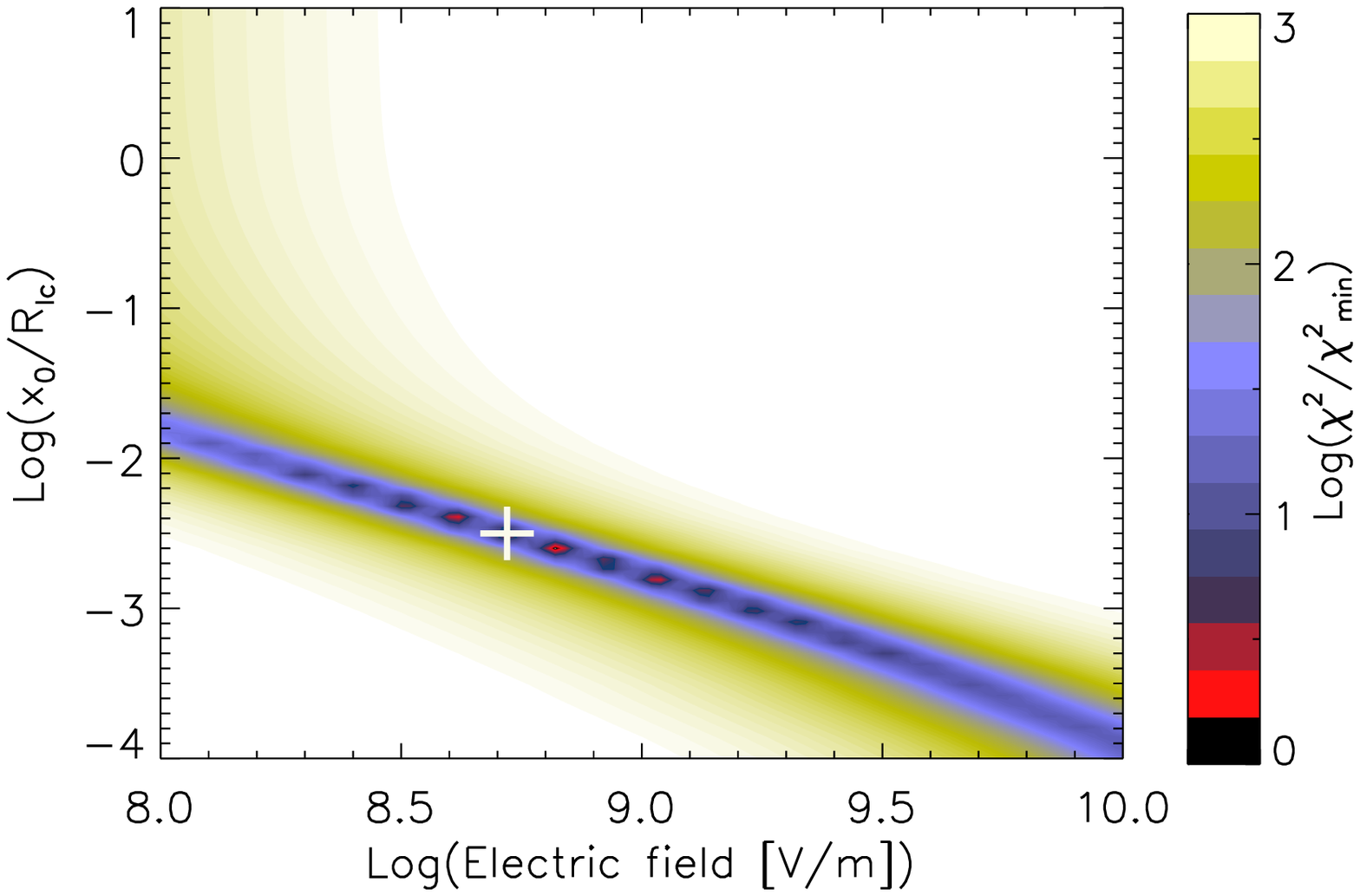}
\put(20,15){\scriptsize  \red {J0534+2200 (Crab)}}
\end{overpic}
\begin{overpic}[width=0.32\textwidth]{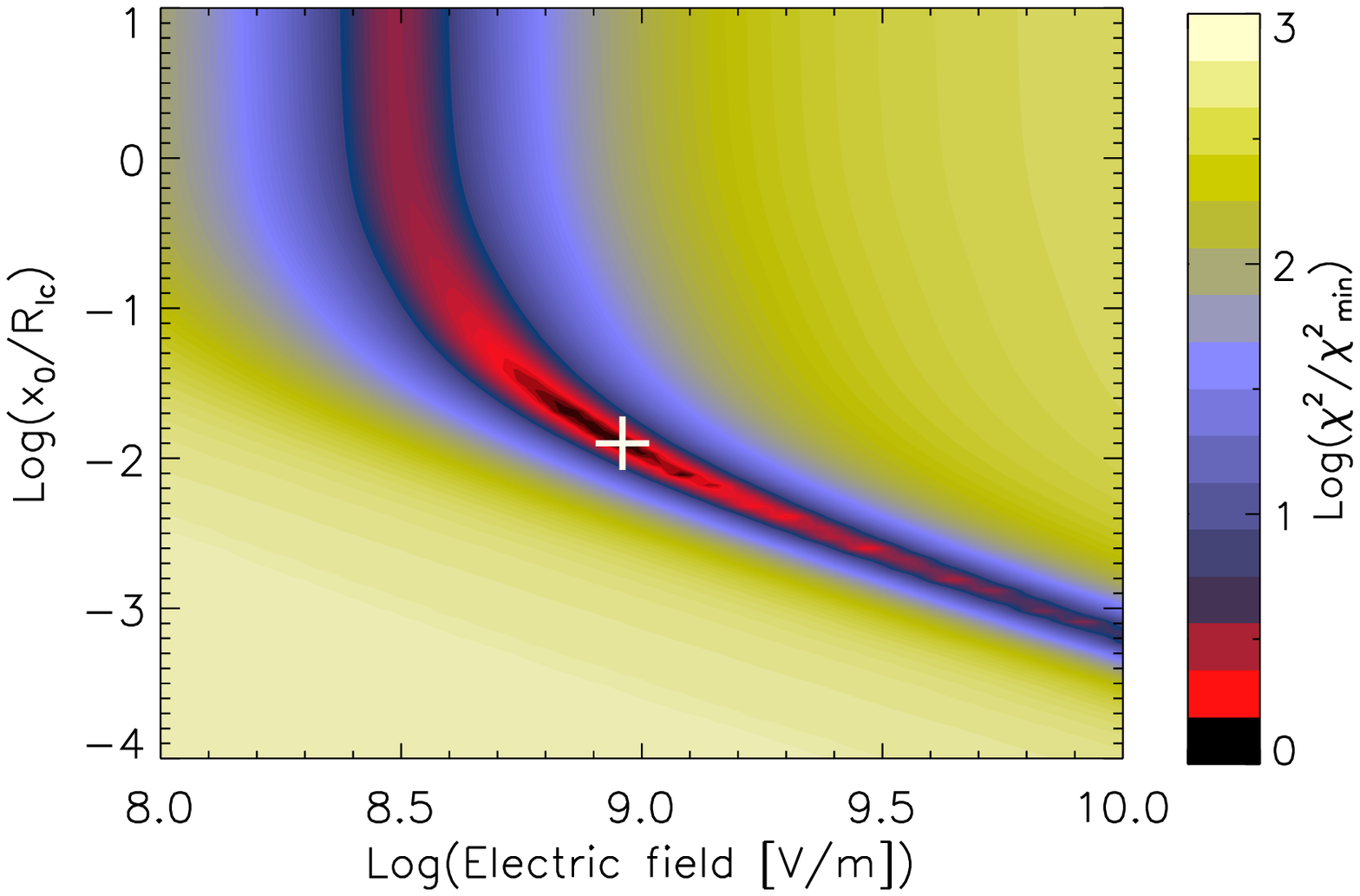}
\put(20,15){\scriptsize \blue {J0613-0200}}
\end{overpic}
\begin{overpic}[width=0.32\textwidth]{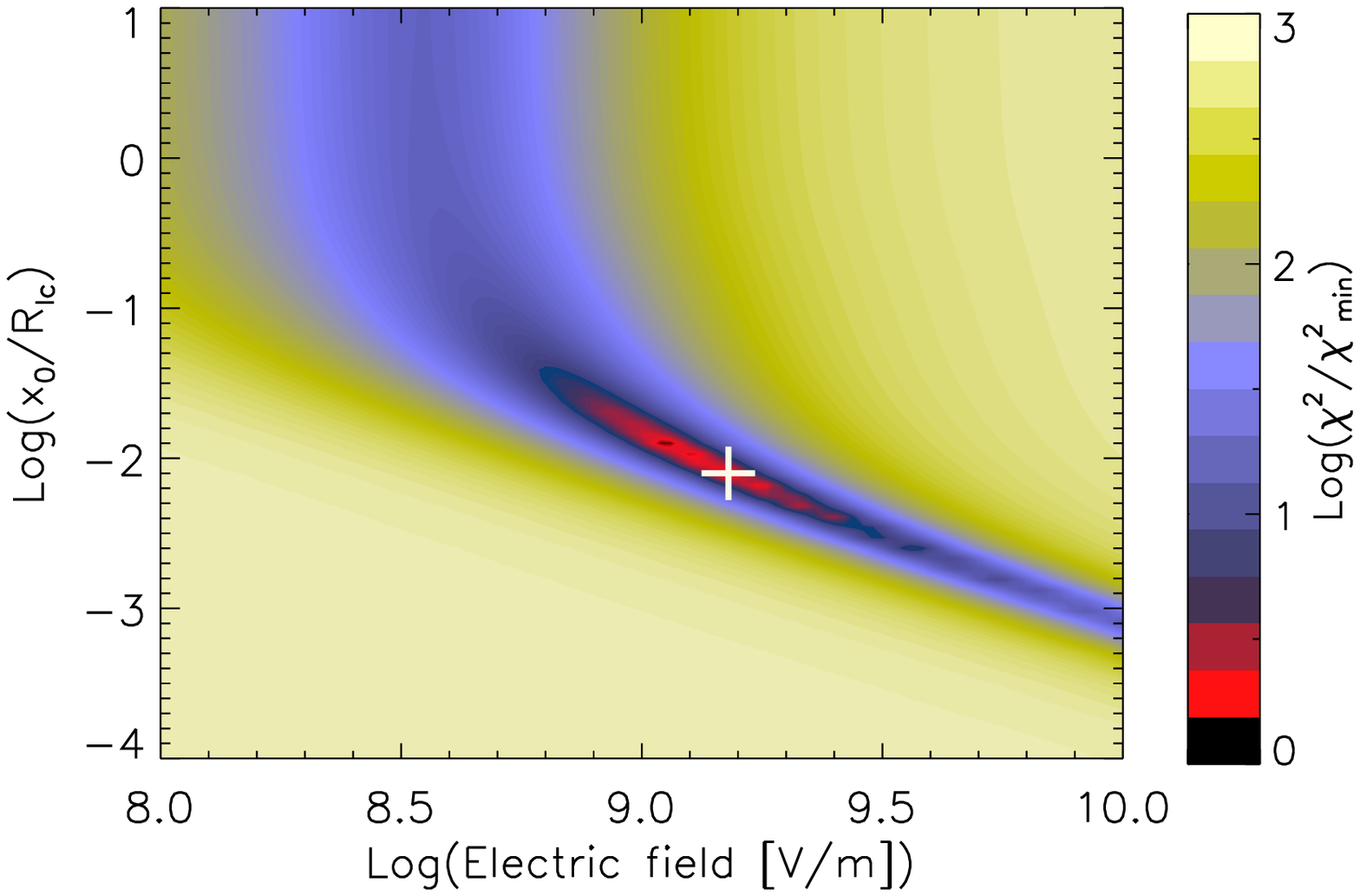}
\put(20,15){\scriptsize \blue {J0614-3329}}
\end{overpic}
\begin{overpic}[width=0.32\textwidth]{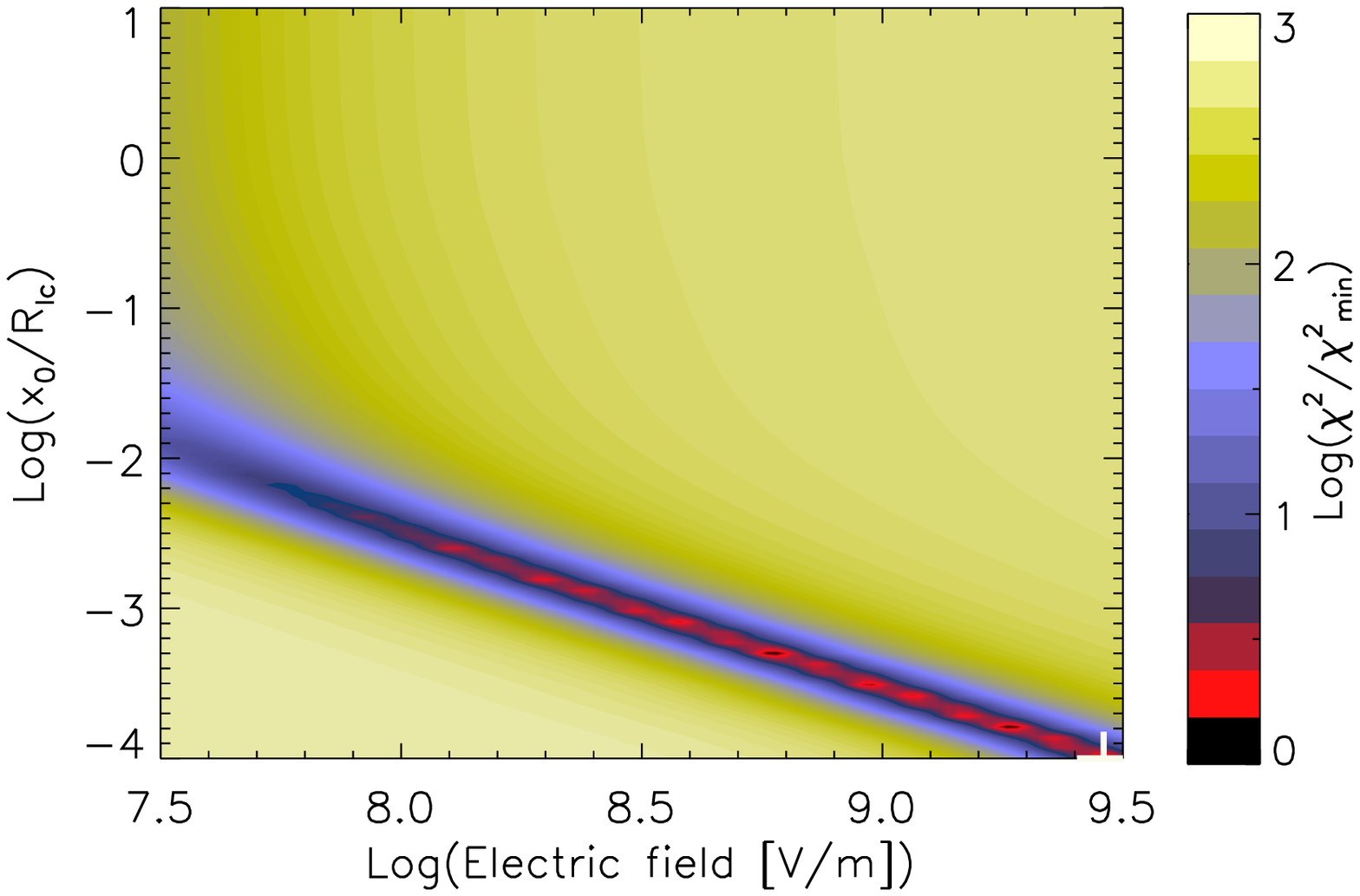}
\put(20,15){\scriptsize  \red {J0631+1036}}
\end{overpic}
\begin{overpic}[width=0.32\textwidth]{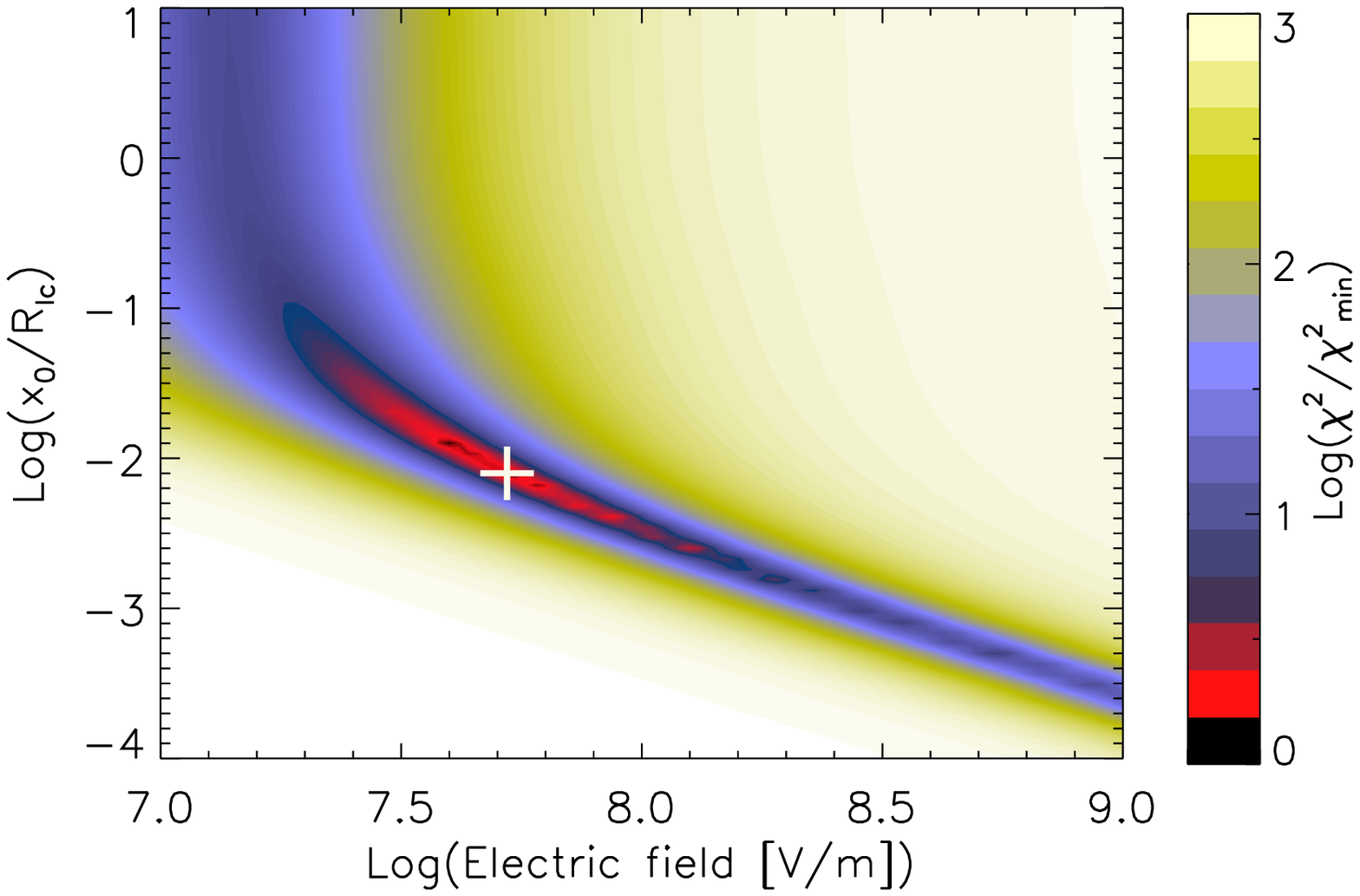}
\put(20,15){\scriptsize  \red {J0633+0632}}
\end{overpic}
\begin{overpic}[width=0.32\textwidth]{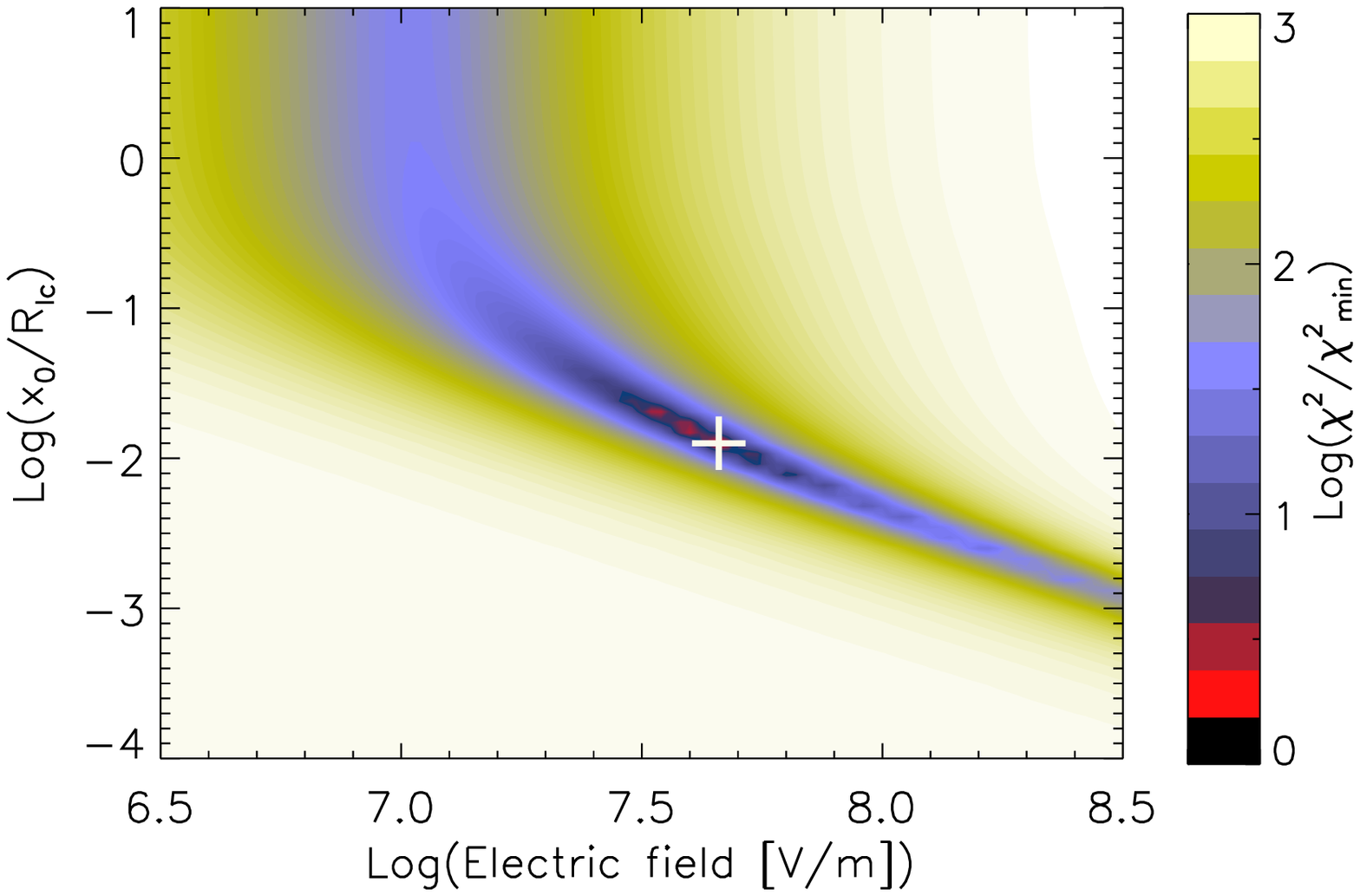}
\put(20,15){\scriptsize  \red {J0633+1746 (Geminga)}}
\end{overpic}
\begin{overpic}[width=0.32\textwidth]{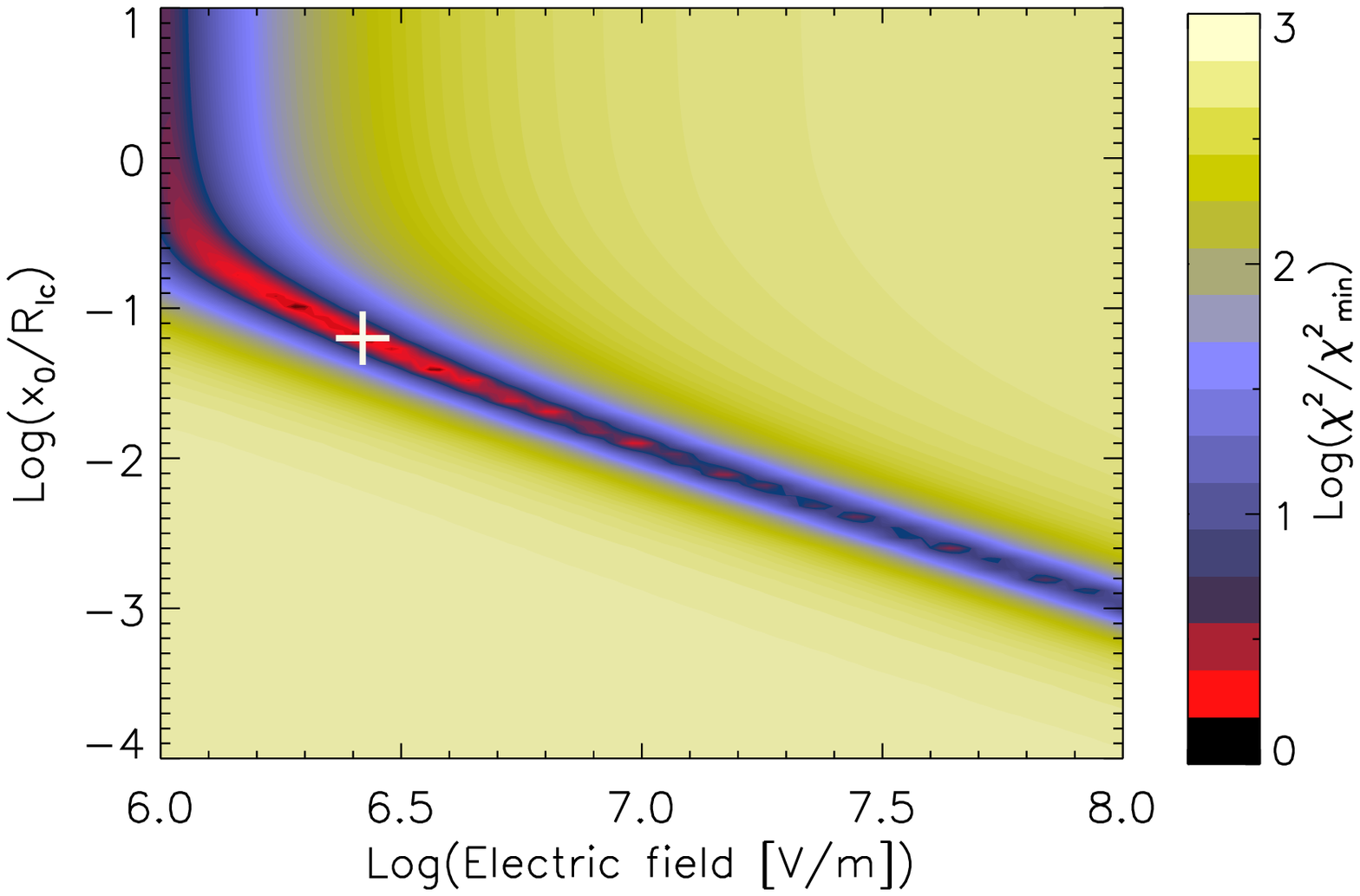}
\put(20,15){\scriptsize  \red {J0659+1414}}
\end{overpic}
\end{center}
\caption{$\chi^2/\chi^2_{\rm min}$ contours on the $\log E_\parallel$--$\log (x_0/R_{\rm lc})$ plane for the pulsars considered in the sample (I). Red and blue labels indicate YPs and MSPs, respectively.}
\label{fig:contours1}
\end{figure*}

\begin{figure*}
\begin{center}
\begin{overpic}[width=0.32\textwidth]{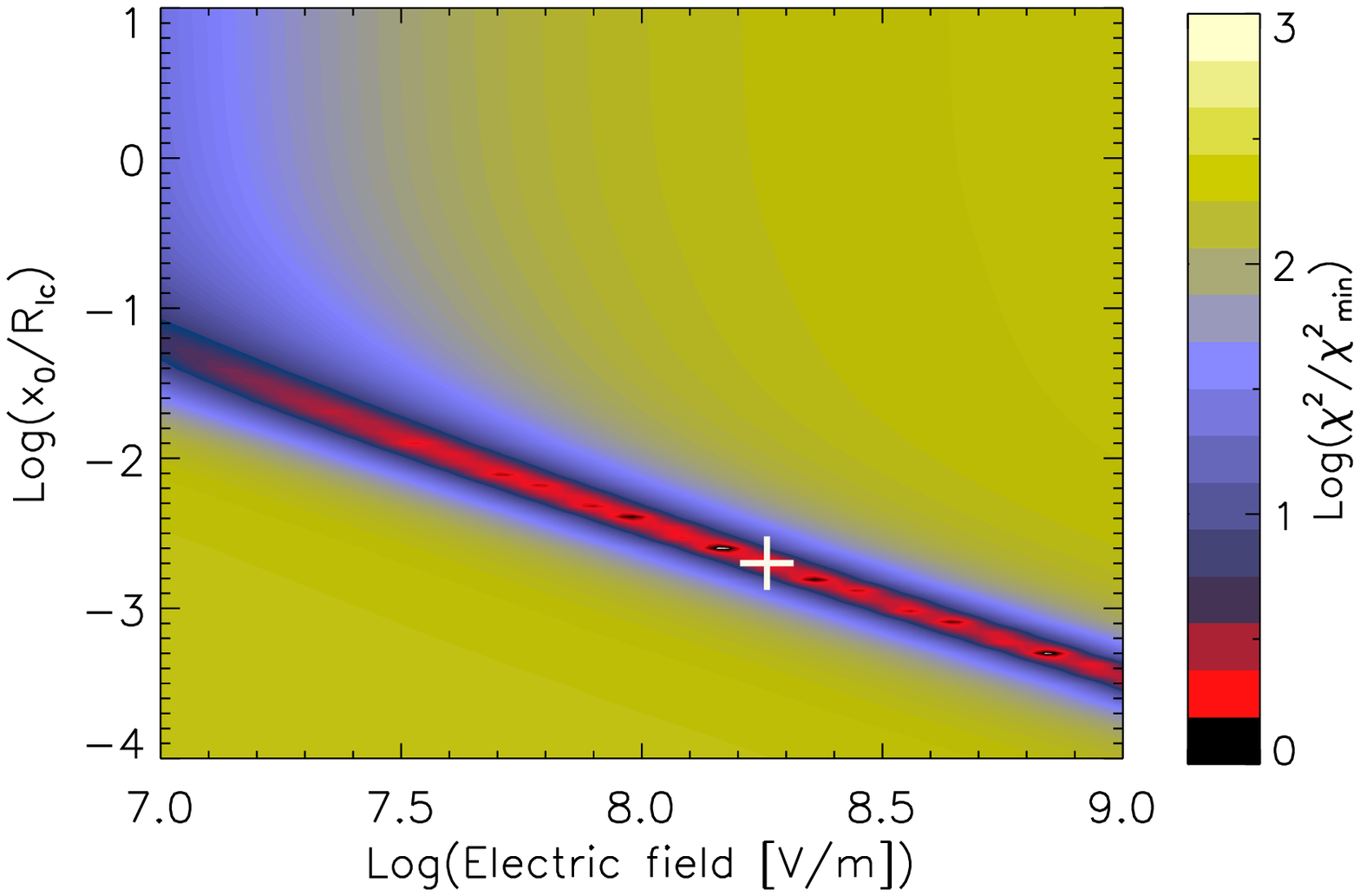}
\put(20,15){\scriptsize  \red {J0734-1559}}
\end{overpic}
\begin{overpic}[width=0.32\textwidth]{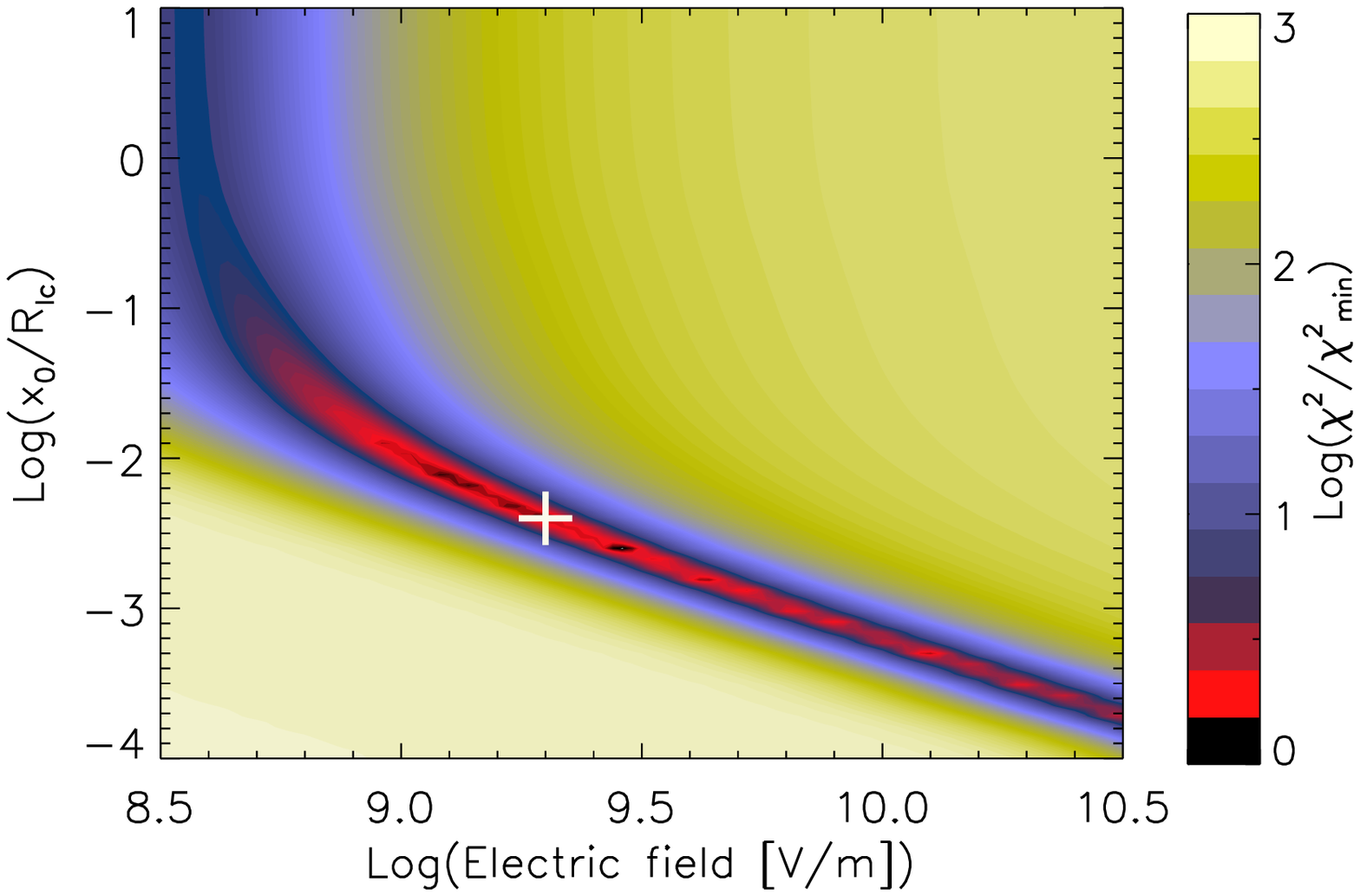}
\put(20,15){\scriptsize \blue {J0751+1807}}
\end{overpic}
\begin{overpic}[width=0.32\textwidth]{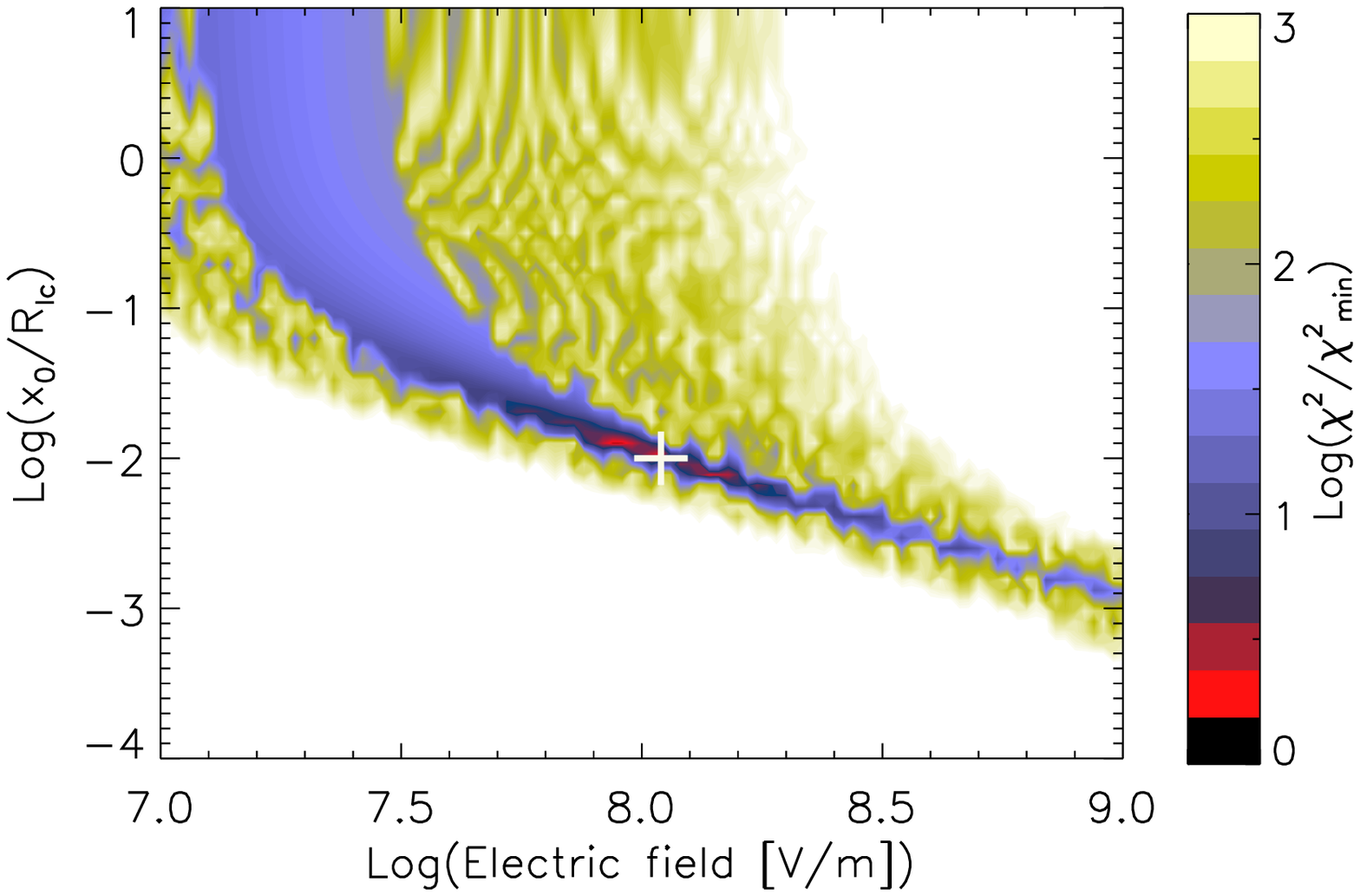}
\put(20,15){\scriptsize  \red {J0835-4510 (Vela)}}
\end{overpic}
\begin{overpic}[width=0.32\textwidth]{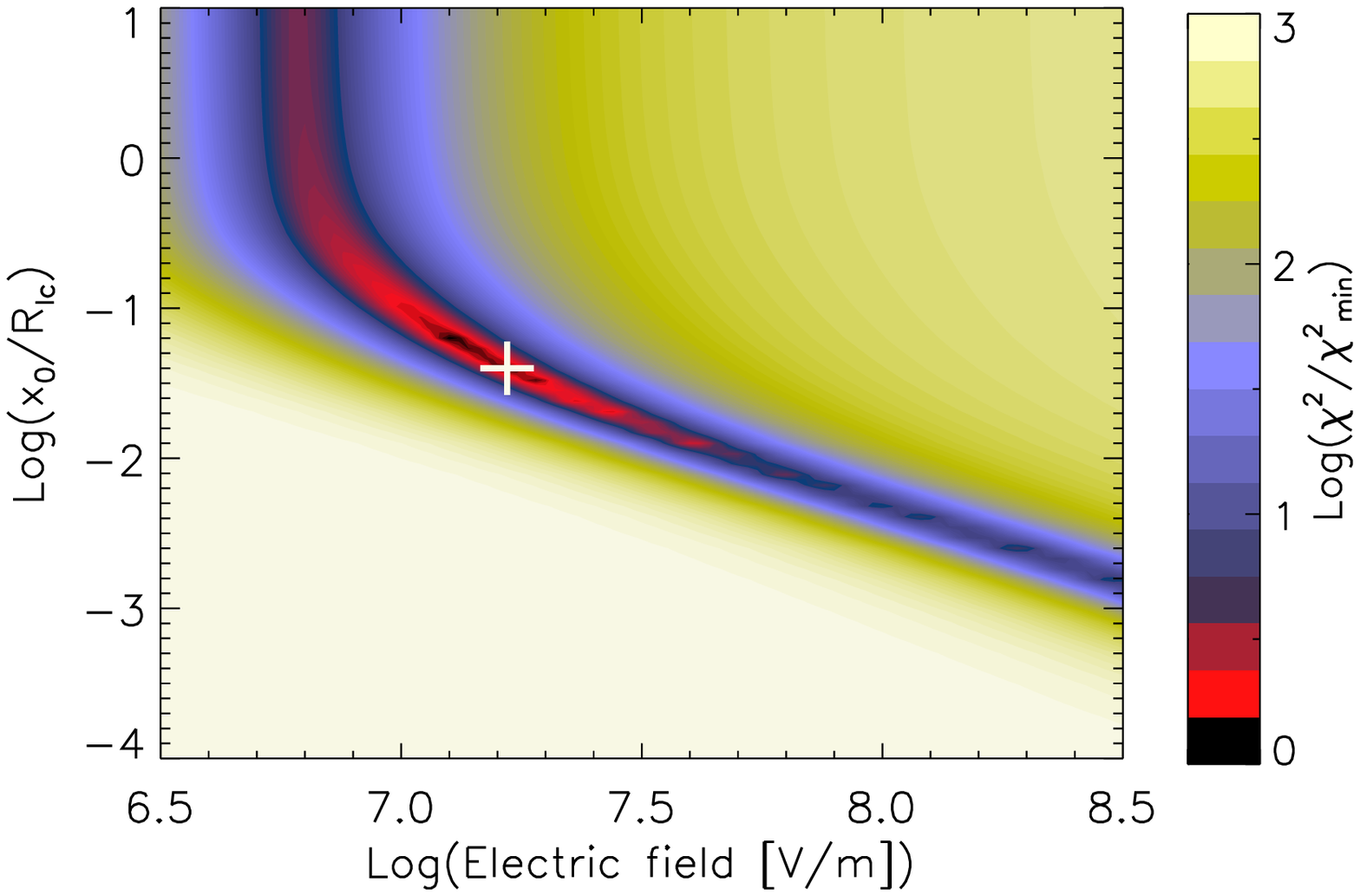}
\put(20,15){\scriptsize  \red {J0908-4913}}
\end{overpic}
\begin{overpic}[width=0.32\textwidth]{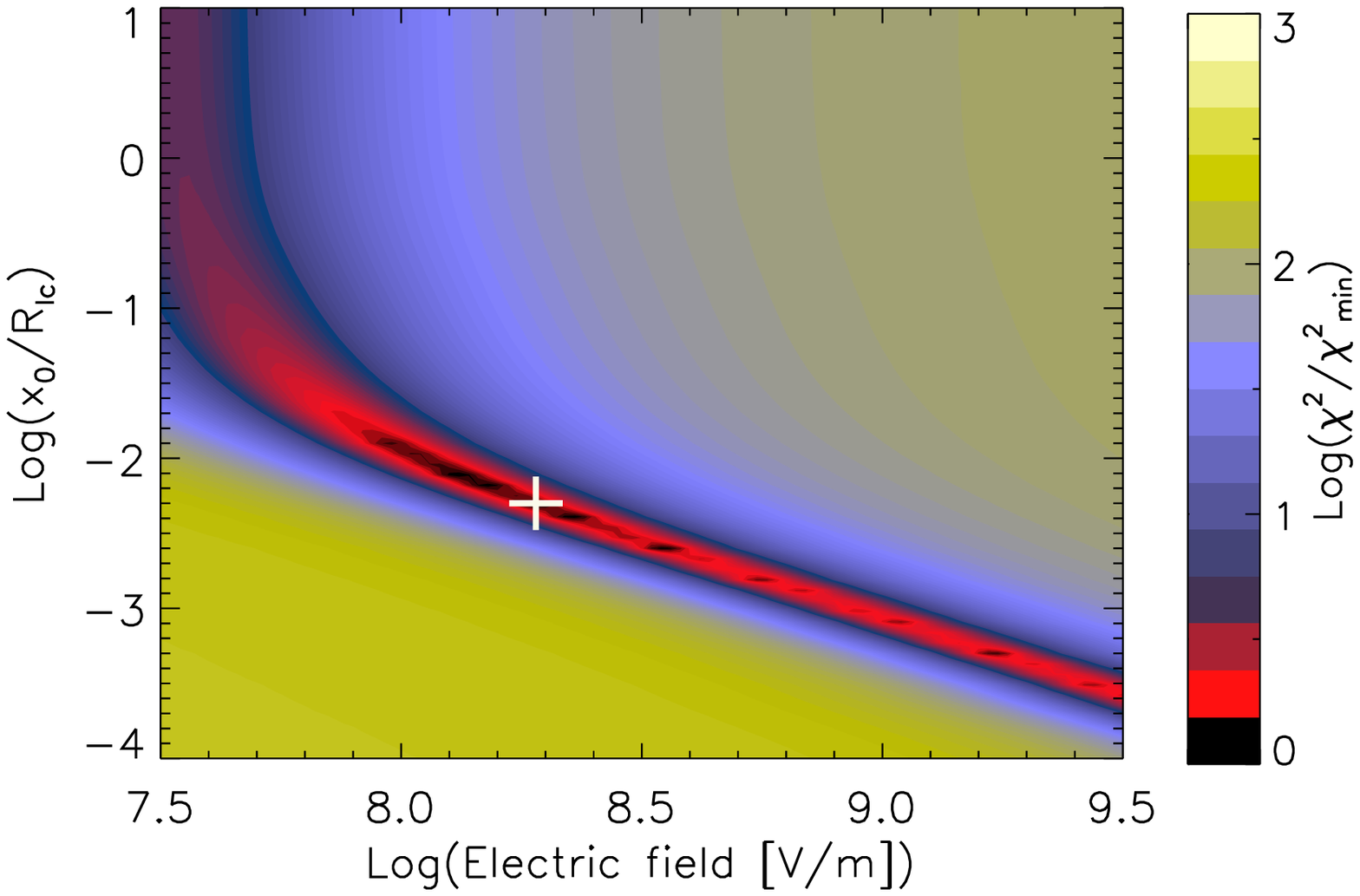}
\put(20,15){\scriptsize  \red {J1016-5857}}
\end{overpic}
\begin{overpic}[width=0.32\textwidth]{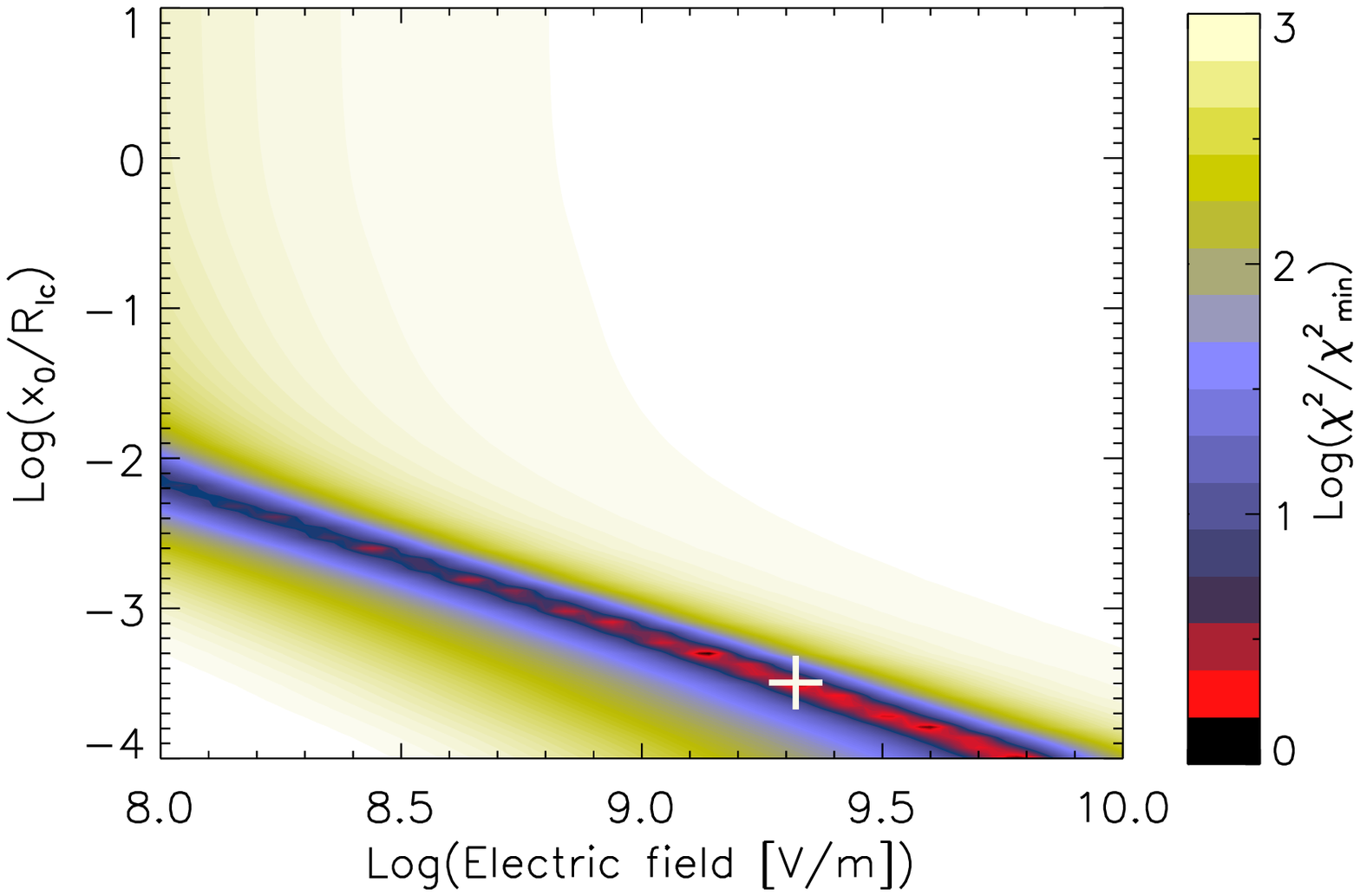}
\put(20,15){\scriptsize  \red {J1023-5746}}
\end{overpic}
\begin{overpic}[width=0.32\textwidth]{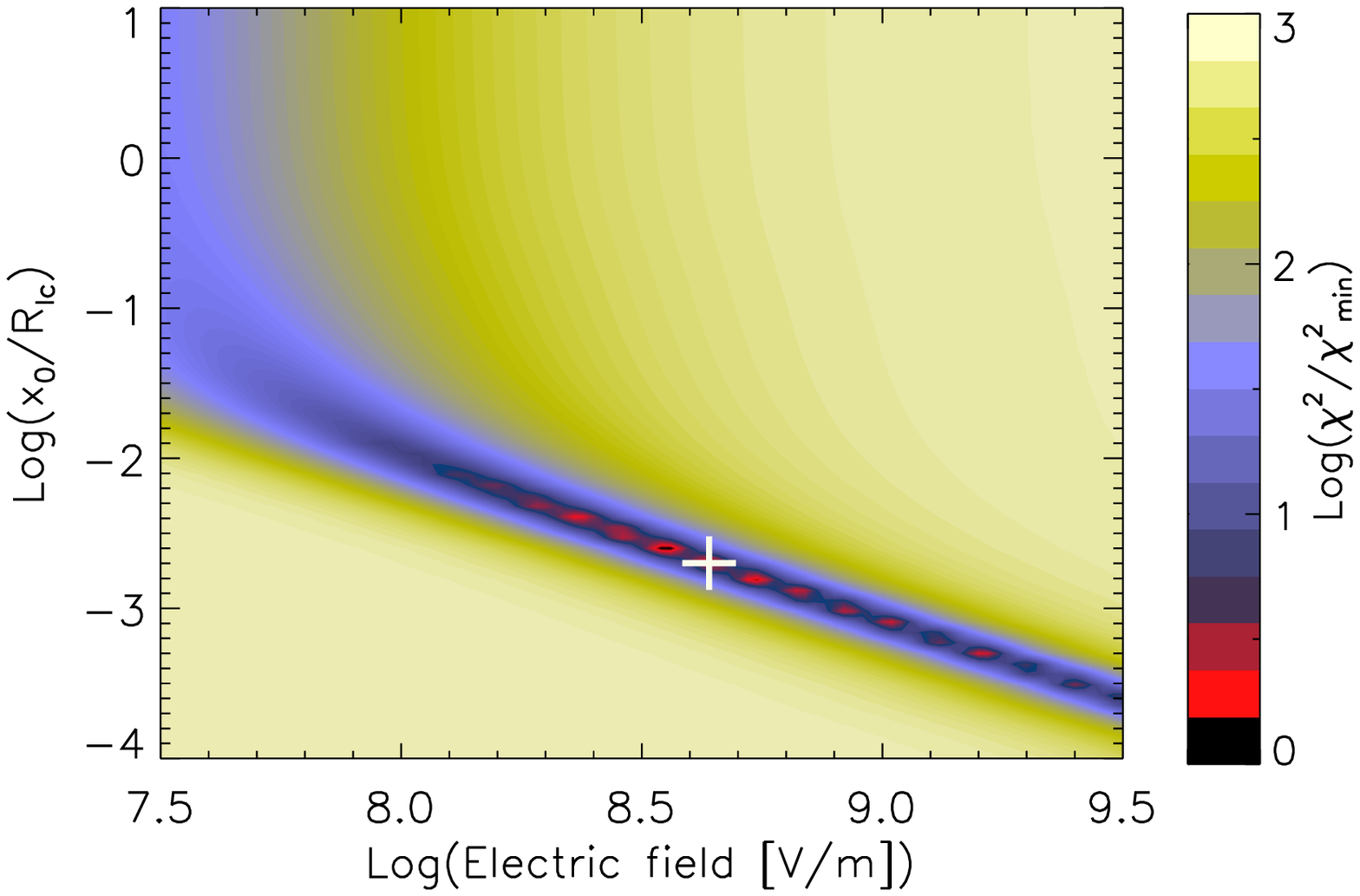}
\put(20,15){\scriptsize  \red {J1028-5819}}
\end{overpic}
\begin{overpic}[width=0.32\textwidth]{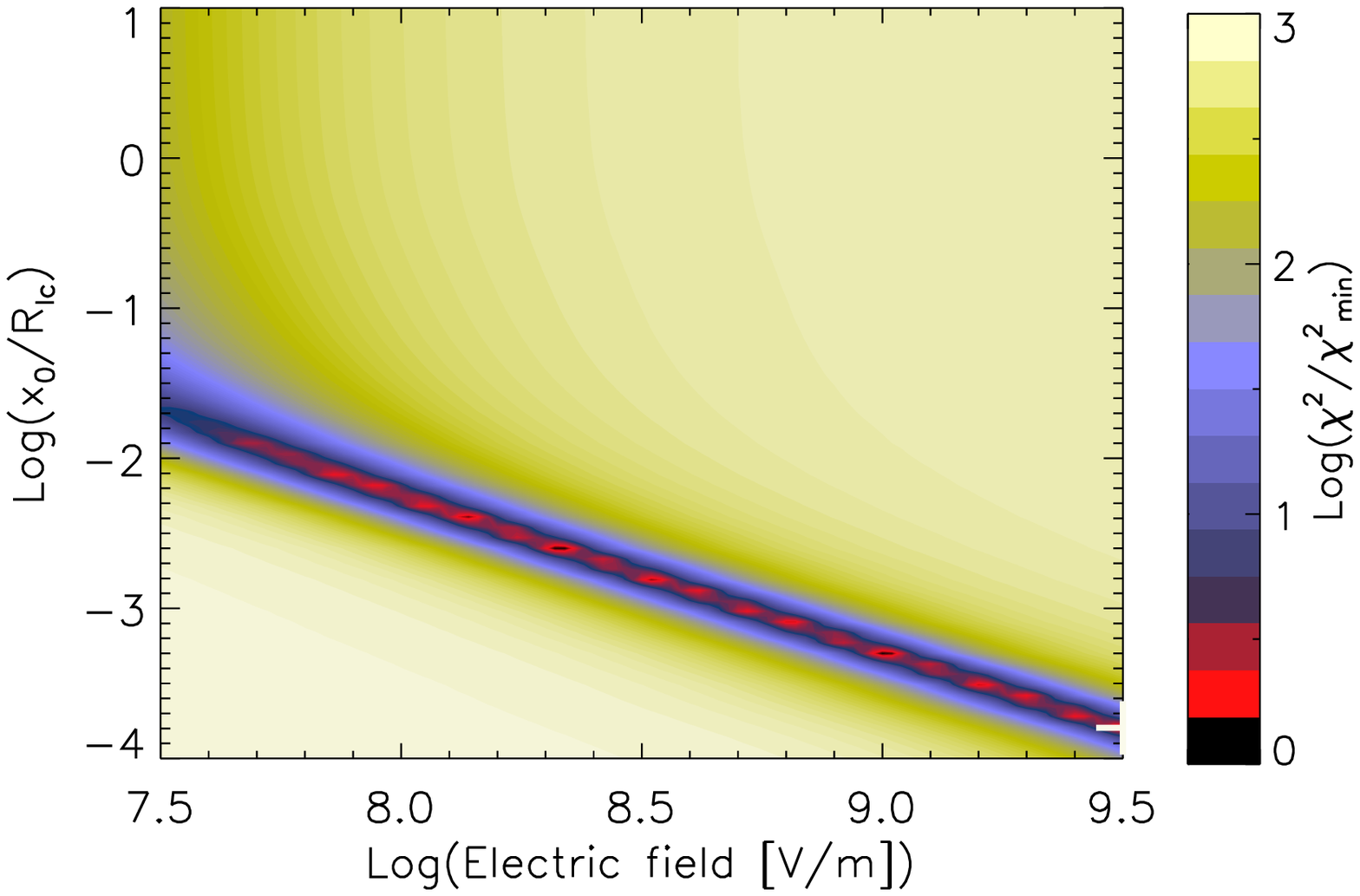}
\put(20,15){\scriptsize  \red {J1044-5737}}
\end{overpic}
\begin{overpic}[width=0.32\textwidth]{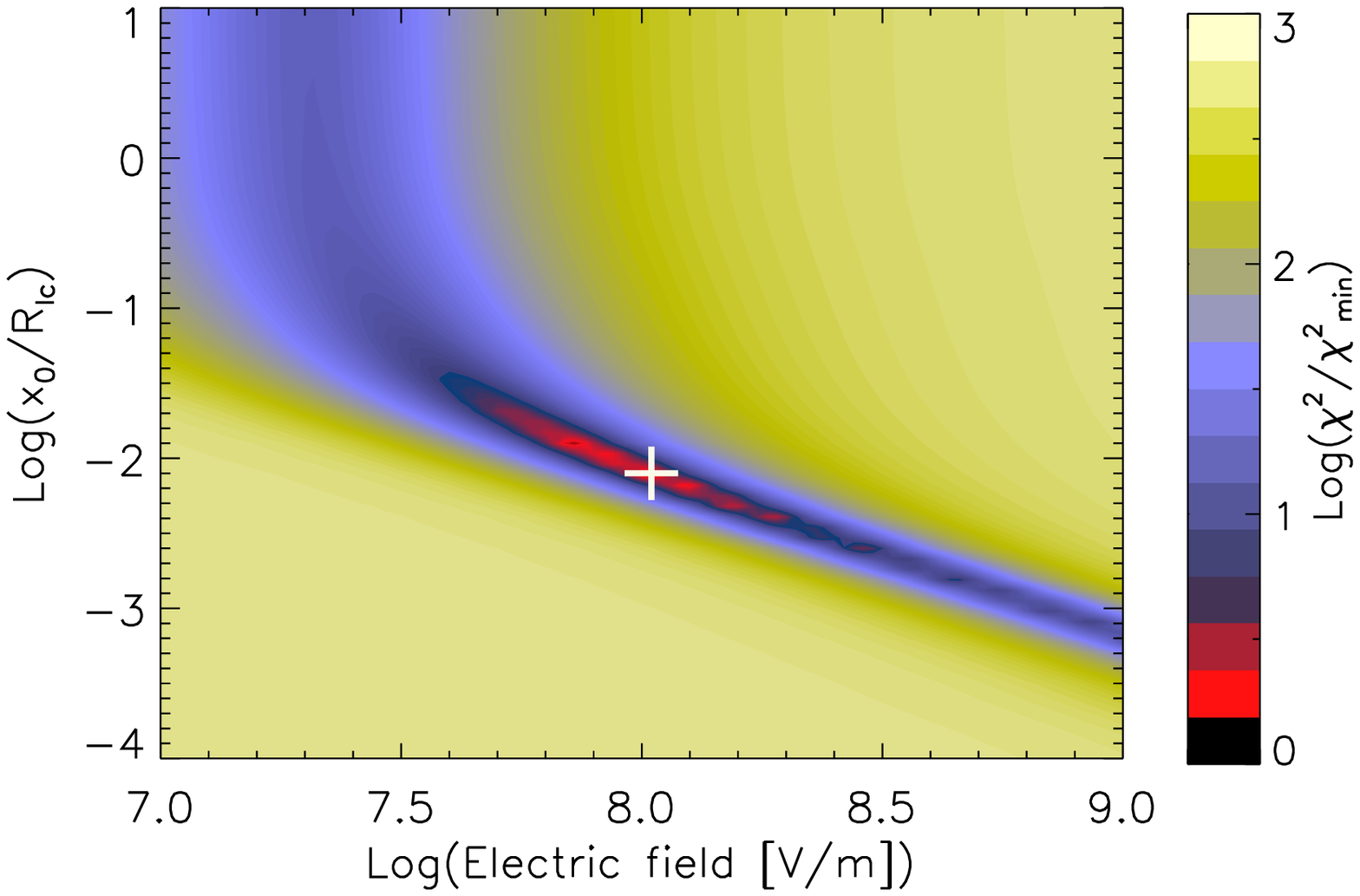}
\put(20,15){\scriptsize  \red {J1048-5832}}
\end{overpic}
\begin{overpic}[width=0.32\textwidth]{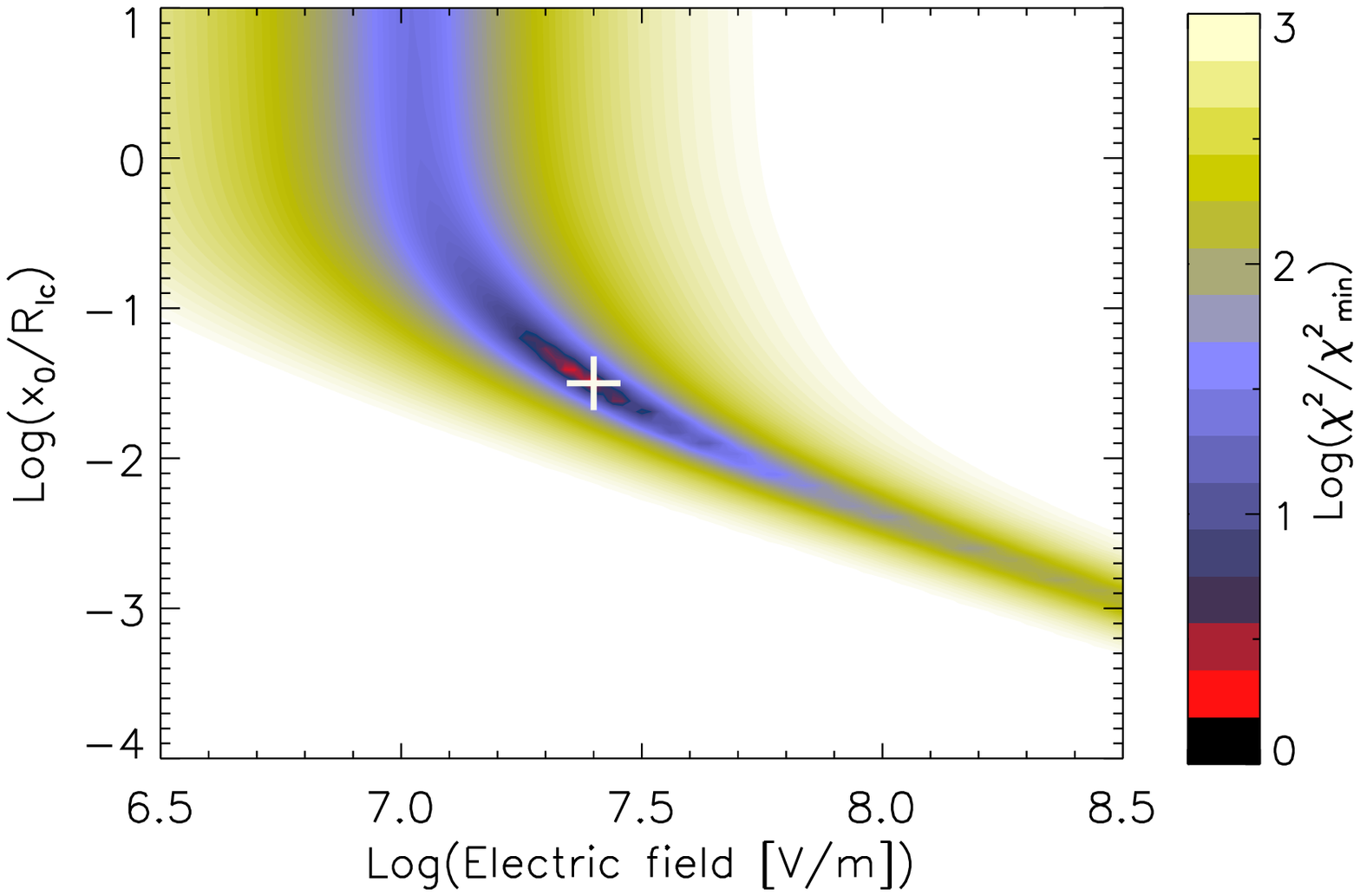}
\put(20,15){\scriptsize  \red {J1057-5226}}
\end{overpic}
\begin{overpic}[width=0.32\textwidth]{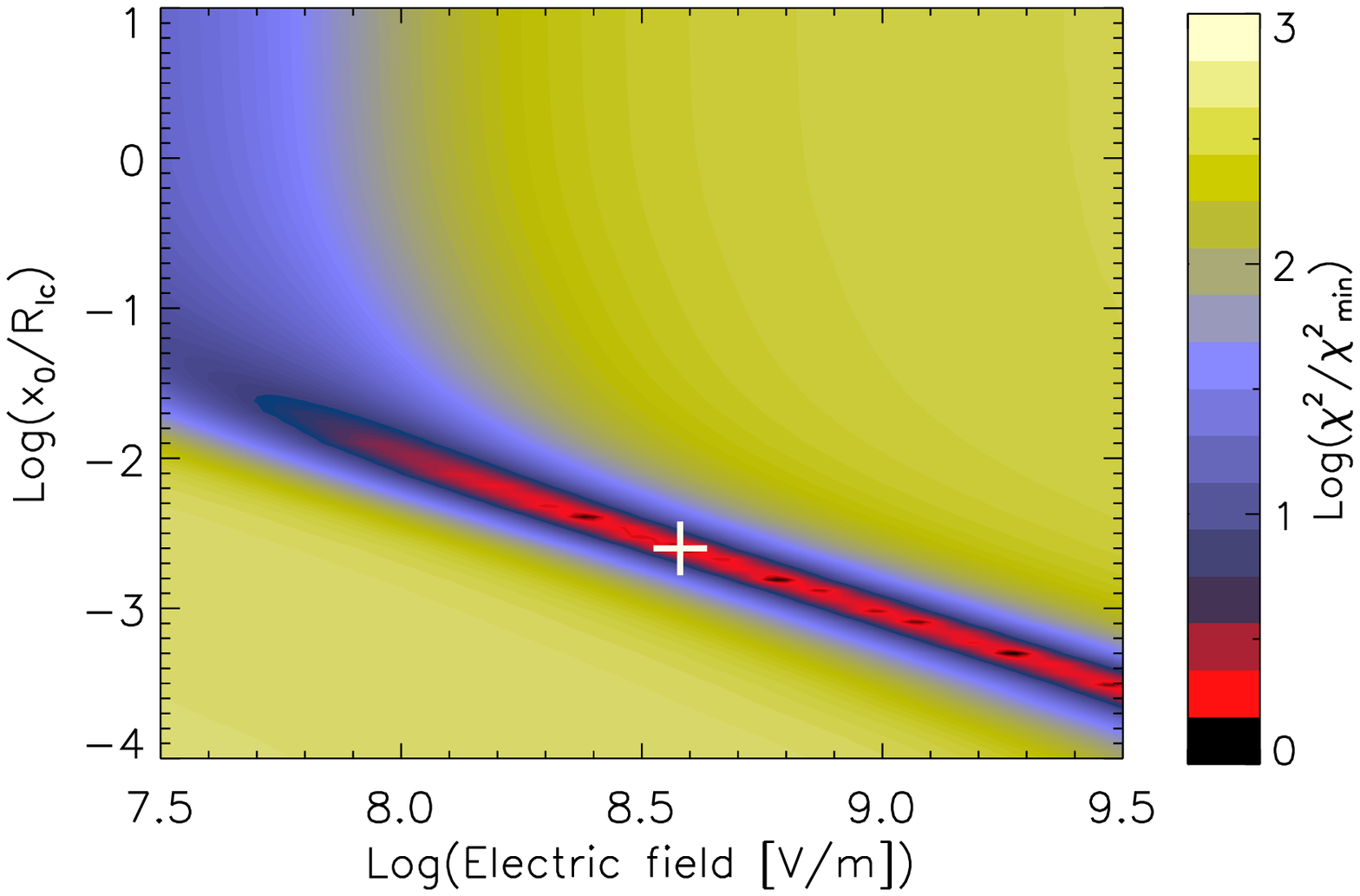}
\put(20,15){\scriptsize  \red {J1105-6107}}
\end{overpic}
\begin{overpic}[width=0.32\textwidth]{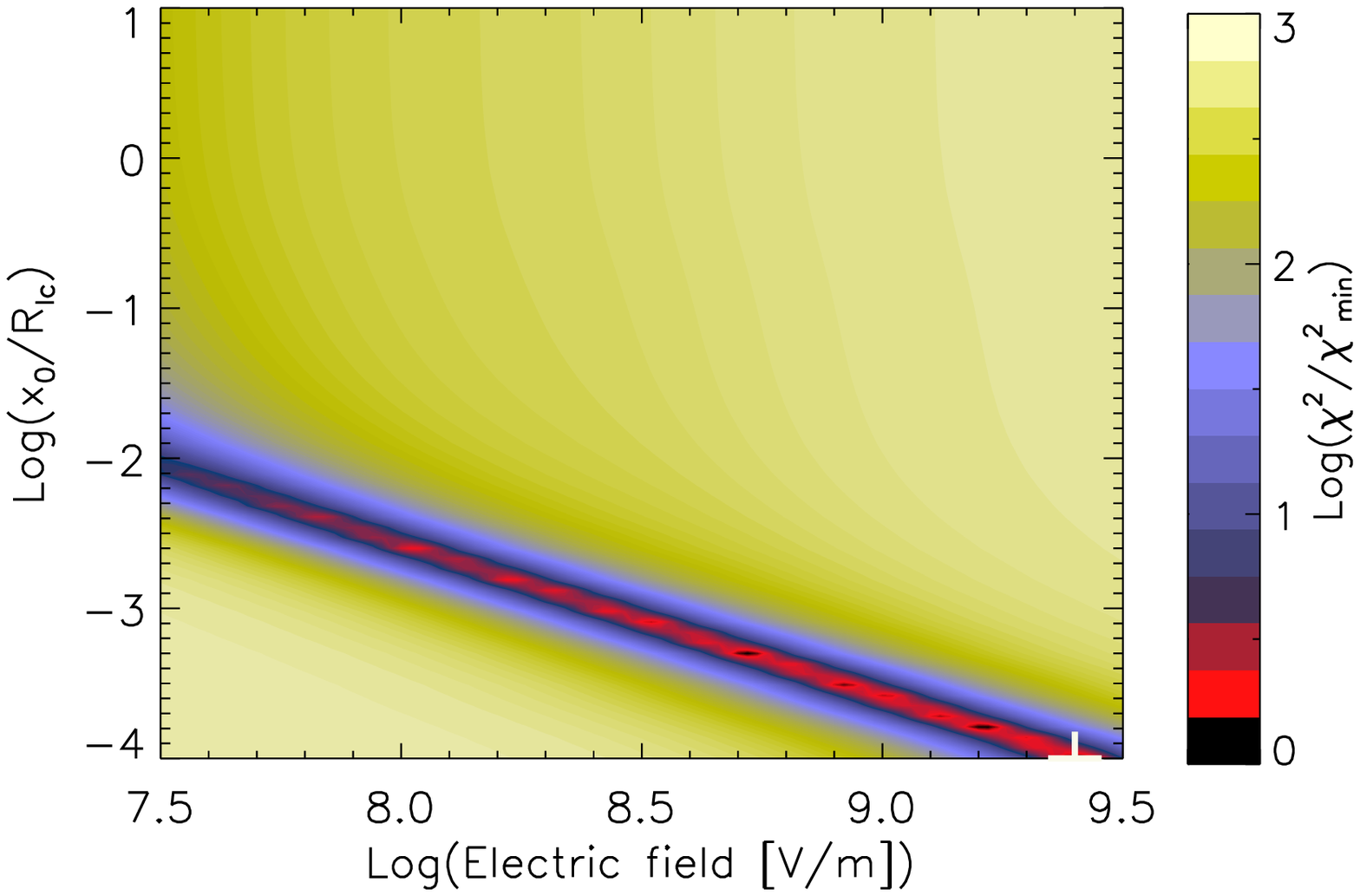}
\put(20,15){\scriptsize  \red {J1119+6127}}
\end{overpic}
\begin{overpic}[width=0.32\textwidth]{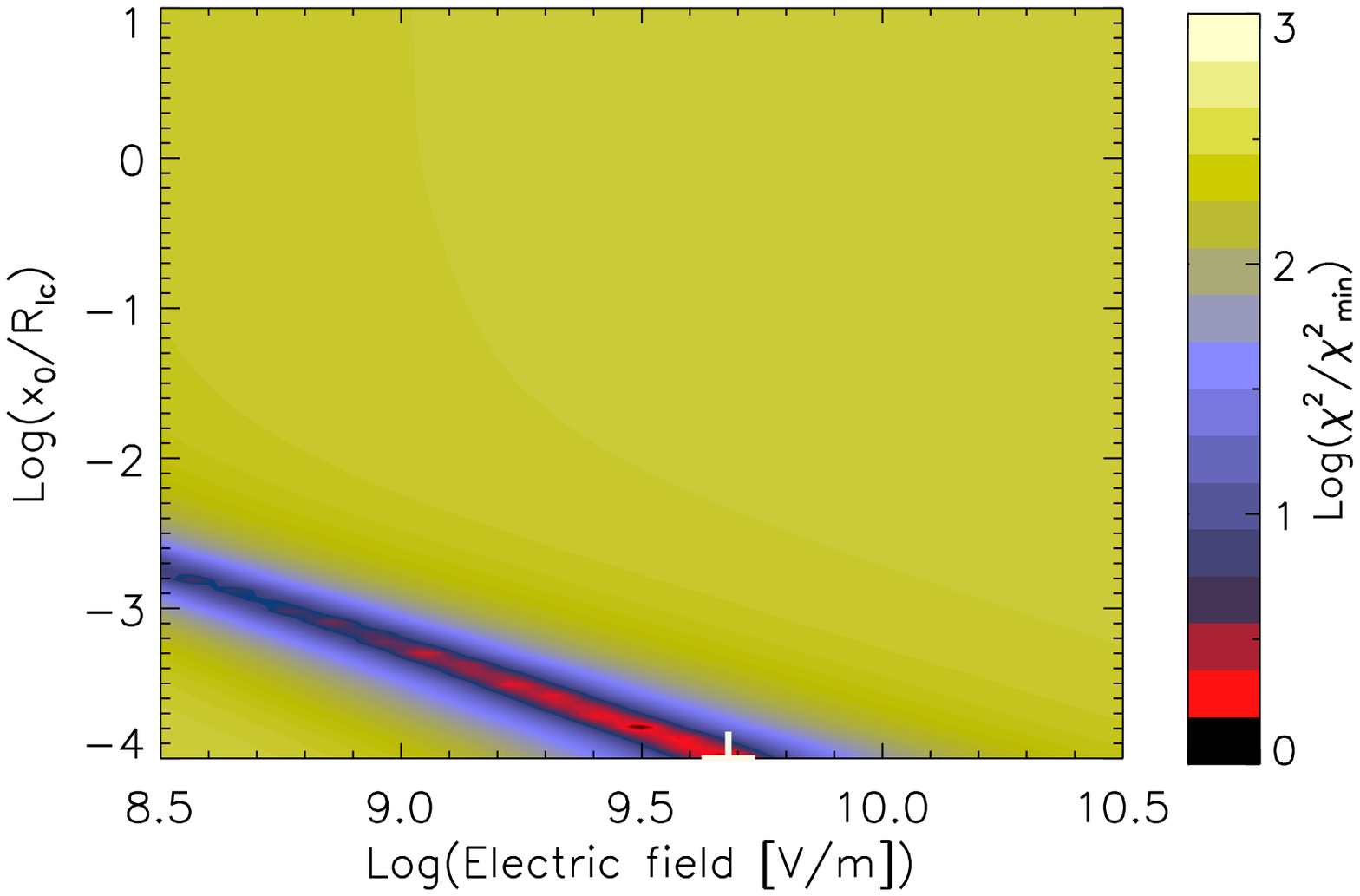}
\put(20,15){\scriptsize \blue {J1124-3653}}
\end{overpic}
\begin{overpic}[width=0.32\textwidth]{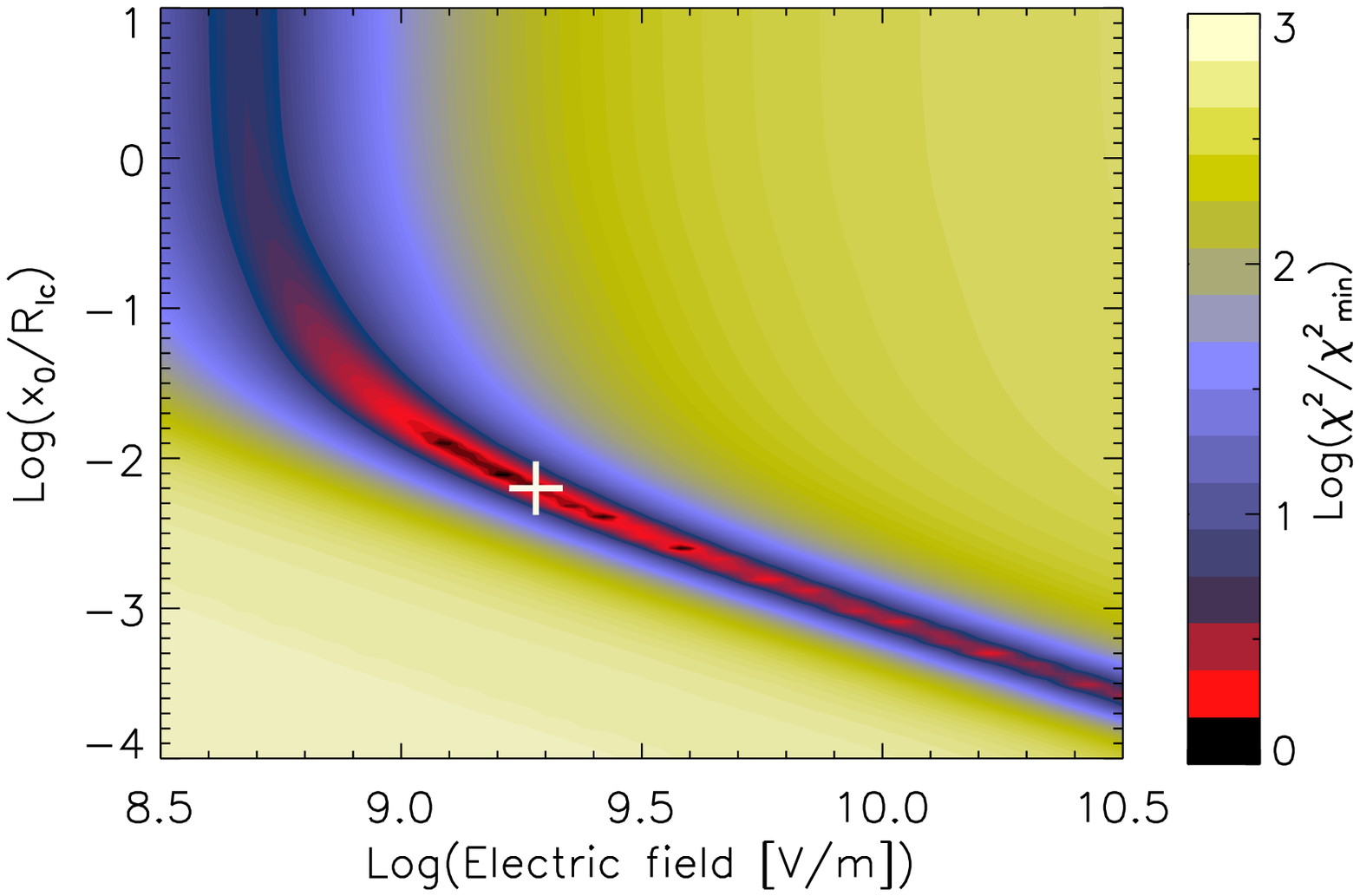}
\put(20,15){\scriptsize  \red {J1124-5916}}
\end{overpic}
\begin{overpic}[width=0.32\textwidth]{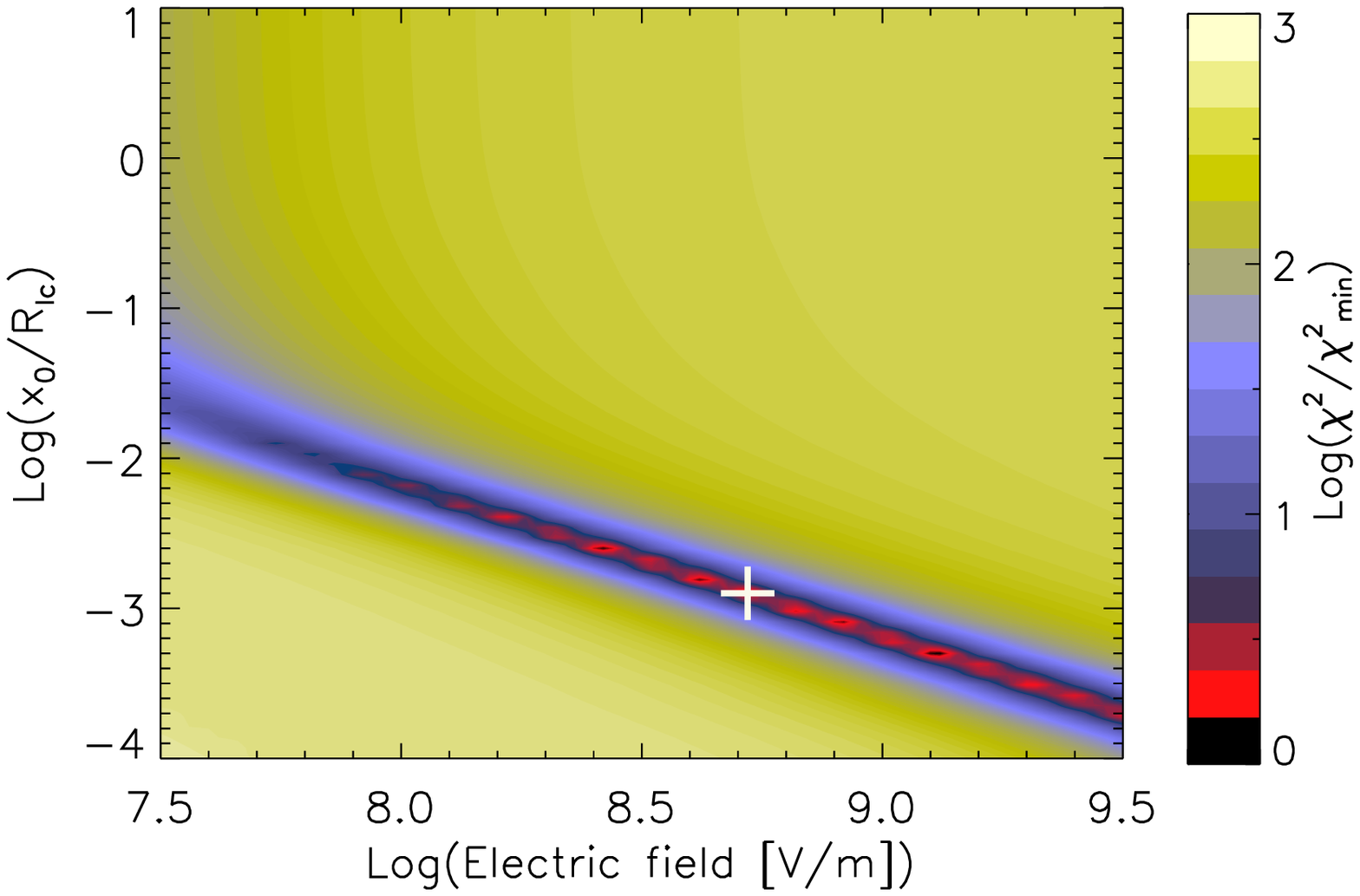}
\put(20,15){\scriptsize  \red {J1135-6055}}
\end{overpic}
\begin{overpic}[width=0.32\textwidth]{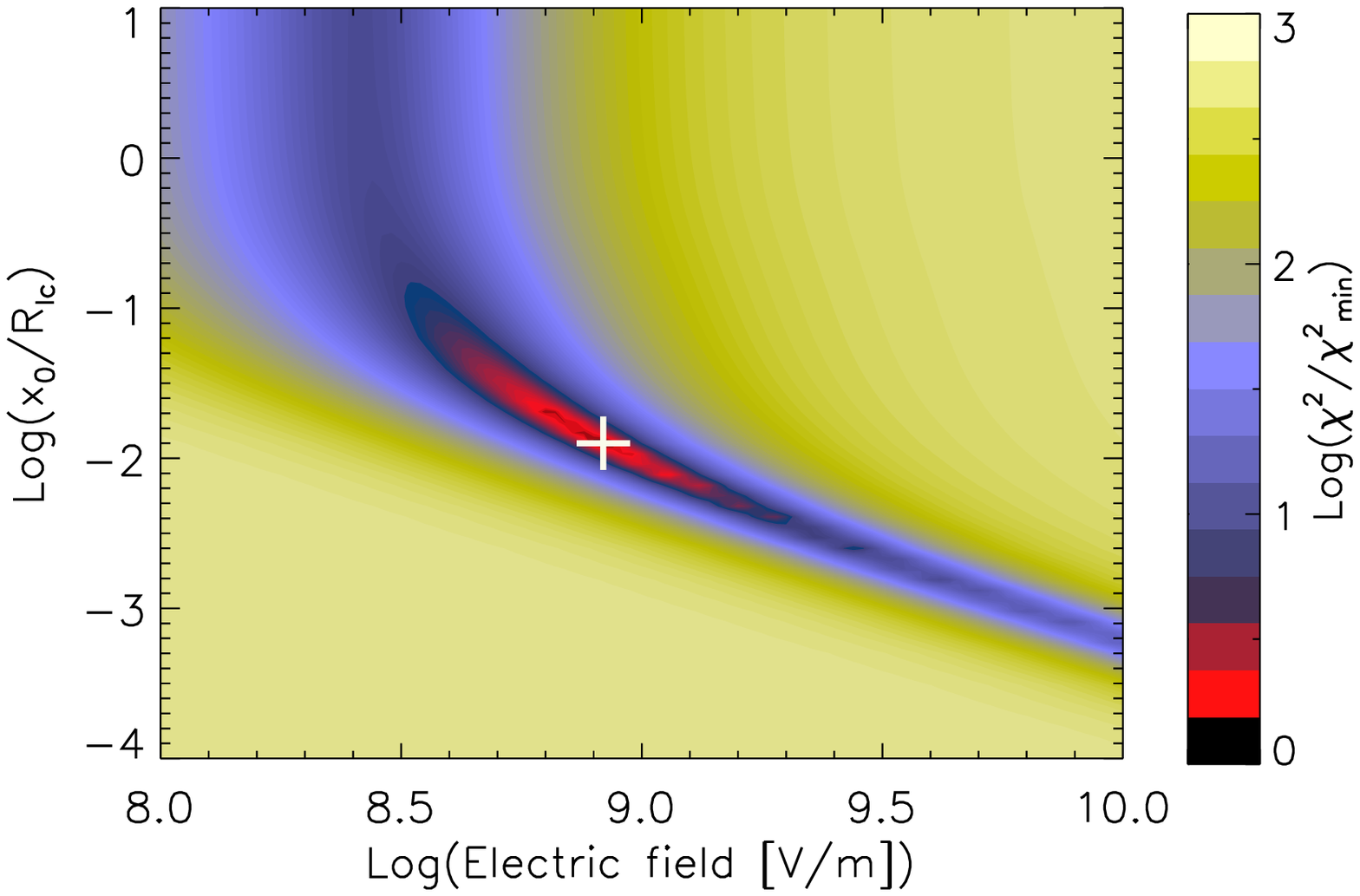}
\put(20,15){\scriptsize \blue{J1231-1411}}
\end{overpic}
\begin{overpic}[width=0.32\textwidth]{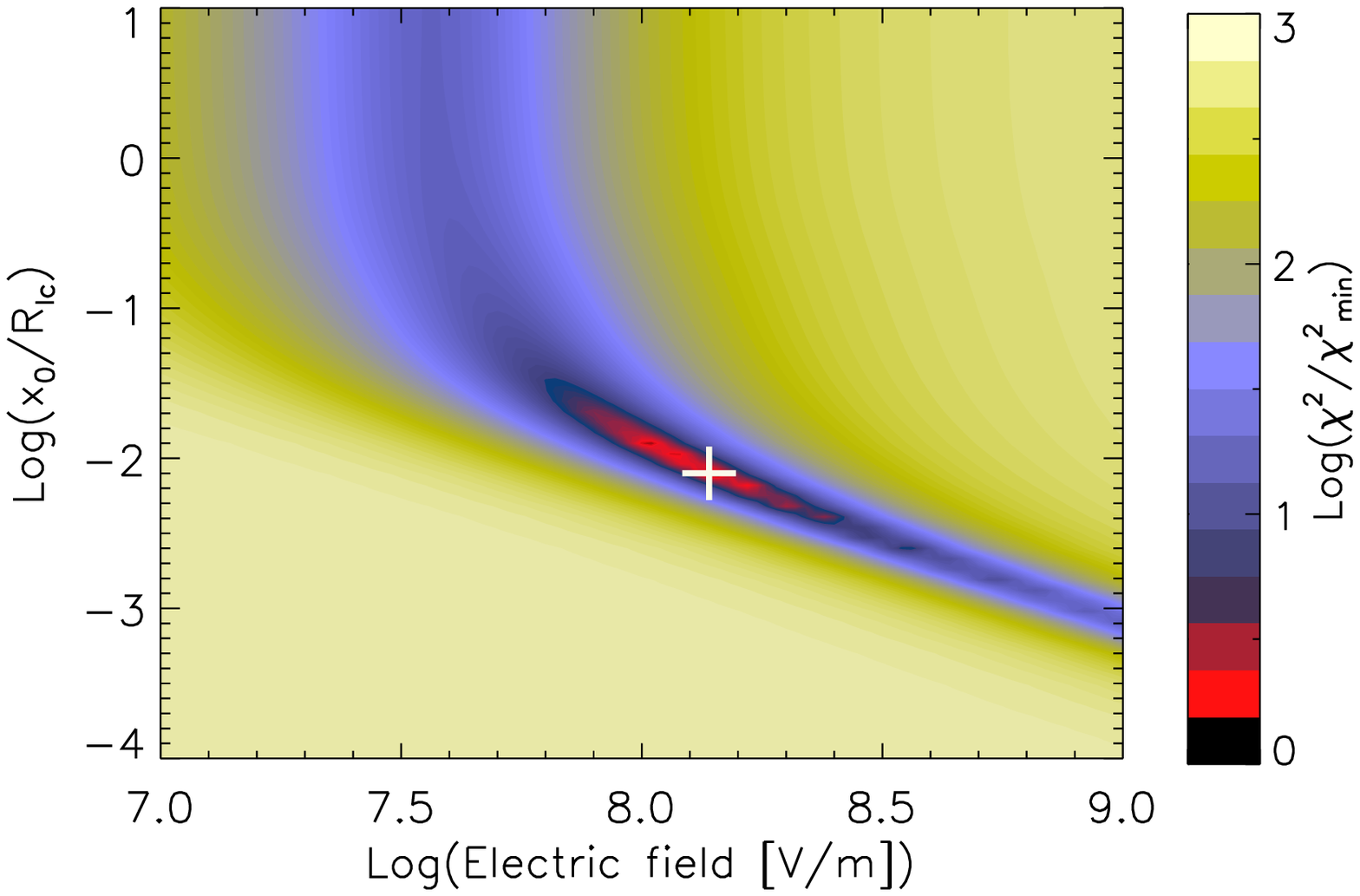}
\put(20,15){\scriptsize  \red {J1413-6205}}
\end{overpic}
\begin{overpic}[width=0.32\textwidth]{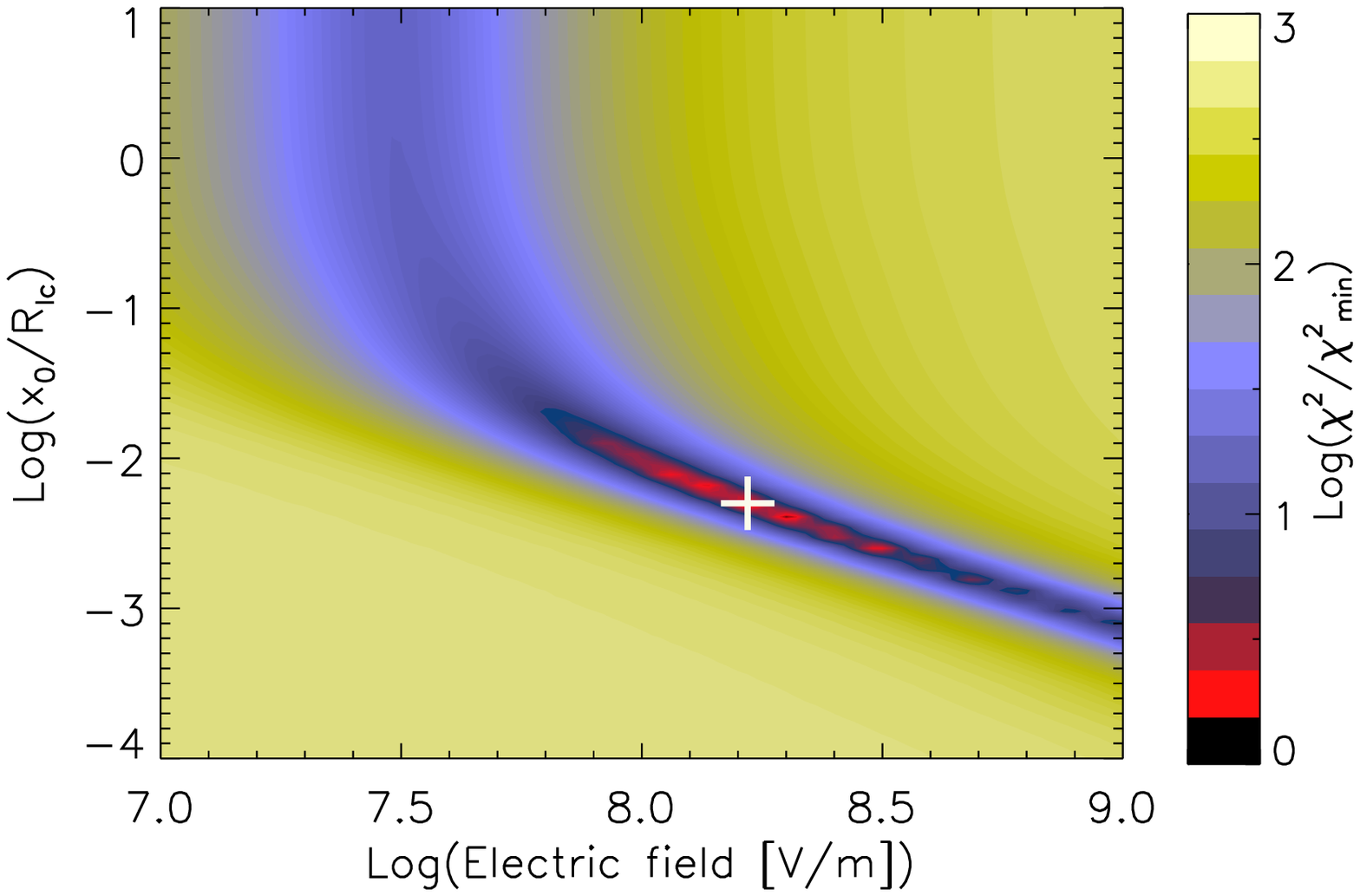}
\put(20,15){\scriptsize  \red {J1418-6058}}
\end{overpic}
\end{center}
\caption{$\chi^2/\chi^2_{\rm min}$ contours on the $\log E_\parallel$--$\log (x_0/R_{\rm lc})$ plane for the pulsars considered in the sample (II).}
\label{fig:contours2}
\end{figure*}

\begin{figure*}
\begin{center}
\begin{overpic}[width=0.32\textwidth]{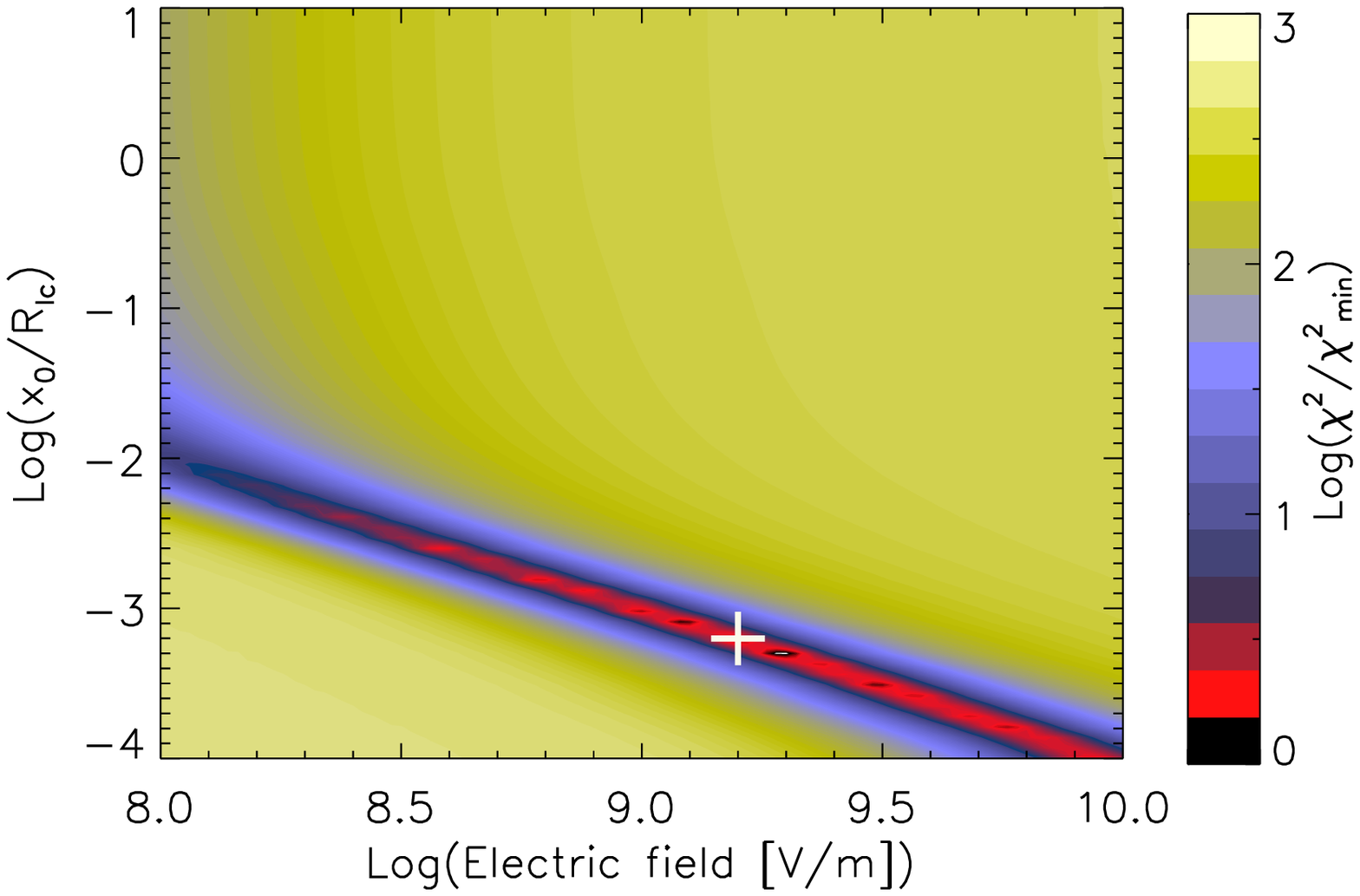}
\put(20,15){\scriptsize  \red {J1420-6048}}
\end{overpic}
\begin{overpic}[width=0.32\textwidth]{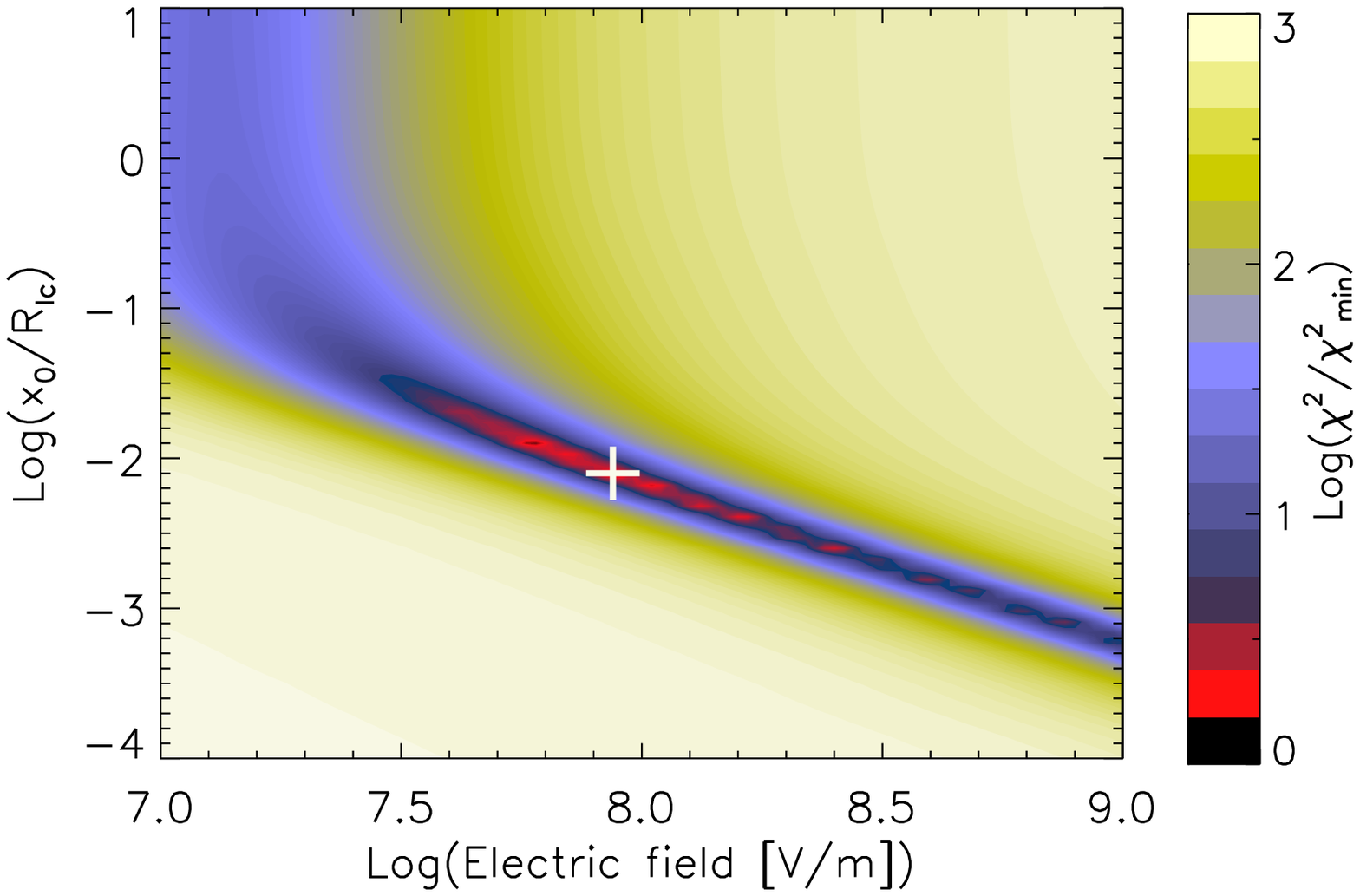}
\put(20,15){\scriptsize  \red {J1429-5911}}
\end{overpic}
\begin{overpic}[width=0.32\textwidth]{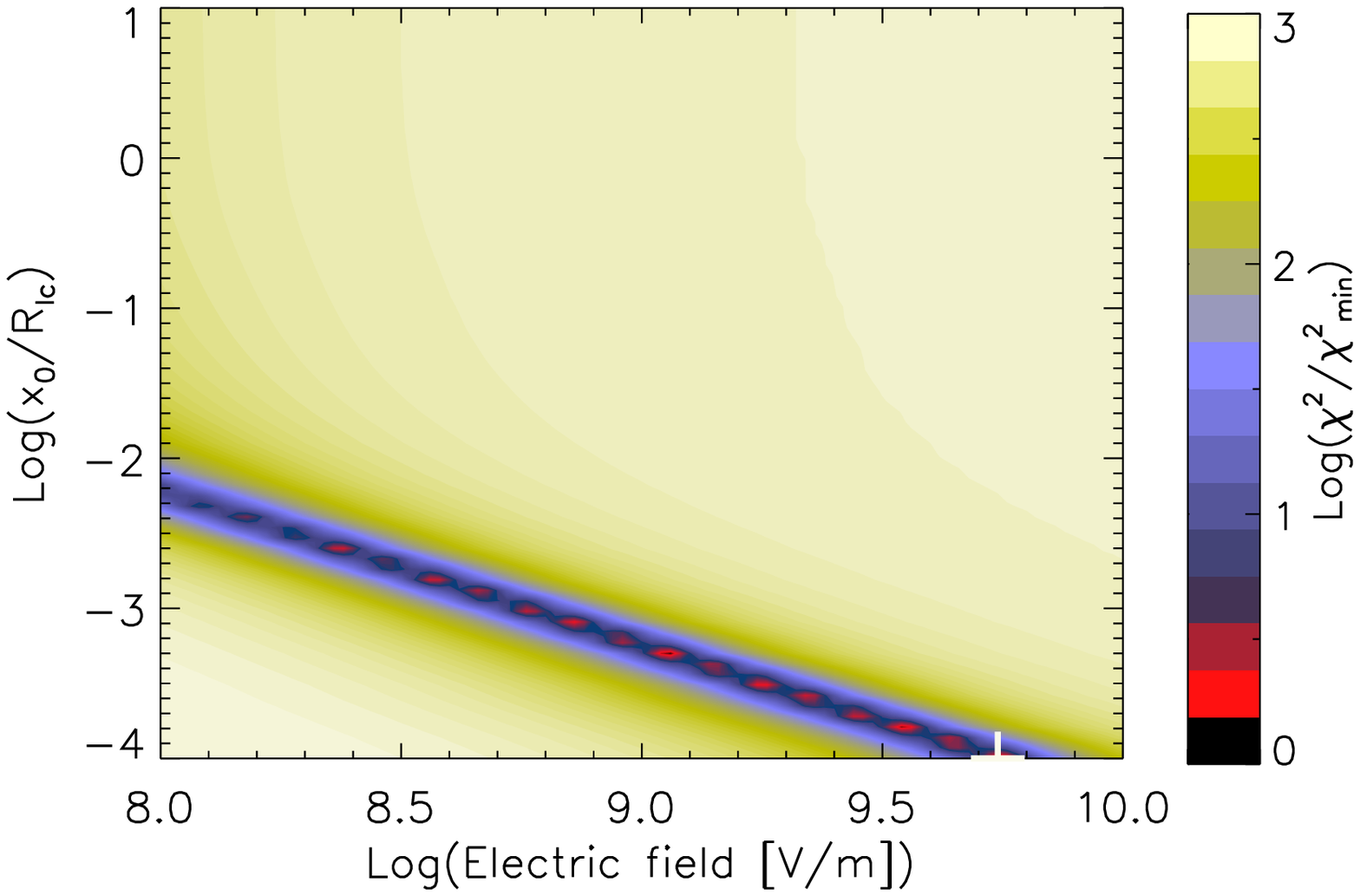}
\put(20,15){\scriptsize  \red {J1459-6053}}
\end{overpic}
\begin{overpic}[width=0.32\textwidth]{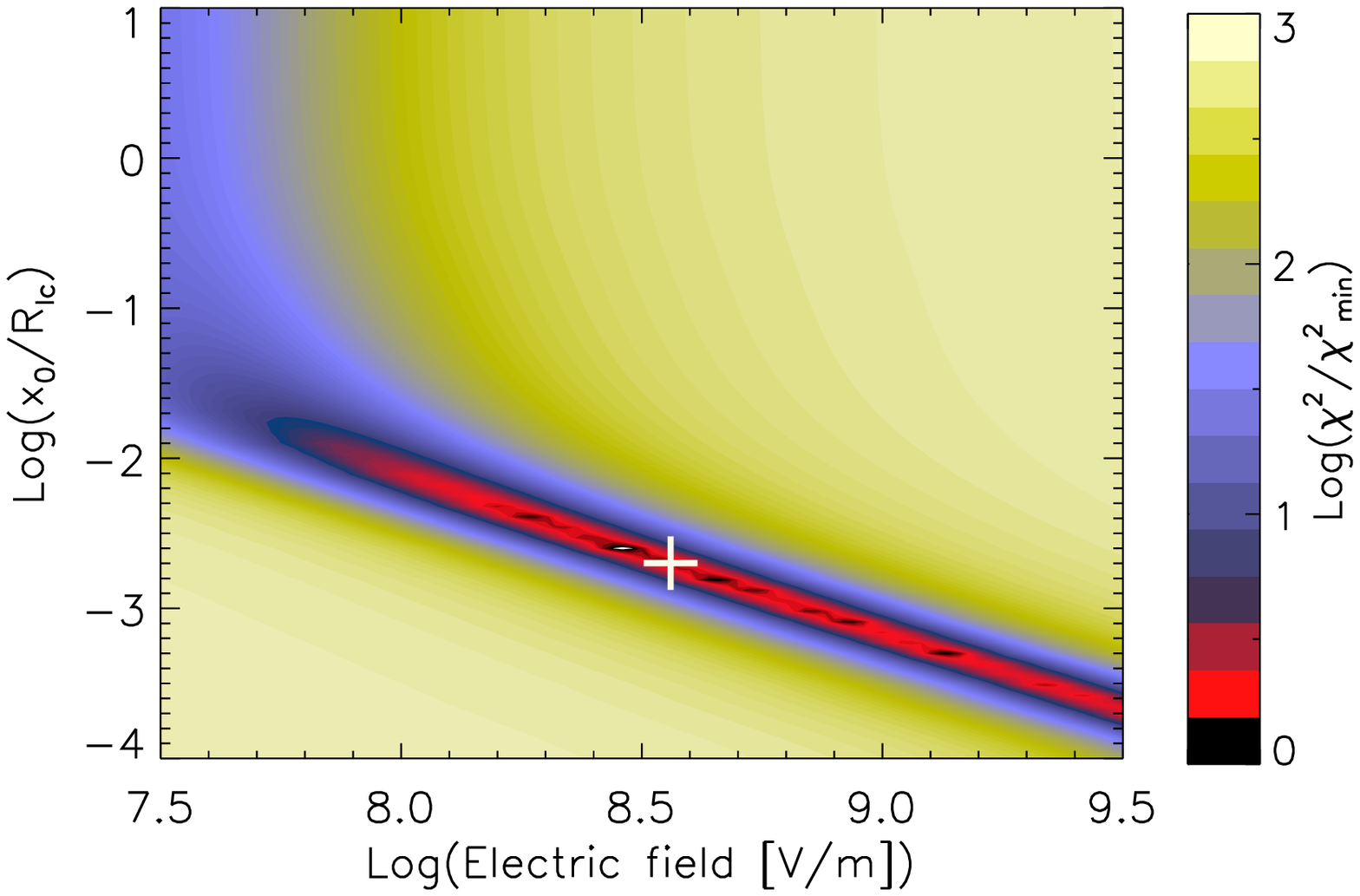}
\put(20,15){\scriptsize  \red {J1509-5850}}
\end{overpic}
\begin{overpic}[width=0.32\textwidth]{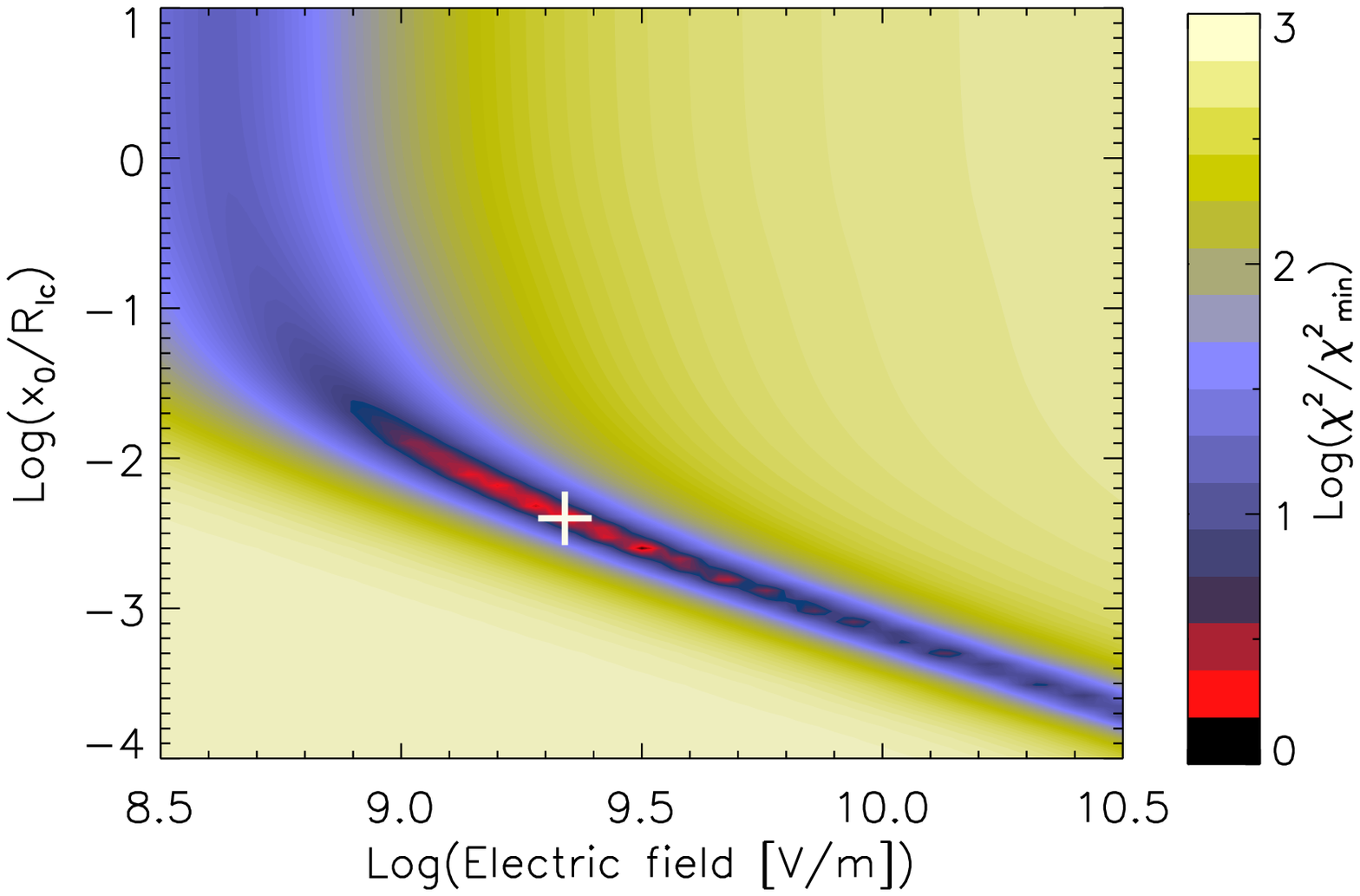}
\put(20,15){\scriptsize \blue {J1514-4946}}
\end{overpic}
\begin{overpic}[width=0.32\textwidth]{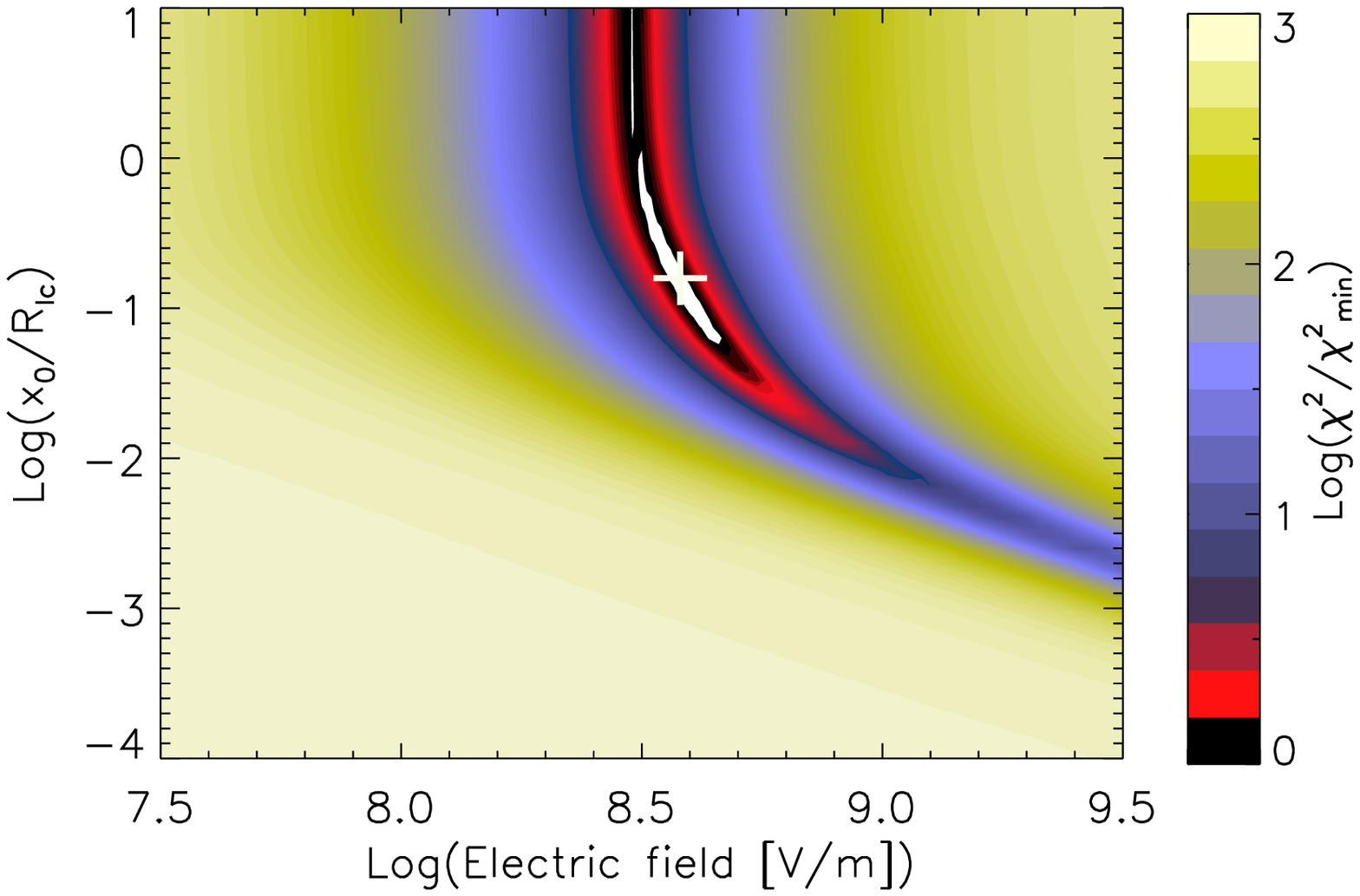}
\put(20,15){\scriptsize  \blue {J1614-2230}}
\end{overpic}
\begin{overpic}[width=0.32\textwidth]{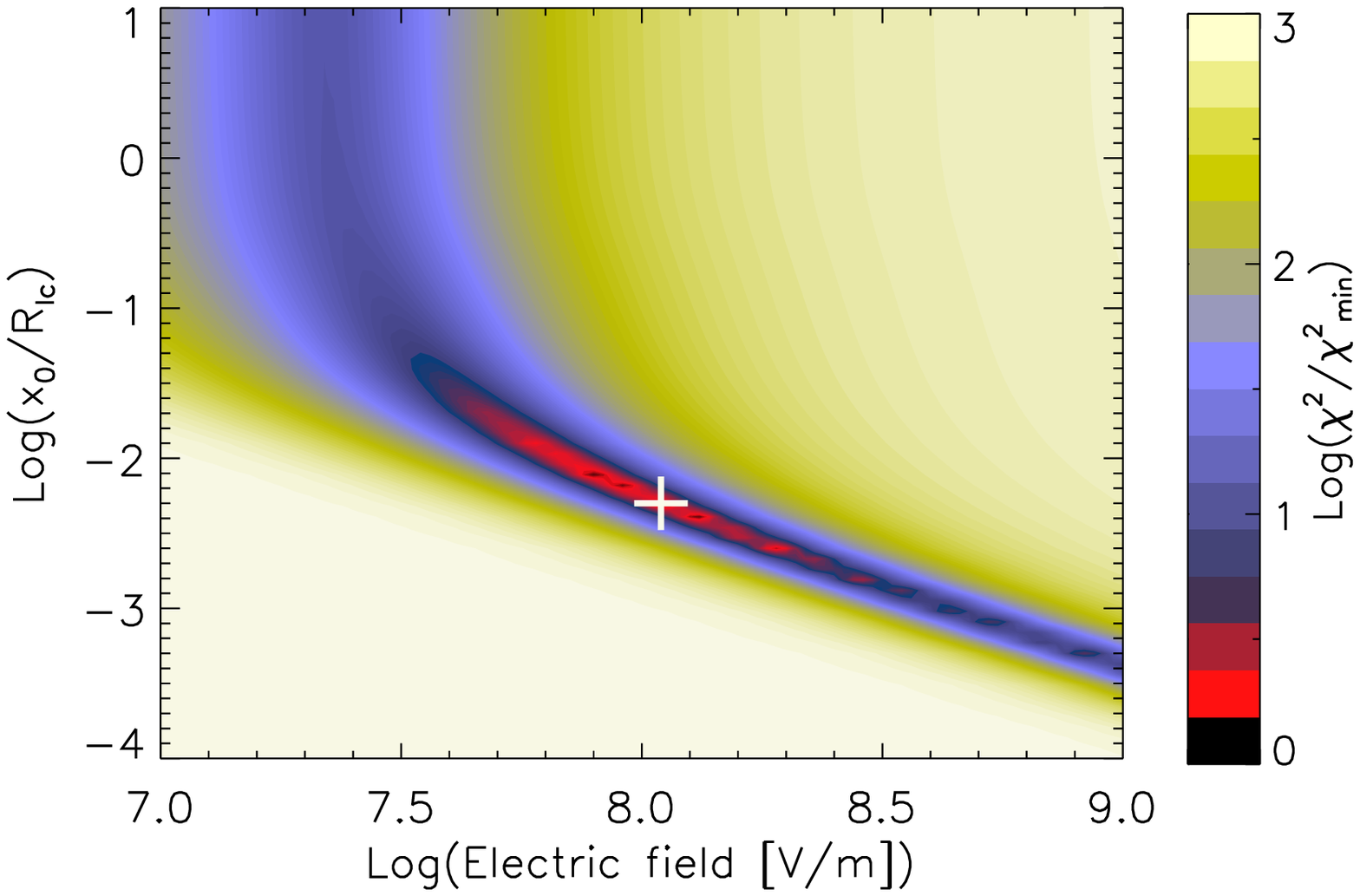}
\put(20,15){\scriptsize  \red {J1620-4927}}
\end{overpic}
\begin{overpic}[width=0.32\textwidth]{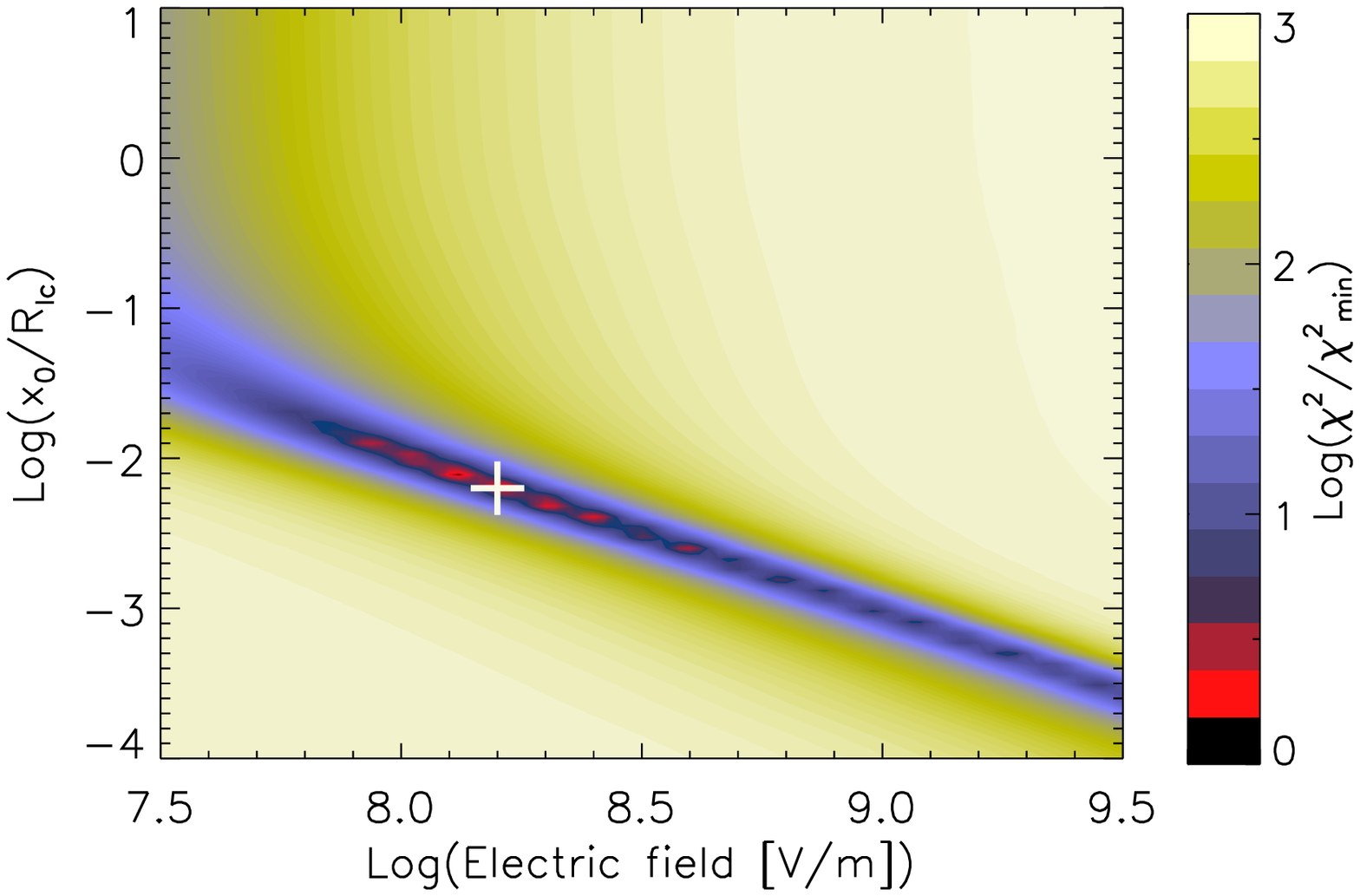}
\put(20,15){\scriptsize  \red {J1709-4429}}
\end{overpic}
\begin{overpic}[width=0.32\textwidth]{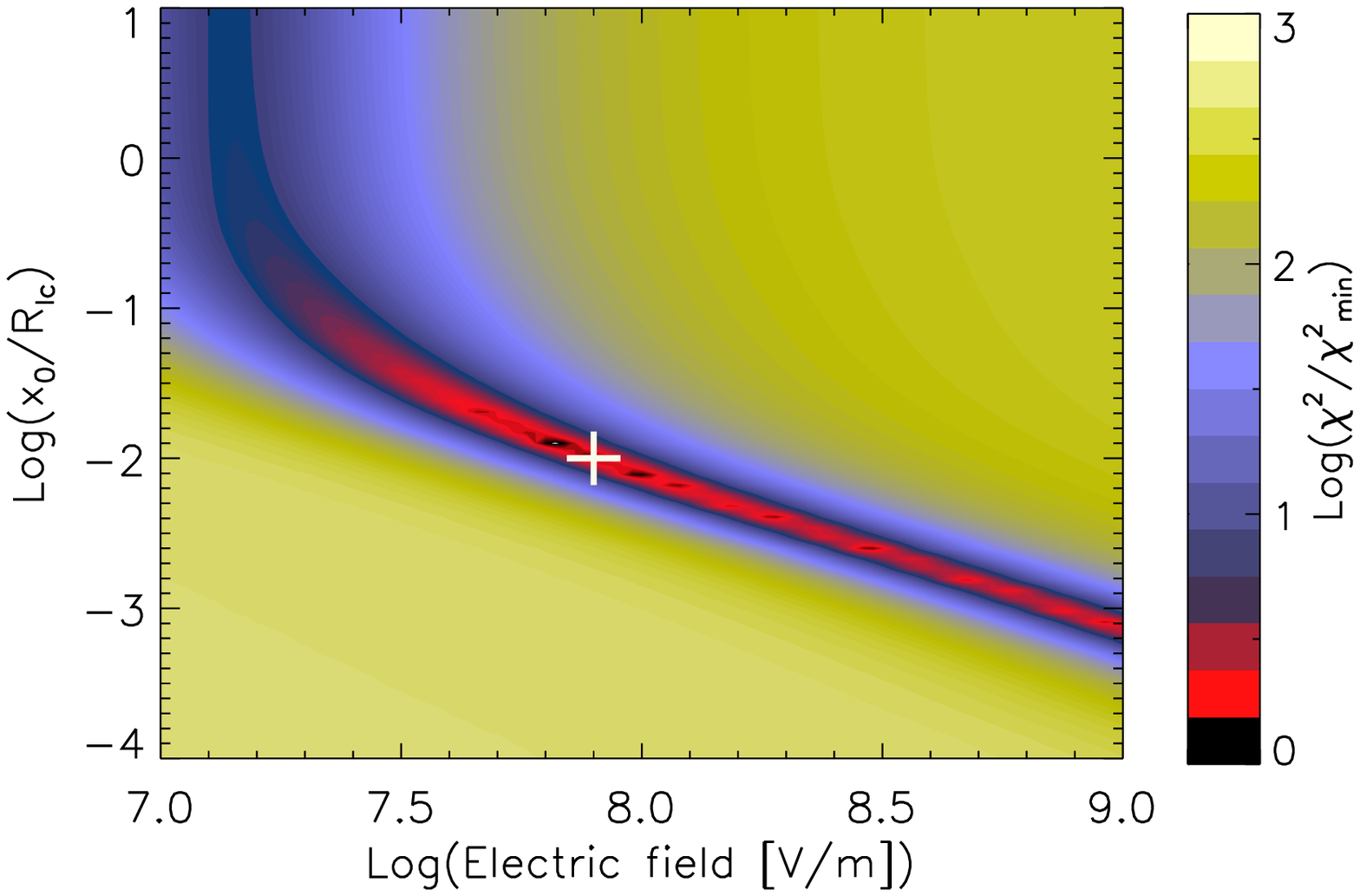}
\put(20,15){\scriptsize  \red {J1718-3825}}
\end{overpic}
\begin{overpic}[width=0.32\textwidth]{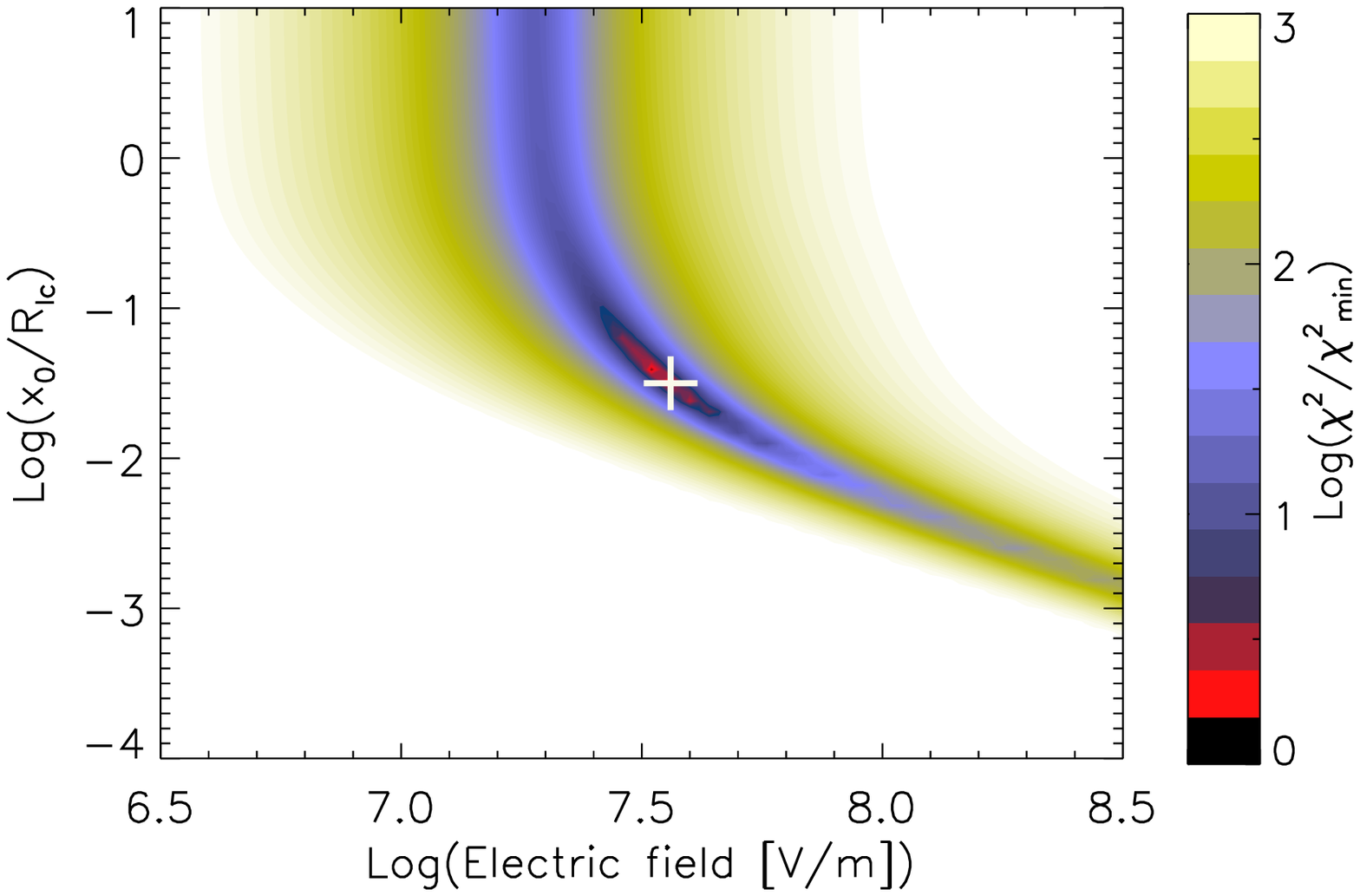}
\put(20,15){\scriptsize  \red {J1732-3131}}
\end{overpic}
\begin{overpic}[width=0.32\textwidth]{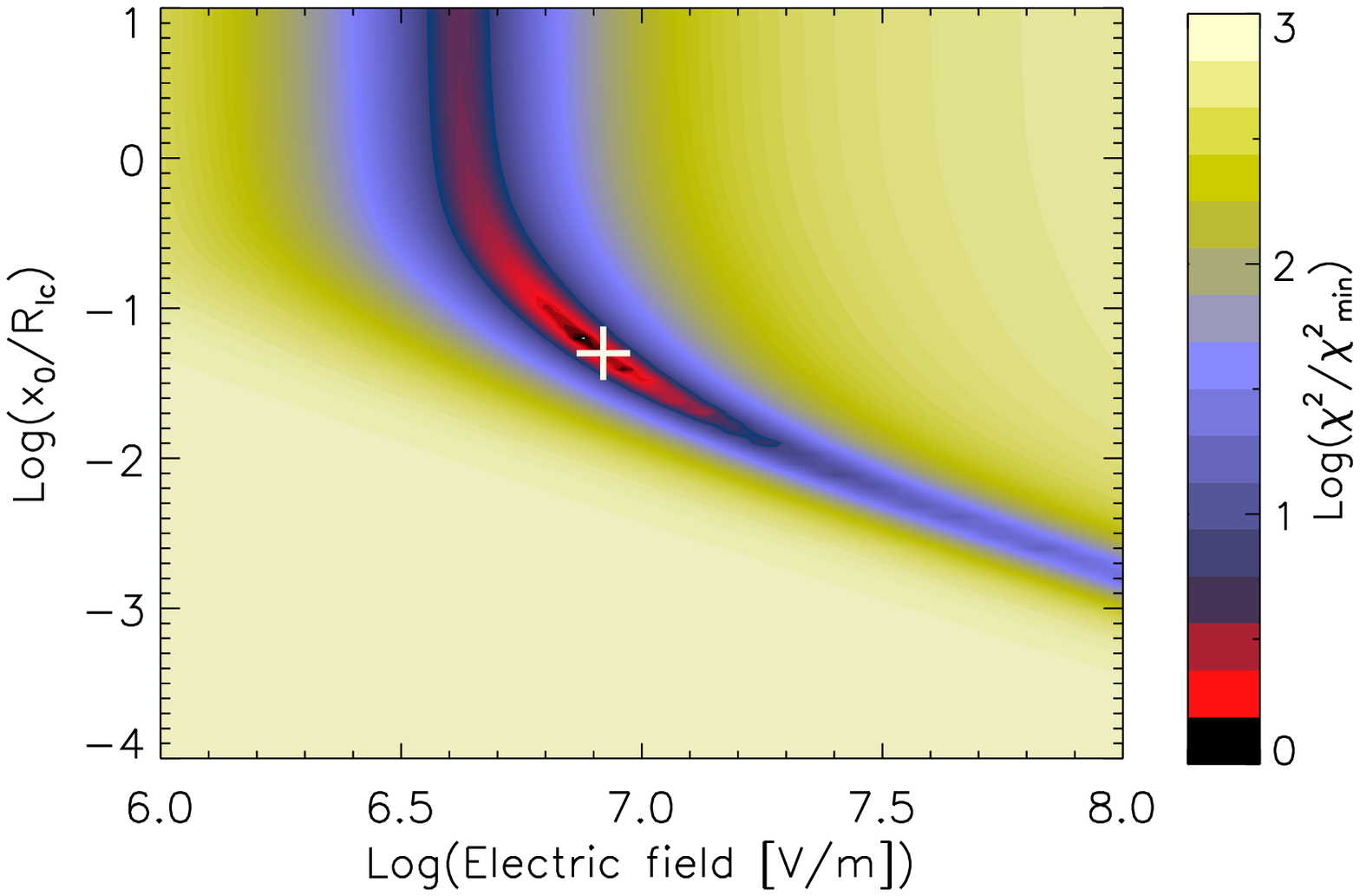}
\put(20,15){\scriptsize  \red {J1741-2054}}
\end{overpic}
\begin{overpic}[width=0.32\textwidth]{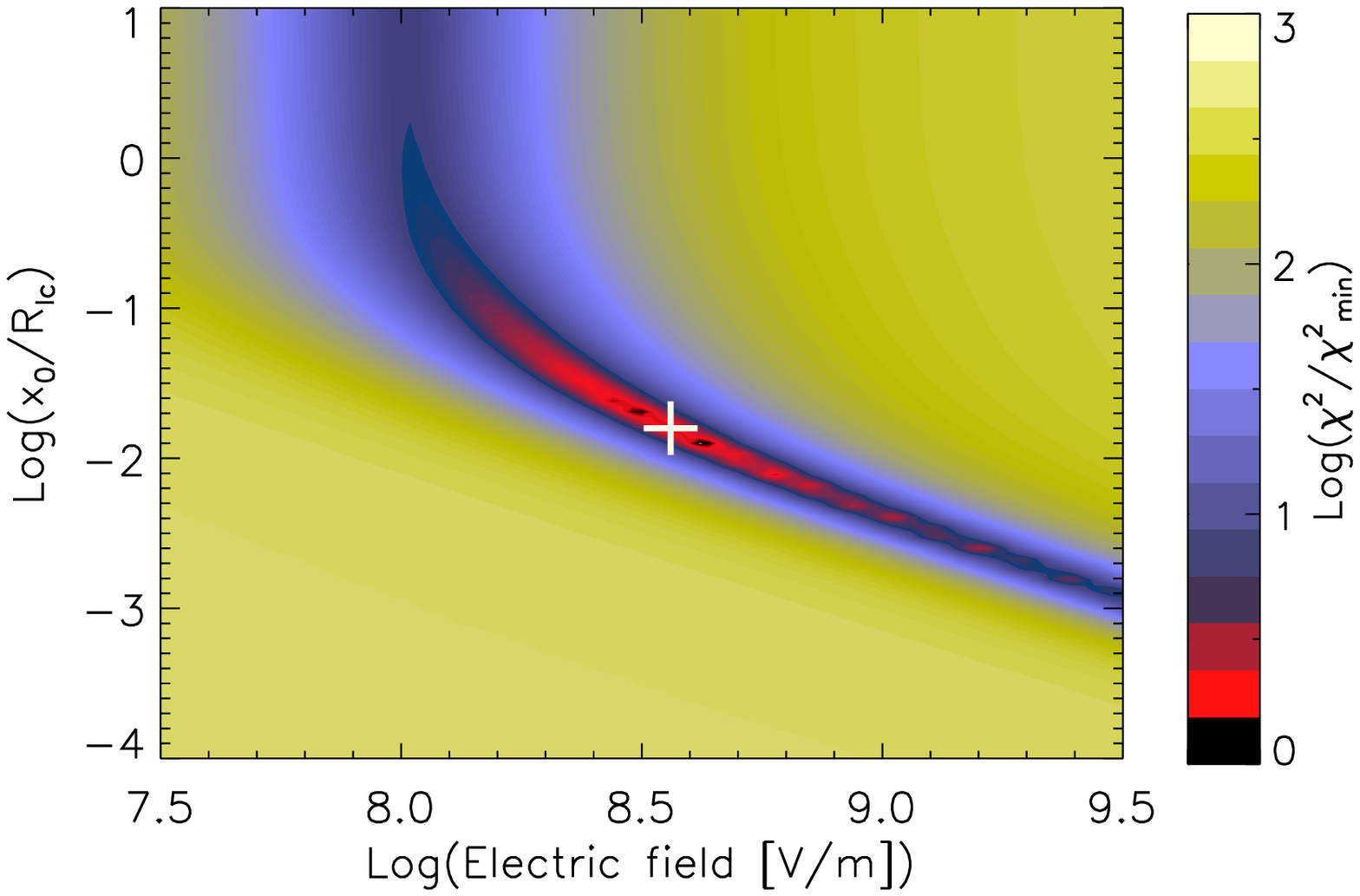}
\put(20,15){\scriptsize \blue {J1744-1134}}
\end{overpic}
\begin{overpic}[width=0.32\textwidth]{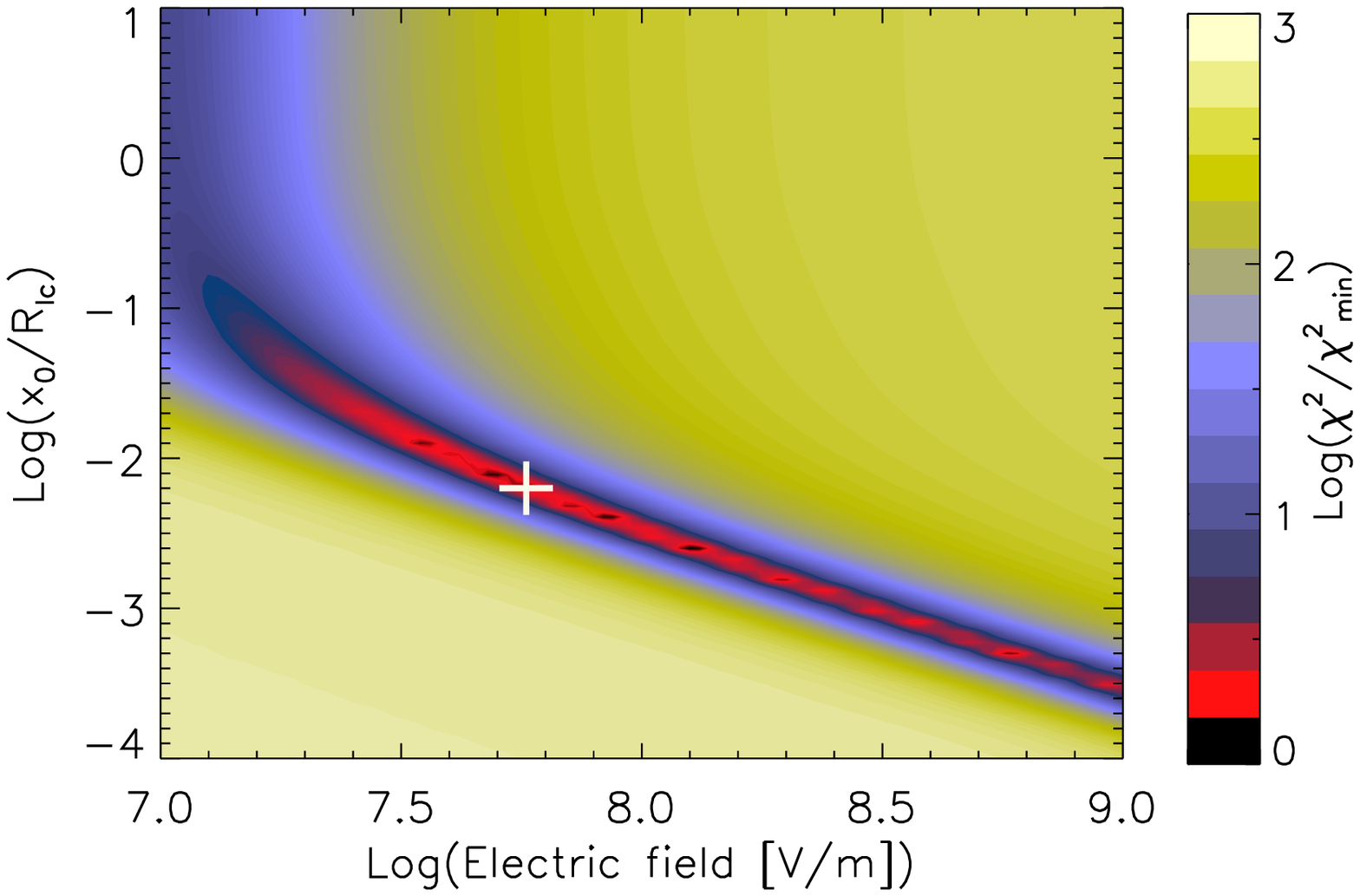}
\put(20,15){\scriptsize  \red {J1746-3239}}
\end{overpic}
\begin{overpic}[width=0.32\textwidth]{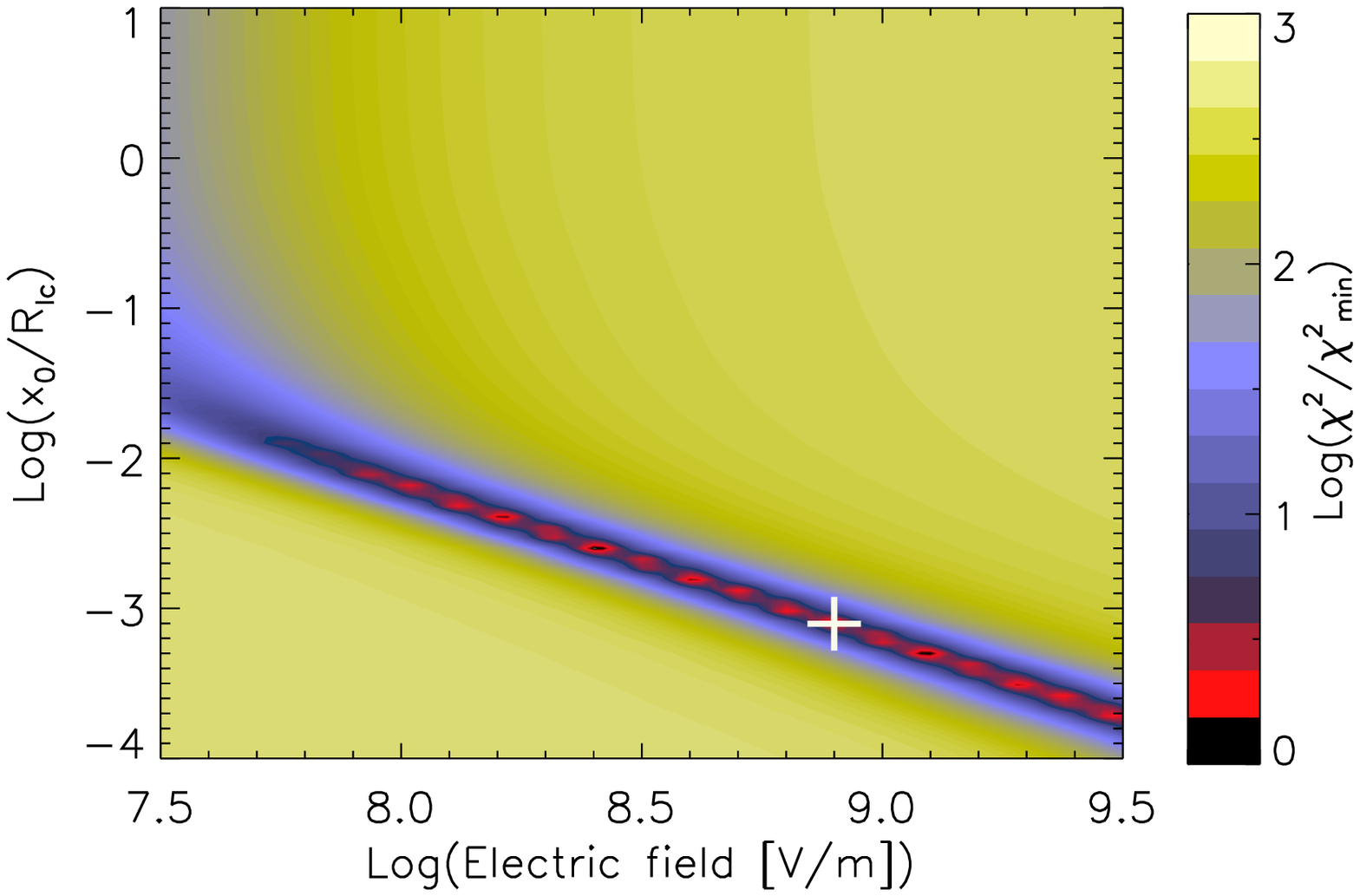}
\put(20,15){\scriptsize  \red {J1747-2958}}
\end{overpic}
\begin{overpic}[width=0.32\textwidth]{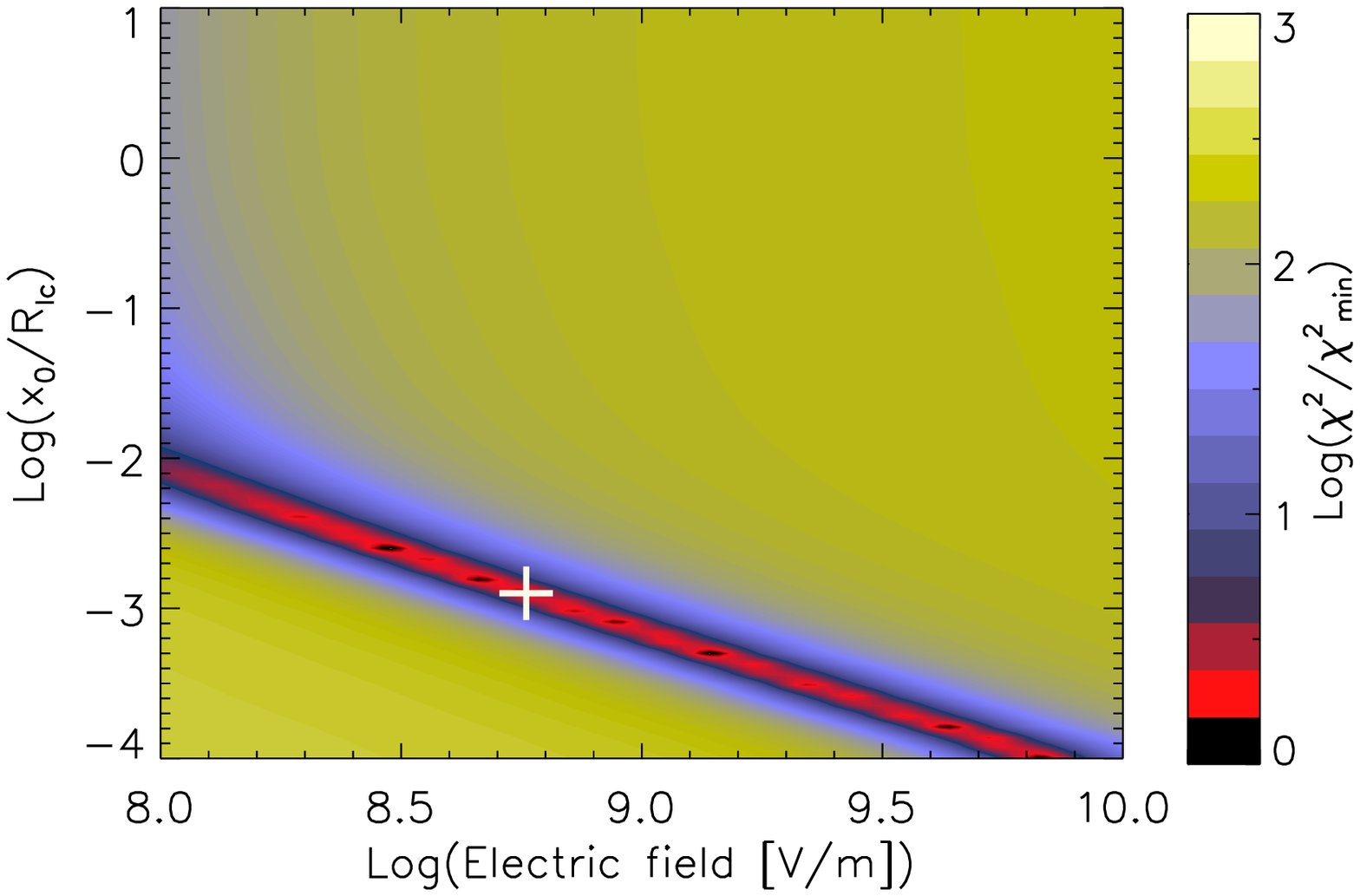}
\put(20,15){\scriptsize \blue{J1803-2149}}
\end{overpic}
\begin{overpic}[width=0.32\textwidth]{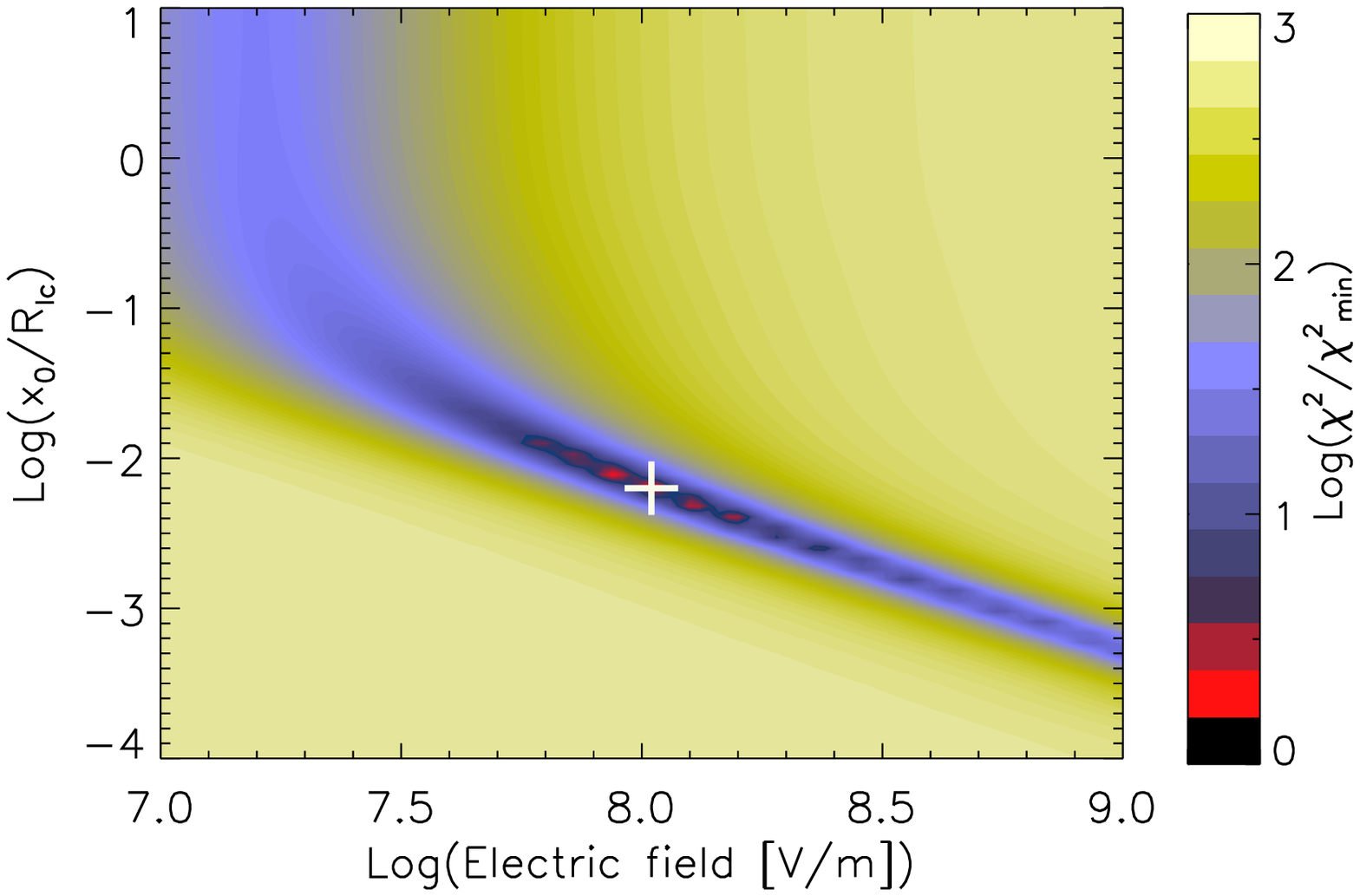}
\put(20,15){\scriptsize  \red {J1809-2332}}
\end{overpic}
\begin{overpic}[width=0.32\textwidth]{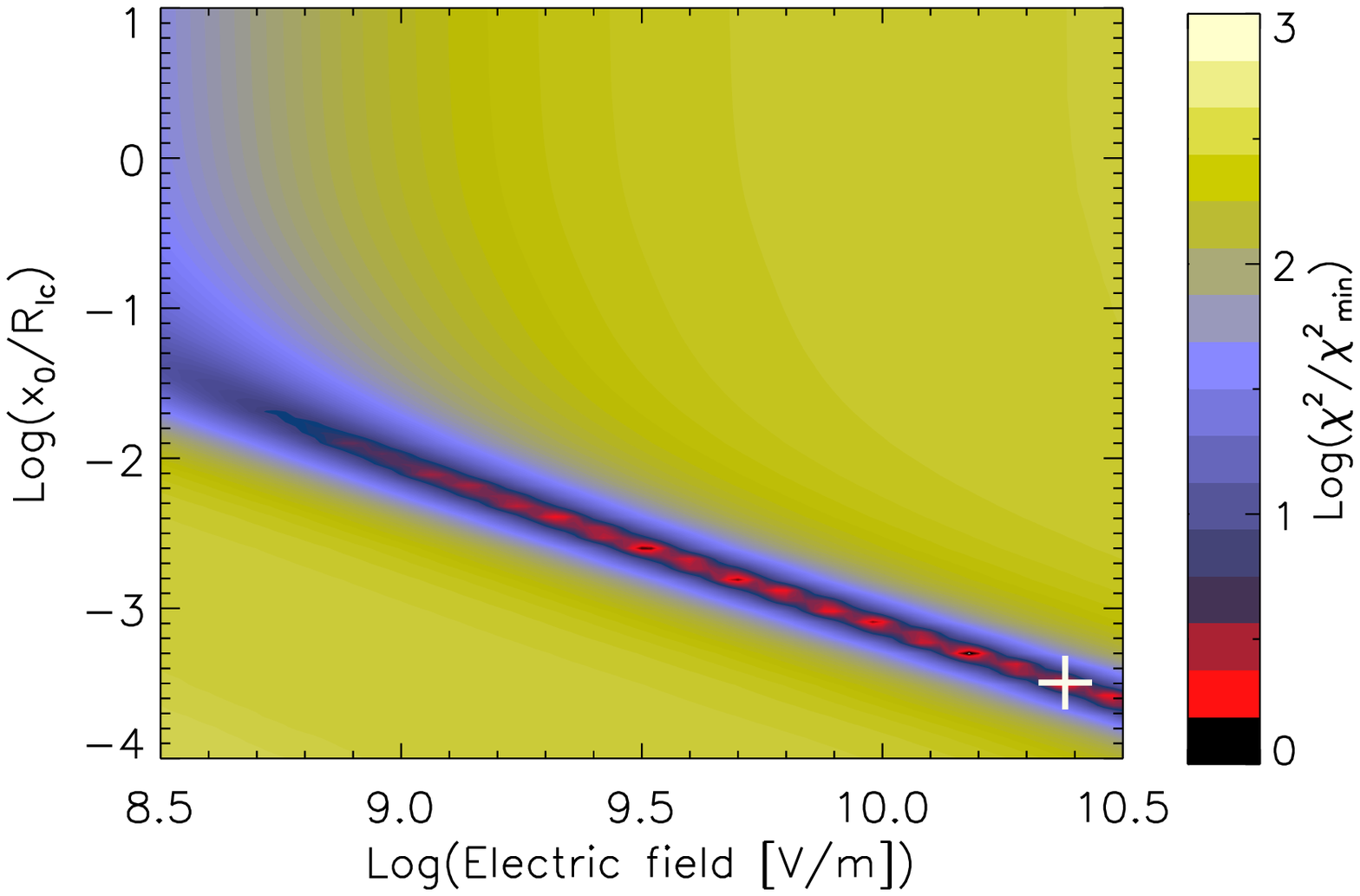}
\put(20,15){\scriptsize \blue{J1810+1744}}
\end{overpic}
\begin{overpic}[width=0.32\textwidth]{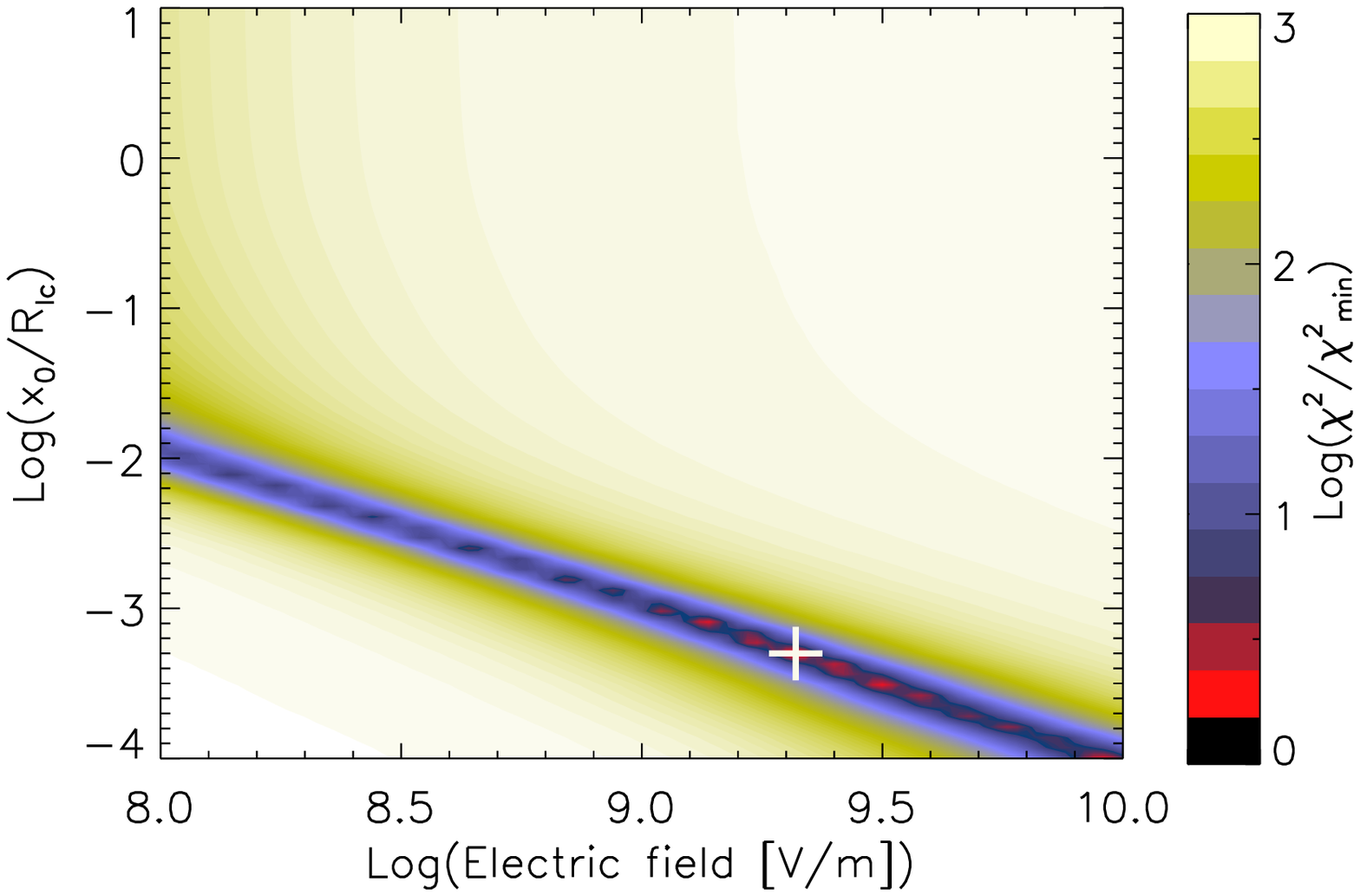}
\put(20,15){\scriptsize  \red {J1813-1246}}
\end{overpic}
\end{center}
\caption{$\chi^2/\chi^2_{\rm min}$ contours on the $\log E_\parallel$--$\log (x_0/R_{\rm lc})$ plane for the pulsars considered in the sample (III).}
\label{fig:contours3}
\end{figure*}

\begin{figure*}
\begin{center}
\begin{overpic}[width=0.32\textwidth]{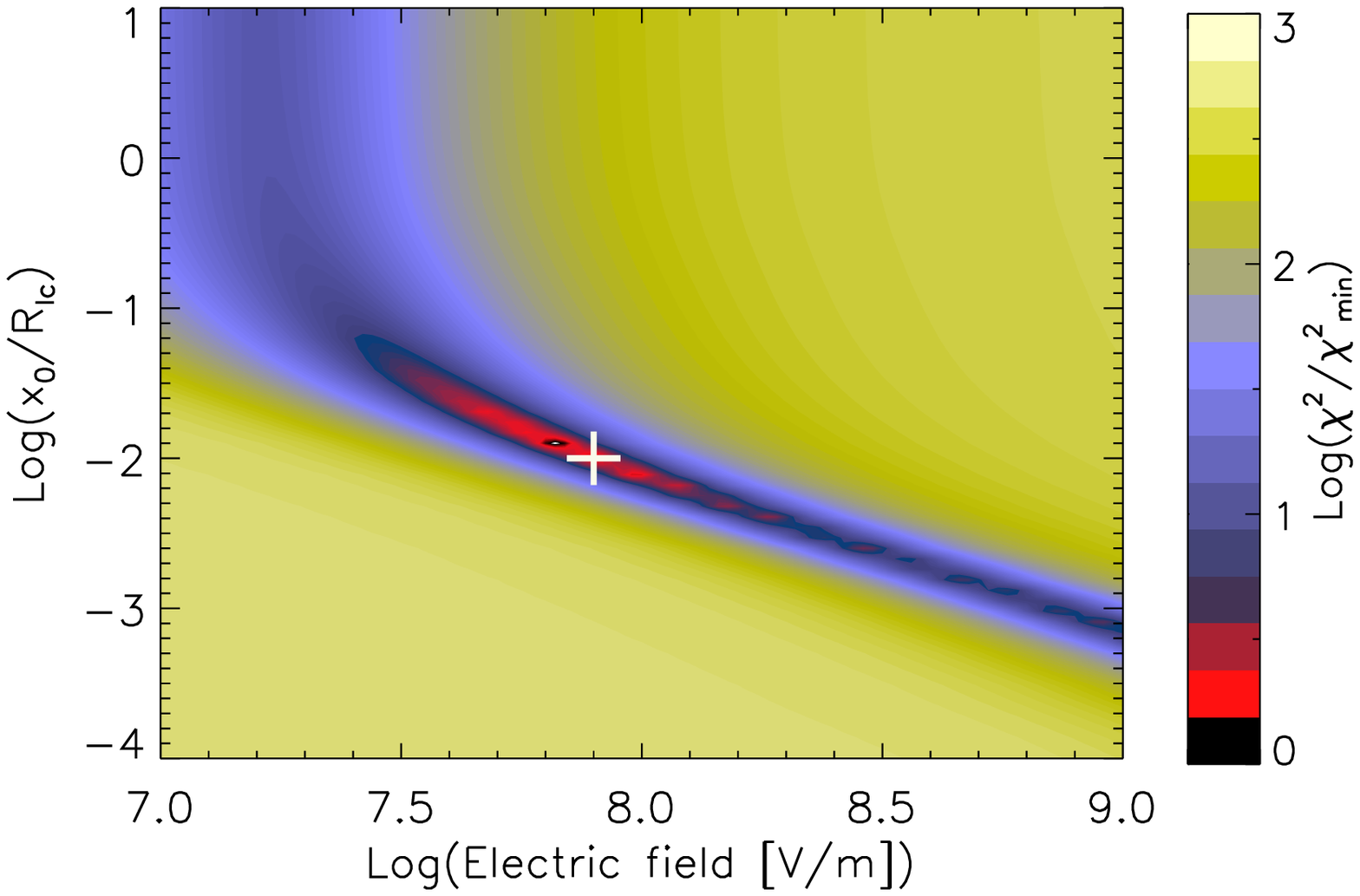}
\put(20,15){\scriptsize  \red {J1826-1256}}
\end{overpic}
\begin{overpic}[width=0.32\textwidth]{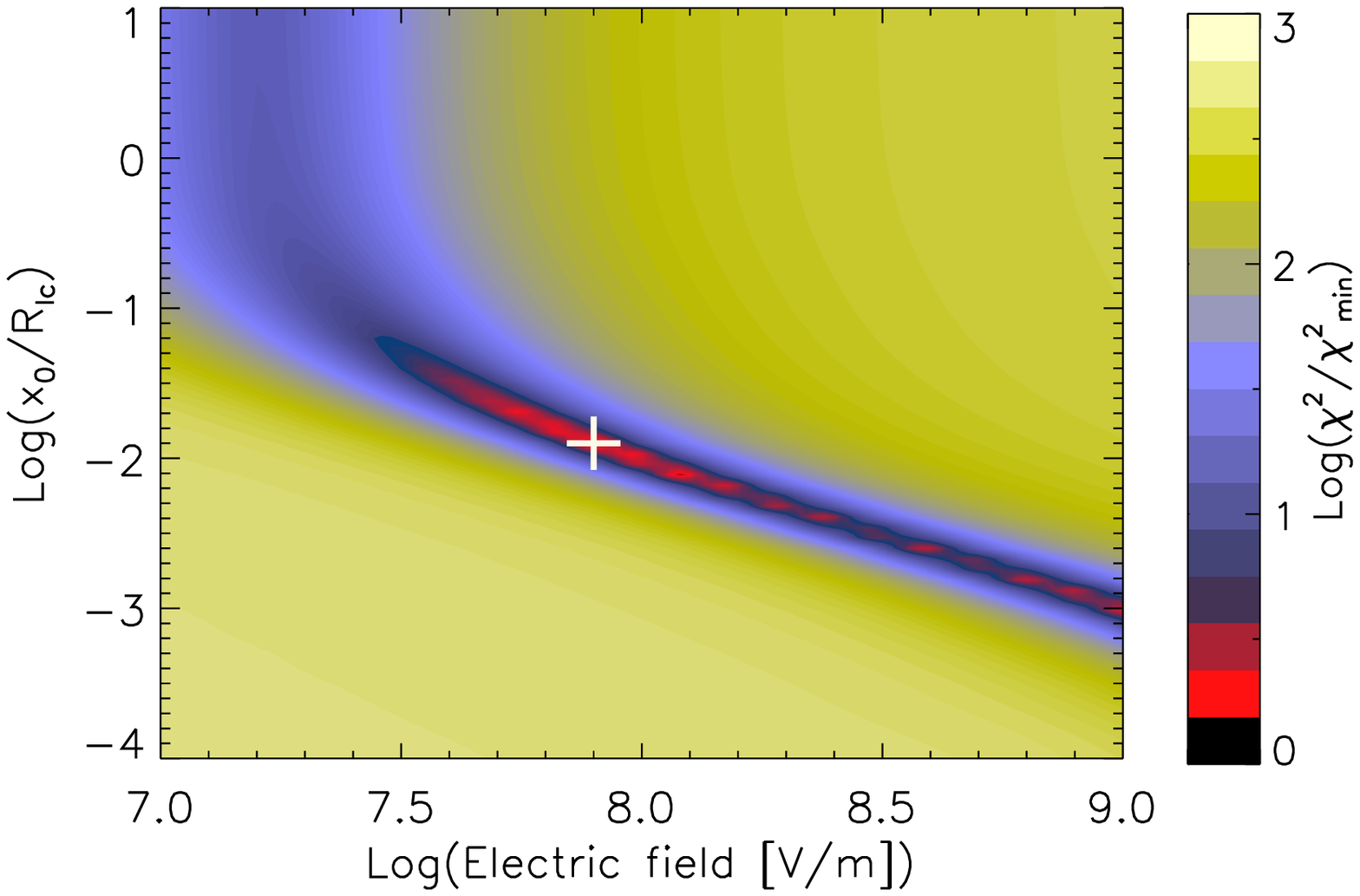}
\put(20,15){\scriptsize  \red {J1833-1034}}
\end{overpic}
\begin{overpic}[width=0.32\textwidth]{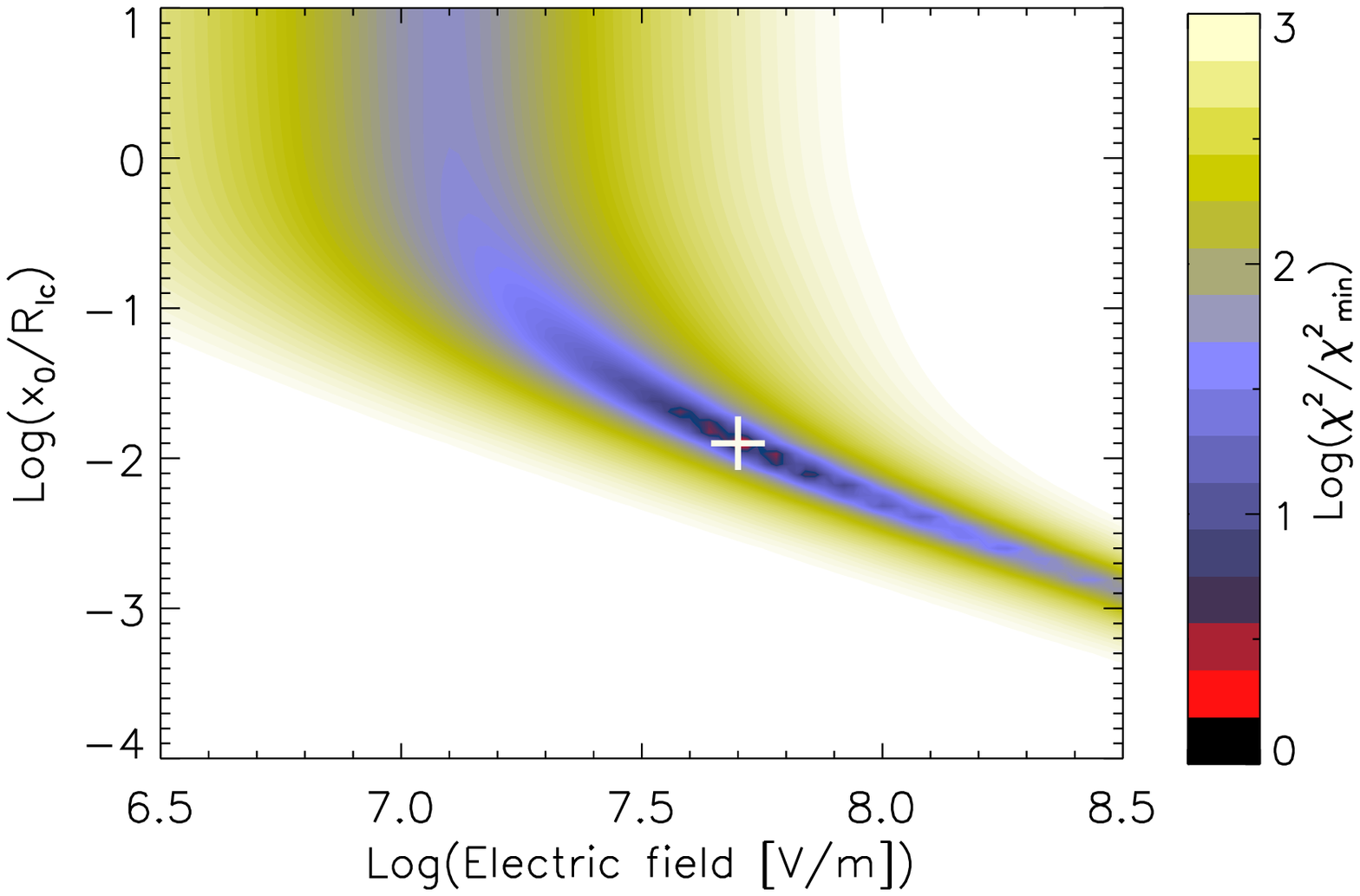}
\put(20,15){\scriptsize  \red {J1836+5925}}
\end{overpic}
\begin{overpic}[width=0.32\textwidth]{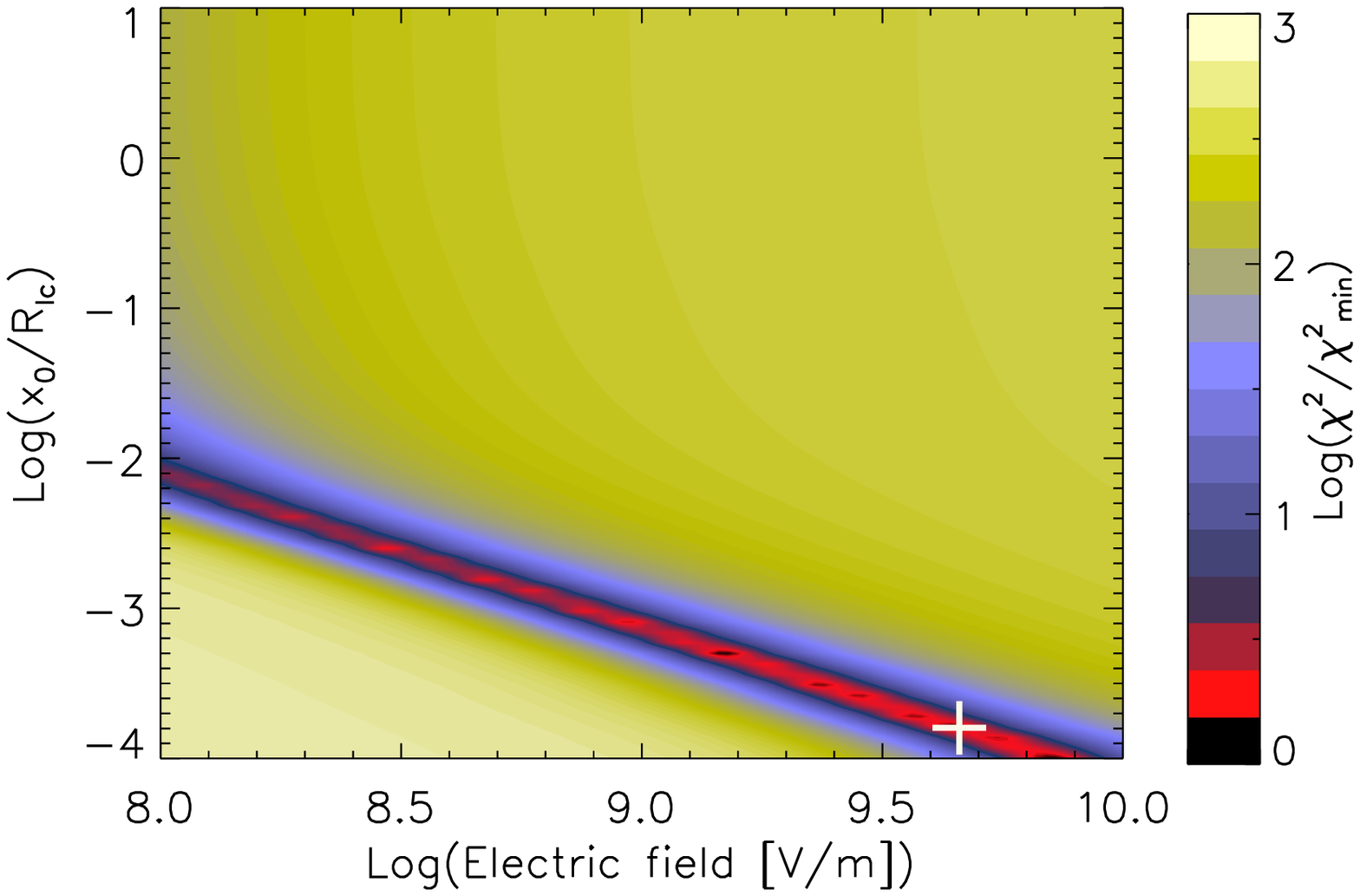}
\put(20,15){\scriptsize  \red {J1838-0537}}
\end{overpic}
\begin{overpic}[width=0.32\textwidth]{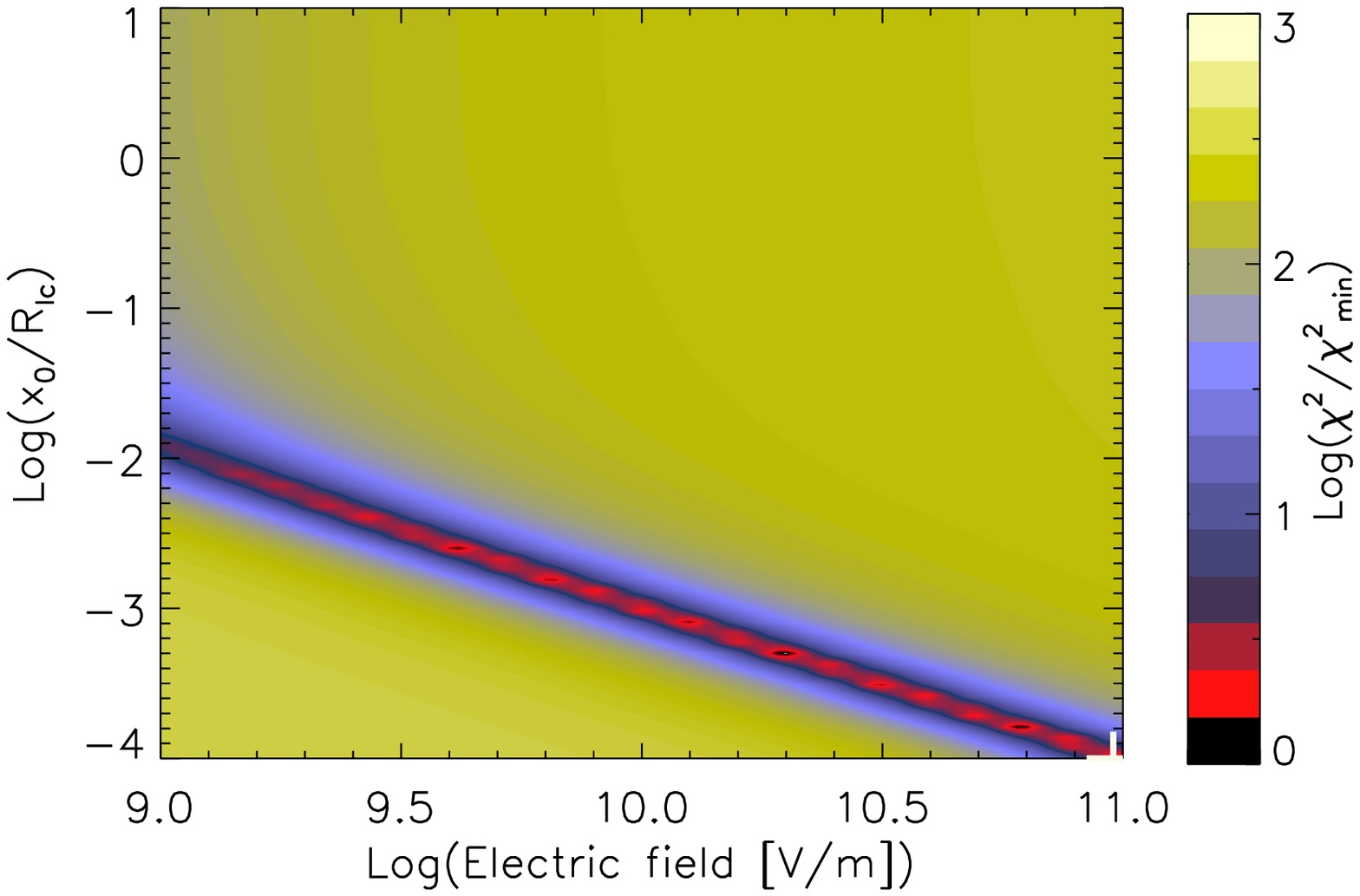}
\put(20,15){\scriptsize \blue {J1902-5105}}
\end{overpic}
\begin{overpic}[width=0.32\textwidth]{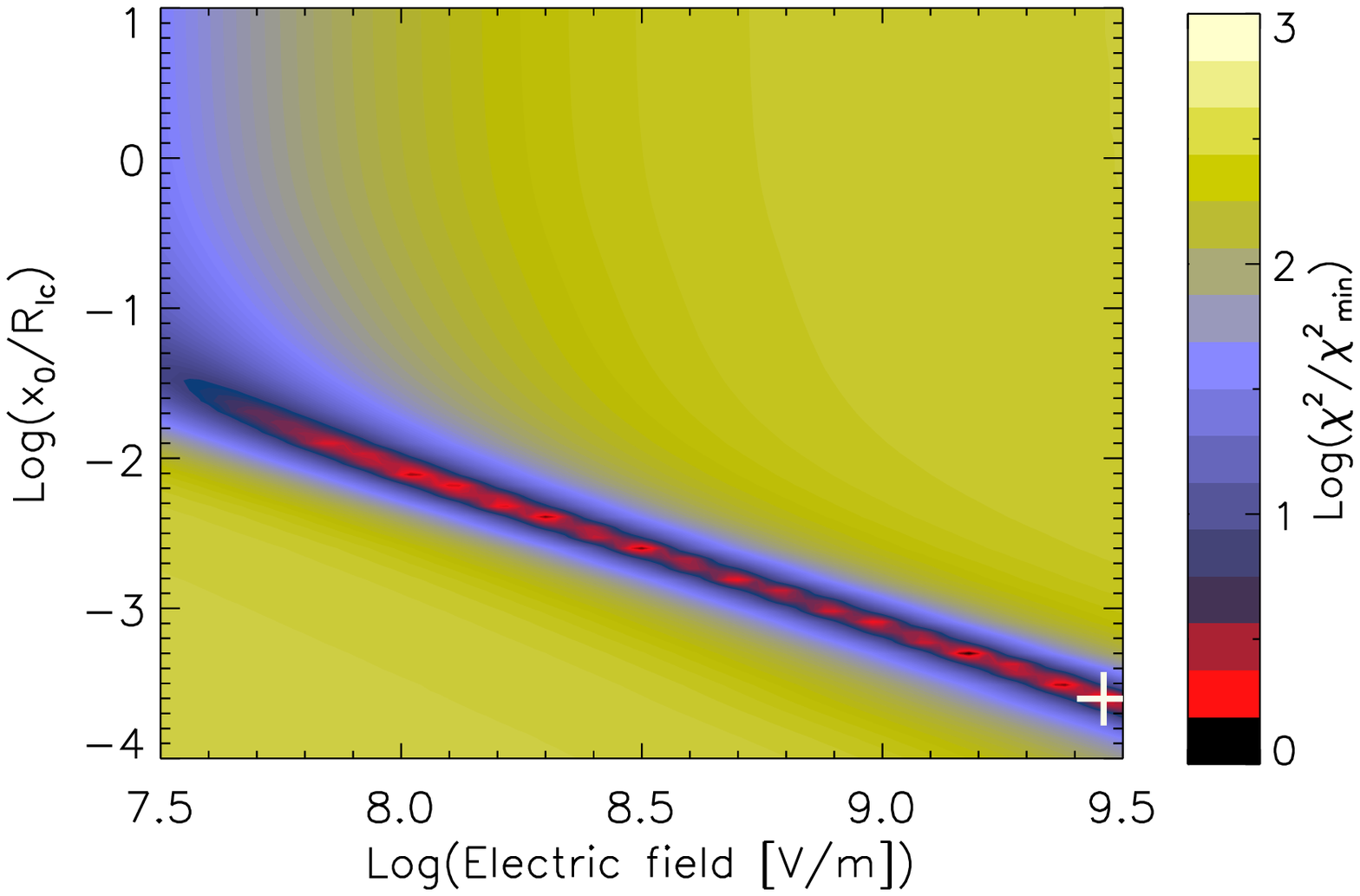}
\put(20,15){\scriptsize  \red {J1907+0602}}
\end{overpic}
\begin{overpic}[width=0.32\textwidth]{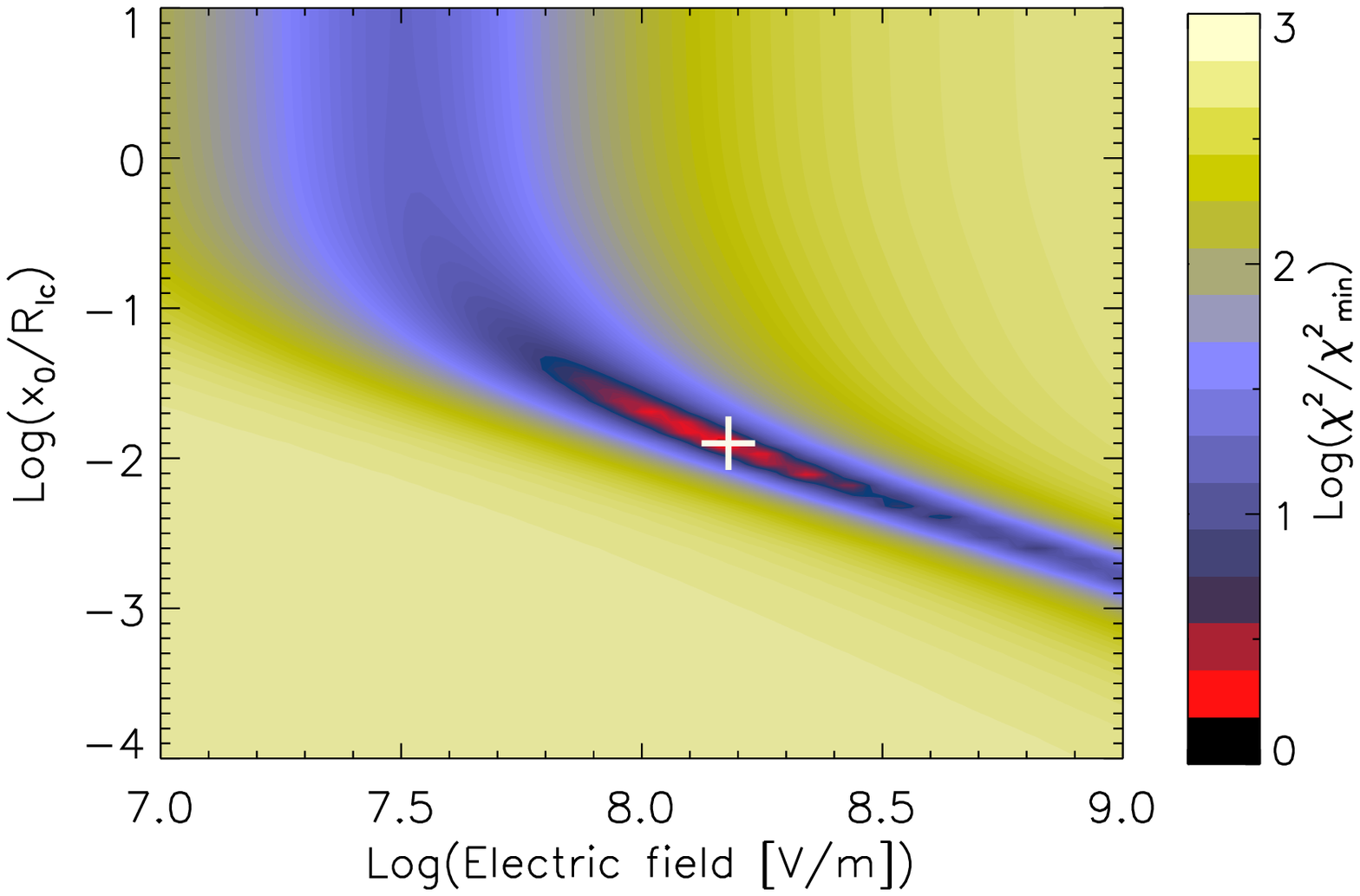}
\put(20,15){\scriptsize  \red {J1952+3252}}
\end{overpic}
\begin{overpic}[width=0.32\textwidth]{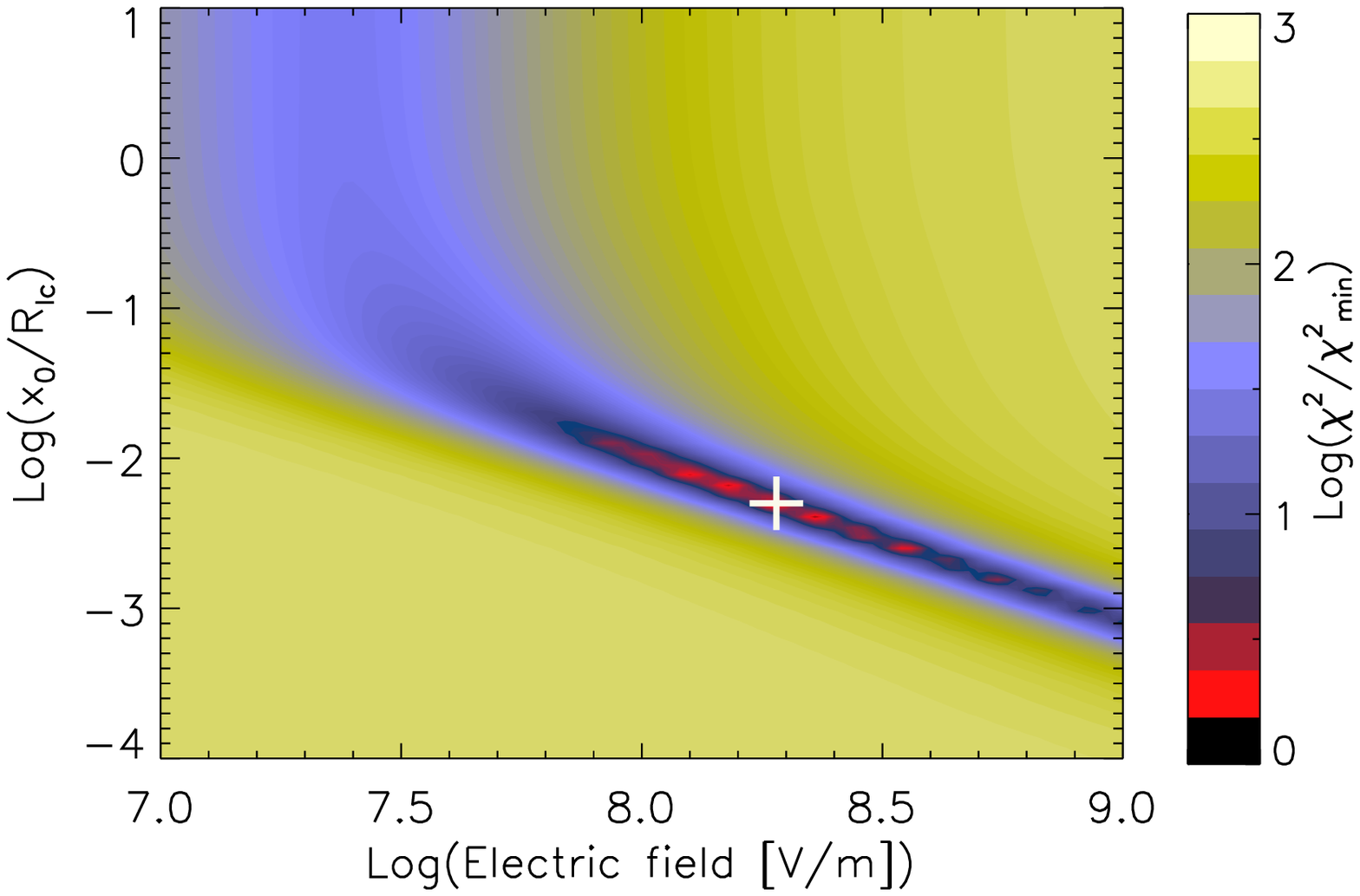}
\put(20,15){\scriptsize  \red {J1954+2836}}
\end{overpic}
\begin{overpic}[width=0.32\textwidth]{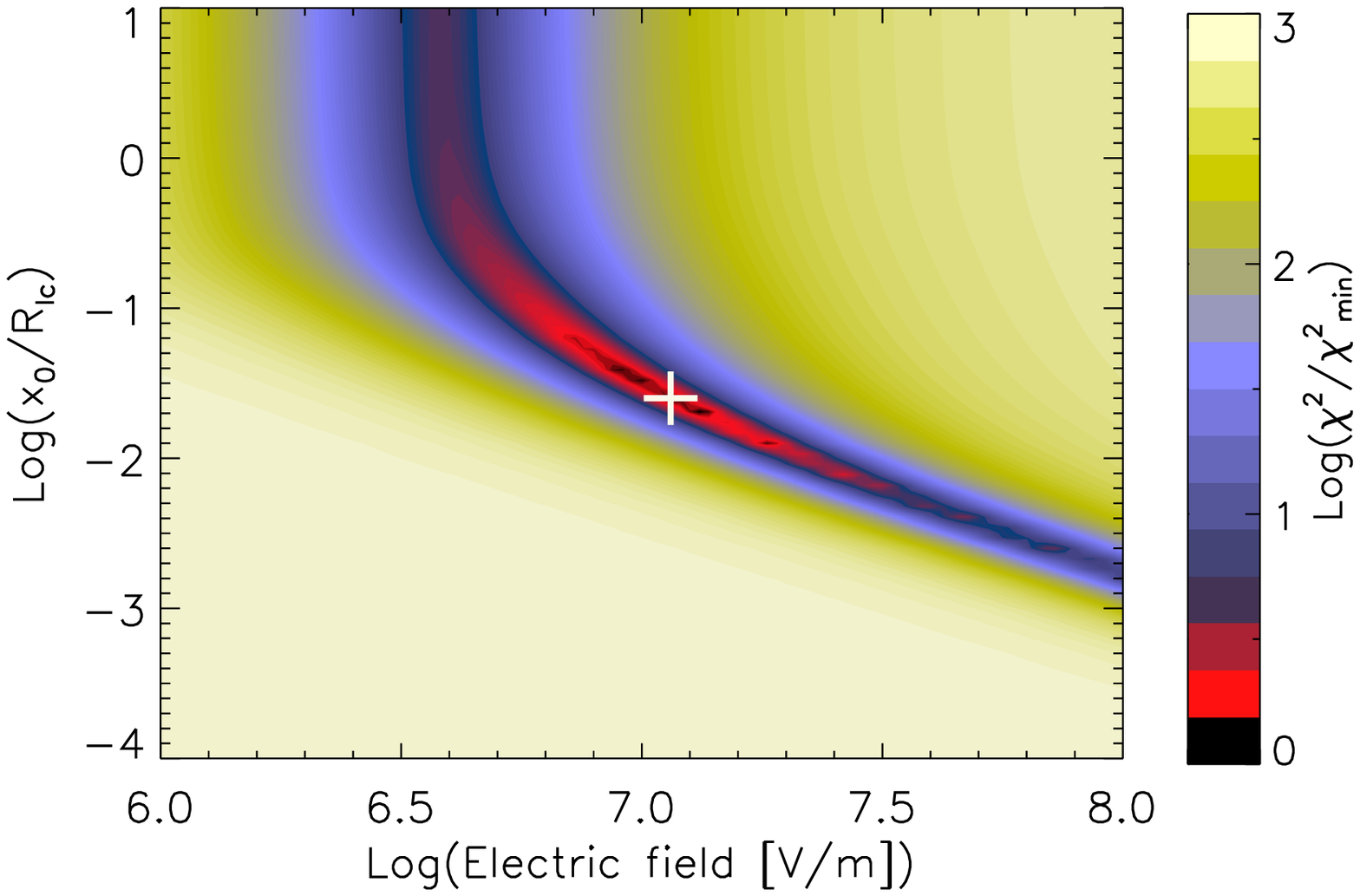}
\put(20,15){\scriptsize  \red {J1957+5033}}
\end{overpic}
\begin{overpic}[width=0.32\textwidth]{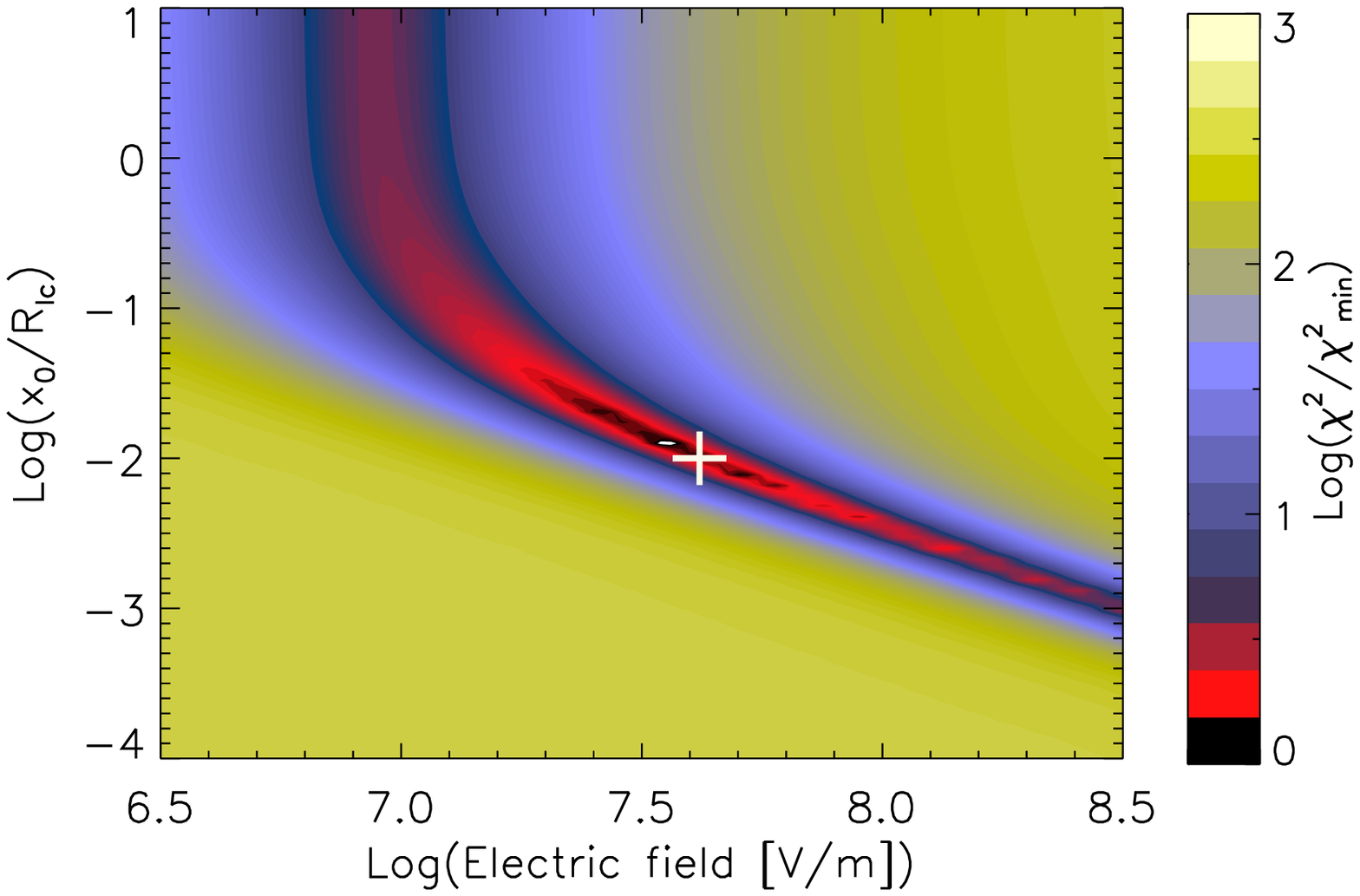}
\put(20,15){\scriptsize  \red {J1958+2846}}
\end{overpic}
\begin{overpic}[width=0.32\textwidth]{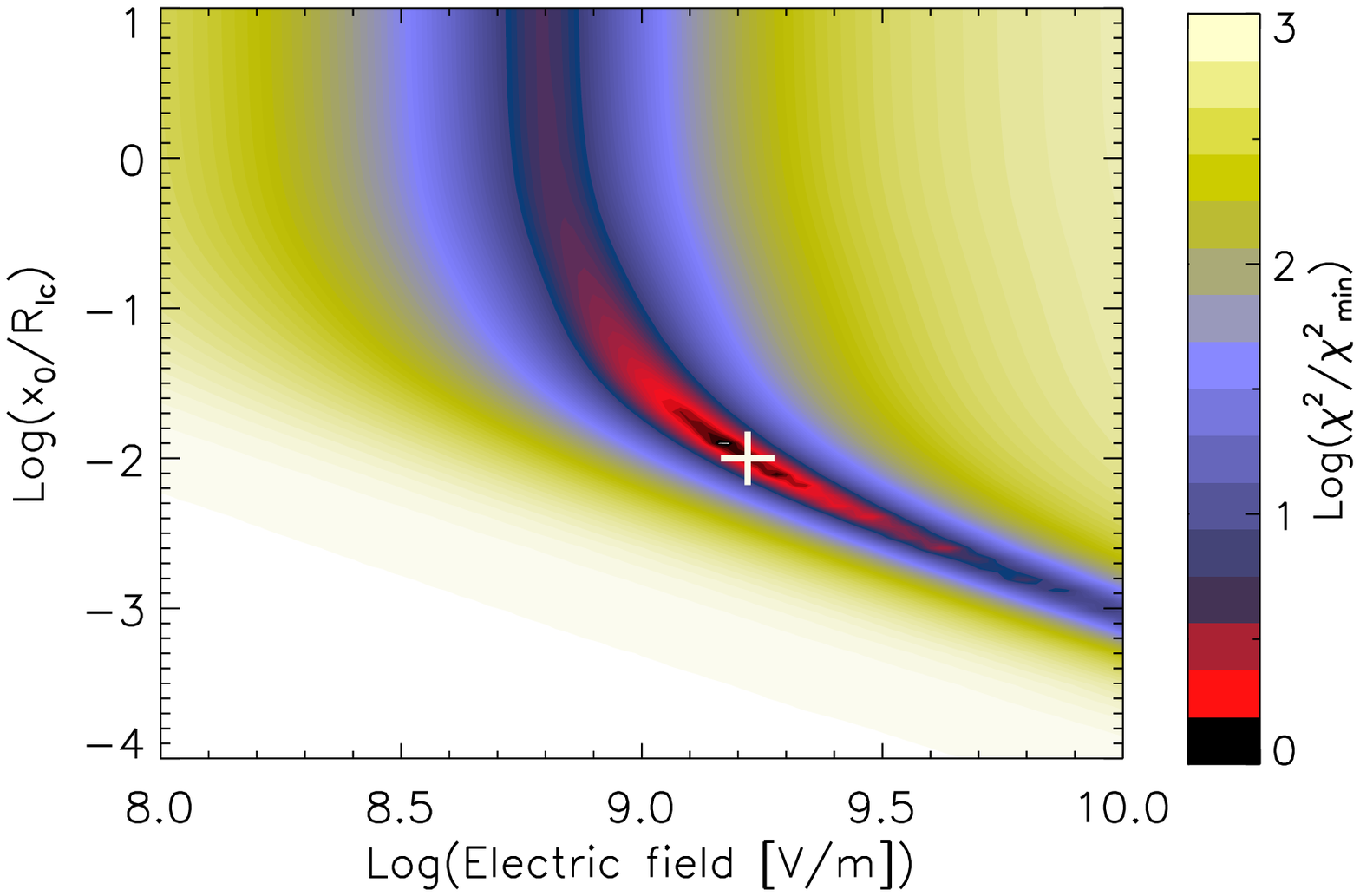}
\put(20,15){\scriptsize \blue {J2017+0603}}
\end{overpic}
\begin{overpic}[width=0.32\textwidth]{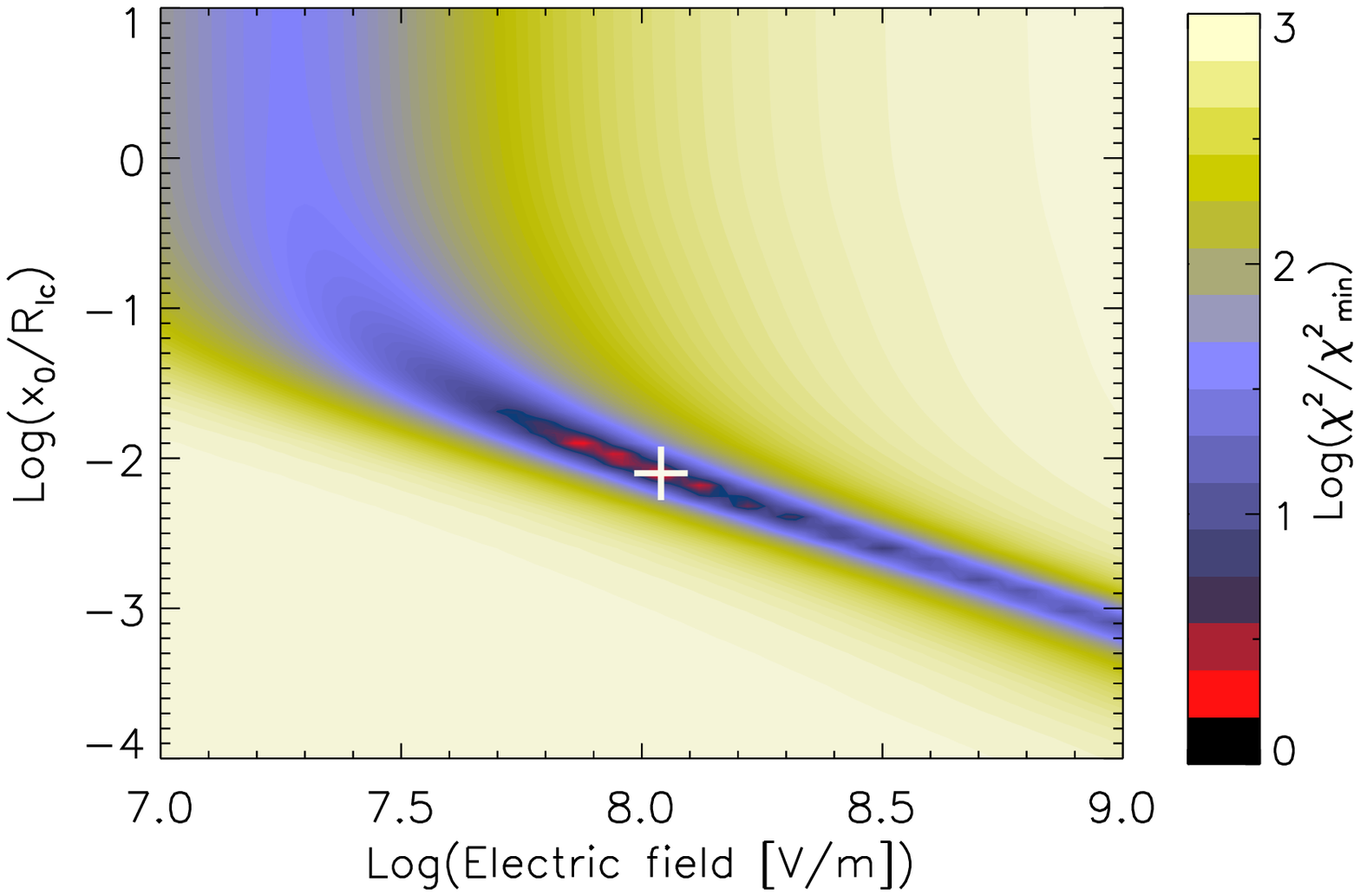}
\put(20,15){\scriptsize  \red {J2021+3651}}
\end{overpic}
\begin{overpic}[width=0.32\textwidth]{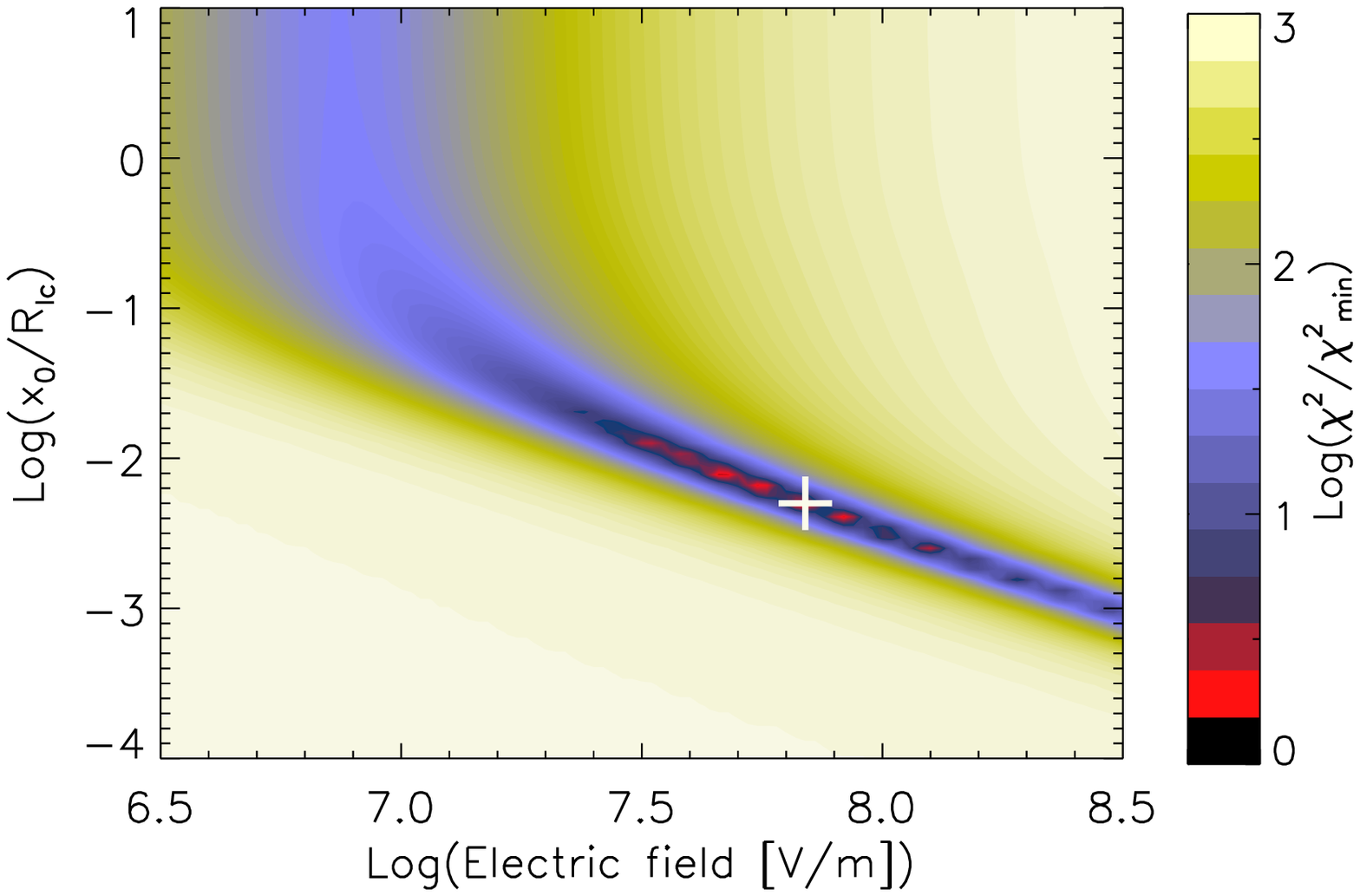}
\put(20,15){\scriptsize  \red {J2021+4026}}
\end{overpic}
\begin{overpic}[width=0.32\textwidth]{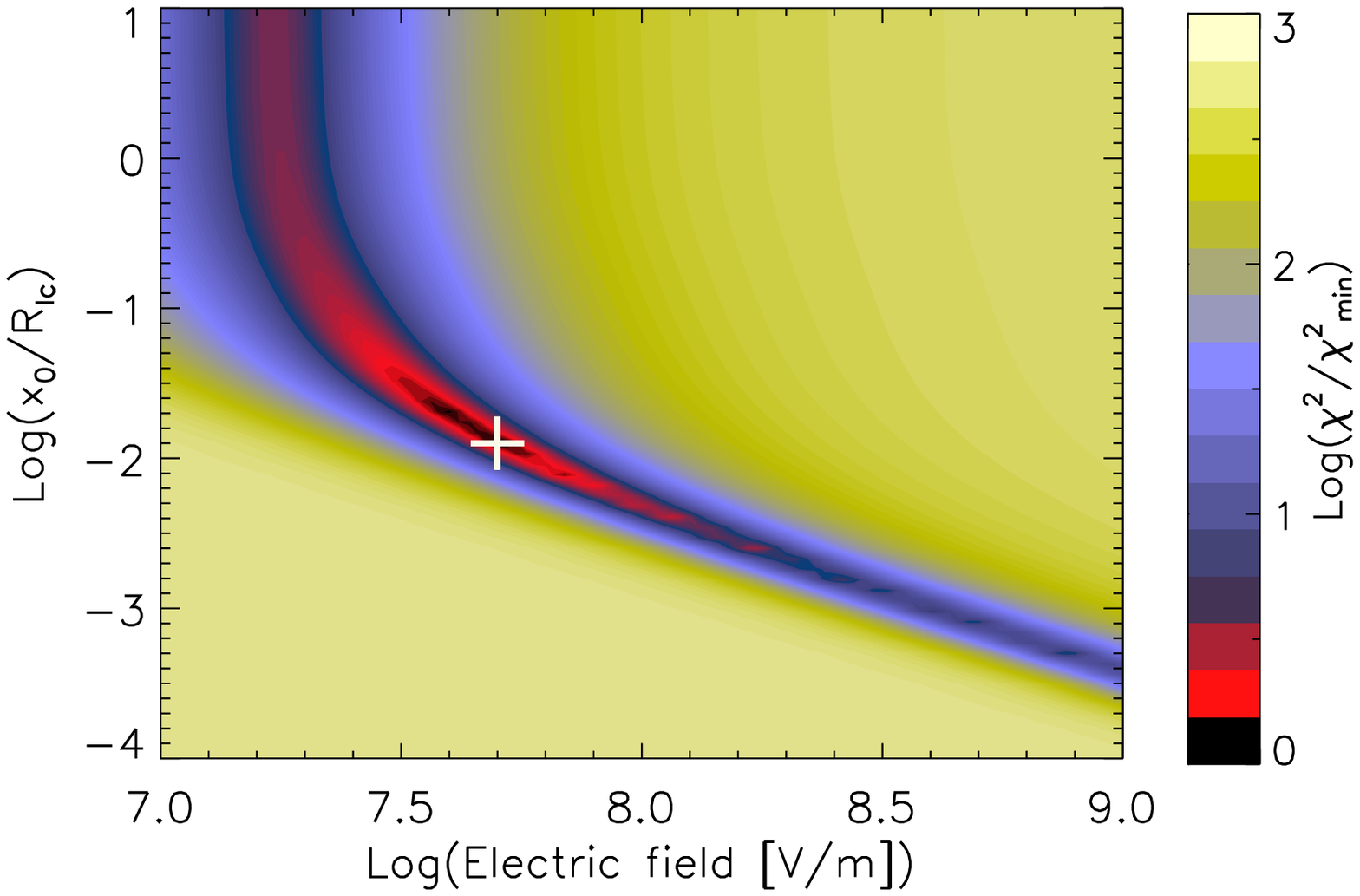}
\put(20,15){\scriptsize  \red {J2028+3332}}
\end{overpic}
\begin{overpic}[width=0.32\textwidth]{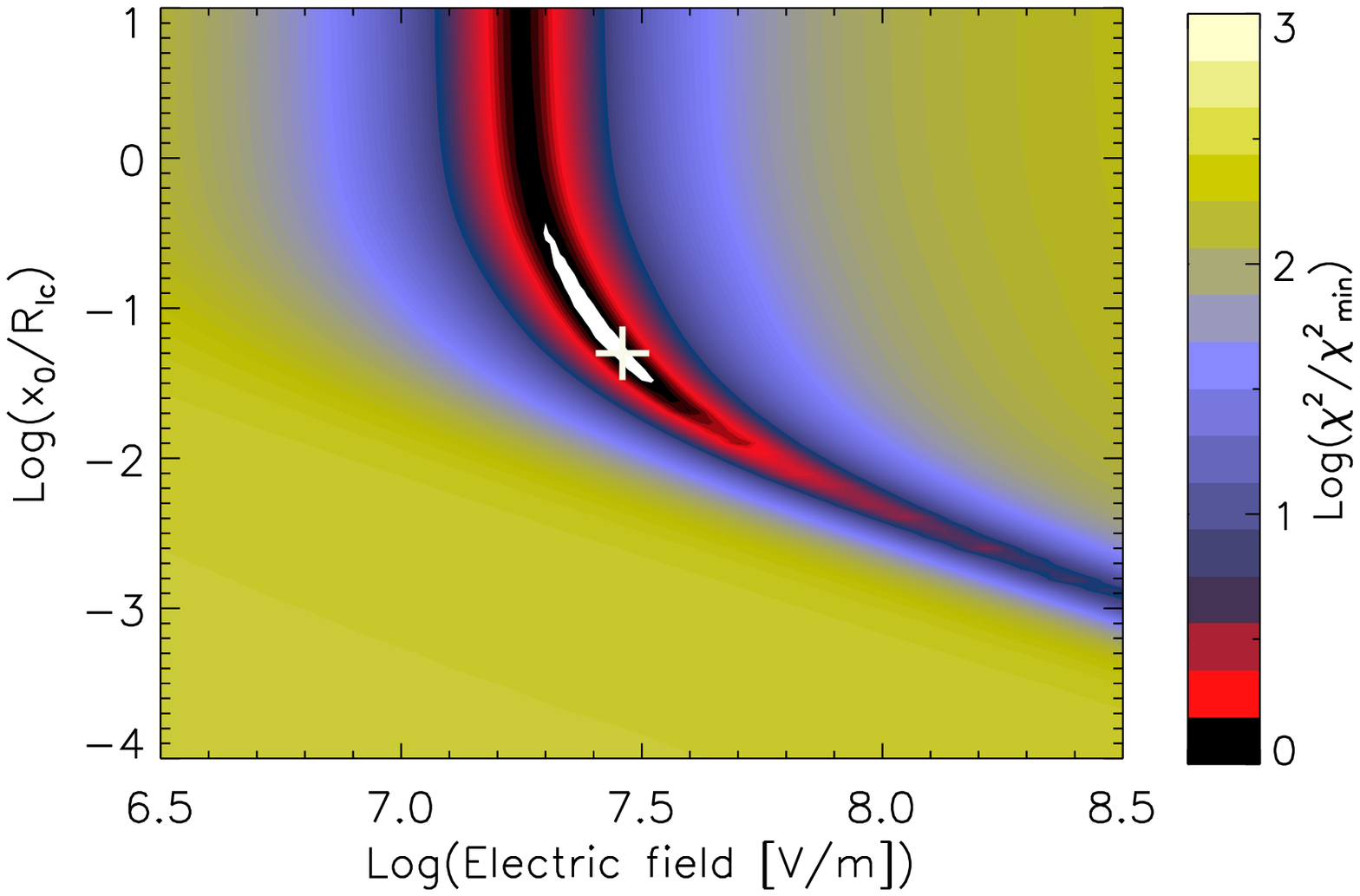}
\put(20,15){\scriptsize  \red {J2030+3641}}
\end{overpic}
\begin{overpic}[width=0.32\textwidth]{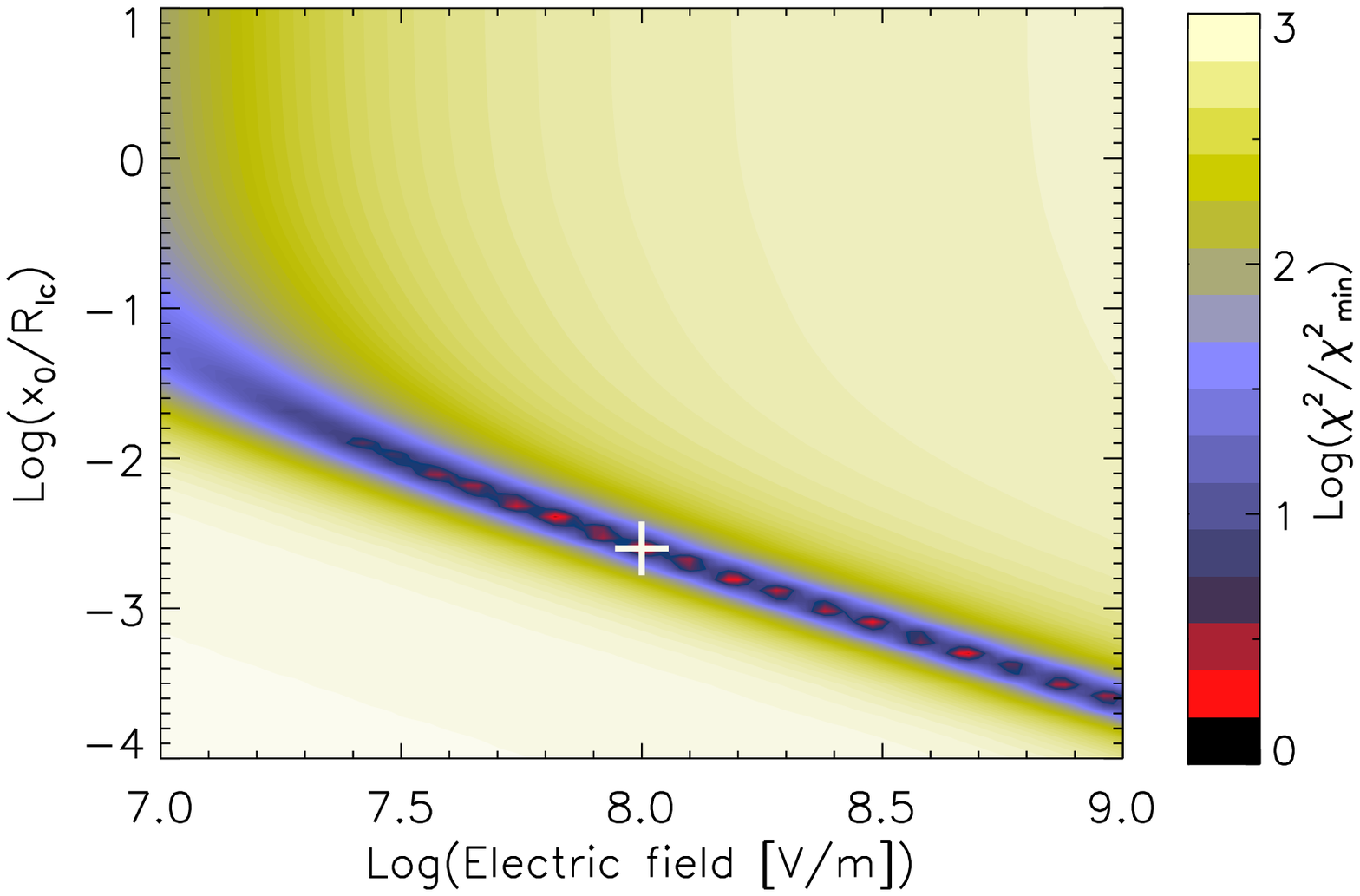}
\put(20,15){\scriptsize  \red {J2030+4415}}
\end{overpic}
\begin{overpic}[width=0.32\textwidth]{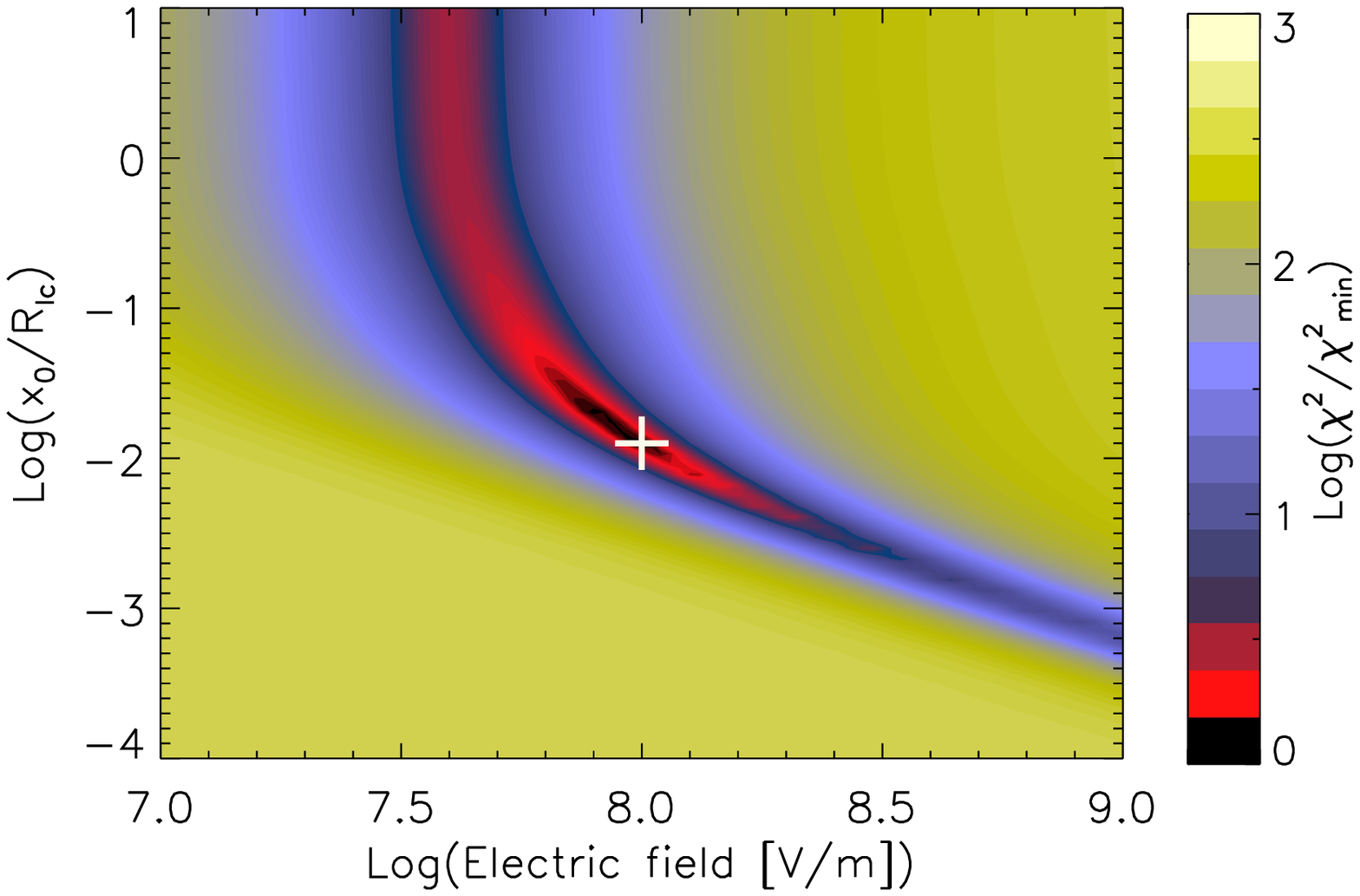}
\put(20,15){\scriptsize  \red {J2032+4127}}
\end{overpic}
\begin{overpic}[width=0.32\textwidth]{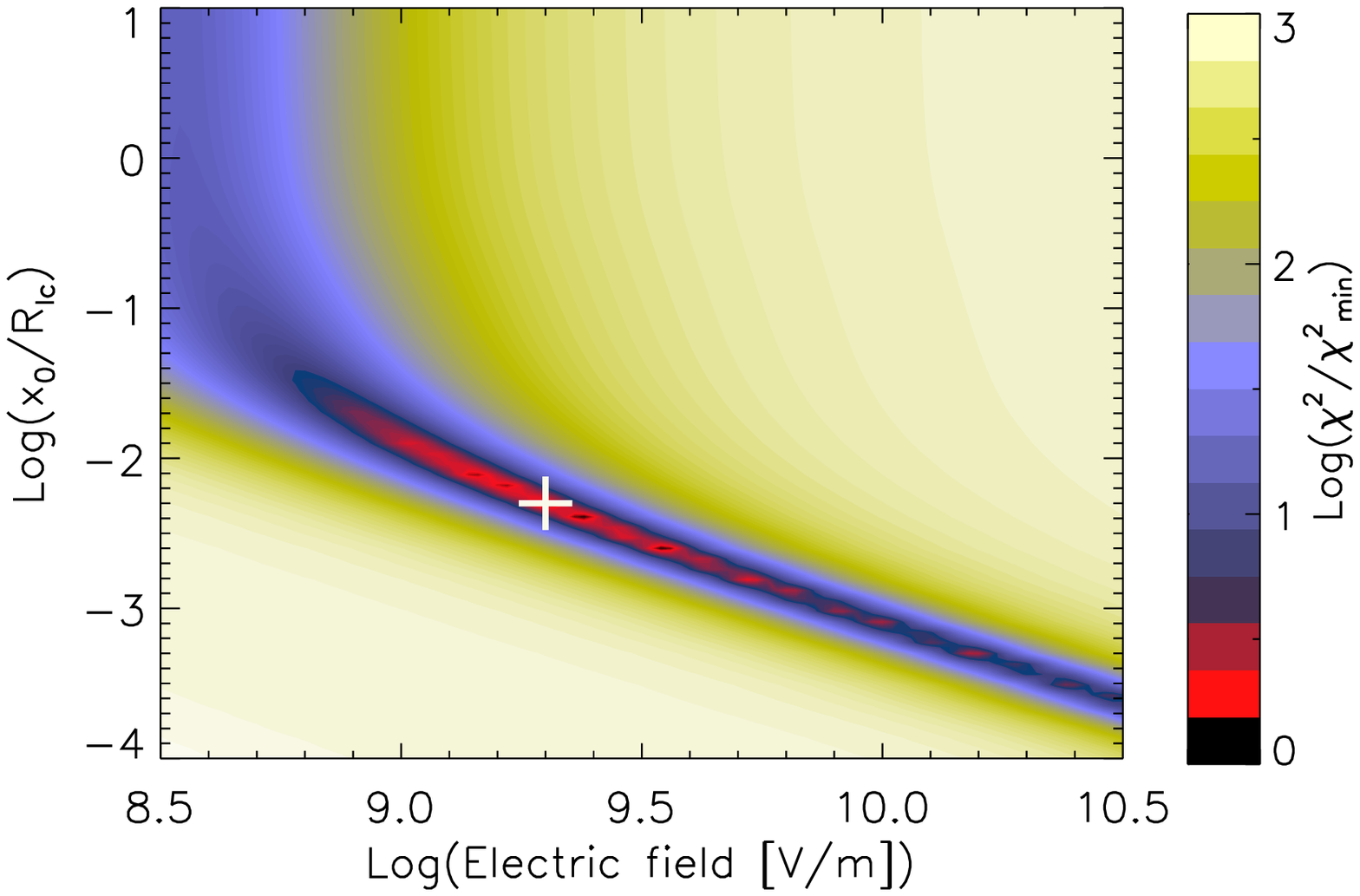}
\put(20,15){\scriptsize \blue {J2043+1711}}
\end{overpic}
\end{center}
\caption{$\chi^2/\chi^2_{\rm min}$ contours on the $\log E_\parallel$--$\log (x_0/R_{\rm lc})$ plane for the pulsars considered in the sample (IV).}
\label{fig:contours4}
\end{figure*}

\begin{figure*}
\begin{center}
\begin{overpic}[width=0.32\textwidth]{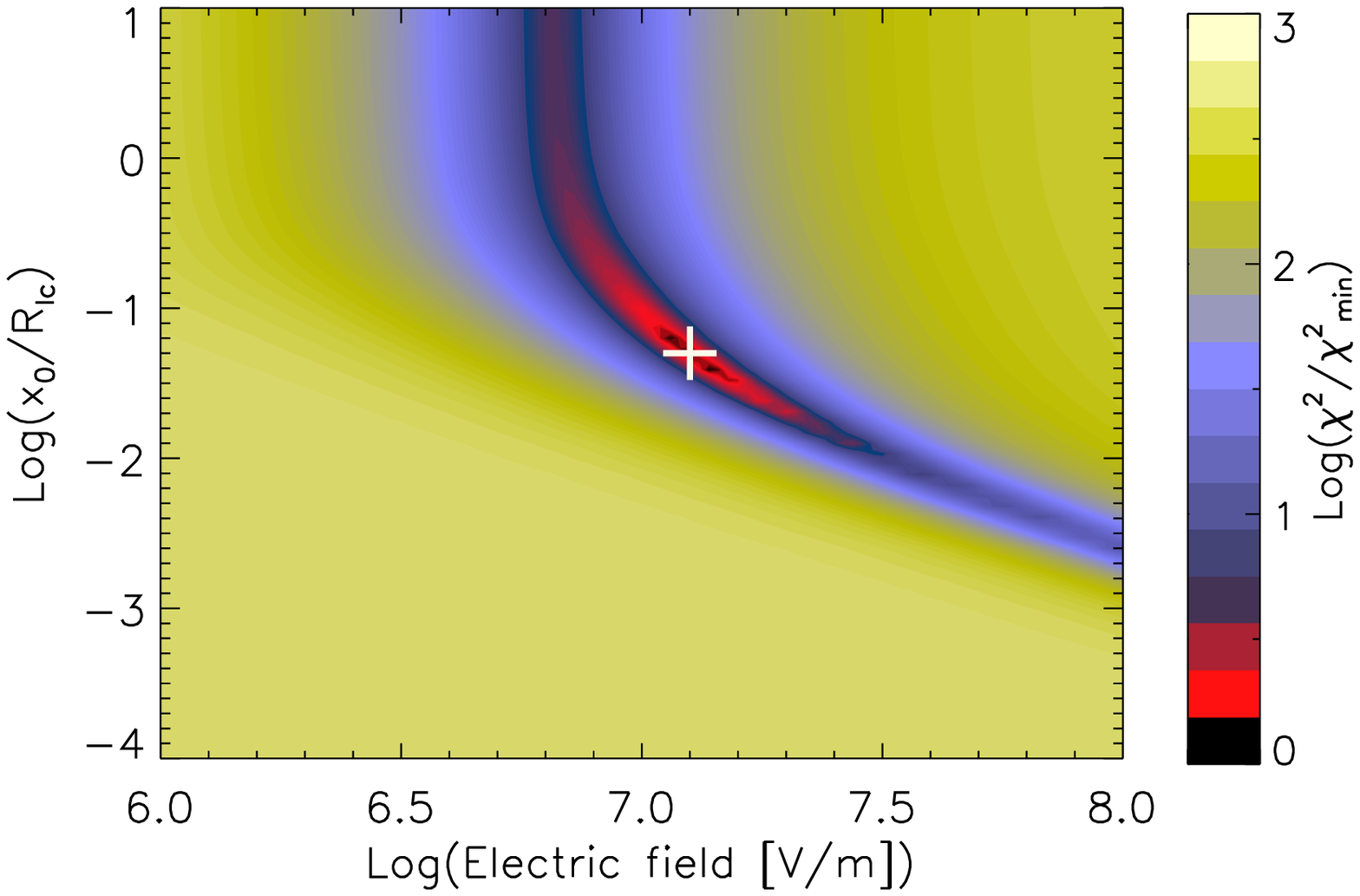}
\put(20,15){\scriptsize  \red {J2055+2539}}
\end{overpic}
\begin{overpic}[width=0.32\textwidth]{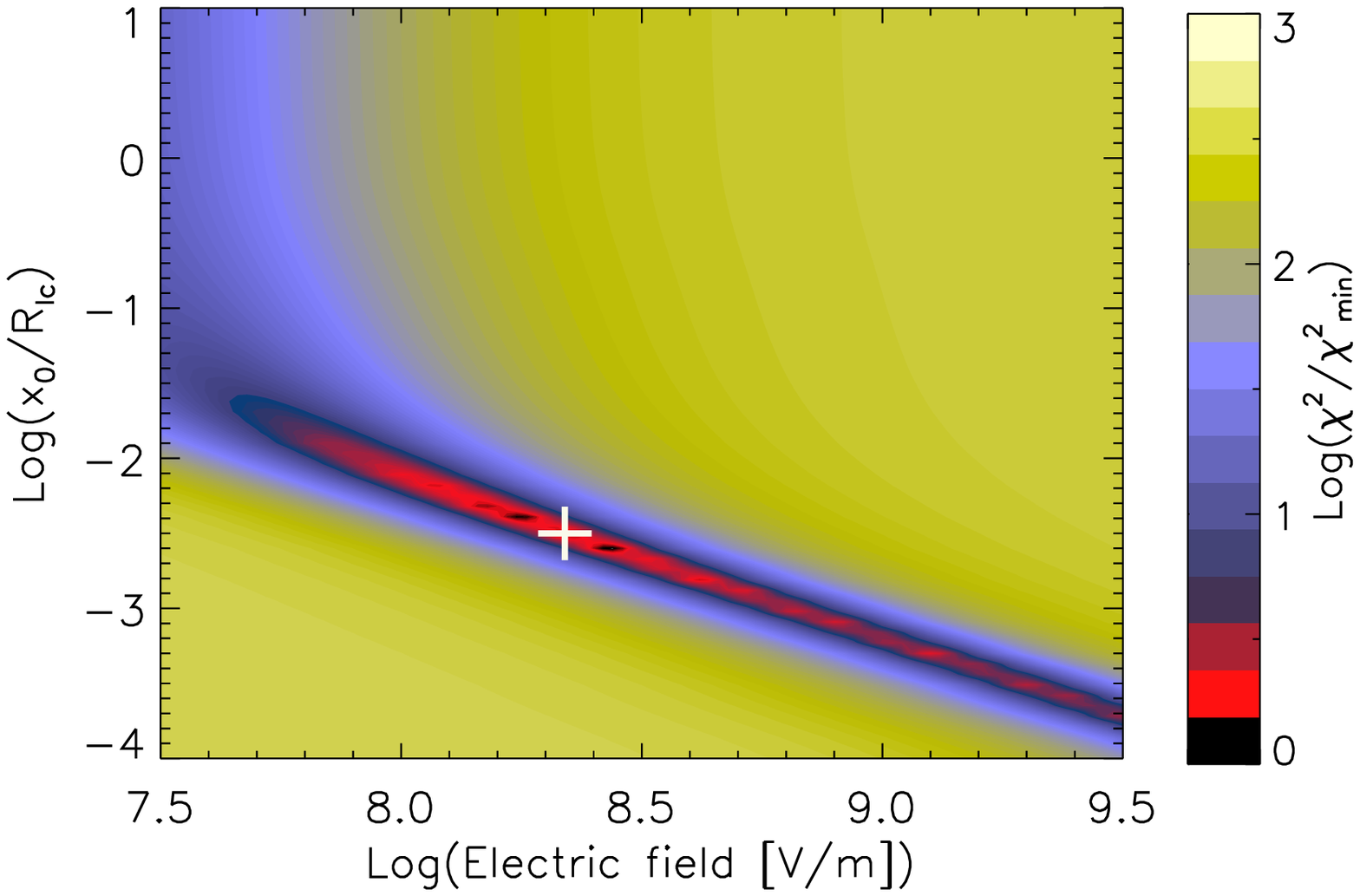}
\put(20,15){\scriptsize  \red {J2111+4606}}
\end{overpic}
\begin{overpic}[width=0.32\textwidth]{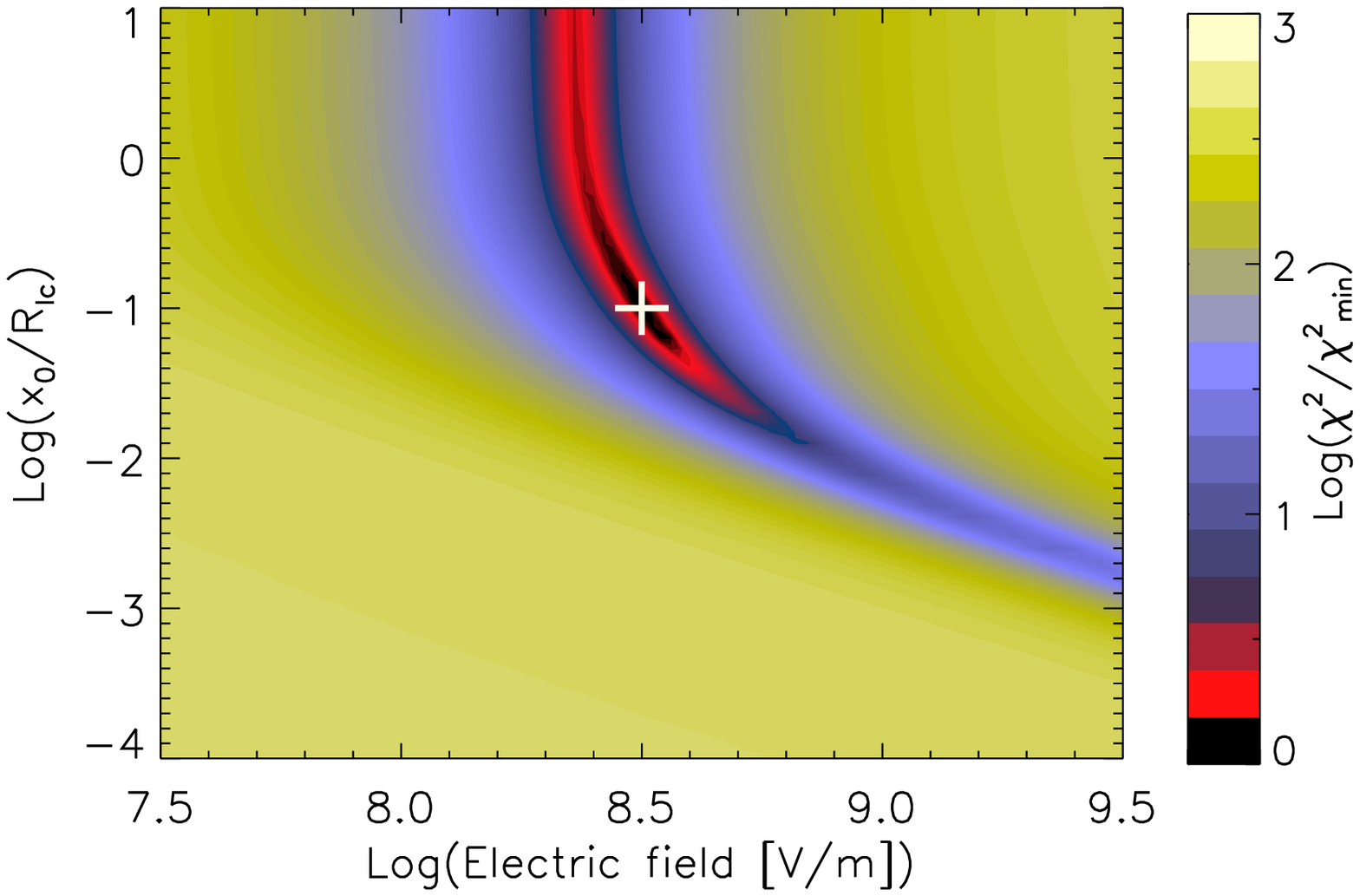}
\put(20,15){\scriptsize \blue {J2124-3358}}
\end{overpic}
\begin{overpic}[width=0.32\textwidth]{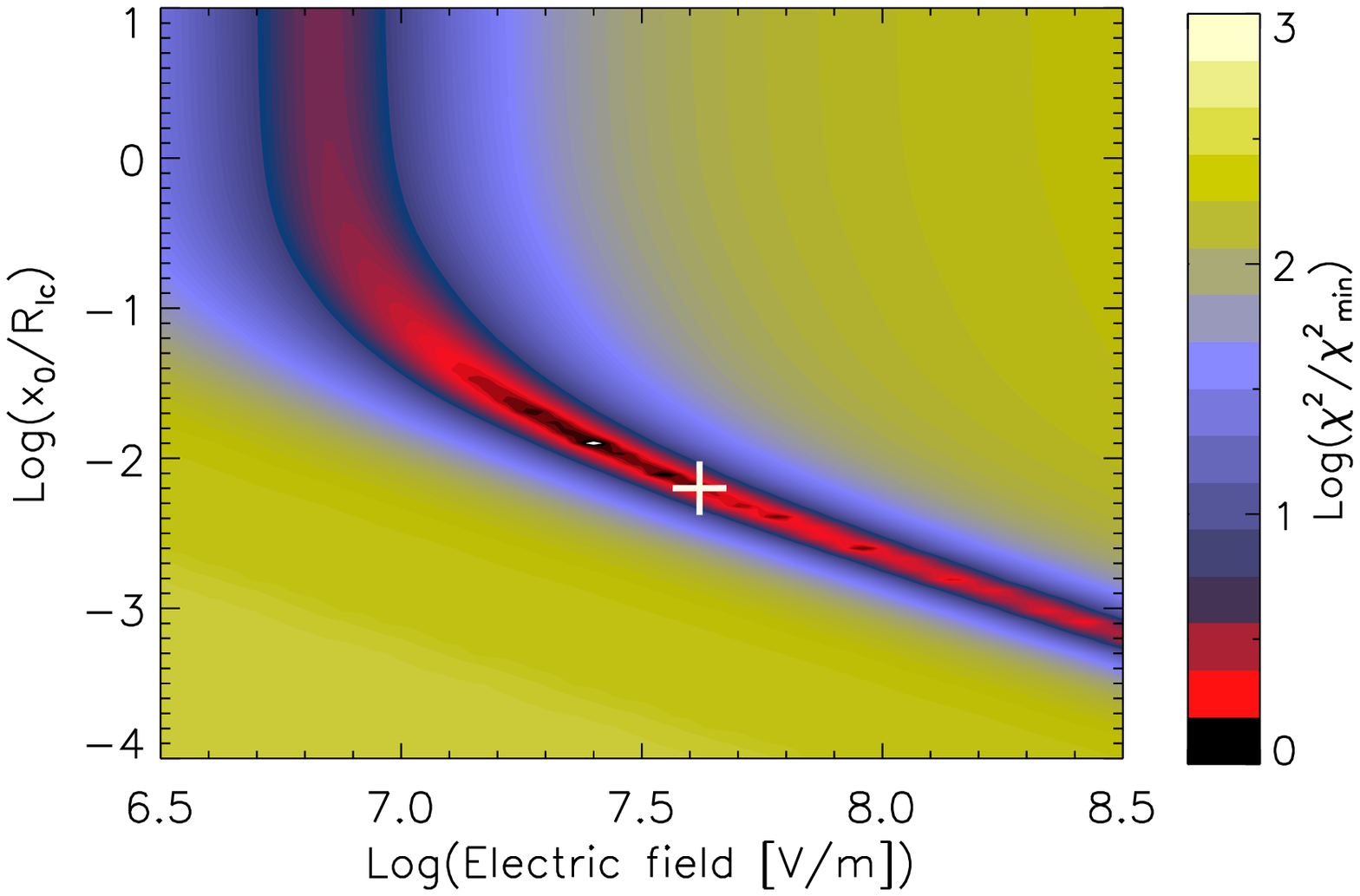}
\put(20,15){\scriptsize  \red {J2139+4716}}
\end{overpic}
\begin{overpic}[width=0.32\textwidth]{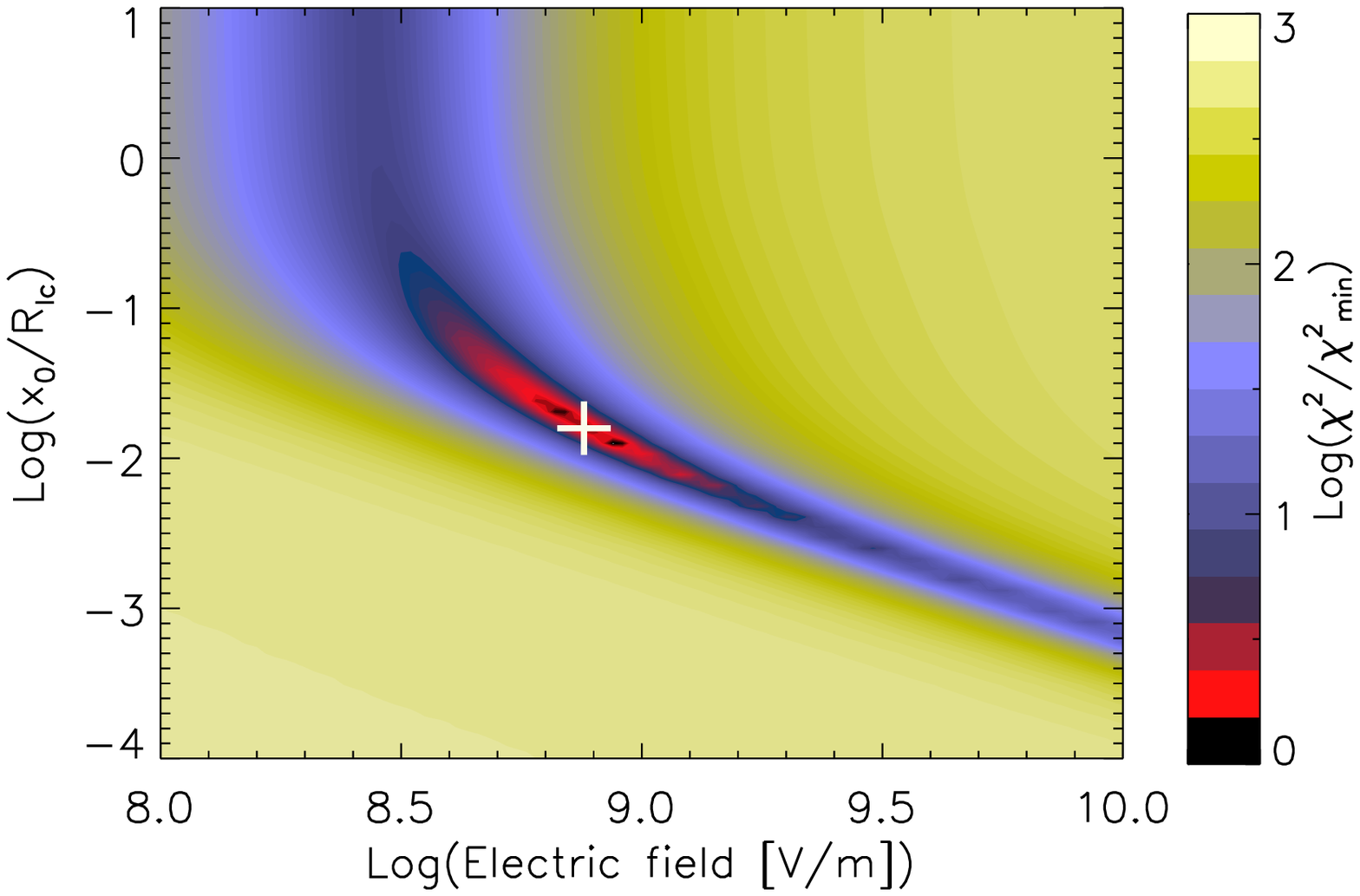}
\put(20,15){\scriptsize  \blue {J2214+3000}}
\end{overpic}
\begin{overpic}[width=0.32\textwidth]{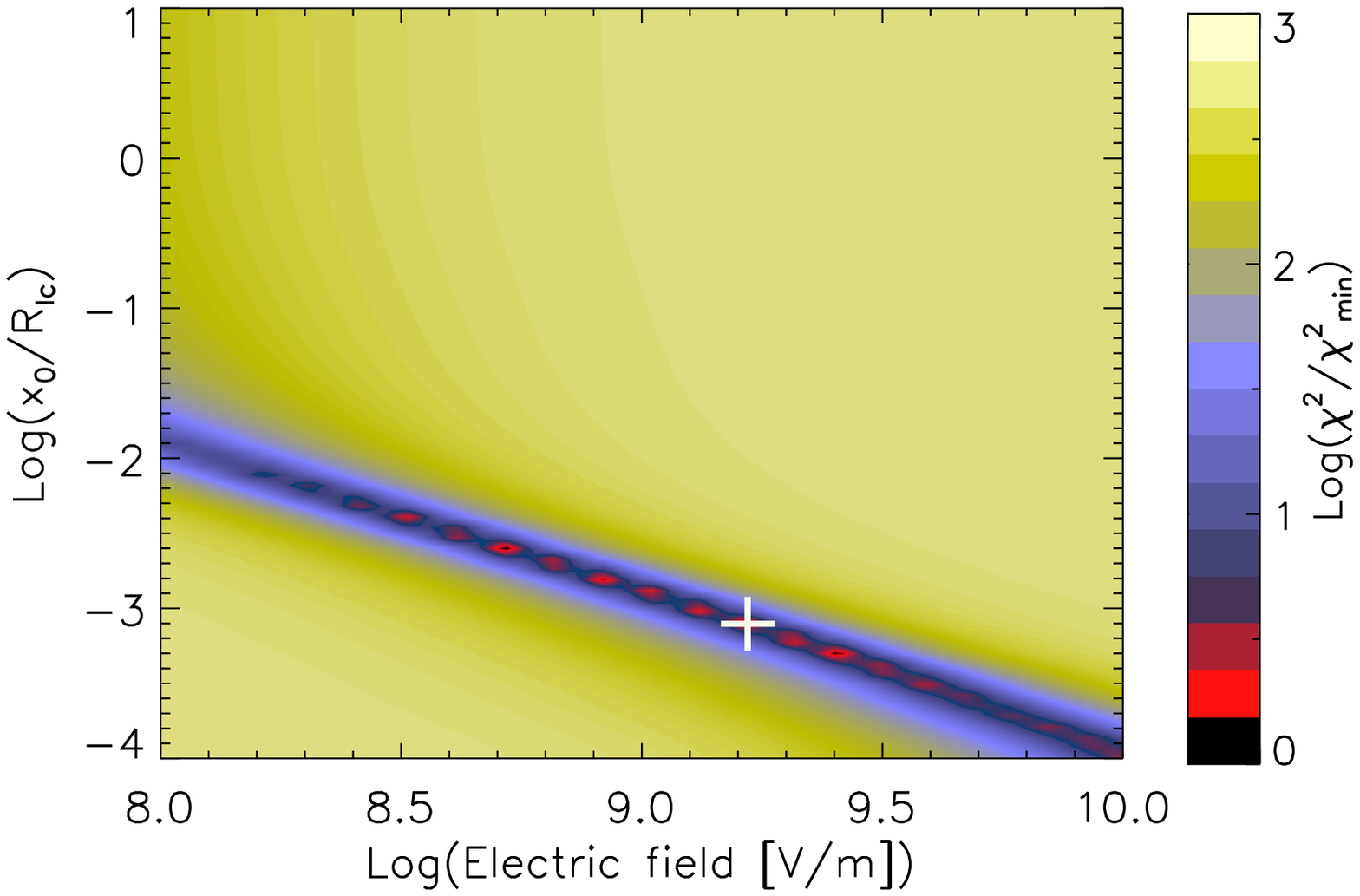}
\put(20,15){\scriptsize  \red {J2229+6114}}
\end{overpic}
\begin{overpic}[width=0.32\textwidth]{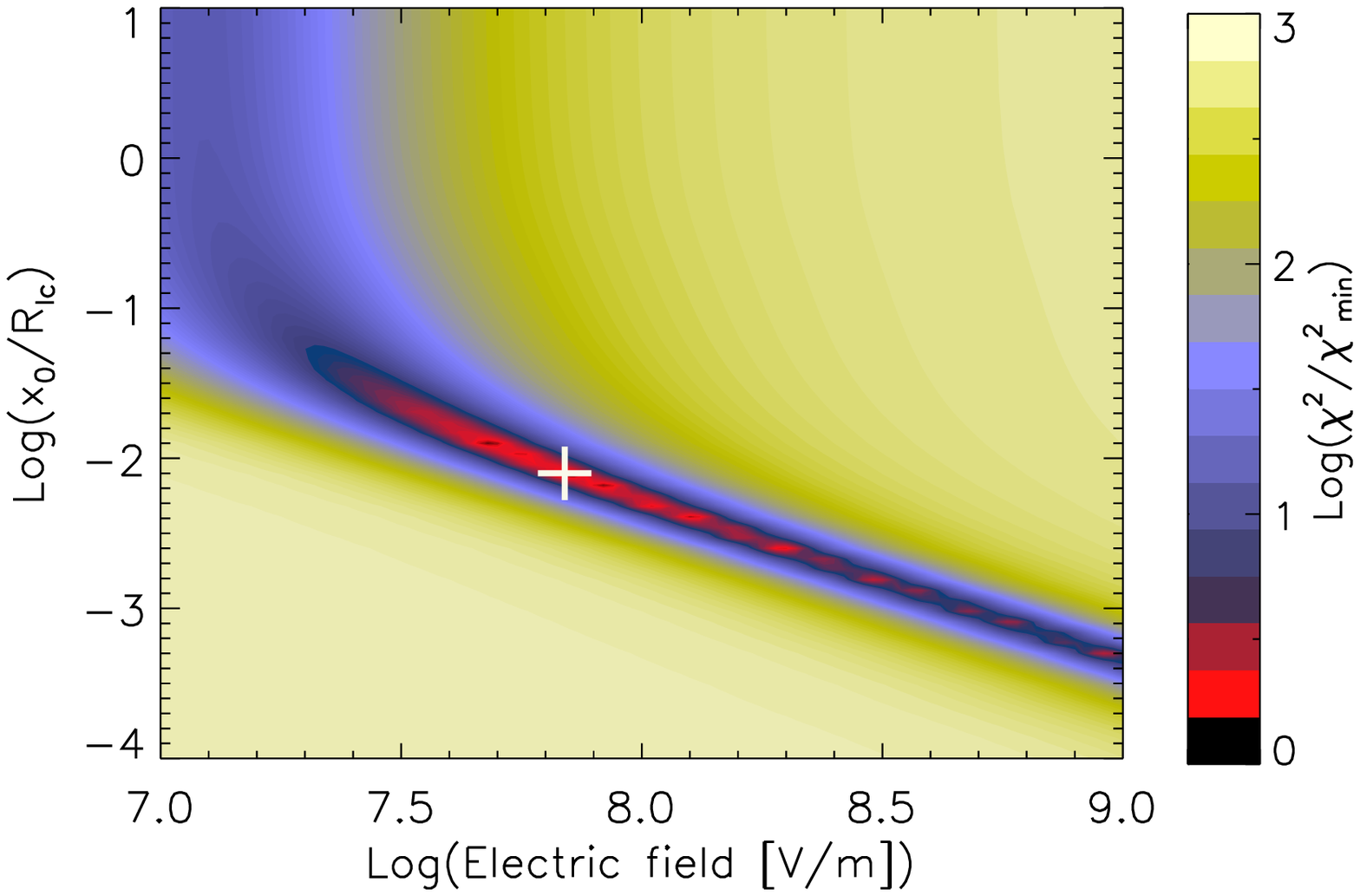}
\put(20,15){\scriptsize  \red {J2238+5903}}
\end{overpic}
\begin{overpic}[width=0.32\textwidth]{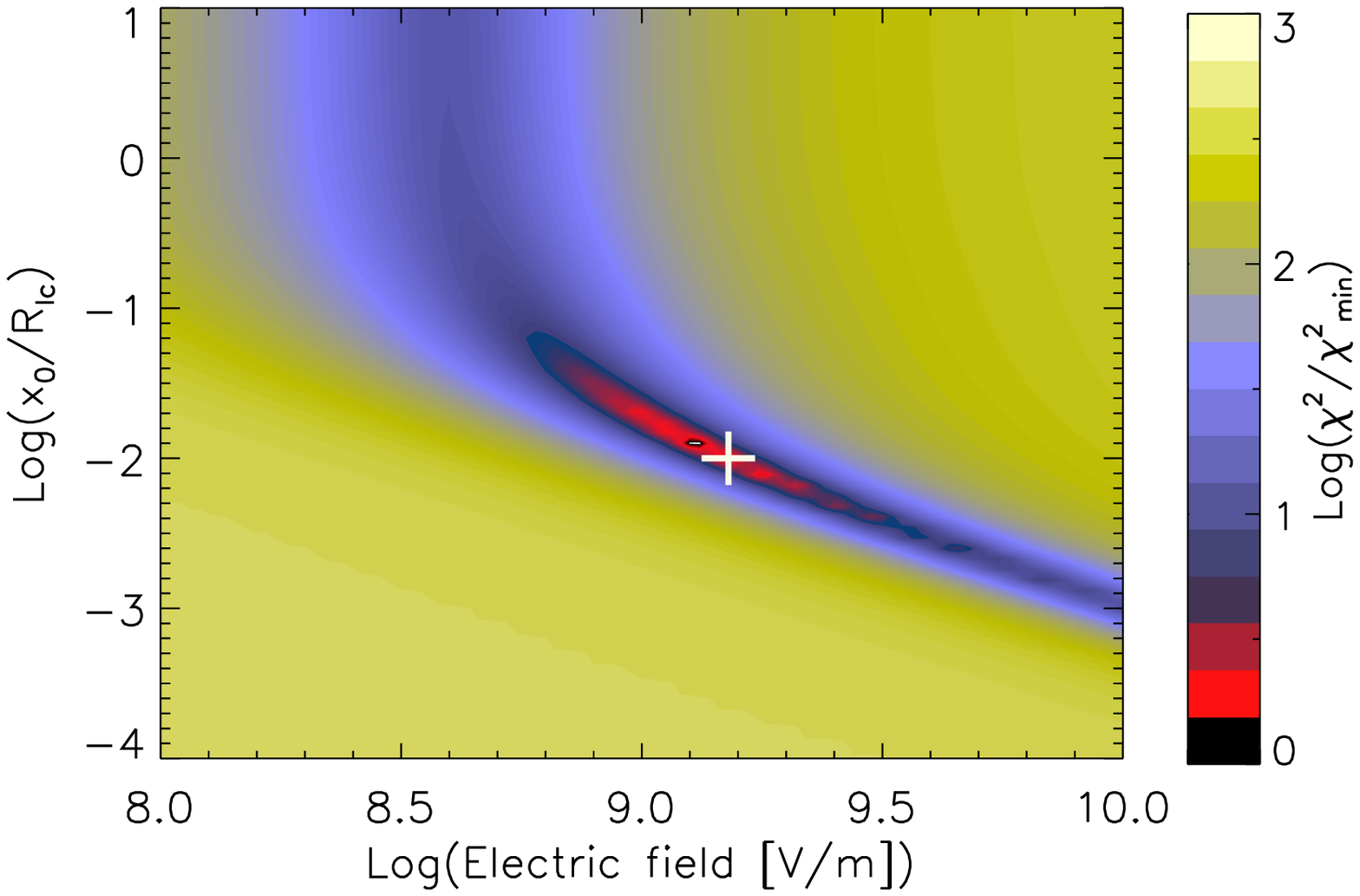}
\put(20,15){\scriptsize \blue {J2241-5236}}
\end{overpic}
\begin{overpic}[width=0.32\textwidth]{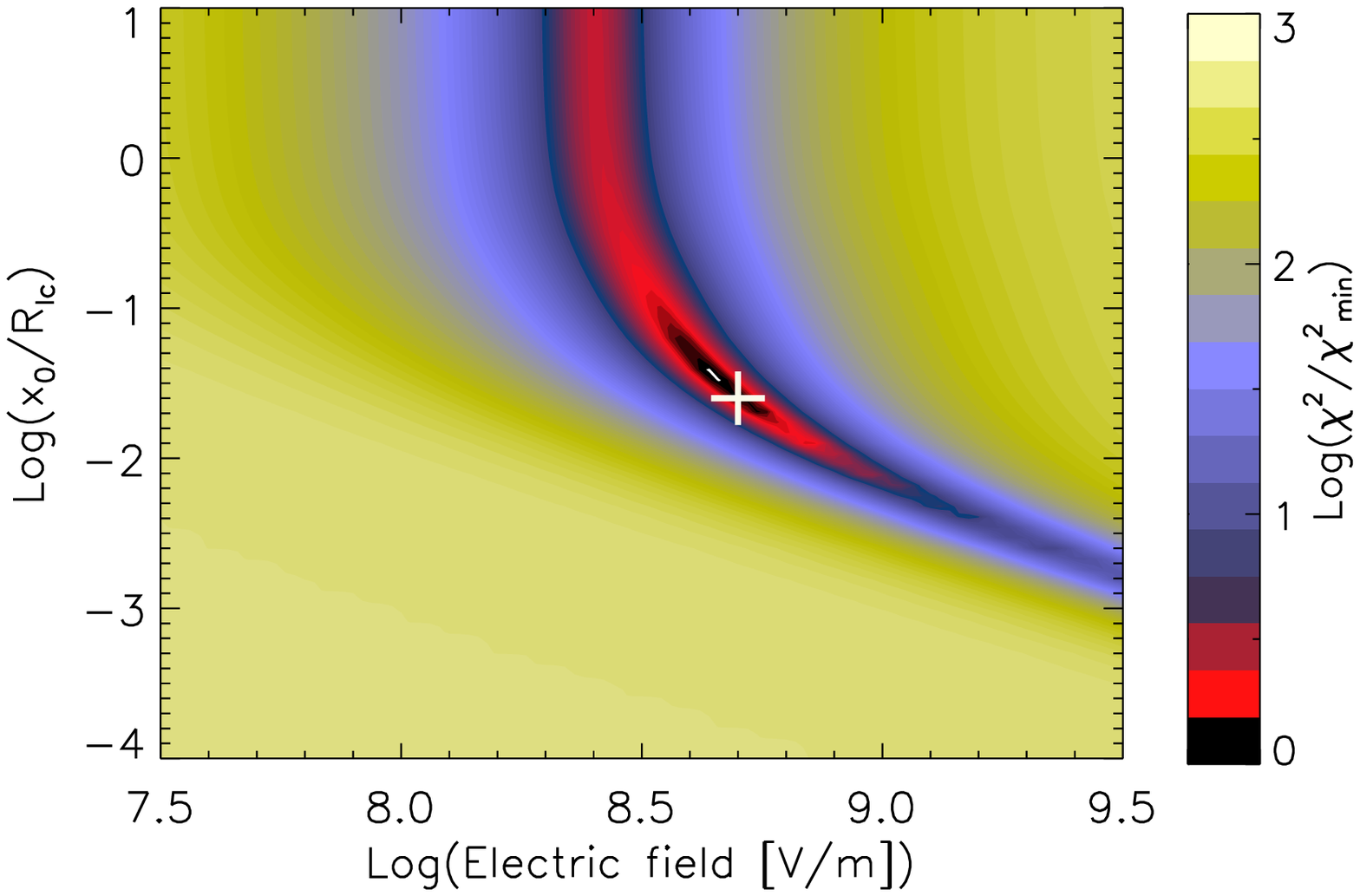}
\put(20,15){\scriptsize \blue {J2302+4442}}
\end{overpic}
\end{center}
\caption{$\chi^2/\chi^2_{\rm min}$ contours on the $\log E_\parallel$--$\log (x_0/R_{\rm lc})$ plane for the pulsars considered in the sample (V).}
\label{fig:contours5}
\end{figure*}

\label{lastpage}
\clearpage

\end{document}